\documentclass[a4paper,fleqn,usenatbib]{mnras}

\usepackage{newtxtext,newtxmath}
\usepackage{longtable}

\usepackage[T1]{fontenc}
\usepackage{ae,aecompl}

\usepackage{float}
\usepackage{placeins}

\newcommand{\msec}[2]{$#1\mbox{$''\mskip-7.6mu.\,$}#2$}
\newcommand{\ms}[2]{$#1\mbox{$^s\mskip-7.6mu.\,$}#2$}

\usepackage{graphicx}	
\usepackage{epstopdf}
\DeclareGraphicsExtensions{.eps}
\usepackage{amsmath}	
\usepackage{amssymb}	
	
\usepackage{pdflscape}
\usepackage{gensymb}

 \usepackage{comment}

\title[Modelling the abundance structure of HNCO toward IRAS~16293$-$2422]{Modelling the abundance structure of isocyanic acid (HNCO) toward the low-mass solar type protostar IRAS~16293$-$2422}

\author[Hern\'andez-G\'omez et al.]{Antonio Hern\'andez-G\'omez,$^{1,2}$\thanks{E-mail: a.hernandez@irya.unam.mx}
Emna Sahnoun$^{3,4}$,
Emmanuel Caux$^{2}$,
\newauthor 
Laurent Wiesenfeld$^{4,5}$,
Laurent Loinard$^{1,6}$,
Sandrine Bottinelli$^{2}$,
\newauthor 
Kamel Hammami$^{3}$,
and Karl M. Menten$^{7}$
\\
$^{1}$Instituto de Radioastronom\'{\i}a y Astrof\'{\i}sica, Universidad Nacional Aut\'onoma de M\'exico, Morelia 58089, Mexico\\
$^{2}$IRAP, Universit\'e de Toulouse, CNRS, UPS, CNES, Toulouse, France\\
$^{3}$Laboratory of Atomic Molecular Spectroscopy and Applications, Department of Physics, Faculty of Science,\\ 
University Tunis El Manar, Campus Universities, 1060 Tunis, Tunisia\\
$^{4}$IPAG, Universit\'e Grenoble Alpes, CNRS,  F-38000 Grenoble, France\\
$^{5}$Laboratoire Aim\'e-Cotton,  Universit\'e Paris-Saclay, CNRS, Orsay, France\\
$^{6}$Instituto de Astronom\'{i}a, Universidad Nacional Aut\'{o}noma de M\'{e}xico, Apartado Postal 70-264, CdMx C.P. 04510, Mexico\\
$^{7}$Max-Planck-Institut f\"ur Radioastronomie, Auf dem H\"ugel 69, D-53121 Bonn, Germany 
}

\date{Accepted XXX. Received YYY; in original form ZZZ}

\pubyear{2018}

\begin{document}
\label{firstpage}
\pagerange{\pageref{firstpage}--\pageref{lastpage}}
\maketitle

\begin{abstract}
Isocyanic acid  (HNCO), the most stable of the simplest molecules containing the four main elements essential for organic chemistry, has been observed in several astrophysical environments such as molecular clouds, star-forming regions, external galaxies and comets. In this work, we model HNCO spectral line profiles toward the low-mass solar type protostar IRAS 16293$-$2422 observed with the ALMA interferometer, the IRAM, JCMT and APEX single-dish radio telescopes, and the HIFI instrument on board the Herschel Space Observatory. In star-forming environments, the HNCO emission is not always in Local Thermodynamical Equilibrium (LTE). A non-LTE radiative transfer approach is necessary to properly interpret the line profiles, and accurate collisional rate coefficients are needed. Here, we used the RADEX package with a completely new set of collisional quenching rates between HNCO and both ortho-H$_2$ and para-H$_2$ obtained from quantum chemical calculations yielding a novel potential energy surface in the rigid rotor approximation. We find that the lines profiles toward IRAS 16293$-$2422 are very well reproduced if we assume that the HNCO emission arises from a compact, dense and hot physical component associated with the hot corino, a warm component associated with the internal part of the protostellar envelope, and a cold and more extended component associated with the outer envelope. The derived HNCO abundances from our model agree well with those computed with the Nautilus chemical code.

\end{abstract}




\section{Introduction}

Isocyanic acid (HNCO) is the most stable of the simplest molecules containing all four atoms essential for life as we know it.\footnote{Cyanic acid  (HOCN), fulminic acid (HCNO), and isofulminic acid (HONC), are all less stable.} In consequence, understanding the formation and evolution of this molecule in star-forming environments might prove relevant to organic chemistry in space. Interstellar HNCO was first reported by \cite{snyder1972} toward the molecular cloud complex Sgr B2, where its emission was found to be abundant and spatially extended. Further studies confirmed the high abundance of HNCO with respect to H$_2$ toward molecular clouds in the direction of the Galactic center \citep[e.g.][]{turner1991, martin2008}, and revealed its presence in various other environments such as hot molecular cores \citep[e.g.][]{blake1987,vandishoeck1995, macdonald1996, helmich1997,bisschop2008}, molecular outflows \citep[e.g.][]{rodriguez-fernandez2010}, external galaxies \citep[e.g.][]{meier2005,martin2006,martin2009} and comets \citep[e.g.][]{lis1997,crovisier1998,biver2006}.  

\noindent HNCO was proposed to be a tracer of dense gas since it has been observed in high density regions \citep[e.g.][]{jackson1984}, as well as a tracer of shocks since its abundance appears to be enhanced toward shocked gas regions \citep[e.g.][]{rodriguez-fernandez2010}. 
Much effort has been devoted to understanding the chemistry of HNCO in a variety of astrophysical environments \citep[e.g.][]{quan2010,marcelino2009,marcelino2010}, but it is not yet fully constrained. In an early study, \cite{iglesias1977} proposed that the formation pathway of HNCO in Sgr B2 could occur in the gas phase through the chemical ion-neutral reactions

\begin{eqnarray}
\mathrm{NCO^+ + H_2 \rightarrow HNCO^+ + H},\\
\mathrm{HNCO^+ + H_2 \rightarrow HNCOH^+ + H},\\
\mathrm{HNCOH^+ + e^- \rightarrow HNCO + H}.
\end{eqnarray}

\noindent \citet{turner2000} also suggested some neutral-neutral reactions to form HNCO:

\begin{eqnarray}
\mathrm{CN +O_2 \rightarrow NCO + O},\\
\mathrm{NCO + H_2 \rightarrow HNCO + H},
\end{eqnarray}

\noindent where reaction (5) has an activation barrier. More recently, \citet{marcelino2009, marcelino2010} proposed a more complete gas phase model to explain the abundance of HNCO and its isomers in cold dense cores.\\ 
Other studies \citep[e.g.][]{garrod2008, tideswell2010} have shown that HNCO could also be formed on dust grain surfaces through the thermal reaction 

\begin{eqnarray}
\mathrm{ NH + CO \rightarrow HNCO },
\end{eqnarray}

\noindent
(which has no activation barrier according to \citealt{garrod2008}) and then be released into the gas phase through desorption. For some time, it was believed that HNCO was directly related with the formation of NH$_2$CHO (formamide, a molecule important in prebiotic chemistry) via hydrogenation on grain surfaces. Recent laboratory experiments, however, have shown that this process is in fact inefficient \citep[e.g.][]{fedoseev2015, noble2015}. Regardless, both gas phase and grain surface chemistry need to be taken into account to correctly model the observed abundances in astrophysical sources of HNCO and other related molecules of potential exobiological interest.

\smallskip

IRAS 16293-2422 (I16293 hereafter), a Class 0 protostar located at $141_{-21}^{+30}$ pc \citep{dzib2018} in the Ophiuchus star-forming region, is particularly interesting in this context. This protostar is often considered a template source for astrochemistry since it has the richest molecular line spectrum known for low-mass protostars, spanning over a wide range of frequencies \citep[e.g.][]{caux2011, jorgensen2016}. In interferometric observations, I16293 is found to be composed of two dense condensations, called A and B, separated by $\sim$705 AU (at a distance of 141 pc), presumably tracing a newborn binary system \citep{wootten1989, mundy1992}. HNCO has been detected toward the compact sources A and B \citep[e.g.][]{bisschop2008} as well as in the surrounding large-scale envelope \citep[e.g.][]{vandishoeck1995}. This makes I16293 an ideal target to study the chemistry of HNCO at multiple scales in a low-mass star-forming environment \citep[e.g.][]{bisschop2008, marcelino2010}. Recently, \citet{lopez-sepulcre2015} modelled the emission of HNCO in I16293 with a radiative transfer code, using the collisional rate coefficients computed by \citet{green1986} for the HNCO-He system. They obtained abundances with respect to H$_2$ of $(5 \pm 4) \times 10^{-12}$ and $(6 \pm 3) \times 10^{-9}$ for the regions where the temperature is, respectively, smaller and higher than 90 K. According to their model, this temperature corresponds to the threshold for thermal desorption of some species from icy dust mantles. Once in the gas phase, HNCO will be subject to collisions with other species. Since H$_2$ is, by far, the most abundant collider in dense astrophysical environments, it would be desirable to use coefficients for the HNCO-H$_2$ system --rather than HNCO-He. Such coefficients recently became available \citep{sahnoun2018} for both forms of H$_2$ (ortho and para) as a result of new quantum chemical calculations (see Section 3.1). 

In this work, we make use this new set of collision coefficients to model the HNCO lines profiles observed at different spatial scales with the Atacama Large Millimeter/submillimeter Array (ALMA) and the single-dish telescopes IRAM, the Atacama Pathfinder Experiment (APEX), the James Clerk Maxwell Telescope (JCMT) and Herschel Heterodyne Instrument for the Far-Infrared (HIFI, \citealt{degraauw2010}), over a wide range of frequencies. In Section 2 we describe the observations in detail, while in Section 3 we describe the new HNCO-H$_2$ collisional coefficients and the adopted physical model. In Section 4 we discuss our findings and compare them with a chemical model and previously published results. Section 5 summarises our results.

\section{Observations}

To study the extended HNCO emission from I16293, we analysed a set of data obtained with the IRAM-30m, JCMT-15m and APEX-12m, ground based single-dish (sub)millimeter wavelength telescopes, as well as from the HIFI instrument on board the Herschel Space Observatory covering a frequency range from 80\,GHz to 1\,THz. The compact emission from the hot corino was studied with interferometric ALMA observations between 329 and 363 GHz. All the observations are described in this section.

\subsection{IRAM-30m and JCMT-15m observations}

First, we use observations that were part of TIMASSS (The IRAS16293-2422 Millimeter And Submillimeter Spectral Survey; \citealt{caux2011}) conducted with the IRAM-30m telescope (Granada, Spain) between 80 and 280 GHz and the JCMT-15m telescope (Mauna Kea, Hawaii) between 328 and 366 GHz with a spectral resolution ranging from 0.51 to 2.25 km s$^{-1}$. These observations were carried out between January 2004 and August 2006. For more details on these observations, see \citet{caux2011}. \\

In addition, higher spectral resolution observations (100\,kHz, 0.13 km s$^{-1}$) of the HNCO $(5_{0\,5}-4_{0\,4})$ transition were carried out with the IRAM-30m telescope between November 1st and 6th 2017 using the broad-band Eight Mixer Receiver (EMIR) receivers connected to a Fast Fourier Transform spectrometer (FFTS). The sky emission was cancelled using the wobbler switching observing mode and a throw of 150$''$. The total observing time for this run was about 5.6 hours and the observed coordinates were $\alpha_{2000}$ = 16$^{\rm h}$ 32$^{\rm m}$ \ms{22}{64}, $\delta_{2000}$ = $-$24$^{\circ}$ 28$'$ \msec{33}{6}.

\subsection{APEX observations}

The observations of I16293 in the frequency range 265--323.5\,GHz were performed with the APEX telescope on the Chajnantor plateau (Chile) during several runs in 2011 and 2012. The observations were carried out using the APEX-1 and APEX-2 receivers in the wobbler switching observing mode, with a throw of 150$''$. A FFTS was connected to the APEX receivers, providing a spectral resolution of 60 kHz and a total bandwidth of about 1.5\,GHz per tuning. The on-source integration time was 30 to 60\,min per setting, depending on the frequency, to reach a similar rms noise level over the complete frequency range observed. The observed coordinates were $\alpha_{2000} = 16^{\rm h}$ 32$^{\rm m}$ \ms{22}{87}, $\delta_{2000}$= $-24^{\circ}$ 28$'$ \msec{36}{6}.\\ 

The transitions between 372 and 462\,GHz were observed during August 2013 under very good weather conditions using a modified version of the First Light Apex Submillimeter Heterodyne receiver \citep[FLASH;][]{heyminck2006}. The spectral resolution delivered by the backends was 38.15 kHz corresponding to a velocity resolution of 0.03 (372\,GHz),  0.027 (418\,GHz) and 0.025 km s$^{-1}$ (462\,GHz). Since I16293 is a very bright sub-millimeter source, the pointing of the telescope was checked regularly using I16293 itself. Thus, the observed position corresponds to the peak of the sub-millimeter emission at $\alpha_{2000}$ = 16$^{\rm h}$ 32$^{\rm m}$ \ms{22}{9}, $\delta_{2000}$ = $-$24$^{\circ}$ 28$'$ \msec{35}{6}.\\

The final velocity resolution of all APEX observations was degraded to $\sim$\,0.6\,km\,s$^{-1}$ to increase the signal to noise ratio without losing much information on the line profile. \\

For both the IRAM November 2017 and the APEX observations, the data reduction was performed using the GILDAS/CLASS90\footnote{http://www.iram.fr/IRAMFR/GILDAS/} package. The telescopes and receivers parameters (main-beam efficiency B$_\text{eff}$, forward efficiency F$_\text{eff}$, half power beam width HPBW) were taken from the IRAM and the APEX webpages. The rms noise achieved is typically 10\,mK (T$_\text{mb}$) per 0.6\,km\,s$^{-1}$ velocity channel for the APEX data, and 3\,mK (T$_\text{mb}$) per 0.6\,km\,s$^{-1}$ velocity channel for IRAM data.\\

\subsection{Herschel-HIFI observations}

Observations with the HIFI instrument onboard the Herschel Space Observatory were conducted as part of the guaranteed-time key program CHESS (Chemical Herschel Surveys of Star-forming regions, \citealt{ceccarelli2010}), whose goal was to perform spectral surveys in the frequency range $480-1790$ GHz with a high spectral resolution ($\sim$ 1.1\,MHz). The data used in this article are part of a full spectral coverage of bands 3b (860-960 GHz ; Obs. Id 1342192330) and 4a (950-1060 GHz ; Obs. Id 1342191619), which were obtained on 19 March, and 3 March 2010, respectively.  The Spectral Scan Double Beam Switch (DBS) with optimization of the continuum observing mode was used. A spectral resolution of 1.1 MHz ($\sim$ 0.3 km s$^{-1}$ at 1 THz) was provided by the HIFI acousto-optic Wide Band Spectrometer (WBS) with an instantaneous bandwidth of 4$\times$1\,GHz \citep{roelfsema2012}. The observed coordinates were $\alpha_{2000} = 16^{\rm h}$ 32$^{\rm m}$ \ms{22}{64}, $\delta_{2000} = -24^{\circ}$ 28$'$ \msec{33}{6}. The DBS reference positions were situated approximately 3$'$ east and west of the source. \\

To carry out a spectral survey, multiple local oscillator tunings are used, in order to cover the required frequency bands. A single local oscillator tuning spectrum consists of eight separate spectra: four per polarization (horizontal and vertical) in 4 sub-bands of $\sim$1\,GHz each. Using the HIPE \citep{ott2010} ``flagTool" task, we removed the spurs not automatically eliminated by the pipeline. Standing waves and baselines removal on each sub-band were performed with the HIPE tasks ``fitHifiFringe" and ``fitBaseline". The sideband deconvolution was performed using the HIPE task  ``doDeconvolution", and the resulting deconvolved spectra observed in both polarizations were averaged to improve the noise in the final spectra, given that they had similar quality. The task ``fitBaseline" was then ran to obtain the continuum values which are well fitted by order 3 polynomials over the frequency range of the whole sub-bands. These single side band continuum values were then added to the spectra at the considered frequencies. Finally, we used the forward efficiency of 0.96 and the (frequency-dependent) beam-efficiency given in Table 1 of \citet{roelfsema2012} to convert the intensities from antenna to main-beam temperature scale.

\subsection{ALMA observations}
 \label{ALMA observations}

\begin{figure}
\includegraphics[width=0.45\textwidth]{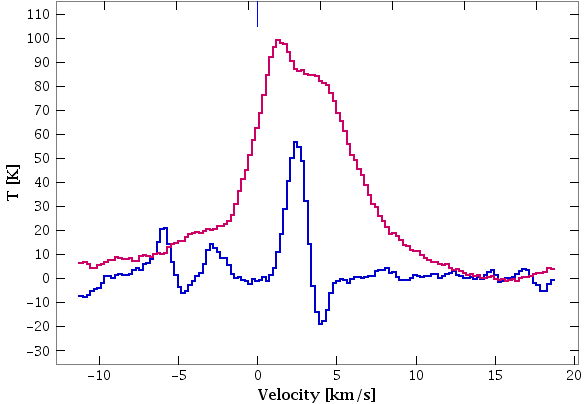}
  \caption{HNCO $15_{0\,,15}-14_{0 \,,14}$ line emission as observed towards IRAS16293 in the PILS survey toward source A (red line), and source B (blue line).}
  \label{fig:ALMA_obs}
\end{figure}

PILS, the ALMA Protostellar Interferometric Line Survey (\citealt{jorgensen2016}), reported ALMA observations in the frequency range [329.15, 362.90] GHz with a $\sim $0.2 km s$^{-1}$ channel spacing and a $\sim$0.5$''$ spatial resolution. Five HNCO lines used in this work were observed in this survey, HNCO ($15_{0\,15}-14_{0 \,14}$), ($15_{1\,14}-14_{1 \,13}$), ($16_{0\,16}-15_{0 \,15}$), ($16_{1\,16}-15_{1 \,14}$) and ($16_{1\,16}-15_{1 \,15}$). These interferometric observations, not sensible to the extended emission, were only used to determine some physical parameters of the hot corino component. As can be seen in Figure \ref{fig:ALMA_obs}, the HNCO ($15_{0\,15}-14_{0 \,14}$) line integrated intensity is 9 times larger for A than for B, allowing us to assume in this work that source A is solely responsible of the hot corino emission.\\

Table \ref{tab:example_table} summarises the observation parameters as well as spectroscopic information for the observed lines from the Cologne Database for Molecular Spectroscopy\footnote{https://www.astro.uni-koeln.de/cdms} \citep[CDMS,][]{muller2001,muller2005}. This database makes use of spectroscopic data from \citet{kukolich1971}, \citet{hocking1975}, \citet{niedenhoff1995} and \citet{lapinov2007}. All line intensities are expressed in main beam brightness temperature units (T$_{\text{mb}}$), after correction of rear-ward losses and main-beam efficiency and for the atmospheric attenuation for IRAM, APEX and JCMT observations.\\

\section{Results}

\begin{figure}
\includegraphics[width=8.5cm]{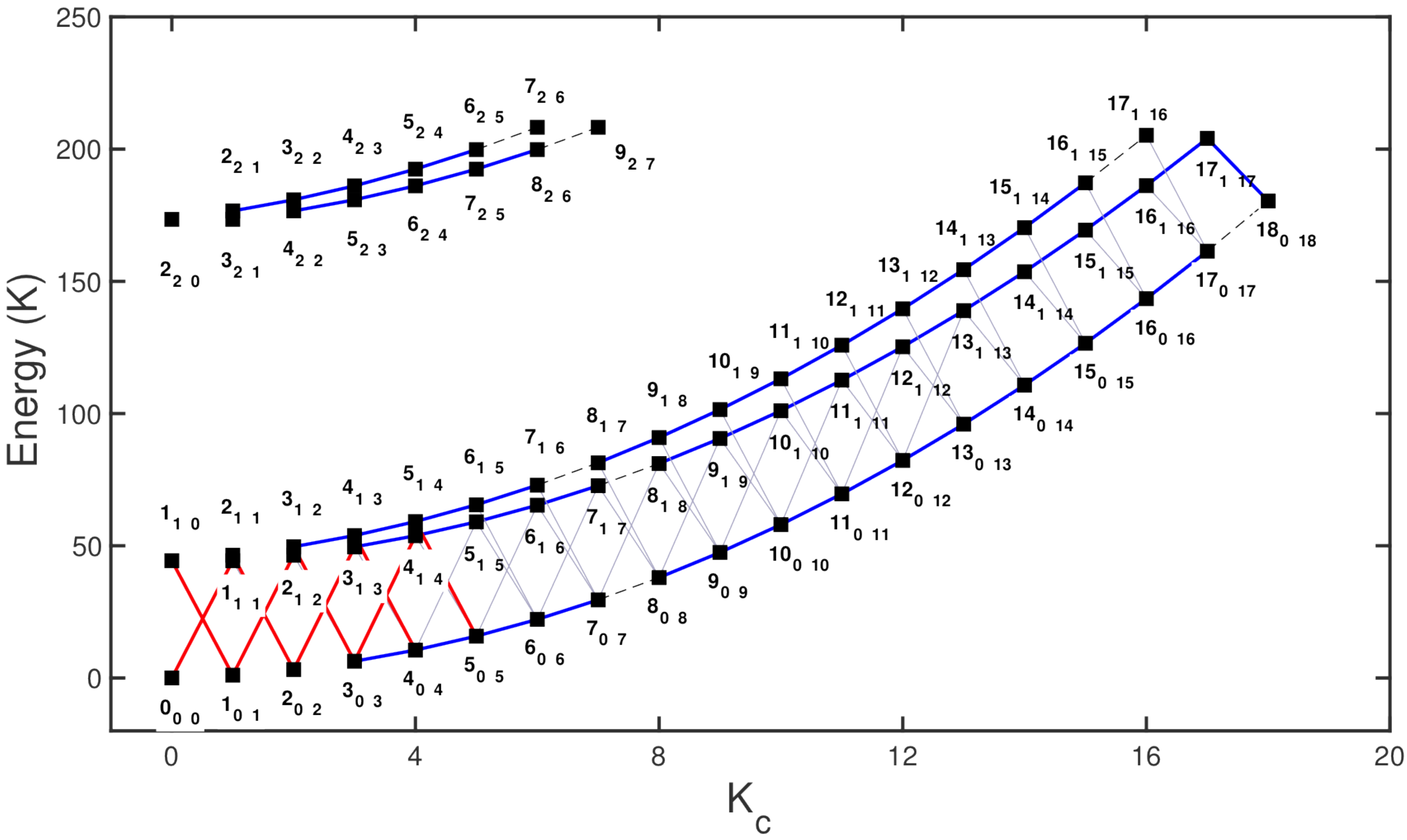}
\caption{Sketch of the rotational levels of HNCO, labelled by $J_{K_aK_c},$ and energy in Kelvin. Lines connect observable transitions, at frequencies $80< \nu < 1200$~GHz. Blue lines indicate transitions detected in emission; red lines indicate transitions detected in absorption. Light grey lines indicate transitions that were looked for but not detected, while dashed lines indicate transitions that were not observed.}
\label{fig1}
\end{figure}

To identify and model the relevant HNCO spectral lines, we used CASSIS\footnote{http://cassis.irap.omp.eu} \citep{2011caux}, a software developed at IRAP-UPS/CNRS which makes use of the CDMS database \citep{muller2001,muller2005}. 35 transitions are present in the IRAM observations, 16 in the APEX ones and 6 in the JCMT ones. We also found 47 transitions falling within the HIFI observations range. Most of these latter lines were not detected (see Appendix), but the corresponding upper limits were used as constraints in the modelling. The details for all the transitions used are given in Table \ref{tab:example_table}. We illustrate in Figure \ref{fig1} the transitions observed and their corresponding levels as a function of the quantum number K$_c$ and the energy above the ground state.

To model the line profiles we used the statistical equilibrium non-LTE radiative transfer code RADEX \citep{vandertak2007} that uses the escape probability formalism. The HNCO collisional coefficients are taken from Sahnoun et al. (2018) and were obtained as we now describe.

\subsection{HNCO collisional coefficients}

The main aspects of the collisional coefficients computation are summarised in this section. The full description can be found in \citet{sahnoun2018}.\\ 

\noindent {\bf Potential Energy Surface}\\

\noindent The Potential Energy Surface (PES) for the HNCO-H$_2$ van der Waals system was recently computed by \citet{sahnoun2018}. This five dimensional PES was computed in the rigid-rotor approximation. The HNCO and H$_2$ internuclear distances are frozen at the experimental average value for the vibrational ground state (distances in bohr, angles in degrees). For H$_2$, we set $r_{\mathrm{HH}}=1.4011$. The planar HNCO parameters are set to $r_{\mathrm{HN}} = 1.9137$, $r_{\mathrm{NC}} = 2.3007$, $r_{\mathrm{CO}} =  2.2028$, $\mathrm{\alpha(HNC)} = 124.0$, $\mathrm{\alpha(NCO) }= 172.1$ \citep{fusina1981harmonic}. We computed the PES for distances $R$ between center of masses from 4.5 to 50~bohr. About 430,000 ab initio points were computed in the C1 symmetry group with the CCSD(T)-F12a method using for atomic bases the standard aug-cc-pVDZ basis set as implemented in the MOLPRO2011 package \citep{werner2012}. The basis set superposition error has been corrected at all geometries with the counterpoise procedure \citep{nizam1988theoretical}. The PES has a global minimum of $V = -235.26\;\mathrm{cm^{-1}}$ located at $R = 7.9$~bohr and angles such that  the H$_2$ molecule is perpendicular to the HNCO plane, and its center is collinear with the NH bond. This PES presents a very large anisotropy because of the rod-like geometry of the HNCO molecule, with the H atom protruding out of the nearly linear NCO arrangement.\\

\noindent {\bf Rotational quenching cross sections and rates}\\

\noindent The PES ab initio points were fit in terms of polyspherical harmonic functions in order to be introduced in the Molscat dynamical code.\footnote{http://ipag.osug.fr/$\sim$afaure/molscat/index.html} Computation of the rotational quenching cross-sections was done within the coupled-states quantum time-independent formalism for collision energies up to $500\, \mathrm{ cm^{-1}}$ (719~K), for both ortho-H$_2$, $J_{\mathrm H_2}=1$ and para-H$_2$,  $J_{\mathrm H_2}=0, \,2$. The cross sections were averaged using the Maxwell-Boltzmann distribution to calculate the rate coefficients as a function of the kinetic temperature. \\

All details about the collisional coefficients computation may be found in a preceding paper, \cite{sahnoun2018}. It was shown that  the quenching rates with ortho-H$_2$ are larger than the corresponding ones with para-H$_2$. Also, the rates connecting the $K_a=0$ levels are somewhat larger than those corresponding to the $K_a=1$ levels. Hence,  quenching redistributes the level populations in a manner far from the black-body distribution.

\subsection{The physical model}

{Several previous studies have shown that I16293 has a hot corino revealed by the emission of numerous complex molecules in both the A and B sources. In addition, a common, extended infalling envelope surrounding the binary system has also been observed by single-dish telescopes \citep[e.g.][]{caux2011}. \citet{crimier2010} determined the physical structure of I16293 from single-dish and interferometric continuum observations, assuming a spherical source, and provided temperature and density profiles up to R\,$\sim$6000\,AU. More recently, \citet{jacobsen2018} studied the inner envelope of I16293 up to R\,$\sim$8000\,AU ($\sim\,113''$ size) with a 3D dust and gas model based on ALMA observations. By comparing those studies, we noted that the model of the envelope in \citet{jacobsen2018} is fully compatible with \citet{crimier2010} in the same radius range. Last, OTF observations of CS \citep{menten1987}, and CN \citep{hernandez-gomez2018}, have shown the presence of an extended cold envelope component (size larger than 100$''$). \\

We therefore modeled the HNCO lines assuming I16293 can be represented by 3 physical components: a single hot corino of size 0.5" (see $\ref{ALMA observations}$), a warm envelope (R\,$\le$\,1000\,AU) and a cold, extended envelope (R\,$\ge$\,1000\,AU), see Figure \ref{fig:structureI16293}. In order to constrain the parameters of each of the three physical components, we have used the 1D physical structure derived by \citet{crimier2010}, fixing thereby the main H$_2$ density for all components from their derived density profile. \\

A fundamental ingredient for the modelling is the continuum level in the spectra. However, because of the observing mode at IRAM and JCMT, no continuum was recovered during those observations. This is not the case for APEX and HIFI observations. To have a consistent model taking into account the continuum level for all lines, we have computed the spectral energy distribution (SED) for the envelope of I16293 with several instruments (such as PACS, SPIRE, MIPS, MAMBO2, LABOCA, IRS, SCUBA2, NIKA2 and HIFI) and derived the expected continuum from $\sim$1 mm to $\sim$70 $\mu$m ($\sim$150 GHz to $\sim$5 THz) (Bottinelli et al., in prep.). We have checked that the predicted continuum level is consistent with the observations within an error of about 15\%. We subtracted the continuum for spectra that show continuum and added the predicted continuum for all lines. \\

\subsection{Model fitting}

In CASSIS, it is possible to model with RADEX \citep{vandertak2007} an observed spectrum with a set of physical components, each of them defined with six physical parameters that will serve as input for RADEX: the size of the component, its density n(H$_2$), its kinetic temperature T$_{\rm kin}$, the Full Width at Half Maximum (FWHM) and the velocity relative to the Local Standard of Rest V$_{\rm LSR}$ of the lines, and the column density N of the studied specie. Since our observations have sufficient spectral resolution, we fixed V$_{\rm LSR}$ of the lines for the warm and cold envelopes to 4.1\,km\,s$^{-1}$. Given that the component associated with the cold envelope has a narrow line width, we have fixed its FWHM to 0.4\,km\,s$^{-1}$. The parameters left to vary during the optimization are therefore the column density and kinetic temperature for all components, FWHM and V$_{\rm LSR}$ for the hot corino, and FWHM and size for the warm envelope. For the hot corino, we adopted an ortho-to-para H$_2$ ratio of 3 but we checked that varying this value did not influence the final results. For both envelope layers, we assumed that para-H$_2$ is the dominant form since para-H$_2$ is more stable at lower temperatures. We have therefore modelled I16293 in terms of the superposition of three physical components represented on Figure 3, with a total of 10 free parameters out of 18 possible.\\

We used the Monte Carlo Markov Chain (MCMC) method inside CASSIS \citep{hastings1970, guan2006} which explores the space of parameters to find the best solution by means of $\chi^2$ minimisation, running 1,000 models with these 10 free parameters randomly chosen. Table \ref{tab:chi2minimisation} shows the best values obtained from the $\chi^2$ minimisation, and Figure \ref{imagenes-cont} shows observed and simulated spectra of some of the HNCO transitions, while all spectra are shown in appendix $\ref{Obs-modHNCO}$. In all cases, the predicted emission has been diluted with the appropriate telescope beam, considering the frequency of the transitions and the assumed size of the given physical component.\\

\begin{figure}
\includegraphics[width=0.5\textwidth]{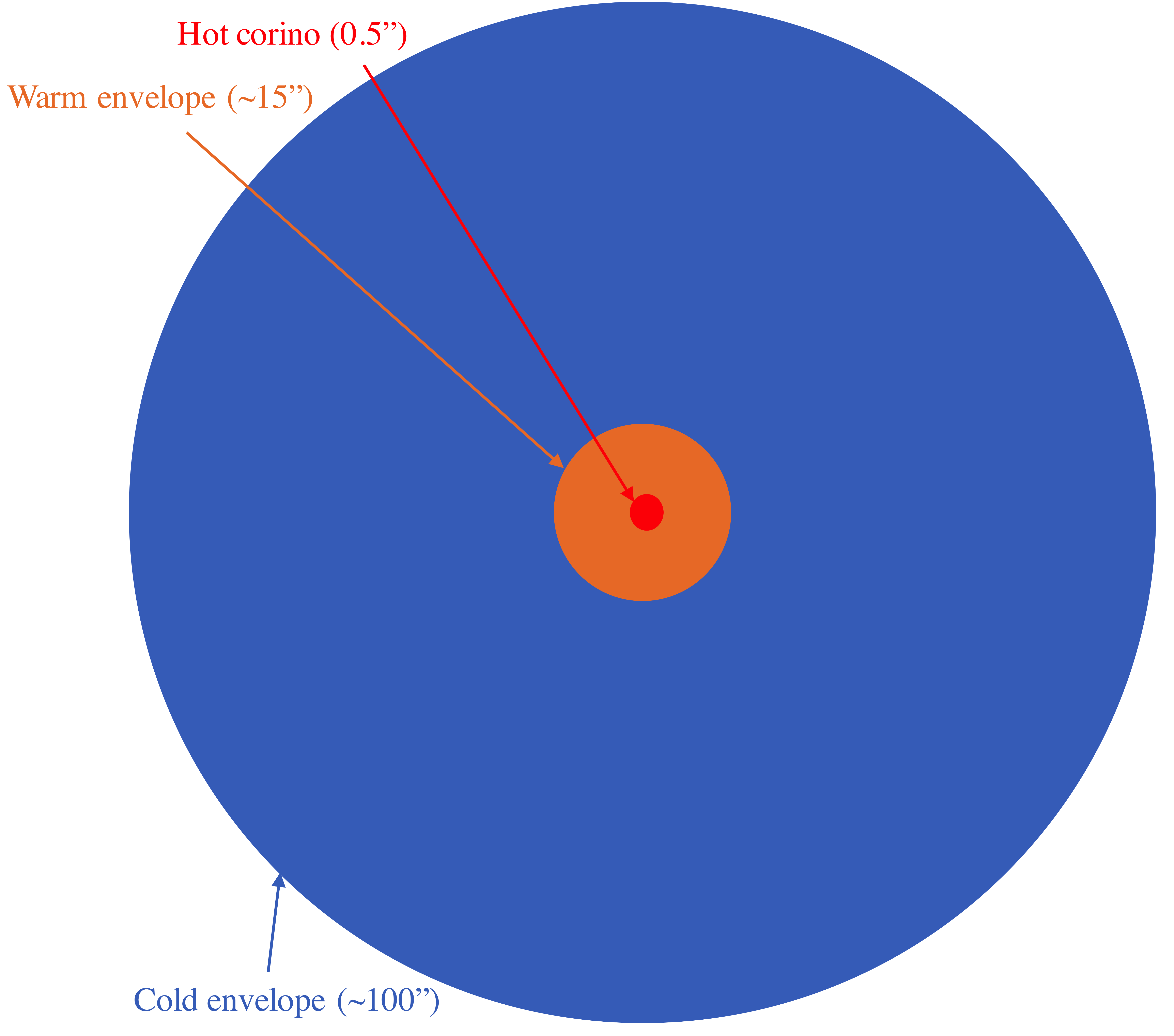}
\caption{Adopted 3 components physical structure to model the HNCO emission in I16293. }
\label{fig:structureI16293}
\end{figure}

It should be noted that the observations used in this study cannot constrain the size of the hot corino, nor that of the extended envelope, the single-dish beams being either too big or too small to provide useful information. The given column-densities and abundances are therefore those corresponding to the adopted sizes of these components (0.5$''$ and 100$''$, respectively), and should be scaled if other sizes are used. Only the size of the HNCO emission in the warm envelope can be correctly constrained with the observations we have in hands. \\

\begin{table*}
\centering
\caption{Best physical parameters (those with $^{*}$ being fixed) obtained with the $\chi^2$ minimization. The abundance was computed with respect to H$_2$.}
\label{tab:chi2minimisation}
\scalebox{1.0}{
\begin{tabular}{lccccccc} 

\hline
Component 	& N 						 & T$_{\rm kin}$ 	& FWHM 		 & V$_{\rm LSR}$	& Size 		& $n$(H$_2$)		& $X(n_{\rm HNCO}/n_{\rm H_2}$)	\\
		 	&(cm$^{-2}$)				 & (K)			& (km s$^{-1}$) & (km s$^{-1}$)	& ($''$)		& (cm$^{-3}$)		&							\\
\hline
Hot corino 	&$(1.9 \pm 0.4 )\times 10^{16}$	 & $190\pm 50$		& $6.2\pm 0.7$	& $2.9\pm 0.1$		& 0.5$^{*}$	& 3$\times 10^{8}$	& $(7.1\pm 0.6) \times 10^{-8}$		\\
Warm envelope 	&$(2.5\pm 0.5)\times 10^{13}$	 & $31\pm 5$		& $4.9\pm 0.8$	& 4.1$^{*}$		& $15.6\pm 2.4$& 5$\times 10^{7}$	& $(1.8\pm 0.4) \times 10^{-11}$	\\
Cold envelope 	&$(5.3\pm 0.6)\times 10^{12}$	 & $8.7\pm 0.8$		& 0.4$^{*}$	& 4.1$^{*}$		& 100.0$^{*}$	& 3$\times 10^{6}$	& $(9.8\pm 0.3) \times 10^{-12}$	\\
\hline
\end{tabular}
}
\end{table*}

\begin{figure*}
\centering
\setlength\tabcolsep{3.7pt}
\begin{tabular}{c c c}
\includegraphics[width=0.315\textwidth, trim= 0 0 0 0, clip]{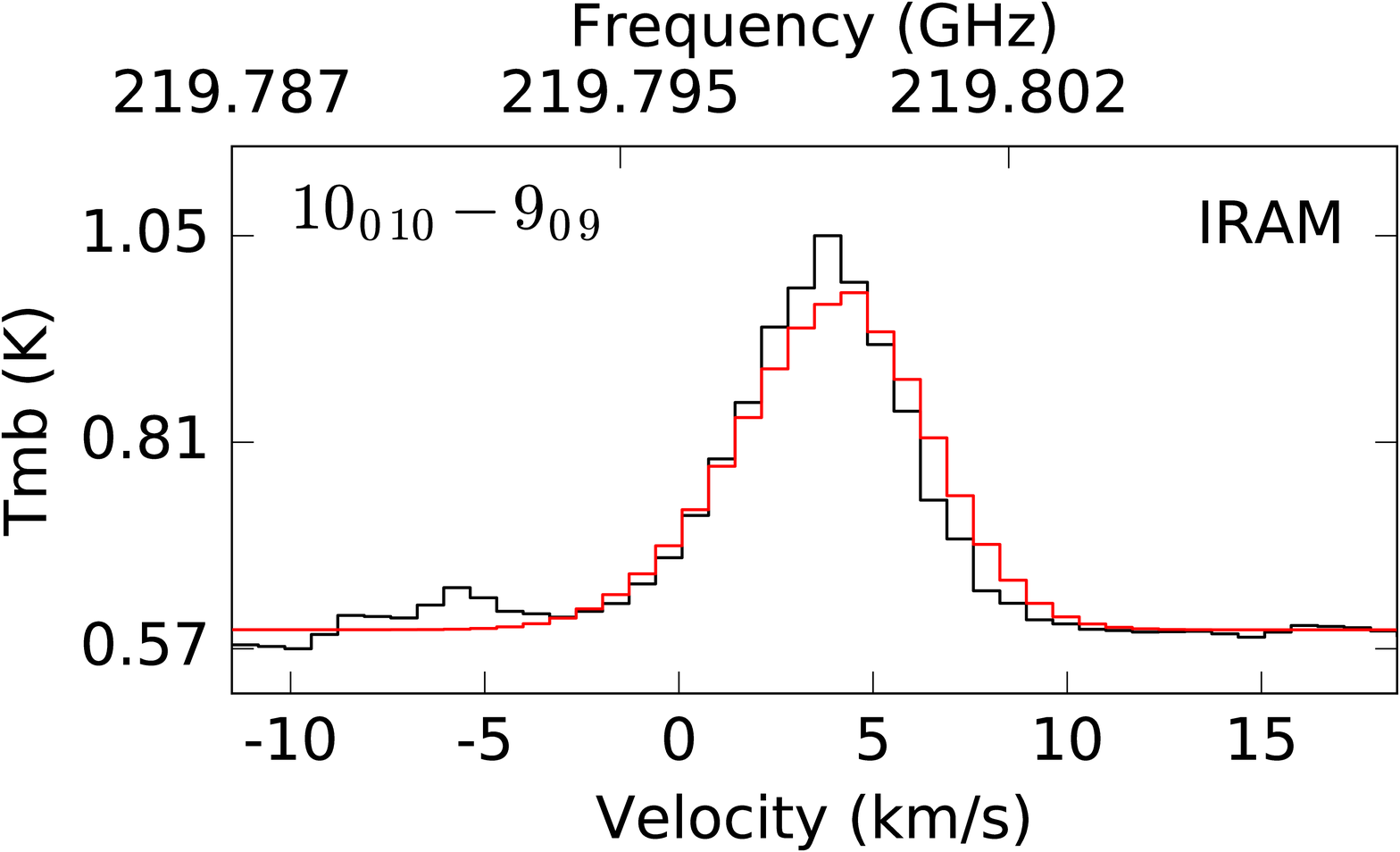} &\includegraphics[width=0.315\textwidth,trim = 0 0 0 0,clip]{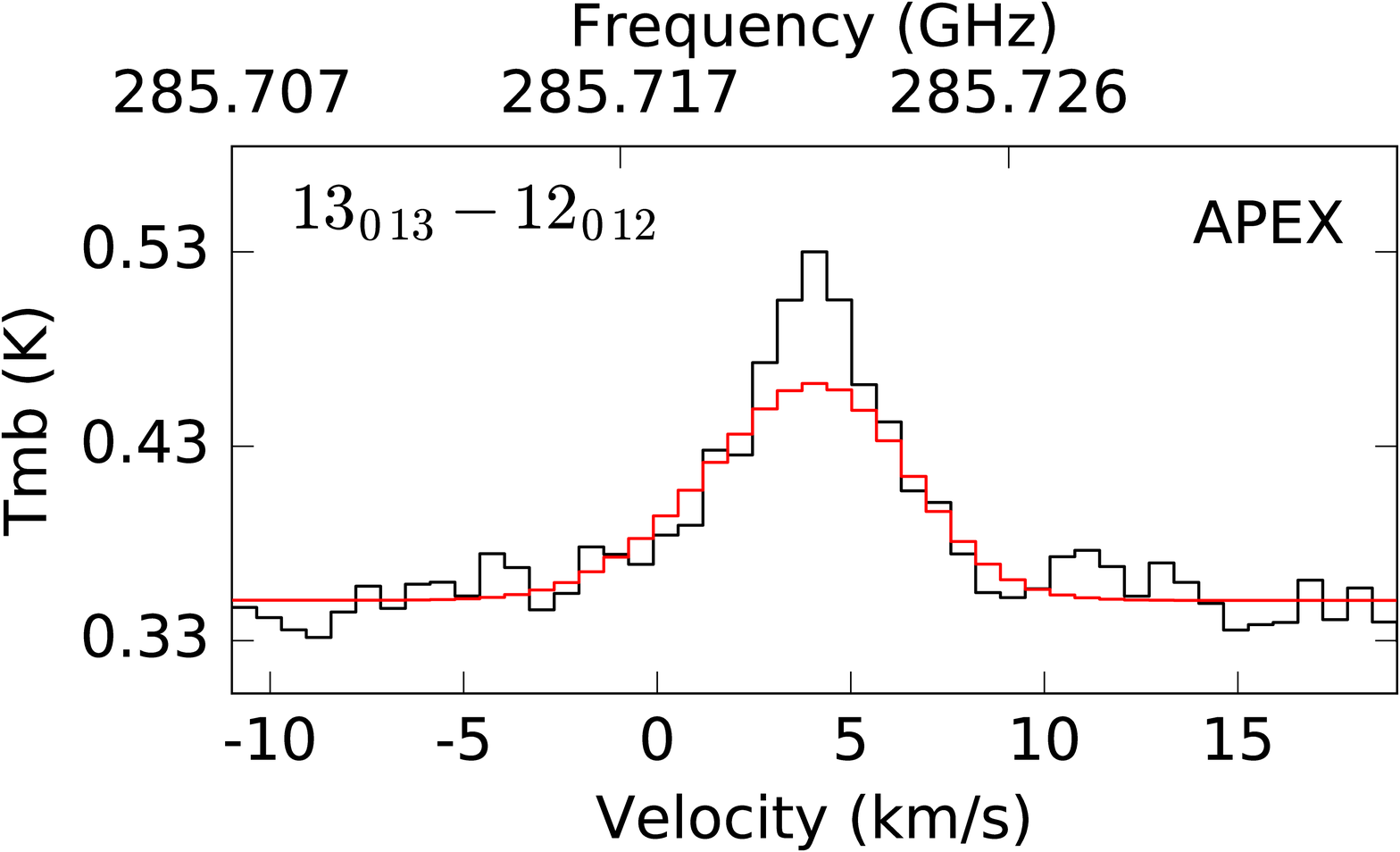}  &\includegraphics[width=0.315\textwidth,trim = 0 0 0 0,clip]{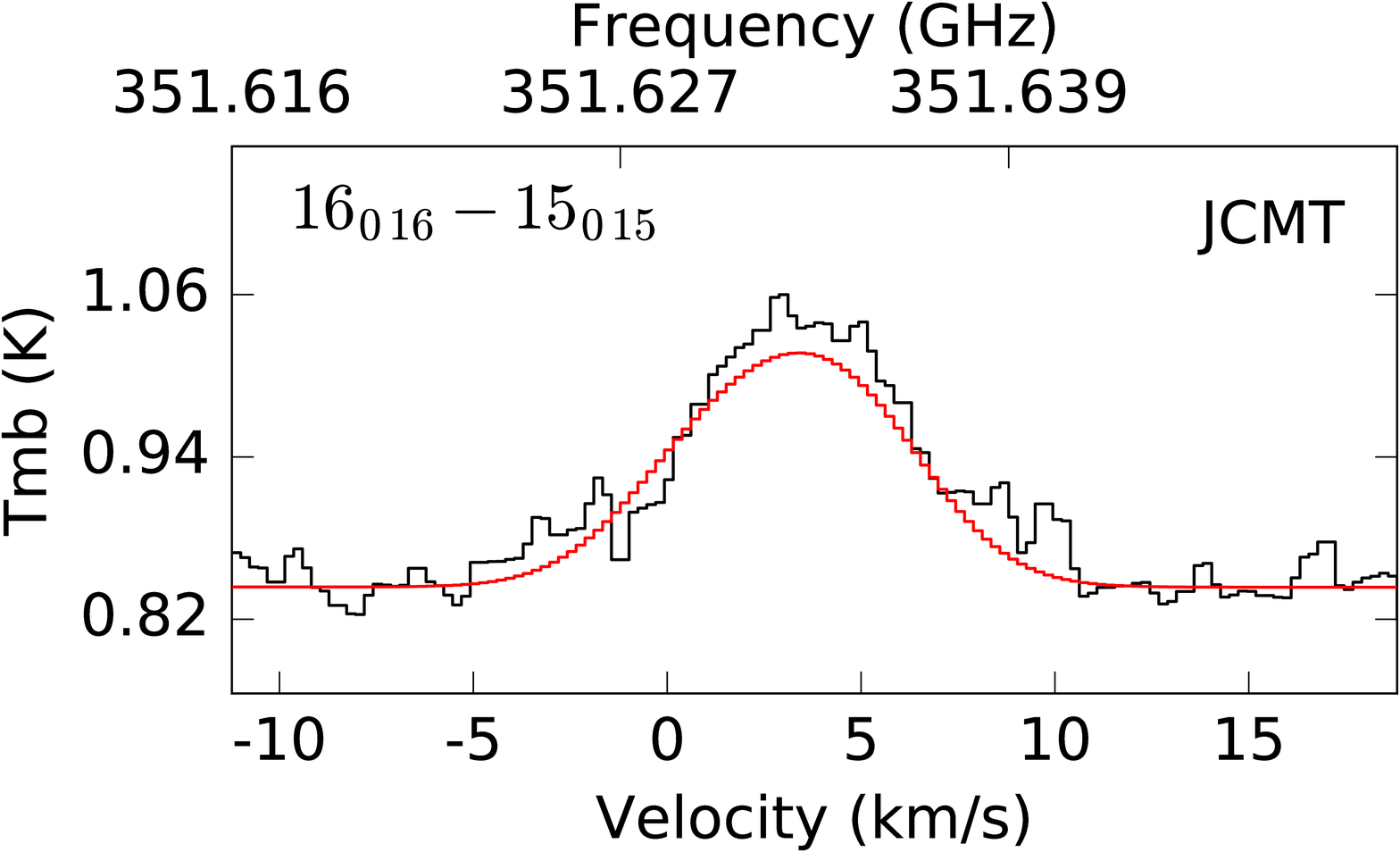}  \\

\includegraphics[width=0.315\textwidth, trim= 0 0 0 0, clip]{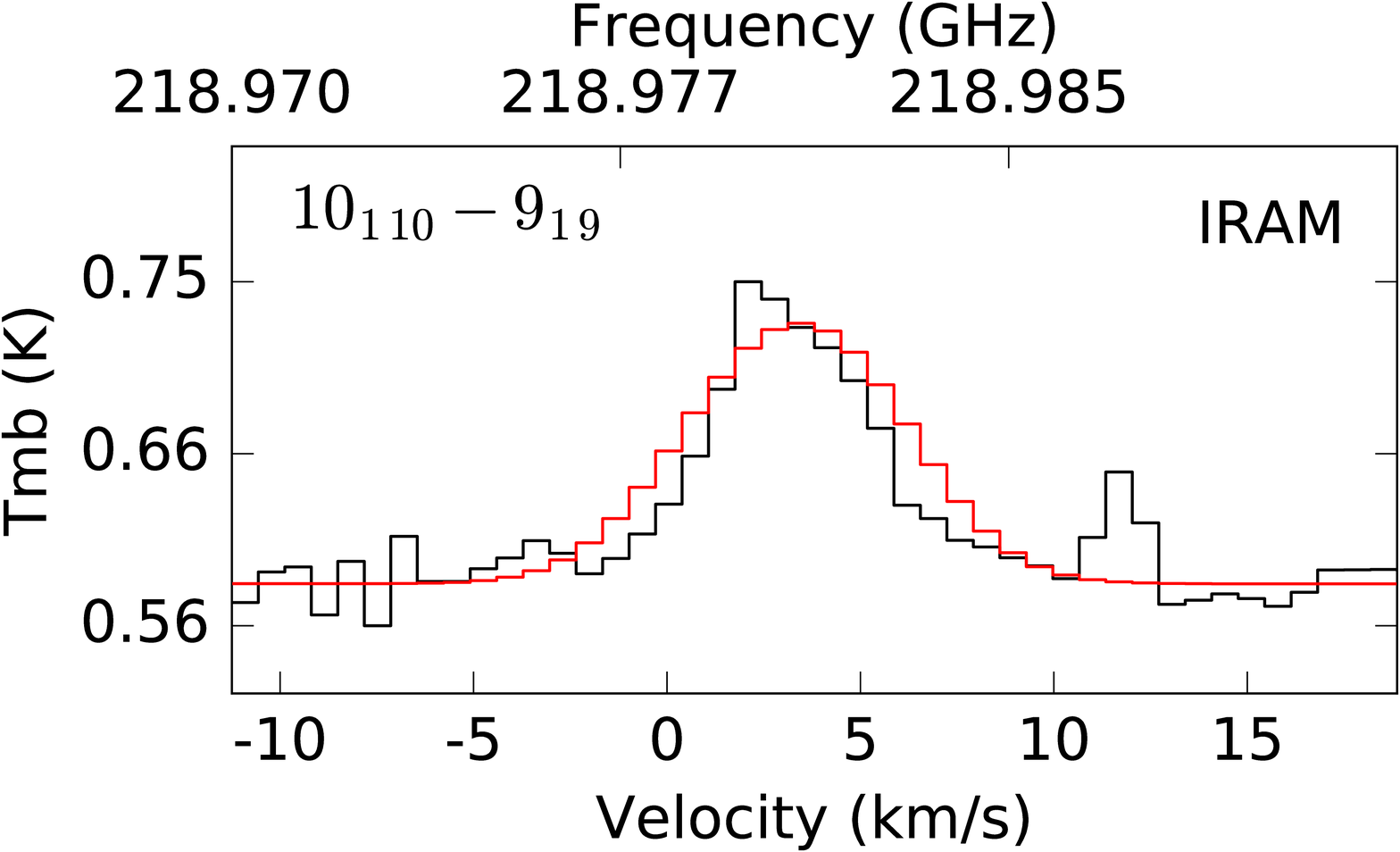} &\includegraphics[width=0.315\textwidth,trim = 0 0 0 0,clip]{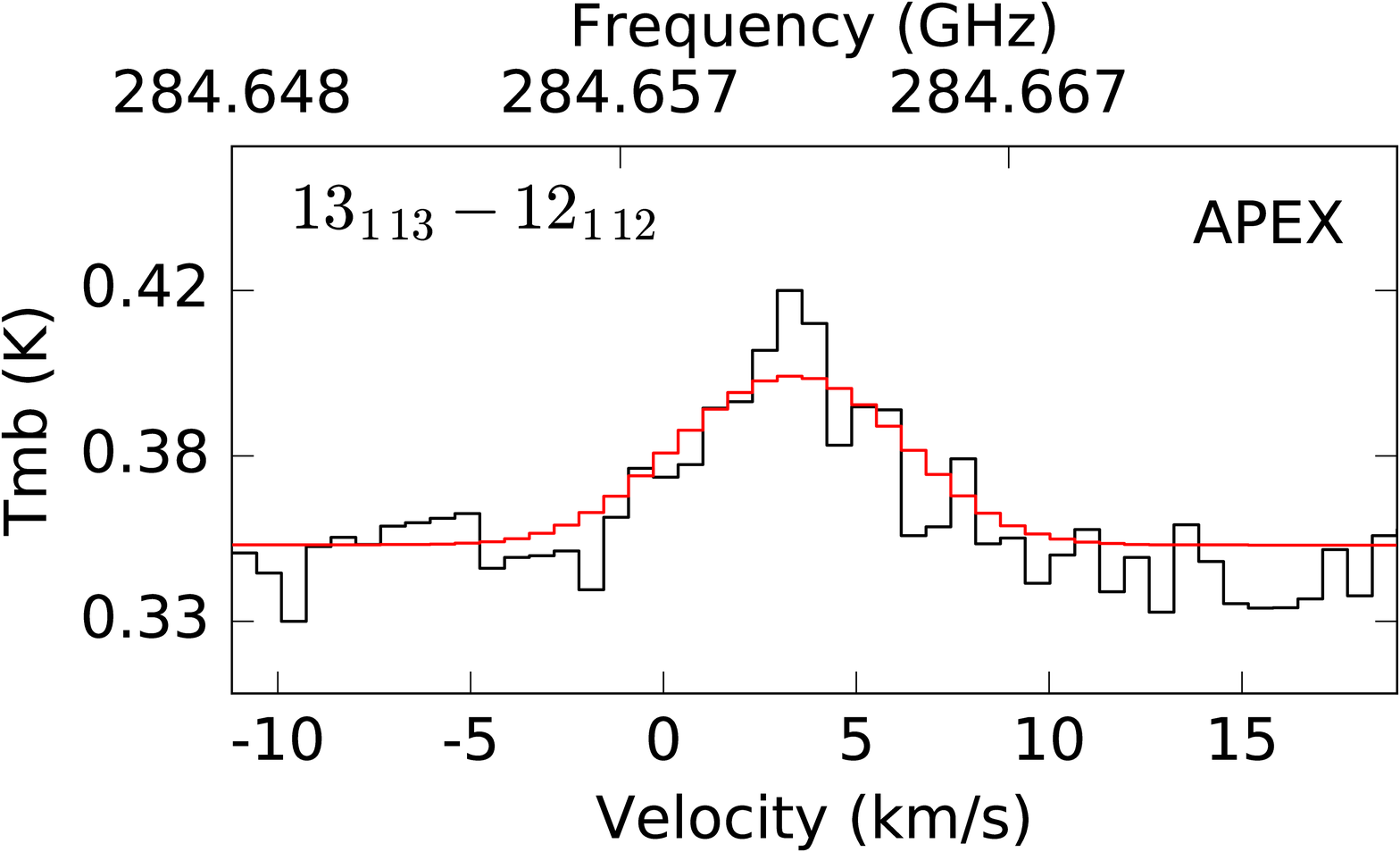}  &\includegraphics[width=0.315\textwidth,trim = 0 0 0 0,clip]{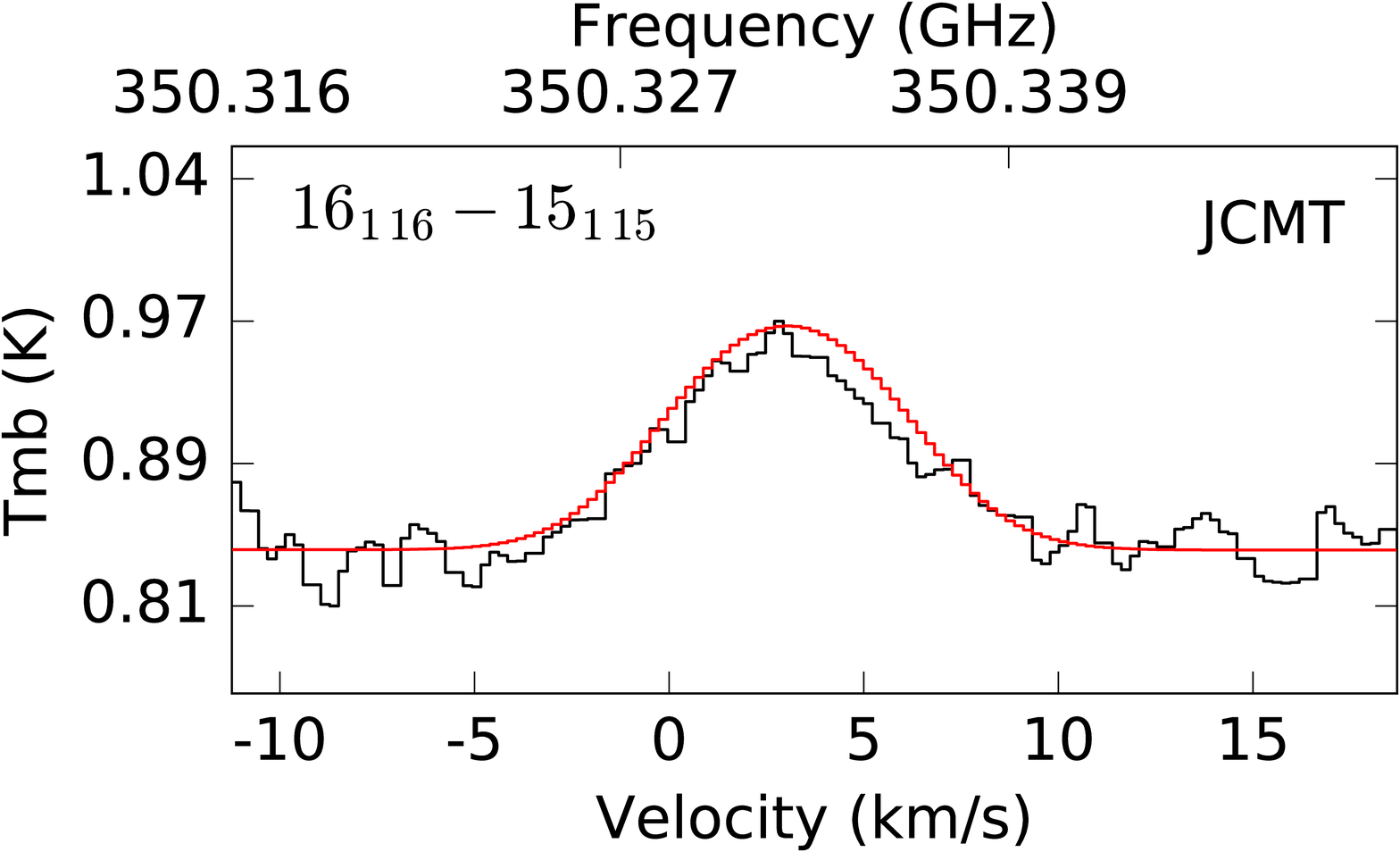}  \\

\includegraphics[width=0.315\textwidth, trim= 0 0 0 0, clip]{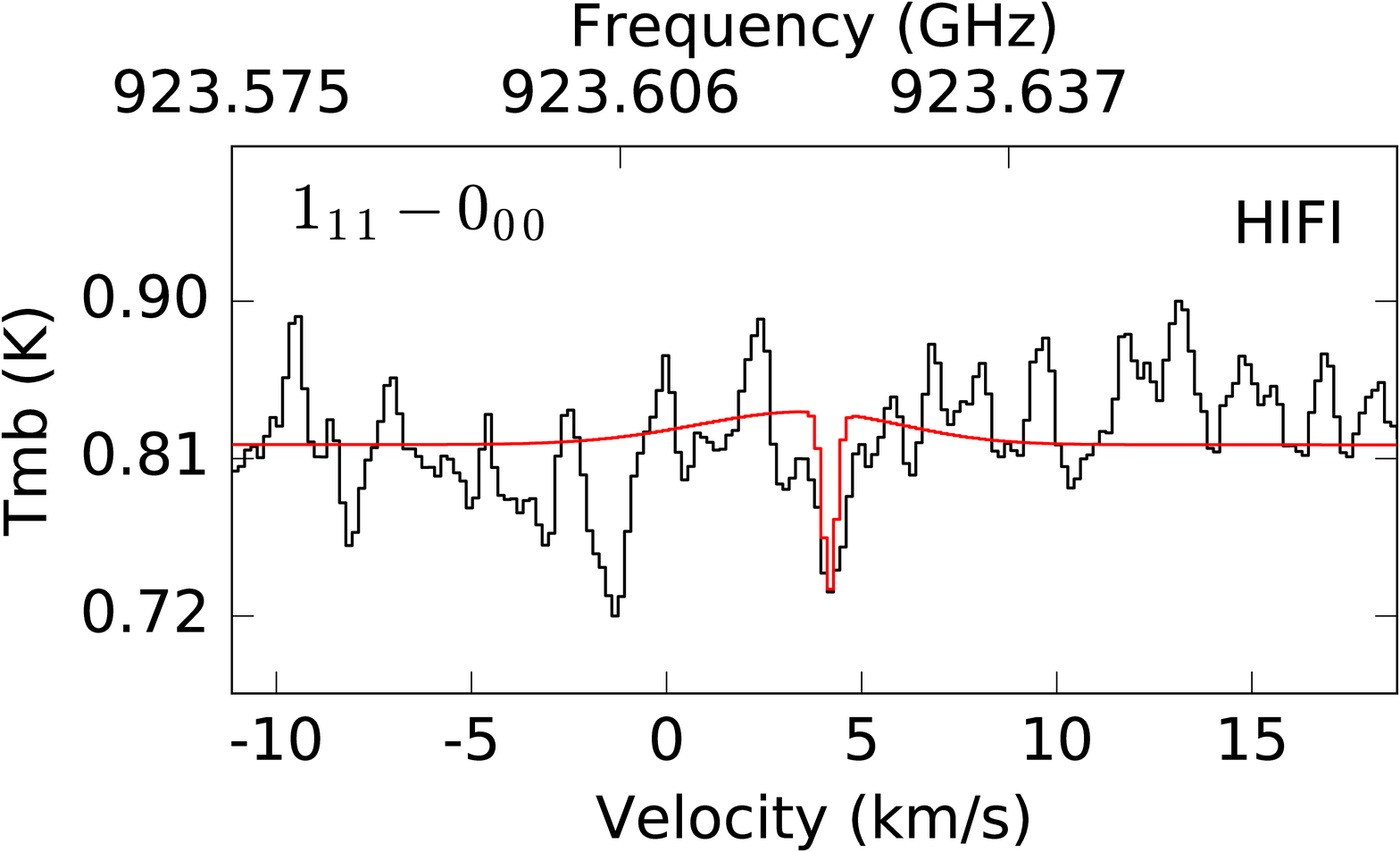} &\includegraphics[width=0.315\textwidth,trim = 0 0 0 0,clip]{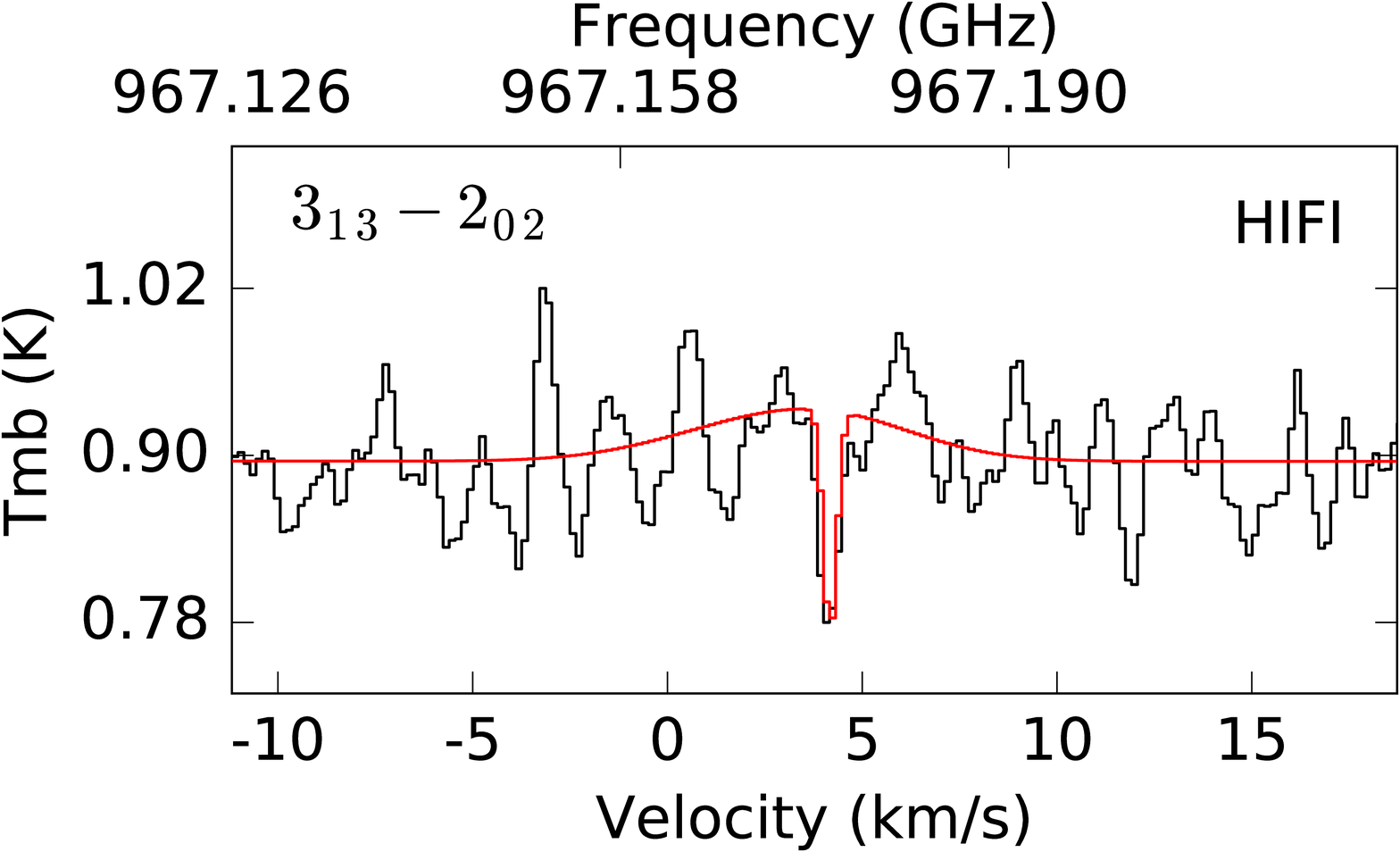}  &\includegraphics[width=0.315\textwidth,trim = 0 0 0 0,clip]{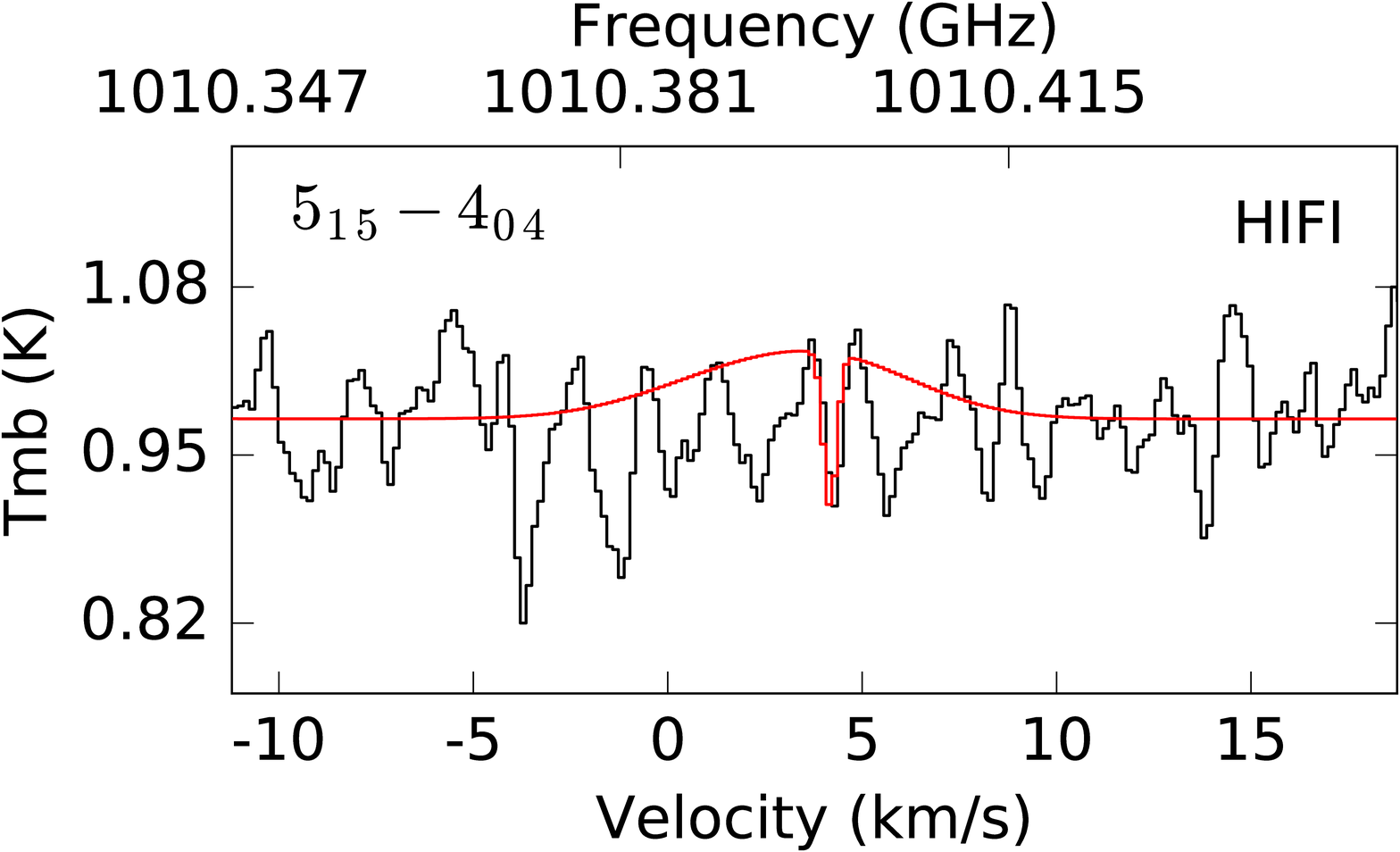}  \\

\end{tabular}
\caption{In black, we show some of the observed HNCO line profiles as seen by different telescopes, while in red we show the predicted profiles from our model with the parameters reported in Table 2.}
\label{imagenes-cont}
\end{figure*}

\section{Discussion}
\subsection{HNCO physical parameters}

The best results from our model give a temperature for the  HNCO-bearing gas associated with the hot corino of about 200 K and a column density of ($1.9\pm 0.4) \times 10^{16}$ cm$^{-2}$, which corresponds to an abundance relative to H$_2$ of $(7.1\pm 0.6) \times 10^{-8}$ for a fixed size of 0.5$''$. This abundance is in agreement with typical values measured for other hot cores \citep[e.g.][]{bisschop2007}. \citet{martin-domenech2017} derived the column density of HNCO towards the hot corino associated with source B in I16293 measuring the HNC$^{18}$O column density and adopting an isotopic ratio of $\mathrm{^{16}O/^{18}O=500}$. They obtained $N$(HNCO)$=(4.9\pm 1.9)\times 10^{16}$ cm$^{-2}$ and an abundance of $(1.8\pm 0.7) \times 10^{-9}$. Although  the column density for source A computed with our model is similar to the column density derived by \citet{martin-domenech2017} for source B, we predict a higher abundance for source A.
For the warm envelope, we found a column density of $(2.5\pm 0.5)\times 10^{13}$ cm$^{-2}$, a temperature of $(31\pm 5$) K, a size of $(15.6\pm 2.4)''$ and an abundance of ($1.8 \pm 0.4) \times 10^{-11}$. The line widths are smaller (4.9 km s$^{-1}$) than those found for the hot corino (6.2 km s$^{-1}$). This behaviour is expected if we consider that the infall and rotation speeds are larger toward the center of the envelope. For the cold outer layer of the envelope, we derive a temperature of about 9 K and a column density of ($5.3\pm 0.6) \times 10^{12}$ cm$^{-2}$. The resulting abundance is ($9.8 \pm 0.3) \times 10^{-12}$, which is very similar to that of the warm envelope.\\

In a previous study, \citet{vandishoeck1995} derived a value for the HNCO column density in the envelope of I16293 based on data obtained with the JCMT and the CSO (Caltech Submillimeter Observatory located on Mauna Kea, Hawaii) of ($3.4\pm 1.5) \times 10^{13}$ cm$^{-2}$ and an abundance with respect to H$_2$ of 1.7$\times 10^{-10}$ by using the rotational diagram technique, which assumes LTE conditions (although no source size for the envelope was derived from their observed HNCO transitions). The column density derived by us is consistent with that reported by \citet{vandishoeck1995} but our abundance is lower by a factor of 10. This difference could be related with the different assumed H$_2$ column densities. Moreover, assuming LTE conditions might not be appropriate for the envelope of I16293. Indeed, \citet{lopez-sepulcre2015} already pointed out this problem. These latter authors computed the abundances for HNCO using a non-LTE radiative transfer model together with collision coefficients from \citet{green1986} and found abundances between (6$\pm$3) $\times$ 10$^{-9}$ for T$>90$ K in the warm envelope and (5$\pm$4) $\times$ 10$^{-12}$ for T$<90$ K corresponding to the cold envelope. We argue that the differences between our abundances and those of \citet{lopez-sepulcre2015} reflect, in part, the different number of components considered in the different models: while ours considers three distinct components, \citet{lopez-sepulcre2015} only includes two. Our hot corino component has a size of 0.5$''$, corresponding to a radius of 35 AU. This is significantly more compact that the region (of radius, fortuitously, 90 AU) where the temperature is higher than 90 K. As a consequence, this latter region in the model of \citet{lopez-sepulcre2015} would correspond in our model to a mixture of the hot corino region and the warm envelope component. It is, therefore, not surprising that the abundance, (6$\pm$3) $\times$ 10$^{-9}$, reported by \citet{lopez-sepulcre2015} for this region is intermediate between the abundances we derive here for the hot corino and the warm envelope. For the region outside of 90 AU, we derive an abundance of 1 to 2 $\times$ 10$^{-11}$, which is within a factor of two of that derived by \citet{lopez-sepulcre2015}.\\

\subsection{Spectral modeling}

We see that most of the lines are very well reproduced by the model based on the physical parameters discussed above (see Figure 3). However, some of them are underestimated (see the predicted line profiles by the model in the Appendix; note that in some cases, the predicted emission could not be computed for some HNCO transitions since their corresponding collisional rate coefficient was lacking in the computation of \citet{sahnoun2018}). For the spectroscopic branch $K_a=1$, the agreement between the model and the data is less good for transitions at lowest frequencies (e.g. $J=4,5$). We have searched for line blending with other species for these transitions in particular to verify if the line intensities were affected by this problem but we did not found a clear contribution from other molecules. For the transitions belonging to the $K_a=2$ levels, the problem is more severe. It could be argued that collision coefficients are particularly inaccurate for such lines. However, examining the spectra of similar molecules, like $\mathrm{H_2CO}$,  one sees the same type of problems. A thorough discussion is given by \cite{mangum93, mangum15}. Formaldehyde has a rotational spectra similar to HNCO, with a heavy rod or rod-like part (NCO vs. CO), and only  light H atom(s)  breaking the rod-like symmetry. Both molecules have thus very distinct $K_a=0,1,2, \ldots$ branches well separated one from the other. Other molecules, less abundant, present a very similar type of spectroscopy (e.g. $\mathrm{H_2CS}$ and l-$\mathrm{C_3H_2}$).\\ 

While for H$_2$CO the intensities and line-shapes are well understood for $K_a\leq 2$ (see methods proposed by \citealt{mangum93}), this is not always true for the higher lying $K_a>2$ branch, very similarly to our case of HNCO. The higher $K_a=2$ transitions we observe connect levels which are not correctly modelled. That is, the levels with high-$J$ and $K_a=2$ or any level with $K_a>2$ are not considered in the computation of the collisional coefficients used in this paper. These high levels could be populated by specific excitation, perhaps originating in some hotter photonic bath or some specific excitation, not properly taken into account.\\

Note that we have not considered infrared (IR) pumping to vibrationally excited states of HNCO. For this molecule, the $\nu_4$, $\nu_6$ and $\nu_5$ fundamental bending vibrations have energies above ground by 1118, 949, and 831 K, respectively \citep[see, e.g., ][]{Yamada1977}. Various rotational lines with $J=4$ and $5$ from within the $\nu_5 = 1$ and $\nu_6 = 1$ (and possibly the $\nu_4 = 1$) states have been detected by \citet{Belloche2017} with ALMA toward the hot core Sgr B2 (N2). These lines intensities are characterized by a source model whose angular source size and rotational temperature are not too different from the values used by us for I16293. Some of the lines had even been detected in an earlier survey with the IRAM 30-m telescope \citet{Belloche2013}. While quite weak for detection by the single dishes employed by us, future ALMA data could address HNCO excitation in the hot corino in I16293. Deciding whether vibrational excitation by IR photons or collisions are both feasible or whether such excitation and re-decay to the ground state could influence the intensities of the $K_a = 2$ lines lies beyond the scope of the present paper.


\subsection{Comparison with Green collisional coefficients}

The new collisional coefficients with respect to those computed by \citet{green1986} bring three advantages : 

\begin{itemize}
\item[(a)] More temperatures are considered, 17 with the new set of coefficients (7,  10,  and 20 to 300 K with a step of 20 K) with respect to 5 for Green coefficients (20, 40, 80, 160 and 320\,K).
\item[(b)] More collisional transitions are considered (2272 against 2254).
\item[(c)] The coefficients are computed separately for collisions with p-H$_2$ and o-H$_2$, allowing to deal with the o/p ratio, an important factor in cold environments.
\end{itemize}

While the computed spectra using both set of collisional coefficients are similar for our I16293 three components model at frequencies $\le$\,500\,GHz, there are some differences at higher frequencies, as can be seen on Figure \ref{fig:lw-vs-green}, justifying the use of this new set for an optimal modelisation.

\begin{figure}
\includegraphics[width=0.48\textwidth]{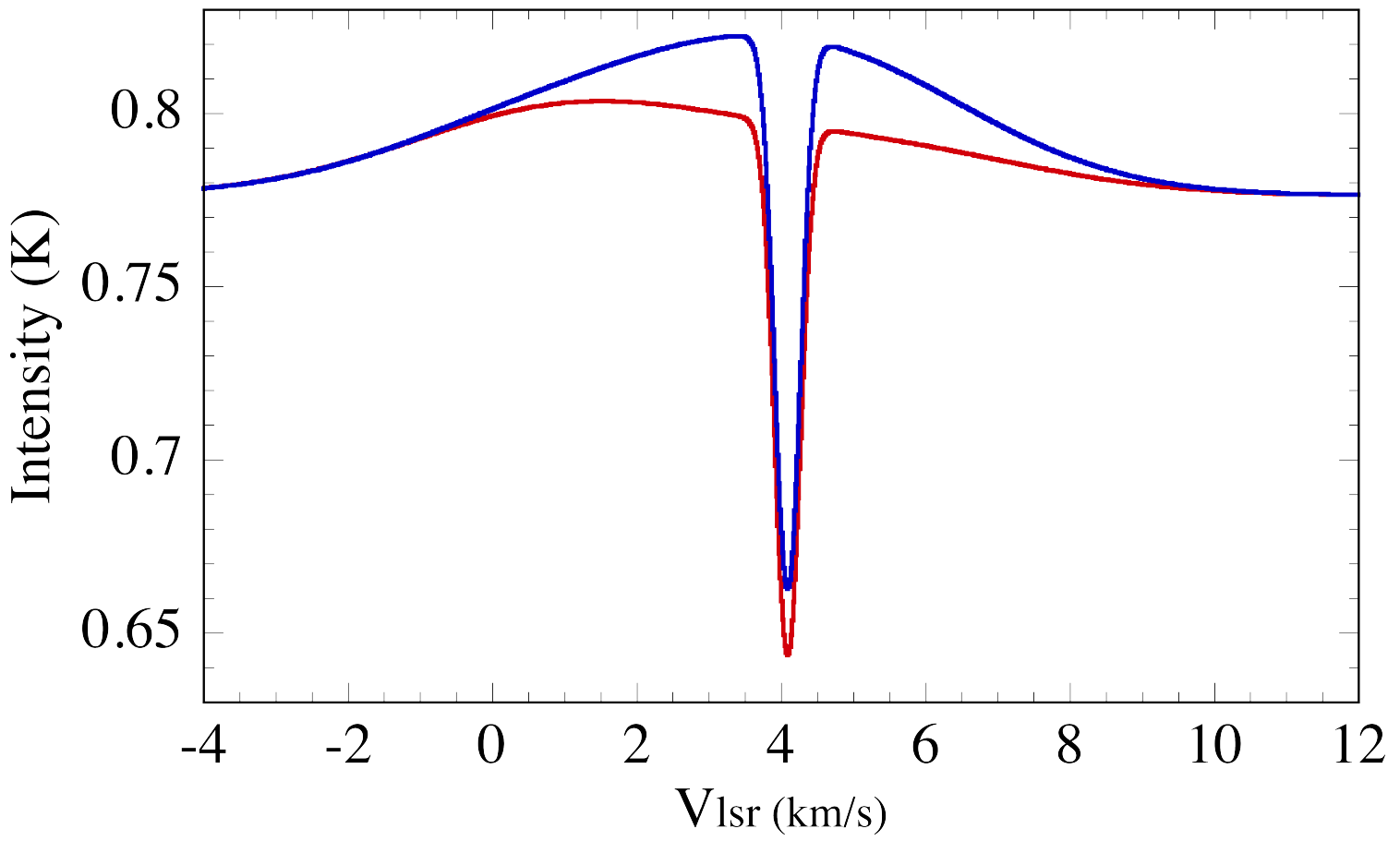}
  \caption{Comparison of the I16293 computed spectrum for the HNCO (4$_{1,3}-4_{0,4}$) transition ($\nu$ = 902505.9937\,MHz, E$_\text{up}$ = 53.86\,K A$_\text{ij}$ = 7.85 $\times 10^{-3}$) with our 3 components model. Blue : Green coefficients, Red : new set of coefficients. }
  \label{fig:lw-vs-green}
\end{figure}

\subsection{Chemical modelling with Nautilus}

To further investigate the chemistry of HNCO in the envelope of I16293 and to probe the abundances derived from our full non-LTE radiative transfer model, we computed the abundance profiles using the Nautilus chemical code. Nautilus is a three phase gas-grain chemical code that allows us to compute the chemical composition as a function of time. The details of the chemical reactions included in the code are explained in \citet{ruaud2016}. The gas phase network used by Nautilus is based on kida.uva.2014 \citep{wakelam2015}, while the grain chemistry is presented in \citet{ruaud2015}. The modelling is made in two steps: first, we run a simulation for the parental cloud where the protostar I16293 was formed, and then we take the resulting abundances as input for the protostar phase, where we use the 1D structure for the envelope defined by \citet[][i.e., the temperature and density profiles]{crimier2010} to compute the corresponding abundance profile.

For the parental cloud, as \citet{hincelin2011}, we have used the atomic initial abundances and the following set of physical parameters, which are typical for cold dense clouds in the Solar neighbourhood: $n = 3\times 10^4$ cm$^{-3}$, T $ = 10$ K, C/O ratio $= 0.7$, gas to dust ratio $ =  100$ and grain size $ = 0.1$ $\mu$m, UV field $G_0 = 1$ Habing $(1.6 \times 10^{-3}$ erg cm$^{-2}$ s$^{-1}$), visual extinction A$\mathrm{_v} = 3$, cosmic ray ionization rate $\zeta = 1.3 \times 10^{-17}$ s$^{-1}$, evolution time for the cloud to reach the pre-stellar phase $ = 2.5 \times 10^5$ years. 
We ran several simulations changing the value of A$\mathrm{_v}$ (from 3 to 30), density (from $3\times10^3$ to $3\times10^5$ cm$^{-3}$), evolution times (from $10^5$ to $5\times10^5$ years), C/O ratio (0.7 and 1.2) and cosmic ray ionization rate (from $10^{-17}$ to $10^{-16}$  s$^{-1}$), and found that the only parameters changing the abundance profiles are the visual extinction A$\mathrm{_v}$ and the cosmic ray ionization rate $\zeta$. The abundance profiles that best reproduces the observations are obtained with A$\mathrm{_v} = 4$ and  $\zeta = 8.0 \times10^{-17}$ s$^{-1}$. Although this value of $\zeta$ is higher that the commonly assumed standard value for the Solar neighbourhood $(1.3\times10^{-17}$ s$^{-1})$, the Ophiuchus star-forming region is known to have a high cosmic ionization rate \citep[e.g.][]{hunter1994}.

\subsection{Results of the Nautilus chemical model for HNCO}

Once the final abundances are computed, we use them as input and run a simulation taking into account the density and temperature profiles defined by \citet{crimier2010} for the envelope of I16293. To do that, we kept the same visual extinction and cosmic ionization rate used for the initial cloud. The resulting HNCO radial abundance profile for different ages of the protostar is shown in Figure \ref{fig:nautilus}. The model for the warm envelope predicts a high abundance (larger than $\sim 10^{-9}$) close to the hot corino value (in agreement with the model described above), while for the warm envelope radius derived from our non-LTE radiative transfer model of 7.8$''$, the abundance varies from $10^{-12}$ to $10^{-10}$, depending on the age of the protostar. For the external envelope, the abundance profile seems to change more rapidly with time. At a radius of 30$''$, where the difference between the predicted abundances seems to be larger, the model predicts a smaller abundance for largest age ($\sim 10^{-12}$ for $1.4\times 10^5$ yr) and vice-versa  ($\sim 10^{-9}$ for $1.0\times 10^4$ yr). The final values for the abundance predicted with Nautilus (although smaller) are closer to the abundance predicted by the radiative transfer model for younger ages. In fact, the age of I16293 in the literature is reported to be between $\sim 10^4- 10^{5}$ years.  For instance, \citet{quenard2018} studied the emission of HNCO towards the hot corino associated with source B and the cold envelope of I16293 using the chemical code UCLCHEM. They compared the abundance for source B obtained by \citet[][$1.8\times 10^{-9}$]{martin-domenech2017} and the abundance for the cold envelope derived by \citet[][$1.7\times 10^{-10}$]{vandishoeck1995} and found an age for the protostar close to $\sim (2-4) \times 10^4$ years.\\

\begin{figure*}\center
\includegraphics[width=0.9\textwidth]{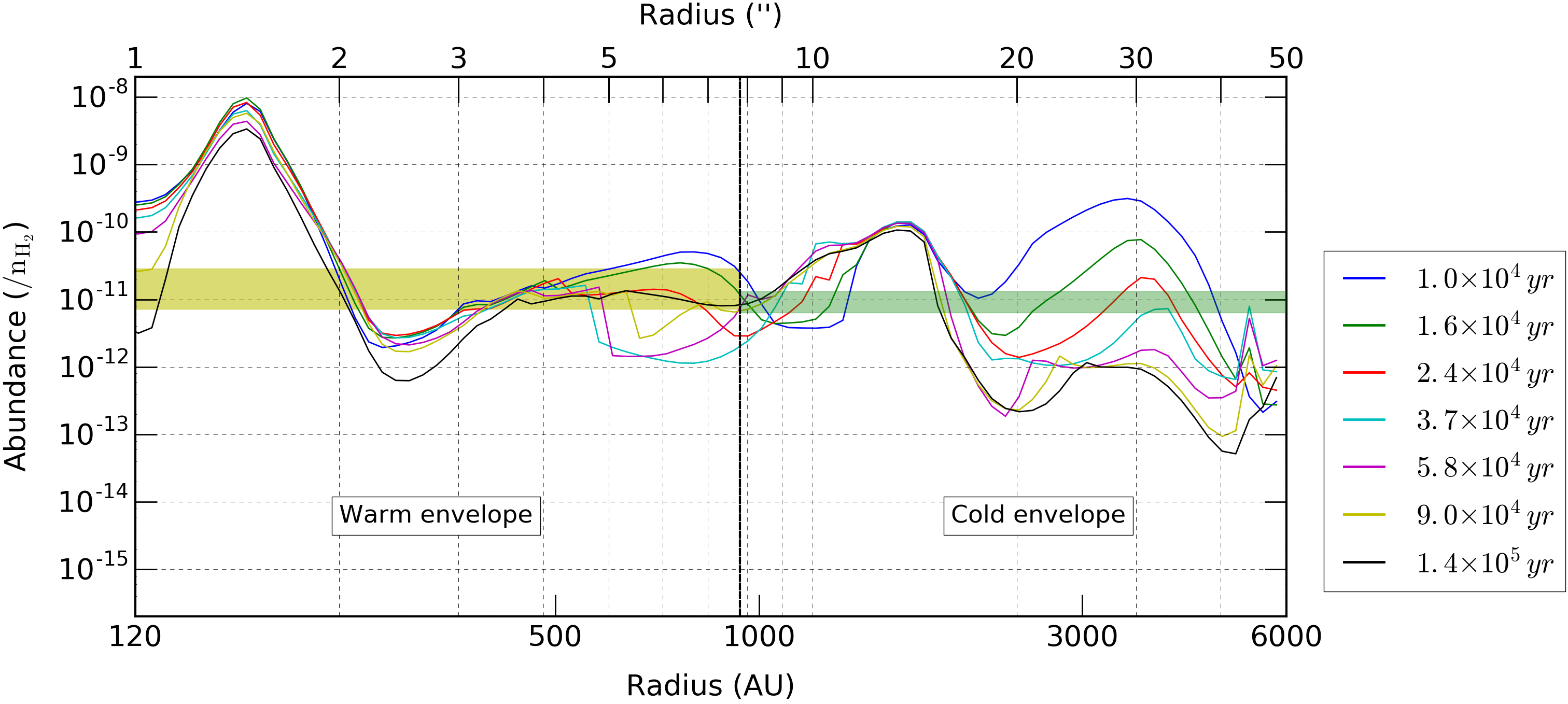}
  \caption{HNCO abundance profile computed with Nautilus for the envelope of I16293 assuming a cosmic ionization rate of $8.0\times 10^{-17}$ s$^{-1}$. The colours of the lines indicate the age of the protostar in years. The light yellow horizontal band represent the predicted abundance for the warm envelope from our CASSIS-RADEX model ($1.8\pm 0.4 \times 10^{-11}$) , while the light green colour band represent the predicted abundance for the cold envelope ($9.8\pm 0.3 \times 10^{-12}$). Both bands have a 3$\sigma$ error width. The black vertical line represents the radius of the warm envelope of $7.8''$. }
  \label{fig:nautilus}
\end{figure*}

From the simulations, we conclude that we cannot use HNCO as a chemical clock to constrain the age of the protostar due to the strong dependence on the initial parameters for the model. If we take an age of about 5.8$\times 10^4$ years, we observe that the abundance profile, although not constant, is in good agreement with the abundances derived directly from the observations with our radiative transfer model. We also found that the Nautilus input parameters used to reproduce the HNCO observed abundances are very similar to those derived by other authors using Nautilus and single-dish data for other molecules such as HDO \citep{coutens2012}, CH \citep{bottinelli2014}, CH$_3$SH \citep{majumdar2016}, C$_3$H$_2$ \citep{majumdar2017} or HOCO$^+$ \citep{majumdar2018}.\\

Recently, observations of the ortho- and para-ground-state lines of both H$_2$D$^+$ and D$_2$H, deuterated isotopologues of the fundamental H$_3^+$ ion, have been performed toward I16293 with the APEX telescope and the Stratospheric Observatory for Infrared Astronomy (SOFIA) \citep{Bruenken2014, Harju2017}. Given the observed simple line profiles, their narrow width and LSR velocities, the lines from these species originate in the extended envelope of I16293. For both species, their ortho-to-para ratio, OPR, is time-dependent. Modelling the observed OPR for H$_2$D$+$ \citet{Bruenken2014} derive an age of order $10^6$ yr. This value is further constrained to $5\times10^5$ yr by \citet{Harju2017} who combines the observed values for the H$_2$D$^+$ and D$_2$H$^+$ OPR. This age is considerably older that the values from Nautilus modelling. While this might be understandable for the age of the hot corino, the discrepancy between the age derived from the H$_2$D$^+$ and HNCO in the extended envelope is difficult to explain.

\section{Conclusions}

In this work, we have presented an analysis of isocyanic acid (HNCO) line emission towards IRAS 16293-2422 as observed with single-dish telescopes (IRAM, APEX, JCMT and Herschel/HIFI) over a wide range of frequencies. The HNCO line profiles are interpreted as the superposition of three physical components along the line of sight: a dense, warm and compact source associated with the hot corino in I16293, a more extended component associated with the warm part of the envelope, and a cold and extended component corresponding to the outer layer of the envelope. We have constrained most of the parameters for each physical component based on the structure derived by \citet{crimier2010}.\\  

We used a full non-LTE radiative transfer model in CASSIS-RADEX to predict the line emission profiles by using the new HNCO collisional rate coefficients computed by \citet{sahnoun2018} from a set of rotational excitation quenching rates between HNCO and both ortho and para H$_2$. We obtain physical values for the hot corino that are in very good agreement with what has been found for other hot cores \citep[e.g.][]{bisschop2007}. From the derived column densities, we found that the abundances in both warm and cold layers of the envelope are very similar (close to $\sim10^{-11}$).\\ 

 While the transitions on the $K_a=0,1$ bands are well reproduced, the transitions belonging to the upper rotational band $K_a=2$ levels are not. We argue that those levels could be populated by specific excitation due to a hotter photonic or collisional bath and that the modelling is not able to take these excitations into account.\\
 
We used the three phase chemical gas-grain code Nautilus to compute the chemical composition of the envelope of I16293 and produce an HNCO radial abundance profile.  We found that the younger ages for the protostar are in better agreement with our radiative transfer model results. However, HNCO cannot be used as a reliable chemical clock due to the high dependence on the initial parameters in our simulations.

\section*{Acknowledgements}
HIFI has been designed and built by a consortium of institutes and university departments from across Europe, Canada and the United States under the leadership of SRON Netherlands Institute for Space Research, Groningen, The Netherlands and with major contributions from Germany, France and the US. Consortium members are: Canada: CSA, U.Waterloo; France: CESR, LAB, LERMA, IRAM; Germany: KOSMA, MPIfR, MPS; Ireland, NUI Maynooth; Italy: ASI, IFSI-INAF, Osservatorio Astrofisico di Arcetri-INAF; Netherlands: SRON, TUD; Poland: CAMK, CBK; Spain: Observatorio Astron\'{o}mico Nacional (IGN), Centro de Astrobiolog\'{i}a (CSIC-INTA). Sweden: Chalmers University of Technology - MC2, RSS \& GARD; Onsala Space Observatory; Swedish National Space Board, Stockholm University - Stockholm Observatory; Switzerland: ETH Zurich, FHNW; USA: Caltech, JPL, NHSC.\\
This work is based on observations carried out under project number [014-17] with the IRAM 30m telescope. IRAM is supported by INSU/CNRS (France), MPG (Germany) and IGN (Spain).\\
APEX is a collaboration between the Max-Planck-Institut fur Radioastronomie, the European Southern Observatory, and the Onsala Space Observatory.\\
The James Clerk Maxwell Telescope is operated by the East Asian Observatory on behalf of The National Astronomical Observatory of Japan; Academia Sinica Institute of Astronomy and Astrophysics; the Korea Astronomy and Space Science Institute; the Operation, Maintenance and Upgrading Fund for Astronomical Telescopes and Facility Instruments, budgeted from the Ministry of Finance (MOF) of China and administrated by the Chinese Academy of Sciences (CAS), as well as the National Key R\&D Program of China (No. 2017YFA0402700). Additional funding support is provided by the Science and Technology Facilities Council of the United Kingdom and participating universities in the United Kingdom and Canada.\\
This paper makes use of the following ALMA data: ADS/JAO.ALMA\#2013.1.00278.S. ALMA is a partnership of ESO (representing its member states), NSF (USA) and NINS (Japan), together with NRC (Canada), MOST and ASIAA (Taiwan), and KASI (Republic of Korea), in cooperation with the Republic of Chile. The Joint ALMA Observatory is operated by ESO, AUI/NRAO and NAOJ.\\
A.H-G., E.C., L.L and S.B. acknowledge the financial support of the French/Mexico CONACyT -- ECOS-Nord Project ECOS-nord n$^{\circ}$ : M14U01 : SPECIMEN : {\bf S}tructure {\bf P}hysiqu{\bf E} et {\bf CI}n\'{e}{\bf M}atiqu{\bf E} d'IRAS 16293 : mol\'{e}cules et conti{\bf N}uum. A.H-G. and L.L. acknowledge the financial support of DGAPA, UNAM (project IN112417), and CONACyT, Mexico. L.W. and E.S. thank the COST action CM1401, ``Our Astrochemical History" for some travel support.
 





\begin{thebibliography}{199}
\bibitem[\protect\citeauthoryear{Belloche et al.}{2013}]{Belloche2013} Belloche A., M{\"u}ller H.~S.~P., Menten K.~M., Schilke P., Comito C., 2013, A\&A, 559, A47  
\bibitem[\protect\citeauthoryear{Belloche et al.}{2017}]{Belloche2017} Belloche A., et al., 2017, A\&A, 601, A49 
\bibitem[Bisschop et al.(2007)]{bisschop2007} Bisschop, S.~E., J{\o}rgensen, J.~K., van Dishoeck, E.~F., \& de Wachter, E.~B.~M.\ 2007, \aap, 465, 913 
\bibitem[Bisschop et al.(2008)]{bisschop2008} Bisschop, S.~E., J{\o}rgensen, J.~K., Bourke, T.~L., Bottinelli, S., \& van Dishoeck, E.~F.\ 2008, \aap, 488, 959
\bibitem[Biver et al.(2006)]{biver2006} Biver, N., Bockel{\'e}e-Morvan, D., Crovisier, J., et al.\ 2006, \aap, 449, 1255 
\bibitem[Blake et al.(1987)]{blake1987} Blake, G.~A., Sutton, E.~C., Masson, C.~R., \& Phillips, T.~G.\ 1987, \apj, 315, 621  
\bibitem[Bottinelli et al.(2014)]{bottinelli2014} Bottinelli, S., Wakelam, V., Caux, E., et al.\ 2014, \mnras, 441, 1964 
\bibitem[\protect\citeauthoryear{Br{\"u}nken et al.}{2014}]{Bruenken2014} Br{\"u}nken S., et al., 2014, Natur, 516, 219 
\bibitem[Caux et al.(2011)]{caux2011} Caux, E., Kahane, C., Castets, A., et al.\ 2011, \aap, 532, A23 
\bibitem[Caux et al.(2011)]{2011caux} Caux, E., Bottinelli, S., Vastel, C., \& Glorian, J.~M.\ 2011, The Molecular Universe, 280, 120 
\bibitem[Ceccarelli(2010)]{ceccarelli2010}Ceccarelli C.,  et al. 2010, A$\&$A, 521, L22
\bibitem[Coutens et al.(2012)]{coutens2012} Coutens, A., Vastel, C., Caux, E., et al.\ 2012, \aap, 539, A132 
\bibitem[Ceccarelli et al.(2000)]{ceccarelli2000} Ceccarelli, C., Loinard, L., Castets, A., Tielens, A.~G.~G.~M., \& Caux, E.\ 2000, \aap, 357, L9 
\bibitem[Crovisier(1998)]{crovisier1998} Crovisier, J.\ 1998, Faraday Discussions, 109, 437 
\bibitem[Crimier et al.(2010)]{crimier2010} Crimier, N., Ceccarelli, C., Maret, S., et al.\ 2010, \aap, 519, A65 
\bibitem[de Graauw et al.(2010)]{degraauw2010} de Graauw, T., Helmich, F.~P., Phillips, T.~G., et al.\ 2010, \aap, 518, L6 
\bibitem[Dzib et al.(2018)]{dzib2018} Dzib, S.~A., Ortiz-Le{\'o}n, G.~N., Hern{\'a}ndez-G{\'o}mez, A., et al.\ 2018, arXiv:1802.03234 
\bibitem[Fedoseev et al.(2015)]{fedoseev2015} Fedoseev, G., Ioppolo, S., Zhao, D., Lamberts, T., \& Linnartz, H.\ 2015, \mnras, 446, 439 
\bibitem[Fusina \& Mills(1981)]{fusina1981harmonic} Fusina, L., \& Mills, I.~M.\ 1981, Journal of Molecular Spectroscopy, 86, 488 
\bibitem[Garrod et al.(2008)]{garrod2008} Garrod, R.~T., Widicus Weaver, S.~L., \& Herbst, E.\ 2008, \apj, 682, 283-302 
\bibitem[Green(1986)]{green1986}Green, S. 1986, NASA Technical Memorandum 87791
\bibitem[Guan et al.(2006)]{guan2006} Guan Y., Flei{\ss}ner R., Joyce P., Krone S. M., 2006, Stat Comput,16, 193
\bibitem[Hastings(1970)]{hastings1970} Hastings W. K., 1970, Biometrika, 57, 97
\bibitem[\protect\citeauthoryear{Harju et al.}{2017}]{Harju2017} Harju J., et al., 2017, ApJ, 840, 63  
\bibitem[Helmich \& van Dishoeck(1997)]{helmich1997} Helmich, F.~P., \& van Dishoeck, E.~F.\ 1997, \aaps, 124, 205 
\bibitem[Hern\'andez-G\'omez et al.(2018)]{hernandez-gomez2018} Hern\'andez-G\'omez A., Caux E., Loinard L., et al. \ 2018, submitted
\bibitem[Heyminck et al.(2006)]{heyminck2006} Heyminck, S., Kasemann, C., G{\"u}sten, R., de Lange, G., \& Graf, U.~U.\ 2006, \aap, 454, L21 
\bibitem[Hincelin et al.(2011)]{hincelin2011} Hincelin, U., Wakelam, V., Hersant, F., et al.\ 2011, \aap, 530, A61 
\bibitem[Hocking(1975)]{hocking1975}W. H. Hocking, M. C. L. Gerry, and G. Winnewisser, 1975, Can. J. Phys. 53, 1869
\bibitem[Hunter et al.(1994)]{hunter1994} Hunter, S.~D., Digel, S.~W., de Geus, E.~J., \& Kanbach, G.\ 1994, \apj, 436, 216 
\bibitem[Iglesias(1977)]{iglesias1977} Iglesias, E.\ 1977, \apj, 218, 697 
\bibitem[Jackson et al.(1984)]{jackson1984} Jackson, J.~M., Armstrong, J.~T., \& Barrett, A.~H.\ 1984, \apj, 280, 608
\bibitem[Jacobsen et al.(2018)]{jacobsen2018} Jacobsen, S.~K., J{\o}rgensen, J.~K., van der Wiel, M.~H.~D., et al.\ 2018, \aap, 612, A72  
\bibitem[J{\o}rgensen et al.(2016)]{jorgensen2016} J{\o}rgensen, J.~K., van der Wiel, M.~H.~D., Coutens, A., et al.\ 2016, \aap, 595, A117 
\bibitem[Kukolich(1971)]{kukolich1971}S. G. Kukolich, A. C. Nelson, and B. S. Yamanashi, 1971, J. Am. Chem. Soc. 93, 6769
\bibitem[Lapinov(2007)]{lapinov2007}Lapinov, G. Yu. Golubiatnikov, V. N. Markov, and A. Guarnieri, 2007, Astron. Lett. 33,121
\bibitem[Lis et al.(1997)]{lis1997} Lis, D.~C., Keene, J., Young, K., et al.\ 1997, \icarus, 130, 355 
\bibitem[L{\'o}pez-Sepulcre et al.(2015)]{lopez-sepulcre2015} L{\'o}pez-Sepulcre, A., Jaber, A.~A., Mendoza, E., et al.\ 2015, \mnras, 449, 2438 
\bibitem[MacDonald et al.(1996)]{macdonald1996} MacDonald, G.~H., Gibb, A.~G., Habing, R.~J., \& Millar, T.~J.\ 1996, \aaps, 119, 333 
\bibitem[Majumdar et al.(2016)]{majumdar2016} Majumdar, L., Gratier, P., Vidal, T., et al.\ 2016, \mnras, 458, 1859 
\bibitem[Majumdar et al.(2017)]{majumdar2017} Majumdar, L., Gratier, P., Andron, I., Wakelam, V., \& Caux, E.\ 2017, \mnras, 467, 3525 
\bibitem[Majumdar et al.(2018)]{majumdar2018} Majumdar, L., Gratier, P., Wakelam, V., et al.\ 2018, \mnras, 477, 525 
\bibitem[Mangum et al. (1993)]{mangum93} J.G. Mangum and A. Wootten, 1993, ApJS, 89, 123.
\bibitem[Mangum et al. (2015)]{mangum15} J.G. Mangum and Y.L. Shirley, 2015, PASP, 127, 266.
\bibitem[Marcelino et al.(2009)]{marcelino2009} Marcelino, N., Cernicharo, J., Tercero, B., \& Roueff, E.\ 2009, \apjl, 690, L27 
\bibitem[Marcelino et al.(2010)]{marcelino2010} Marcelino, N., Br{\"u}nken, S., Cernicharo, J., et al.\ 2010, \aap, 516, A105 
\bibitem[Mart{\'{\i}}n et al.(2006)]{martin2006} Mart{\'{\i}}n, S., Mauersberger, R., Mart{\'{\i}}n-Pintado, J., Henkel, C., \& Garc{\'{\i}}a-Burillo, S.\ 2006, \apjs, 164, 450
\bibitem[Mart{\'{\i}}n-Dom{\'e}nech et al.(2017)]{martin-domenech2017} Mart{\'{\i}}n-Dom{\'e}nech, R., Rivilla, V.~M., Jim{\'e}nez-Serra, I., et al.\ 2017, \mnras, 469, 2230 
\bibitem[Mart{\'{\i}}n et al.(2008)]{martin2008} Mart{\'{\i}}n, S., Requena-Torres, M.~A., Mart{\'{\i}}n-Pintado, J., \& Mauersberger, R.\ 2008, \apj, 678, 245 
\bibitem[Mart{\'{\i}}n et al.(2009)]{martin2009} Mart{\'{\i}}n, S., Mart{\'{\i}}n-Pintado, J., \& Mauersberger, R.\ 2009, \apj, 694, 610 
\bibitem[Meier \& Turner(2005)]{meier2005} Meier, D.~S., \& Turner, J.~L.\ 2005, \apj, 618, 259  
\bibitem[Menten et al.(1987)]{menten1987} Menten, K.~M., Serabyn, E., Guesten, R., \& Wilson, T.~L.\ 1987, \aap, 177, L57 
\bibitem[M{\"u}ller et al.(2001)]{muller2001} M{\"u}ller, H.~S.~P., Thorwirth, S., Roth, D.~A., \& Winnewisser, G.\ 2001, \aap, 370, L49 
\bibitem[M{\"u}ller et al.(2005)]{muller2005} M{\"u}ller, H.~S.~P., Schl{\"o}der, F., Stutzki, J., \& Winnewisser, G.\ 2005, Journal of Molecular Structure, 742, 215 
\bibitem[Mundy et al.(1992)]{mundy1992} Mundy, L.~G., Wootten, A., Wilking, B.~A., Blake, G.~A., \& Sargent, A.~I.\ 1992, \apj, 385, 306 
\bibitem[Niedenhoff(1995)]{niedenhoff1995}M. Niedenhoff, K. M. T. Yamada, S. P. Belov, and G. Winnewisser, 1995, J. Mol. Spectrosc. 174, 151
\bibitem[Nizam et al.(1988)]{nizam1988theoretical} Nizam, M., Bouteiller, Y., Silvi, B., et al.\ 1988, Journal of Physics C Solid State Physics, 21, 5351 
\bibitem[Noble et al.(2015)]{noble2015} Noble, J.~A., Theule, P., Congiu, E., et al.\ 2015, \aap, 576, A91 
\bibitem[Ott(2010)]{ott2010}Ott, S., 2010, Astronomical Data Analysis Software and Systems XIX. Proceedings of a conference held October 4-8, 2009 in Sapporo, Japan. Edited by Yoshihiko Mizumoto, Koh-Ichiro Morita, and Masatoshi Ohishi. ASP Conference Series, Vol. 434, 139
\bibitem[Quan et al.(2010)]{quan2010} Quan, D., Herbst, E., Osamura, Y., \& Roueff, E.\ 2010, \apj, 725, 2101 
\bibitem[Qu{\'e}nard et al.(2018)]{quenard2018} Qu{\'e}nard, D., Jim{\'e}nez-Serra, I., Viti, S., Holdship, J., \& Coutens, A.\ 2018, \mnras, 474, 2796 
\bibitem[Roelfsema et al.(2012)]{roelfsema2012} Roelfsema, P.~R., Helmich, F.~P., Teyssier, D., et al.\ 2012, \aap, 537, A17 
\bibitem[Rodr{\'{\i}}guez-Fern{\'a}ndez et al.(2010)]{rodriguez-fernandez2010} Rodr{\'{\i}}guez-Fern{\'a}ndez, N.~J., Tafalla, M., Gueth, F., \& Bachiller, R.\ 2010, \aap, 516, A98 
\bibitem[Ruaud et al.(2015)]{ruaud2015} Ruaud, M., Loison, J.~C., Hickson, K.~M., et al.\ 2015, \mnras, 447, 4004 
\bibitem[Ruaud et al.(2016)]{ruaud2016} Ruaud, M., Wakelam, V., \& Hersant, F.\ 2016, \mnras, 459, 3756 
\bibitem[Sahnoun et al.(2018)]{sahnoun2018} Sahnoun E., Wiesenfeld L., Hammami K., Jaidane N., \ 2018, The Journal of Physical Chemistry A, 122, 3004  
\bibitem[Snyder \& Buhl(1972)]{snyder1972} Snyder, L.~E., \& Buhl, D.\ 1972, \apj, 177, 619 
\bibitem[Tideswell et al.(2010)]{tideswell2010} Tideswell, D.~M., Fuller, G.~A., Millar, T.~J., \& Markwick, A.~J.\ 2010, \aap, 510, A85
\bibitem[Turner(1991)]{turner1991} Turner, B.~E.\ 1991, \apjs, 76, 617 
\bibitem[Turner(2000)]{turner2000} Turner, B.~E.\ 2000, \apj, 542, 837 
\bibitem[van der Tak et al.(2007)]{vandertak2007} van der Tak, F.~F.~S., Black, J.~H., Sch{\"o}ier, F.~L., Jansen, D.~J., \& van Dishoeck, E.~F.\ 2007, \aap, 468, 627  
\bibitem[van Dishoeck et al.(1995)]{vandishoeck1995} van Dishoeck, E.~F., Blake, G.~A., Jansen, D.~J., \& Groesbeck, T.~D.\ 1995, \apj, 447, 760 
\bibitem[Wakelam et al.(2015)]{wakelam2015} Wakelam, V., Loison, J.-C., Herbst, E., et al.\ 2015, \apjs, 217, 20 
\bibitem[Werner et al.(2012)]{werner2012} Werner H., et al.,\ 2012, University College Cardiff Consultants Ltd.: Wales, UK
\bibitem[Wootten(1989)]{wootten1989} Wootten, A.\ 1989, \apj, 337, 858 
\bibitem[\protect\citeauthoryear{Yamada}{1977}]{Yamada1977} Yamada K., 1977, JMoSp, 68, 423   

\bibitem[\protect\citeauthoryear{Author}{2012}]{Author2012}
\end{thebibliography}




\newpage


\appendix

\section{Observed HNCO transitions parameters}

In this section, we give the parameters of all the HNCO observed lines. 

\begin{table}
\caption{Detected HNCO transitions in all observations and their main parameters. The negative sign for some transitions detected with Herschel/HIFI indicate that the line could be in absorption (after subtracting a baseline), although this is within the noise. Columns are frequency, quantum numbers, upper level energy, Einstein A value, value and error of the velocity-integrated main-beam brightness temperature. Telescope used :  [87,\,245]\,GHz : IRAM-30m,  [260,\,320] and [370,\,470]\,GHz : APEX, [328,\,360]\,GHz : JCMT and [>\,490]\,GHz : Herschel-HIFI.}
\label{tab:example_table}
\begin{tabular}{rcrcc} 
\hline
Frequency 		& Transition 					& E$_\text{up}$& A$_{ij}$ 			&	$\int T_\text{mb}dv$	\\
(MHz) 		& ($J_{K_a\,K_c}$)				& (K)		& (s$^{-1}$) 			&	(K km s$^{-1}$)	\\
\hline		
87597.330	&	$4_{1 \, 4} - 3_{1 \, 3}$			&	53.78		&	8.04$\times 10^{-6}$	&	0.29($\pm0.05$)\\
87898.425	&	$4_{2 \, 3} - 3_{2 \, 2}$			&	180.83	&	6.28$\times 10^{-6}$	&	0.18($\pm0.04$)\\
87898.628	&	$4_{ 2 \, 2} - 3_{ 2 \, 1}$			&	180.83	&	6.28$\times 10^{-6}$	&	0.18($\pm0.04$)\\
87925.237	&	$4_{0 \, 4} - 3_{0 \, 3}$			&	10.54		&	8.78$\times 10^{-6}$	&	0.82($\pm0.11$)\\
88239.020	&	$4_{1 \, 3} - 3_{1 \, 2}$			&	53.86		&	8.22$\times 10^{-6}$	&	0.25($\pm0.13$)\\
109495.996	&	$5_{1 \, 5} - 4_{1 \, 4}$		&	59.04		&	1.65$\times 10^{-5}$	&	0.46($\pm0.07$)\\
109872.337	&	$5_{2 \, 4} - 4_{2 \, 3}$		&	186.10	&	1.41$\times 10^{-5}$	&	0.39($\pm0.05$)\\
109872.765	&	$5_{ 2 \, 3} - 4_{ 2 \, 2}$		&	186.10	&	1.41$\times 10^{-5}$	&	0.39($\pm0.05$)\\
109905.749	&	$5_{0 \, 5} - 4_{0 \, 4}$		&	15.82		&	1.75$\times 10^{-5}$&	0.65($\pm0.04$)\\
110298.089	&	$5_{1 \, 4} - 4_{1 \, 3}$		&	59.15		&	1.68$\times 10^{-5}$	&	0.38($\pm0.09$)\\
131394.230	&	$6_{1 \, 6} - 5_{1 \, 5}$		&	65.34		&	2.92$\times 10^{-5}$	&	0.29($\pm0.05$)\\
131845.890	&	$6_{2 \, 5} - 5_{2 \, 4}$		&	192.43	&	2.61$\times 10^{-5}$	&	0.34($\pm0.07$)\\
131846.600	&	$6_{ 2 \, 4} - 5 _{2 \, 3}$		&	192.43	&	2.61$\times 10^{-5}$	&	0.34($\pm0.07$)\\
131885.734	&	$6_{0 \, 6} - 5_{0 \, 5}$		&	22.15		&	3.08$\times 10^{-5}$	&	1.12($\pm0.12$)\\
132356.701	&	$6_{1 \, 5} - 5_{1 \, 4}$		&	65.50		&	2.99$\times 10^{-5}$	&	0.30($\pm0.07$)\\
153291.935	&	$7_{1 \, 7} - 6_{1 \, 6}$		&	72.70		&	4.73$\times 10^{-5}$	&	0.55($\pm0.16$)\\
153818.880	&	$7 _{2 \, 6} - 6_{ 2 \, 5}$		&	199.81	&	4.33$\times 10^{-5}$	&	0.10($\pm0.10$)\\
153820.016	&	$7_{ 2 \, 5} -  6_{ 2 \, 4}$		&	199.81	&	4.33$\times 10^{-5}$	&	0.10($\pm0.10$)\\
153865.086	&	$7_{0 \, 7} - 6_{0 \, 6}$		&	29.53		&	4.94$\times 10^{-5}$	&	1.27($\pm0.20$)\\
154414.765	&	$7_{1 \, 6} - 6_{1 \, 5}$		&	72.91		&	4.84$\times 10^{-5}$	&	0.68($\pm0.24$)\\
197085.416	&	$9_{1 \, 9} - 8_{1 \, 8}$		&	90.57		&	1.03$\times 10^{-4}$	&	0.80($\pm0.13$)\\
197762.939	&	$9 _{2 \, 8} - 8_{ 2 \, 7}$		&	217.74	&	9.66$\times 10^{-5}$	&	0.89($\pm0.43$)\\
197765.372	&	$9 _{2 \, 7} - 8_{ 2 \, 6}$		&	217.74	&	9.66$\times 10^{-5}$	&	0.89($\pm0.43$)\\
197821.461	&	$9_{0 \, 9} - 8_{0 \, 8}$		&	47.47		&	1.07$\times 10^{-4}$	&	2.23($\pm0.32$)\\
198528.881	&	$9_{1 \, 8} - 8_{1 \, 7}$		&	90.91		&	1.05$\times 10^{-4}$	&	0.18($\pm0.03$)\\
218981.009	&	$10_{1 \, 10} - 9_{1 \, 9}$		&	101.07	&	1.42$\times 10^{-4}$	&	0.78($\pm0.07$)\\
219733.850	&	$10_{ 2 \, 9} - 9_{ 2 \, 8}$		&	228.29	&	1.35$\times 10^{-4}$	&	1.33($\pm0.33$)\\
219737.193	&	$10_{ 2 \, 8} - 9_{ 2 \, 7}$		&	228.29	&	1.35$\times 10^{-4}$	&	1.33($\pm0.33$)\\
219798.274	&	$10_{0 \, 10} - 9_{0 \, 9}$		&	58.01		&	1.47$\times 10^{-4}$	&	2.32($\pm0.10$)\\
220584.751	&	$10_{1 \, 9} - 9_{1 \, 8}$		&	101.50	&	1.45$\times 10^{-4}$	&	0.74($\pm0.14$)\\
240875.727	&	$11_{1 \, 11} - 10_{1 \, 10}$	&	112.63	&	1.90$\times 10^{-4}$	&	1.53($\pm0.20$)\\
241703.853	&	$11_{ 2 \, 10} - 10_{ 2 \, 9}$	&	239.89	&	1.81$\times 10^{-4}$&	0.64($\pm0.20$)\\
241708.312	&	$11_{ 2 \, 9} - 10 _{2 \, 8}$		&	239.89	&	1.81$\times 10^{-4}$&	0.64($\pm0.20$)\\
241774.032	&	$11_{0 \, 11} - 10_{0 \, 10}$	&	69.62		&	1.96$\times 10^{-4}$	&	2.67($\pm0.23$)\\
242639.704	&	$11_{1 \, 10} - 10_{1 \, 9}$		&	113.14	&	1.95$\times 10^{-4}$	&	2.25($\pm0.31$)\\
262769.477	&	$12_{1 \, 12} - 11_{1 \, 11}$	&	125.25	&	2.48$\times 10^{-4}$	&	0.75($\pm0.18$)\\
263672.912	&	$12_{ 2 \, 11} - 11_{ 2 \, 10}$	&	252.54	&	2.37$\times 10^{-4}$	&	1.32($\pm0.51$)\\
263678.709	&	$12_{ 2 \, 10} - 11_{ 2 \, 9}$	&	252.54	&	2.37$\times 10^{-4}$	&	1.32($\pm0.51$)\\
263748.625	&	$12_{0 \, 12} - 11_{0 \, 11}$	&	82.28		&	2.56$\times 10^{-4}$	&	0.70($\pm0.07$)\\
264693.655	&	$12_{1 \, 11} - 11_{1 \, 10}$	&	125.85	&	2.54$\times 10^{-4}$	&	0.21($\pm0.06$)\\
284662.172	&	$13_{1 \, 13} - 12_{1 \, 12}$	&	138.91	&	3.17$\times 10^{-4}$	&	0.38($\pm0.08$)\\
285721.951	&	$13_{0 \, 13} - 12_{0 \, 12}$	&	95.99		&	3.26$\times 10^{-4}$	&	0.76($\pm0.09$)\\
286746.514	&	$13_{1 \, 12} - 12_{1 \, 11}$	&	139.61	&	3.24$\times 10^{-4}$	&	0.29($\pm0.11$)\\
\hline
\end{tabular}
\end{table}

\begin{table}
\contcaption{}
\label{tab:continued}
\begin{tabular}{rcrcc} 
\hline
Frequency 		& Transition 					& E$_\text{up}$& A$_{ij}$ 			&	$\int T_\text{mb}dv$	\\
(MHz) 		& ($J_{K_a\,K_c}$)				& (K)		& (s$^{-1}$) 			&	(K km s$^{-1}$)	\\
\hline		
306553.733	&	$14_{1 \, 14} - 13_{1 \, 13}$	&	153.62	&	3.97$\times 10^{-4}$	&	0.42($\pm0.09$)\\
307693.905	&	$14_{0 \, 14} - 13_{0 \, 13}$	&	110.76	&	4.09$\times 10^{-4}$	&	1.08($\pm0.20$)\\
308798.184	&	$14_{1 \, 13} - 13_{1 \, 12}$	&	154.43	&	4.06$\times 10^{-4}$	&	0.72($\pm0.15$)\\
328444.054	&	$15_{1 \, 15} - 14_{1 \, 14}$	&	169.38	&	4.90$\times 10^{-4}$	&	0.68($\pm0.21$)\\
329664.367	&	$15_{0 \, 15} - 14_{0 \, 14}$	&	126.58	&	5.04$\times 10^{-4}$	&	1.79($\pm0.19$)\\
330848.569	&	$15_{1 \, 14} - 14_{1 \, 13}$	&	170.31	&	5.01$\times 10^{-4}$	&	1.63($\pm0.40$)\\
350333.059	&	$16_{1 \, 16} - 15_{1 \, 15}$	&	186.20	&	5.97$\times 10^{-4}$	&	1.01($\pm0.06$)\\
351633.257	&	$16_{0 \, 16} - 15_{0 \, 15}$	&	143.45	&	6.13$\times 10^{-4}$	&	1.63($\pm0.12$)\\
352897.581	&	$16_{1 \, 15} - 15_{1 \, 14}$	&	187.24	&	6.10$\times 10^{-4}$	&	0.88($\pm0.09$)\\
372220.660	&	$17_{1 \, 17} - 16_{1 \, 16}$	&	204.06	&	7.17$\times 10^{-4}$	&	0.56($\pm0.10$)\\
373600.448	&	$17_{0 \, 17} - 16_{0 \, 16}$	&	161.38	&	7.36$\times 10^{-4}$	&	1.64($\pm0.24$)\\
417529.351	&	$19_{0 \, 19} - 18_{0 \, 18}$	&	200.41	&	1.03$\times 10^{-3}$	&	1.37($\pm0.12$)\\
419035.477	&	$19_{ 1 \, 18} - 18_{1 \, 17}$	&	244.41	&	1.03$\times 10^{-3}$&	1.08($\pm0.15$)\\
461450.213	&	$21_{ 0 \, 21} - 20_{ 0 \, 20}$	&	243.65	&	1.39$\times 10^{-3}$&	1.54($\pm0.22$)\\
493675.710	&	$17_{1 \, 17} - 18_{0 \, 18}$	&	204.06	&	6.30$\times 10^{-4}$	&	0.40($\pm0.06$)\\
517020.943	&	$16_{1 \, 16} - 17_{0 \, 17}$	&	186.20	&	7.21$\times 10^{-4}$	&	-0.01($\pm0.001$)\\
540288.323	&	$15_{1 \, 15} - 16_{0 \, 16}$	&	169.38	&	8.21$\times 10^{-4}$	&	-0.01($\pm0.002$)\\
563477.534	&	$14_{1 \, 14} - 15_{0 \, 15}$	&	153.62	&	9.29$\times 10^{-4}$	&	0.05($\pm0.02$)\\
586588.183	&	$13_{1 \, 13} - 14_{0 \, 14}$	&	138.91	&	1.04$\times 10^{-3}$	&	-0.06($\pm0.01$)\\
609619.927	&	$12_{1 \, 12} - 13_{0 \, 13}$	&	125.25	&	1.17$\times 10^{-3}$	&	0.05($\pm0.01$)\\
632572.365	&	$11_{1 \, 11} - 12_{0 \, 12}$	&	112.63	&	1.30$\times 10^{-3}$	&	0.16($\pm0.04$)\\
655445.310	&	$10_{1 \, 10} - 11_{0 \, 11}$	&	101.07	&	1.44$\times 10^{-3}$	&	-0.01($\pm0.004	$)\\
678238.267	&	$9_{1 \, 9} - 10_{0 \, 10}$		&	90.57		&	1.58$\times 10^{-3}$	&	-0.004($\pm0.01$)\\
700951.182	&	$8_{1 \, 8} - 9_{0 \, 9}$		&	81.11		&	1.74$\times 10^{-3}$	&	-0.03($\pm0.01$)\\
723583.635	&	$7_{1 \, 7} - 8_{0 \, 8}$		&	72.70		&	1.89$\times 10^{-3}$	&	0.01($\pm0.01$)\\
746135.382	&	$6_{1 \, 6} - 7_{0 \, 7}$		&	65.34		&	2.05$\times 10^{-3}$	&	-0.04	($\pm0.01$)\\
768606.238	&	$5_{1 \, 5} - 6_{0 \, 6}$		&	59.04		&	2.21$\times 10^{-3}$	&	-0.01($\pm0.01$)\\
790995.975	&	$4_{1 \, 4} - 5_{0 \, 5}$		&	53.78		&	2.35$\times 10^{-3}$	&	-0.08($\pm0.03$)\\
813304.397	&	$3_{1 \, 3} - 4_{0 \, 4}$		&	49.58		&	2.47$\times 10^{-3}$	&	-0.06($\pm0.01$)\\
835531.328	&	$2_{1 \, 2} - 3_{0 \, 3}$		&	46.42		&	2.49$\times 10^{-3}$	&	-0.03($\pm0.003$)\\
857676.609	&	$1_{1 \, 1} - 2_{0 \, 2}$		&	44.32		&	2.25$\times 10^{-3}$	&	-0.04($\pm0.01$)\\	
901800.081	&	$1_{1 \, 0} - 1_{0 \, 1}$		&	44.33		&	7.84$\times 10^{-3}$	&	-0.09($\pm0.01$)\\
901956.917	&	$2_{1 \, 1} - 2_{0 \, 2}$		&	46.45		&	7.84$\times 10^{-3}$	&	-0.03($\pm0.01$)\\
902192.207	&	$3_{1 \, 2} - 3_{0 \, 3}$		&	49.62		&	7.84$\times 10^{-3}$	&	-0.08($\pm0.01$)\\
902505.994	&	$4_{1 \, 3} - 4_{0 \, 4}$		&	53.86		&	7.85$\times 10^{-3}$	&	-0.07($\pm0.01$)\\
902898.335	&	$5_{1 \, 4} - 5_{0 \, 5}$		&	59.15		&	7.86$\times 10^{-3}$	&	-0.05($\pm0.01$)\\
903369.302	&	$6_{1 \, 5} - 6_{0 \, 6}$		&	65.50		&	7.87$\times 10^{-3}$	&	-0.09($\pm0.01$)\\
903918.982	&	$7_{1 \, 6} - 7_{0 \, 7}$		&	72.91		&	7.88$\times 10^{-3}$	&	0.04($\pm0.01$)\\
904547.473	&	$8_{1 \, 7} - 8_{0 \, 8}$		&	81.38		&	7.89$\times 10^{-3}$	&	-0.01($\pm0.004	$)\\
905254.892	&	$9_{1 \, 8} - 9_{0 \, 9}$		&	90.91		&	7.91$\times 10^{-3}$	&	0.04($\pm0.01$)\\
906041.367	&	$10_{1 \, 9} - 10_{0 \, 10}$	&	101.50	&	7.93$\times 10^{-3}$	&	-0.03($\pm0.002	$)\\
906907.042	&	$11_{1 \, 10} - 11_{0 \, 11}$	&	113.14	&	7.95$\times 10^{-3}$	&	0.06($\pm0.01$)\\
907852.072	&	$12_{1 \, 11} - 12_{0 \, 12}$	&	125.85	&	7.97$\times 10^{-3}$	&	-0.004($\pm0.003$) \\
908876.631	&	$13_{1 \, 12} - 13_{0 \, 13}$	&	139.61	&	7.99$\times 10^{-3}$	&	0.04($\pm0.004$)    \\
909980.904	&	$14_{1 \, 13} - 14_{0 \, 14}$	&	154.43	&	8.01$\times 10^{-3}$	&	-0.05($\pm0.010$)   \\
911165.091	&	$15_ {1 \, 14} - 15_{ 0 \, 15}$	&	170.31	&	8.04$\times 10^{-3}$ &	-0.007($\pm0.002$) \\
912429.407	&	$16_{1 \, 15} - 16_{0 \, 16}$	&	187.24	&	8.07$\times 10^{-3}$	&	-0.02($\pm0.003$)   \\
913774.079	&	$17_{1 \, 16} - 17_{0 \, 17}$	&	205.24	&	8.10$\times 10^{-3}$	&	0.01($\pm0.003$)    \\
923621.180	&	$1_{1 \, 1} - 0_{0 \, 0}$		&	44.32		&	5.61$\times 10^{-3}$	&	-0.07($\pm0.01$)     \\
945438.685	&	$2_{1 \, 2} - 1_{0 \, 1}$		&	46.42		&	5.42$\times 10^{-3}$	&	-0.05($\pm0.003$)   \\
967173.925	&	$3_{1 \, 3} - 2_{0 \, 2}$		&	49.58		&	5.52$\times 10^{-3}$	&	-0.06($\pm0.01$)     \\
988826.956	&	$4_{1 \, 4} - 3_{0 \, 3}$		&	53.78		&	5.74$\times 10^{-3}$	&	-0.05($\pm0.01$)     \\
1010397.720	&	$5_{1 \, 5} - 4_{0 \, 4}$		&	59.04		&	6.01$\times 10^{-3}$	&	-0.02($\pm0.007$)   \\
1031886.200	&	$6_{1 \, 6} - 5_{0 \, 5}$		&	65.34		&	6.32$\times 10^{-3}$	&	-0.06($\pm0.001$)   \\
1053292.425	&	$7_{1 \, 7} - 6_{0 \, 6}$		&	72.70		&	6.66$\times 10^{-3}$	&	-0.08($\pm0.01$)     \\
1074616.365	&	$8_{1 \, 8} - 7_{0 \, 7}$		&	81.11		&	7.03$\times 10^{-3}$	&	-0.17($\pm0.01$)     \\
1095858.058	&	$9_{1 \, 9} - 8_{0 \, 8}$		&	90.57		&	7.41$\times 10^{-3}$	&	-0.07($\pm0.01$)      \\
1117017.603	&	$10_{1 \, 10} - 9_{0 \, 9}$		&	101.07	&	7.81$\times 10^{-3}$	&	-0.14($\pm0.01$)      \\
1138095.052	&	$11_{1 \, 11} - 10_{0 \, 10}$	&	112.63	&	8.23$\times 10^{-3}$	&	0.06($\pm0.03$)       \\
1159090.494	&	$12_{ 1 \, 12} - 11_{ 0 \, 11}$	&	125.25	&	8.67$\times 10^{-3}$&	-0.02($\pm-0.003$)  \\
1180004.039	&	$13_{1 \, 13} - 12_{0 \, 12}$	&	138.91	&	9.13$\times 10^{-3}$	&	0.08($\pm0.02$)       \\
\hline
\end{tabular}
\end{table}

\newpage

\section{Observed and modelled HNCO transitions}
\label{Obs-mod HNCO}

In this section, we show all the HNCO observed line profiles individually (in black) and the predicted emission/absorption profiles from our model described in Section 3 (in red). We have separated the observations by the $K_a$ quantum number. Some transitions (indicated in each figure caption) could not be modelled since their collisional rate coefficients were not computed in the quantum chemical calculations of \citet{sahnoun2018}.  

\begin{figure*}
\centering
\setlength\tabcolsep{3.7pt}
\caption{HNCO $K_a=0$ transitions. The collision rate coefficient for the $21_{0 \, 21}-20_{0 \,20}$ transition was lacking and therefore it could not be modelled.}
\begin{tabular}{c c c }
\includegraphics[width=0.315\textwidth, trim= 0 0 0 0, clip]{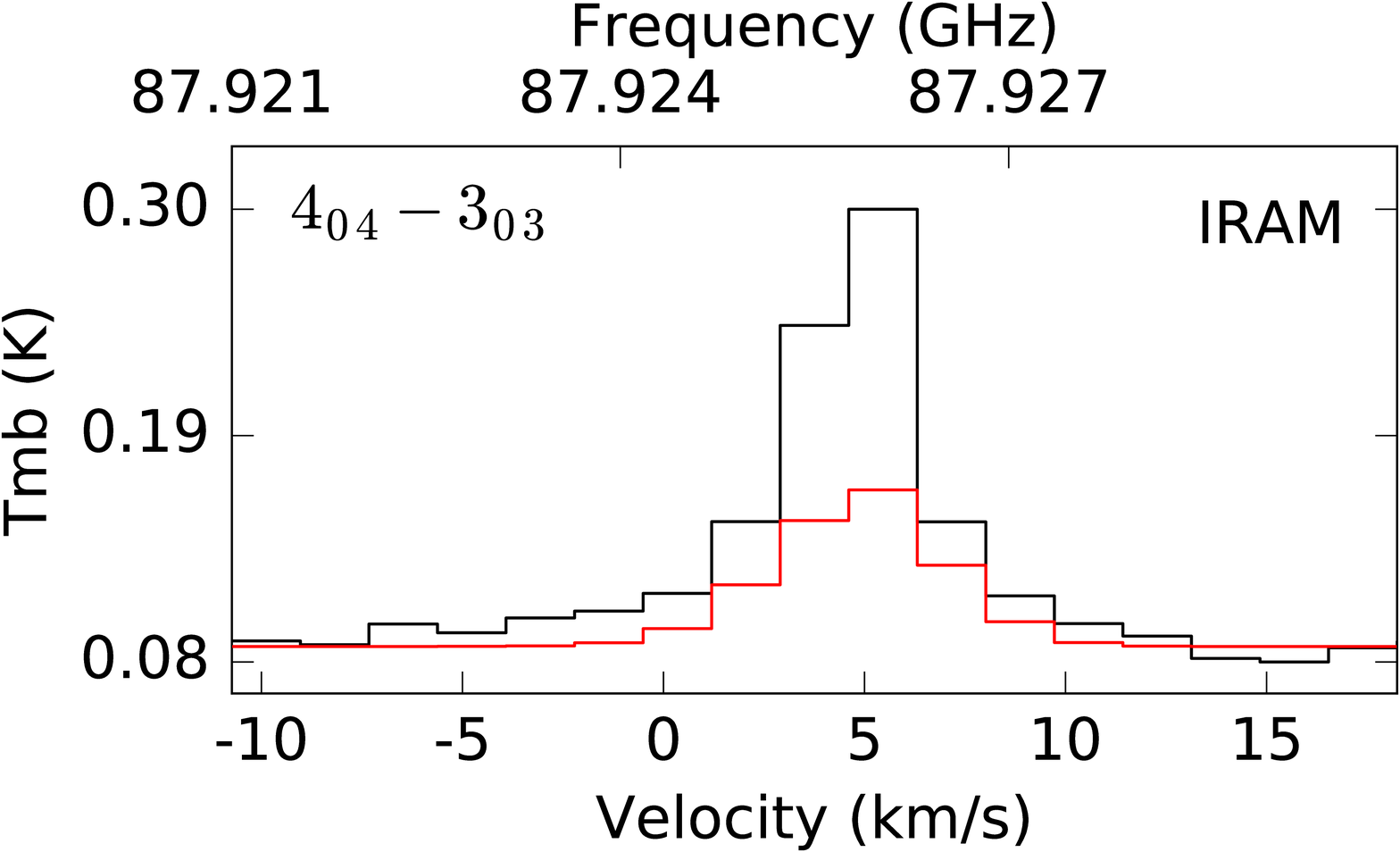} &\includegraphics[width=0.315\textwidth,trim = 0 0 0 0,clip]{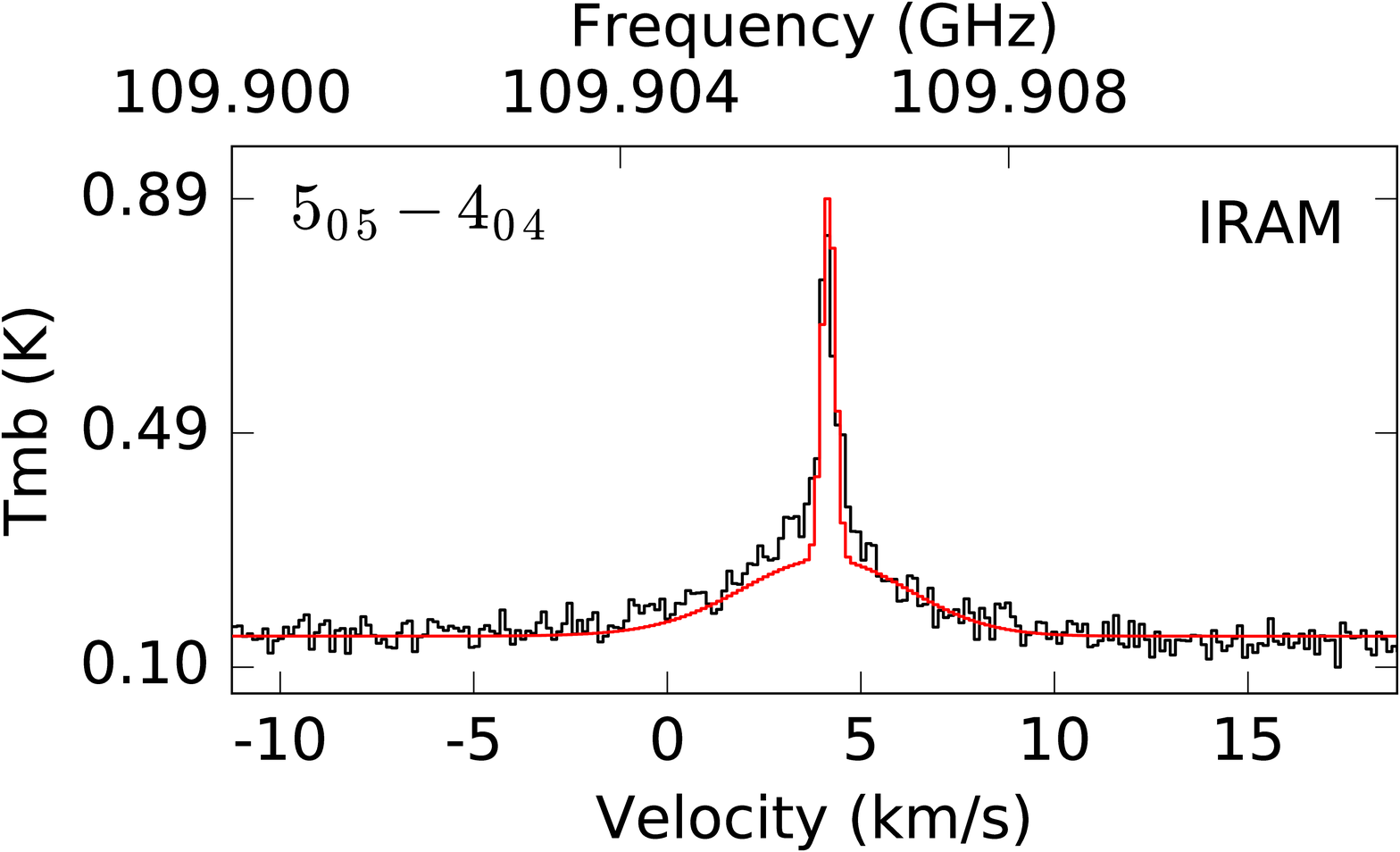}  &\includegraphics[width=0.315\textwidth,trim = 0 0 0 0,clip]{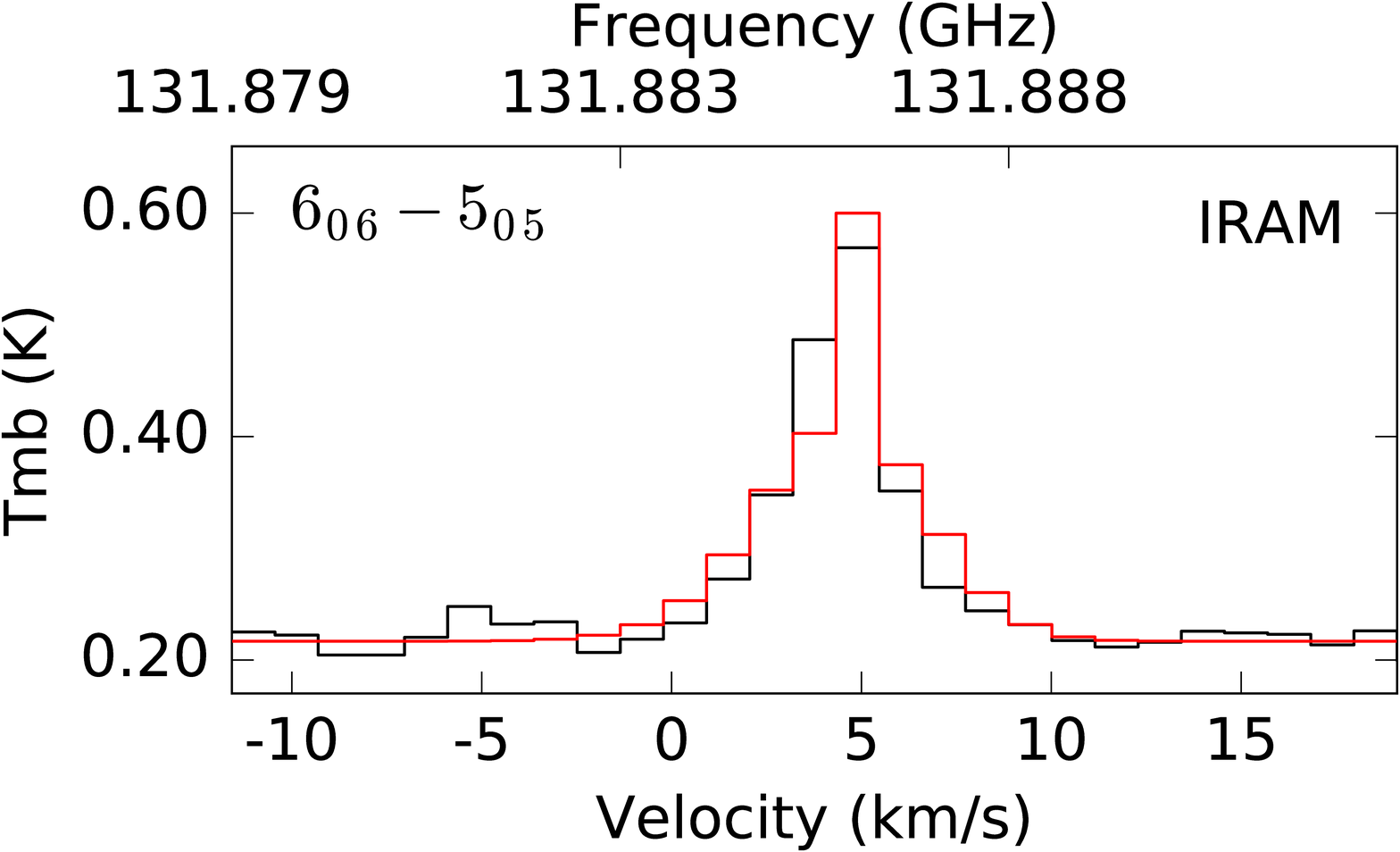} \\
\includegraphics[width=0.315\textwidth,trim = 0 0 0 0,clip]{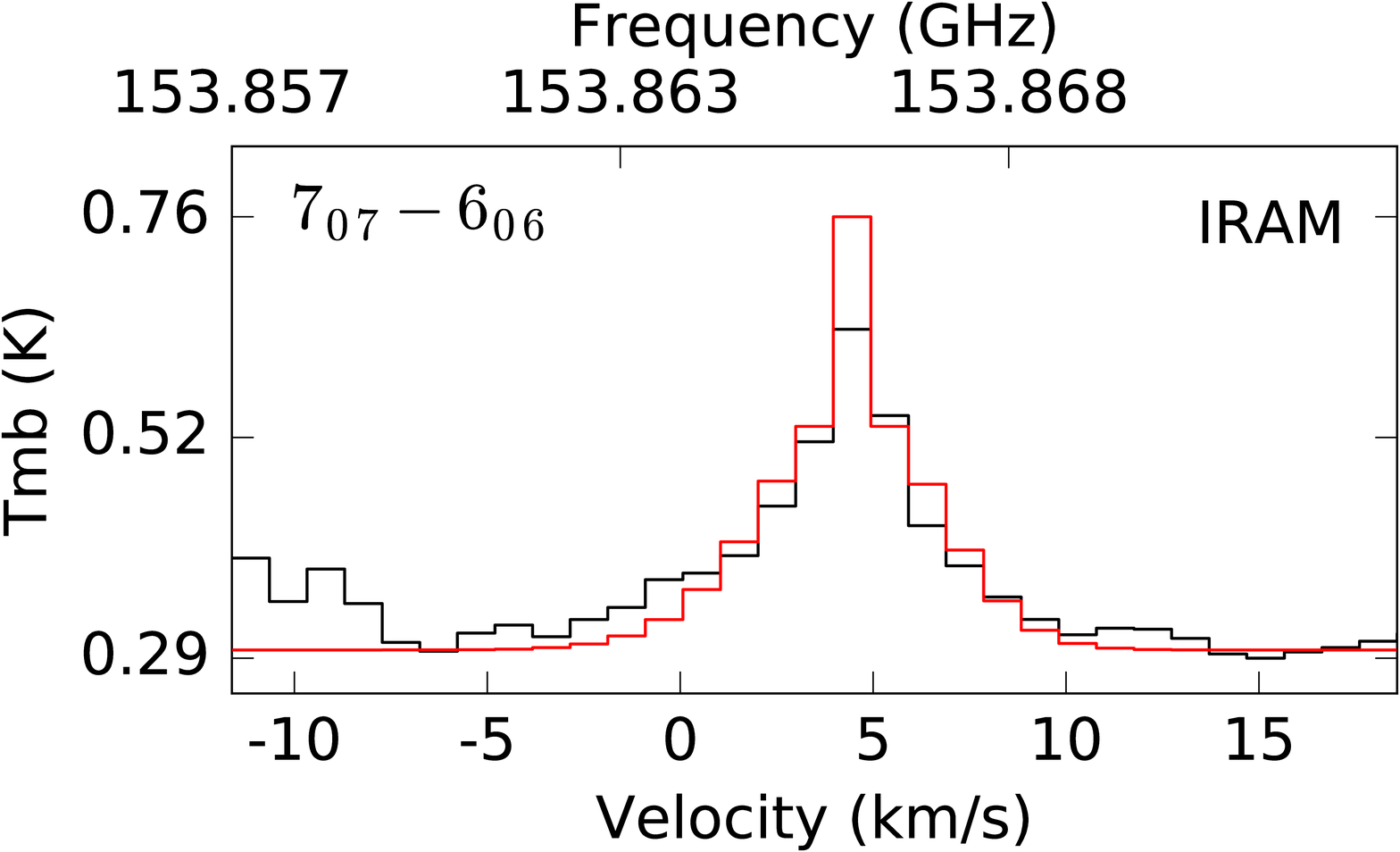}&\includegraphics[width=0.315\textwidth,trim = 0 0 0 0,clip]{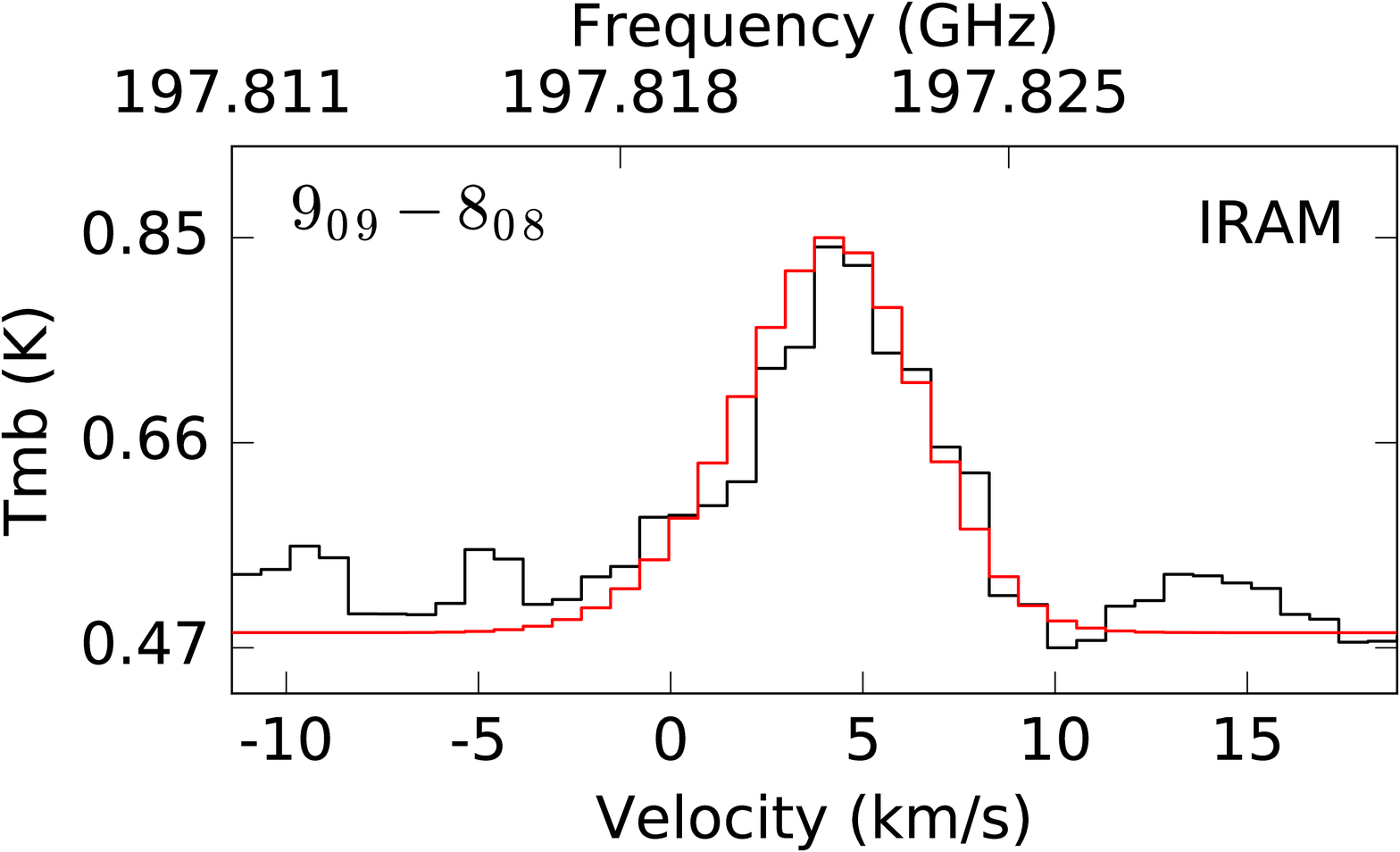}& \includegraphics[width=0.315\textwidth, trim= 0 0 0 0, clip]{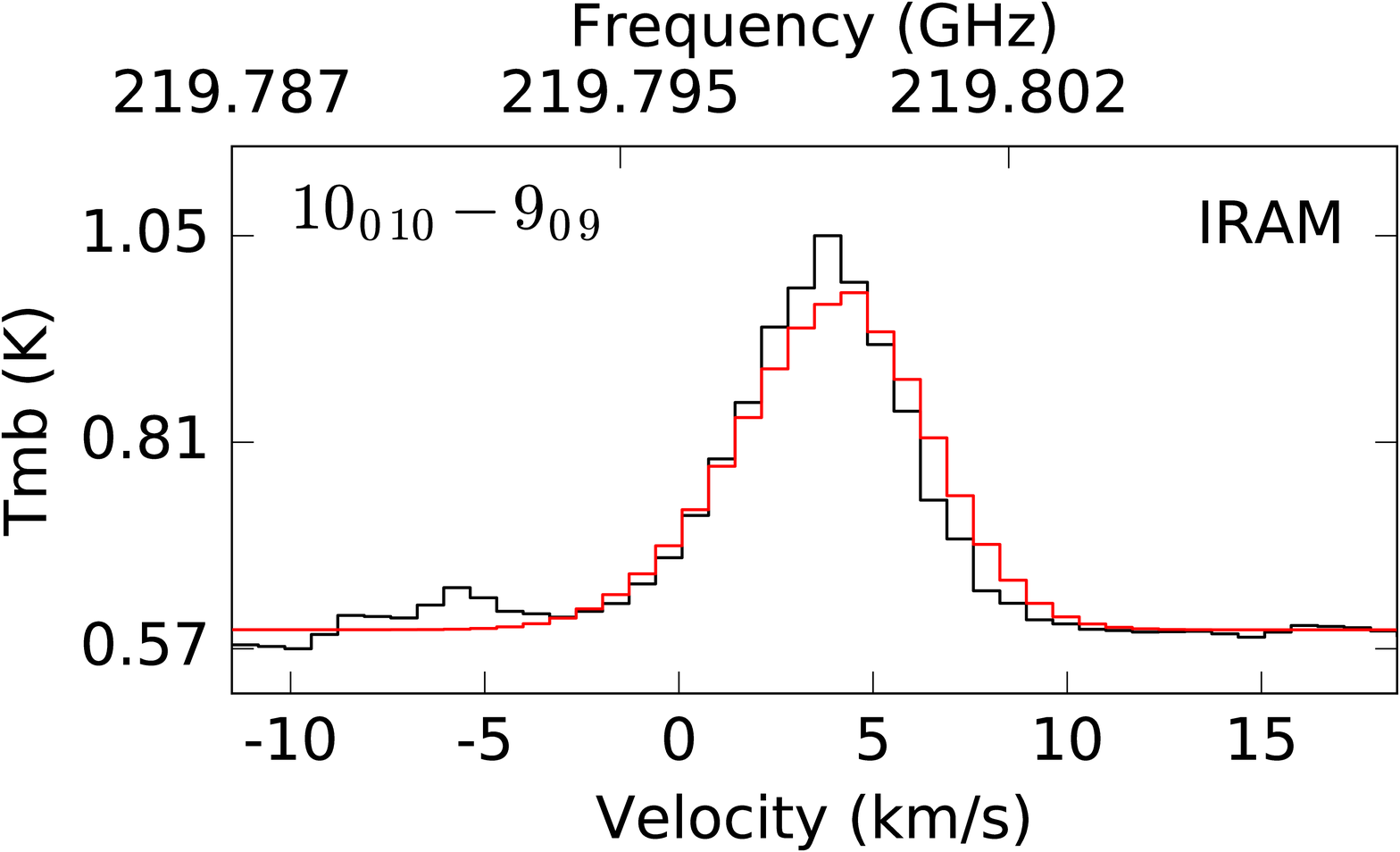} \\
\includegraphics[width=0.315\textwidth,trim = 0 0 0 0,clip]{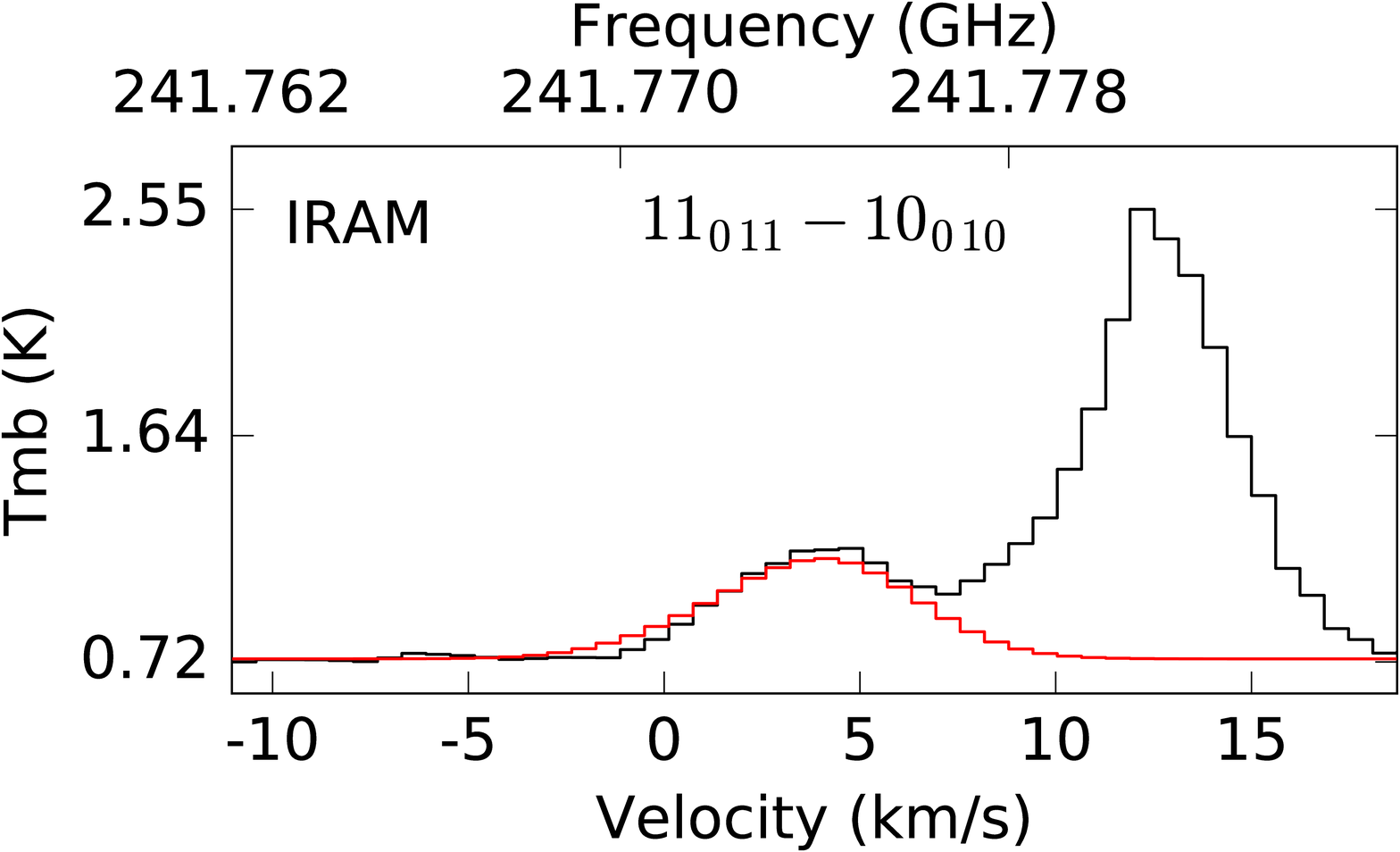}  &\includegraphics[width=0.315\textwidth,trim = 0 0 0 0,clip]{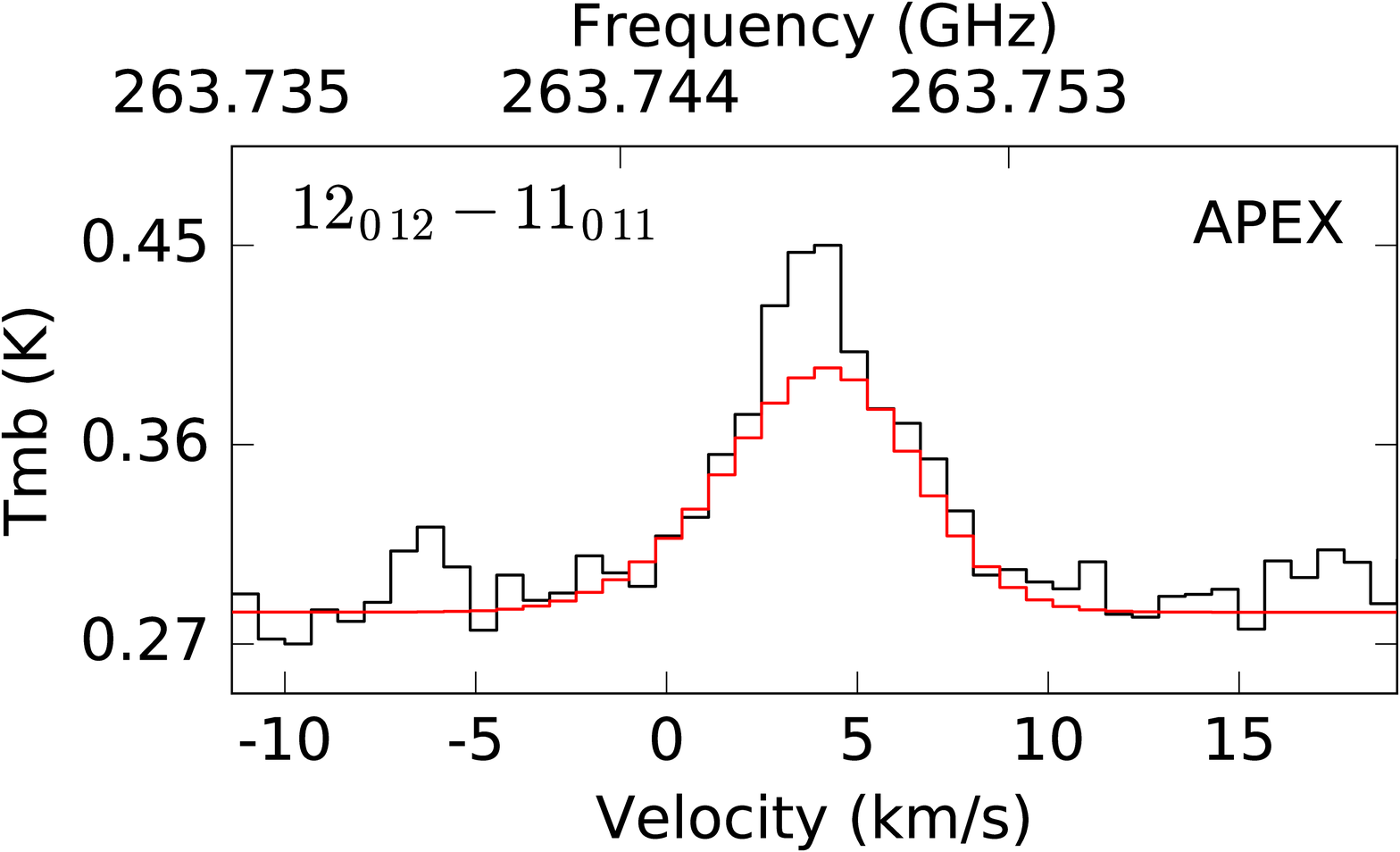} &\includegraphics[width=0.315\textwidth,trim = 0 0 0 0,clip]{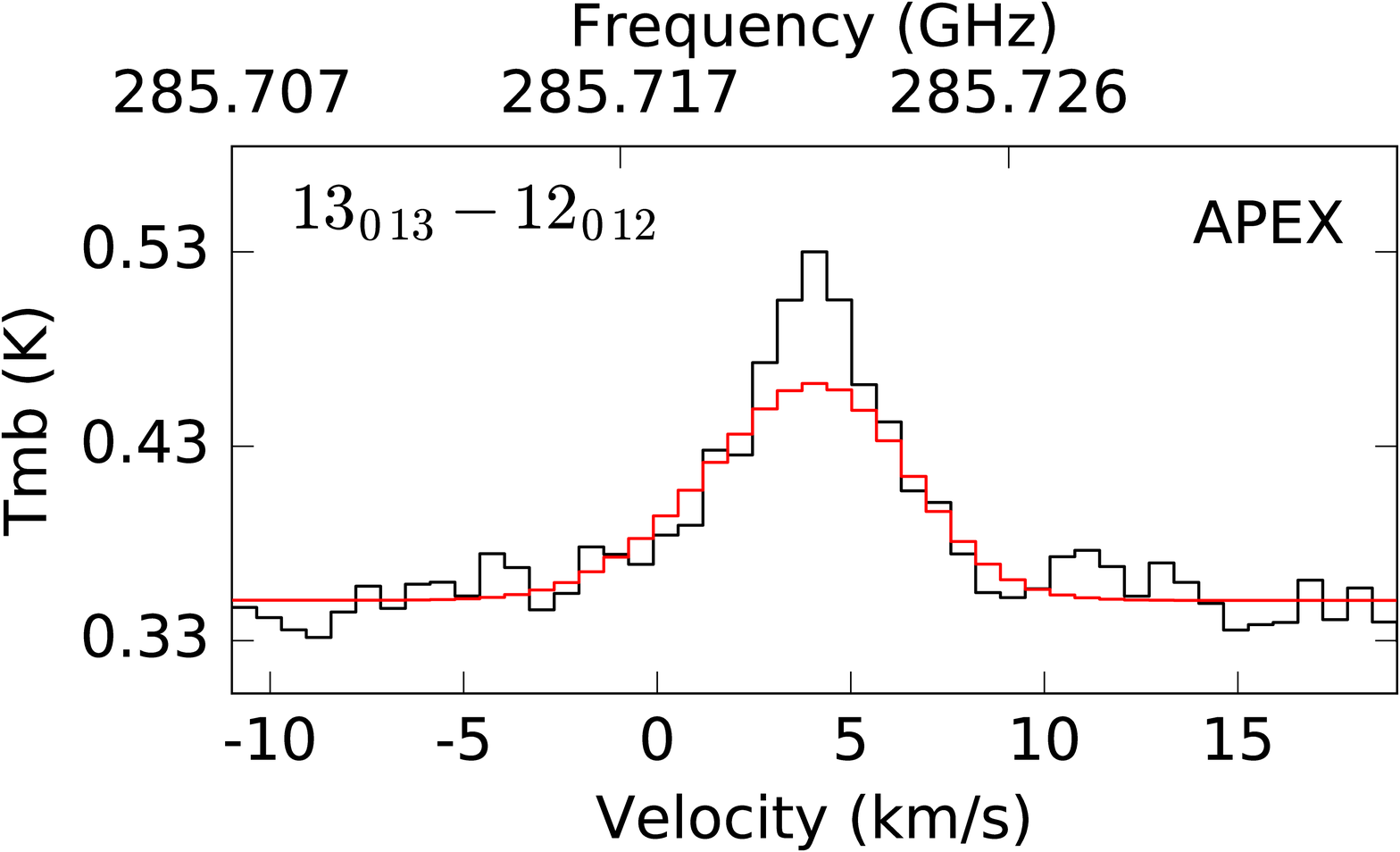}\\
\includegraphics[width=0.315\textwidth,trim = 0 0 0 0,clip]{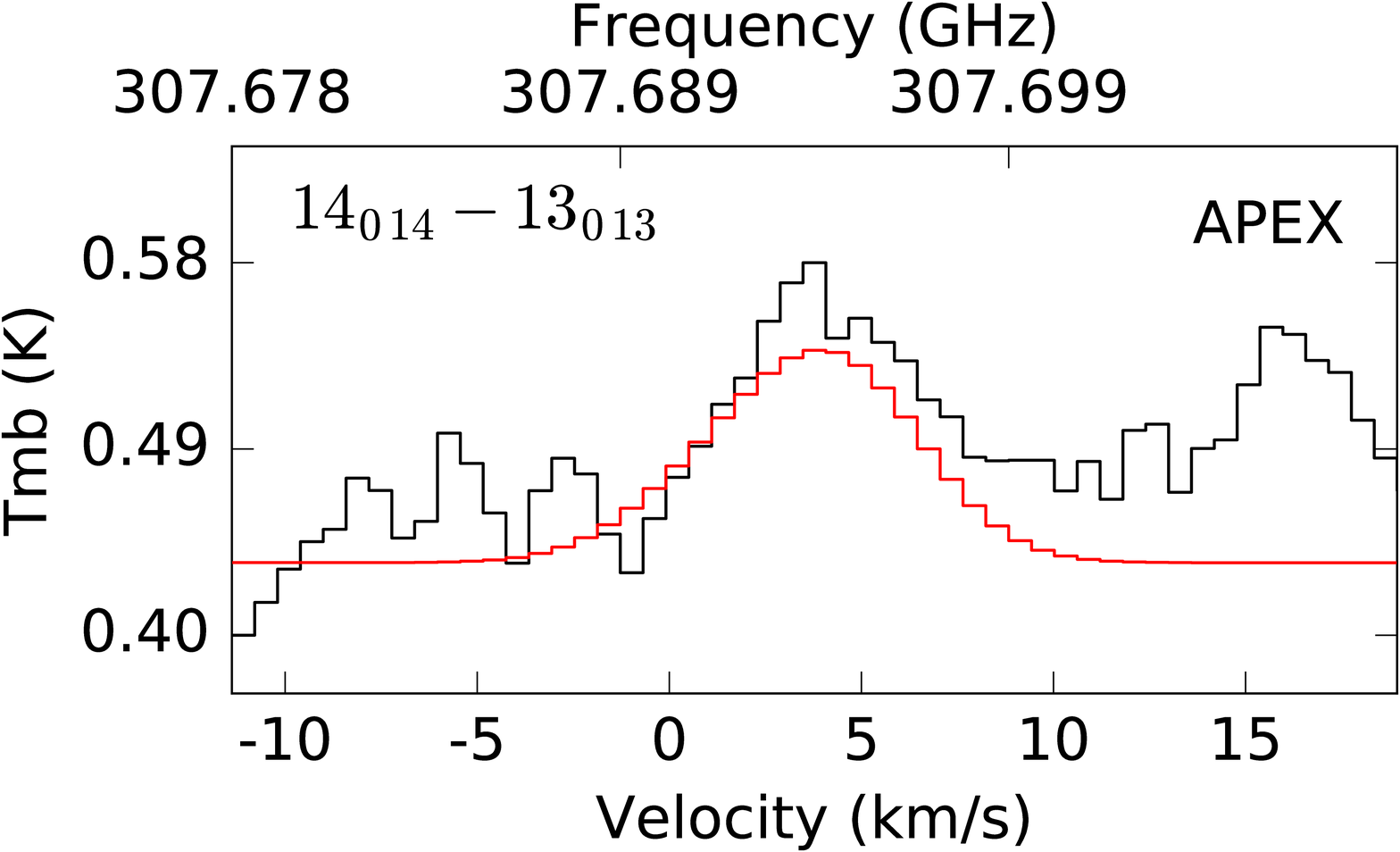} &\includegraphics[width=0.315\textwidth, trim= 0 0 0 0, clip]{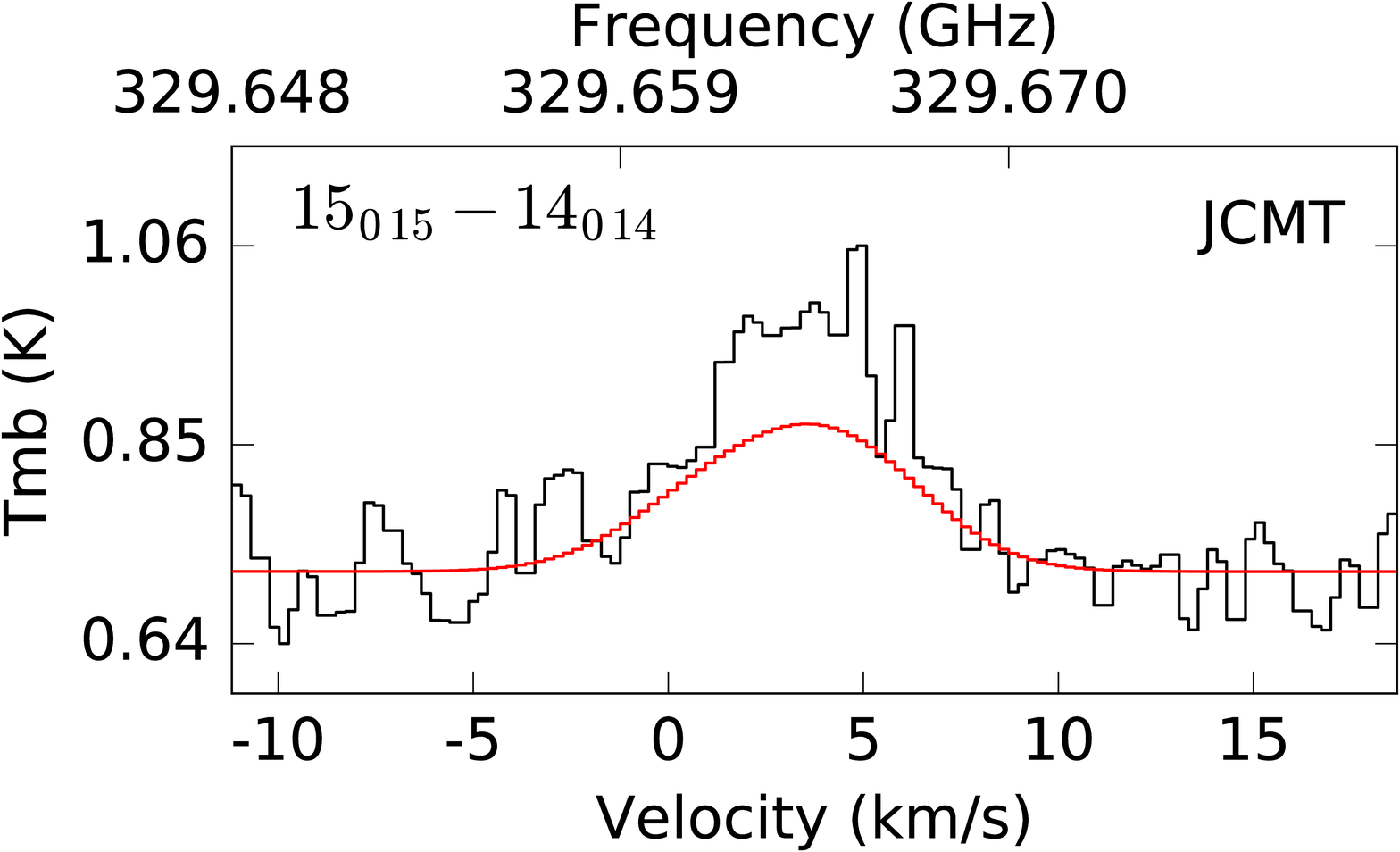} &\includegraphics[width=0.315\textwidth,trim = 0 0 0 0,clip]{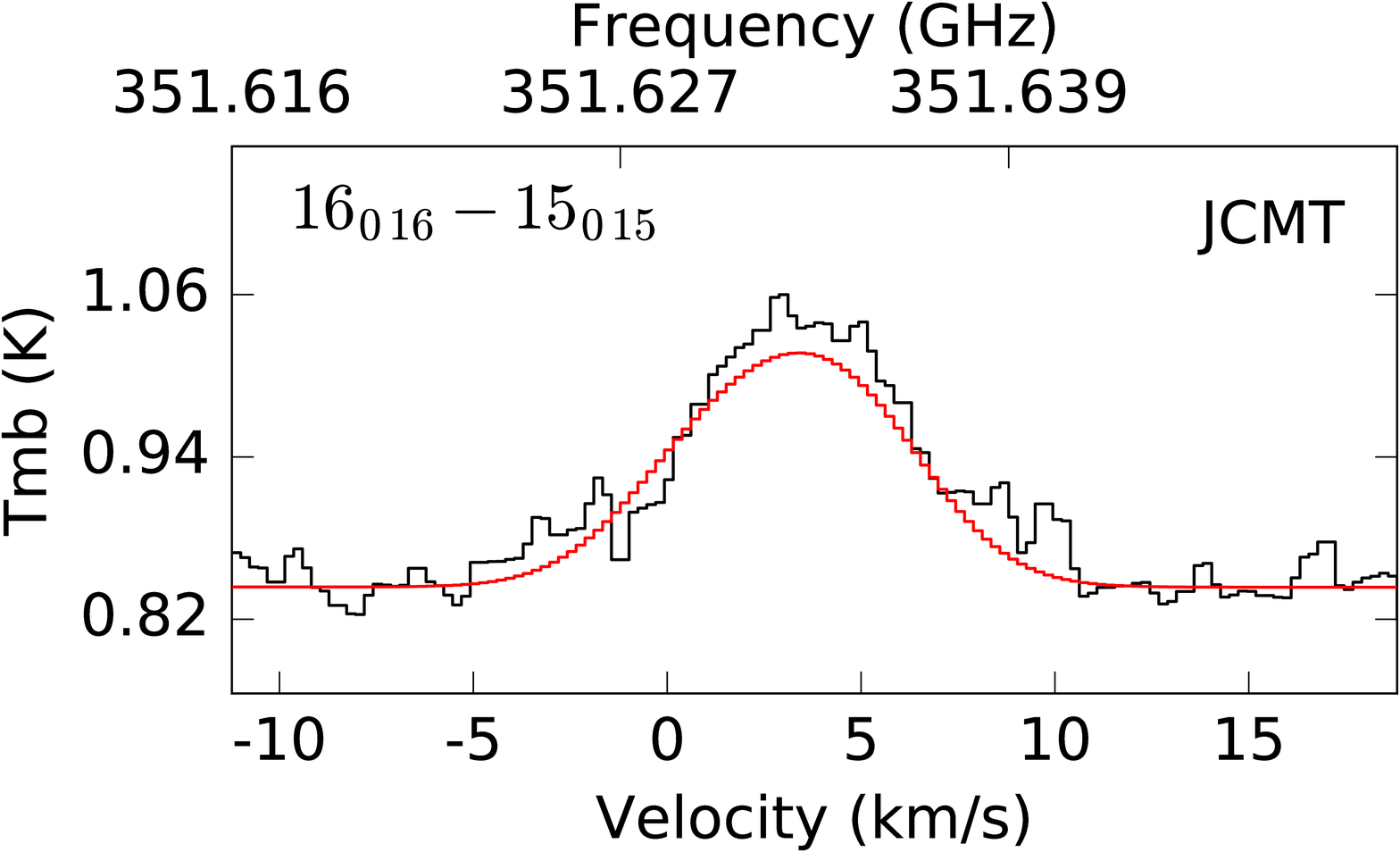} \\ 
\includegraphics[width=0.315\textwidth,trim = 0 0 0 0,clip]{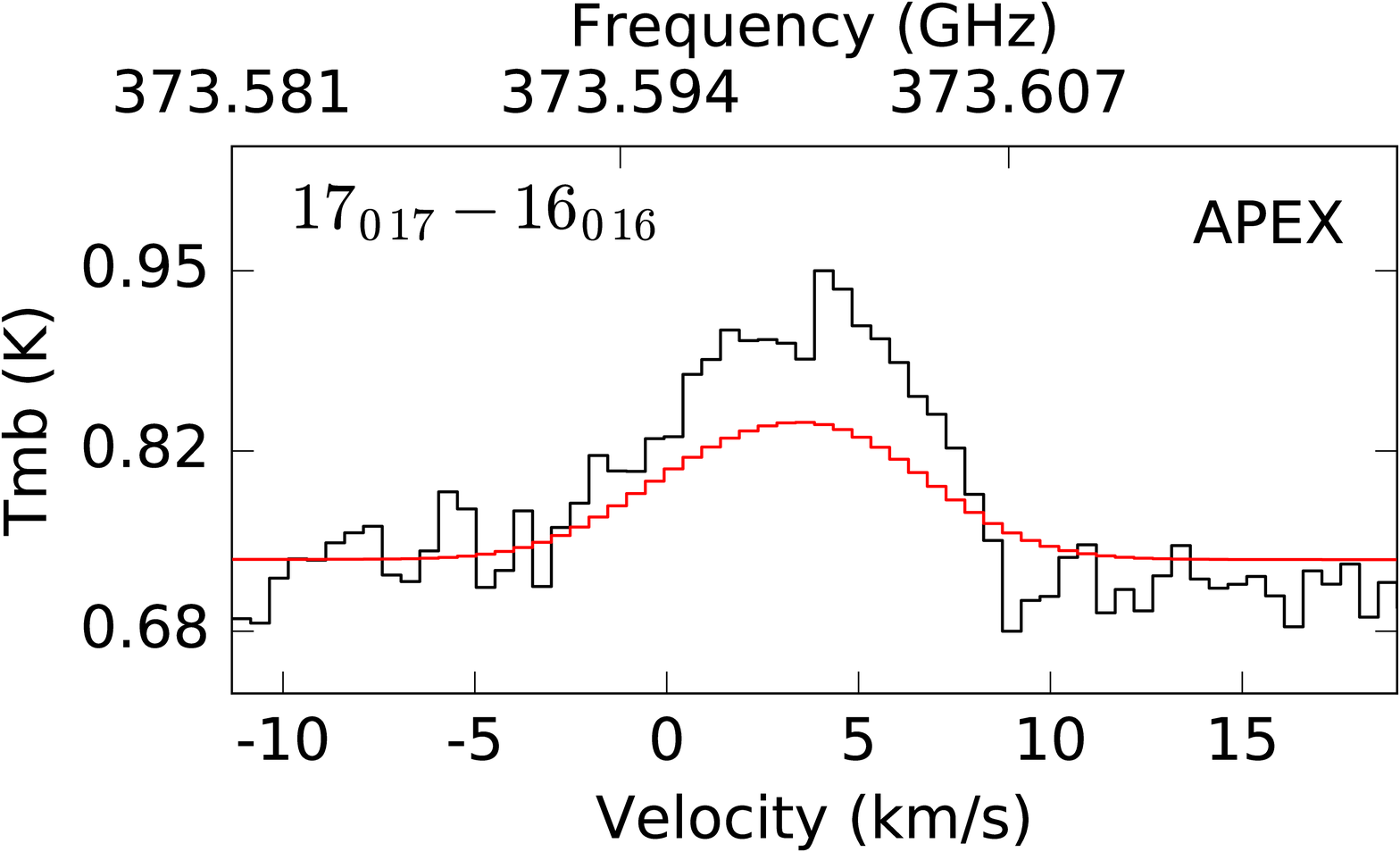} &\includegraphics[width=0.315\textwidth,trim = 0 0 0 0,clip]{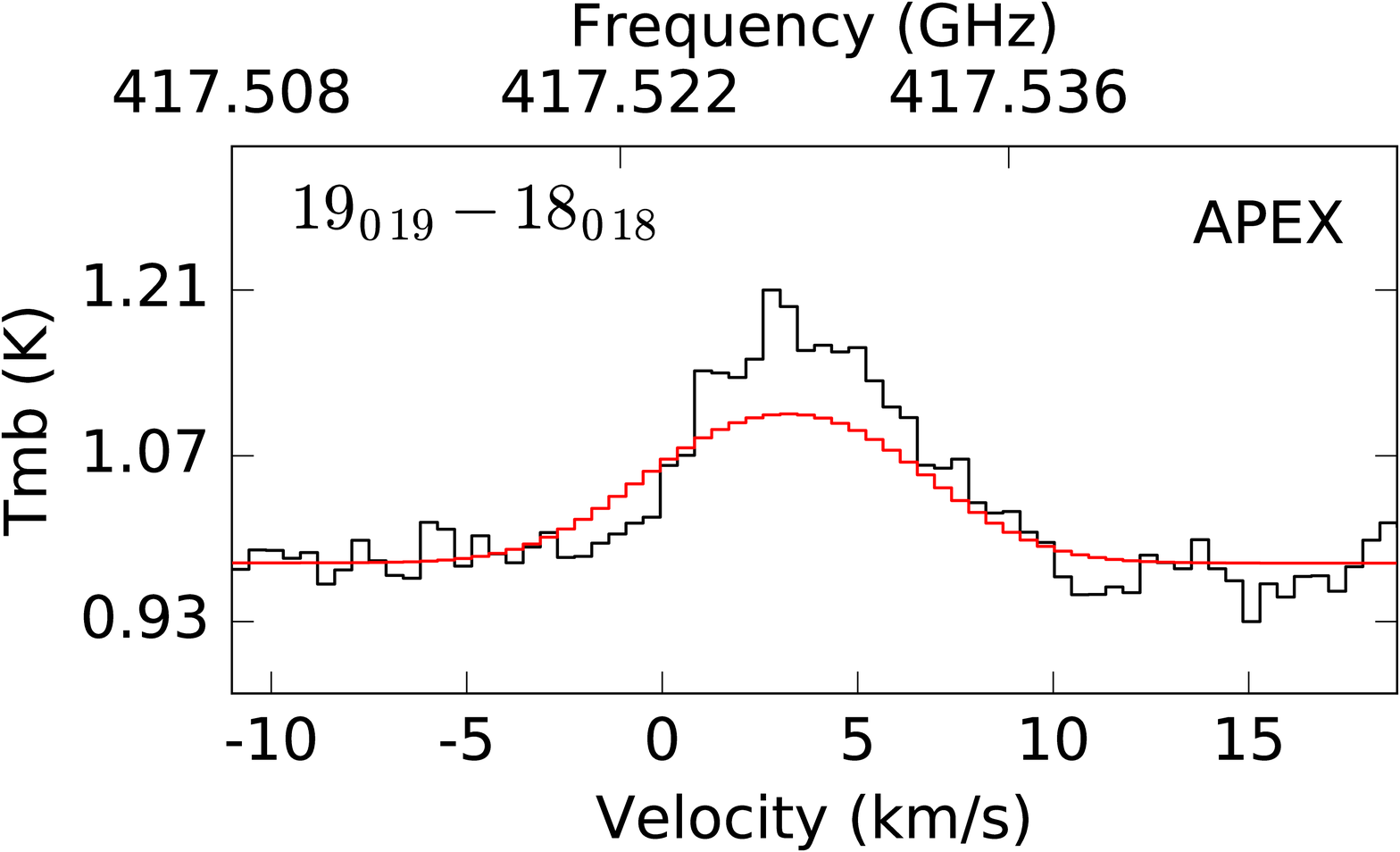}&\includegraphics[width=0.315\textwidth,trim = 0 0 0 0,clip]{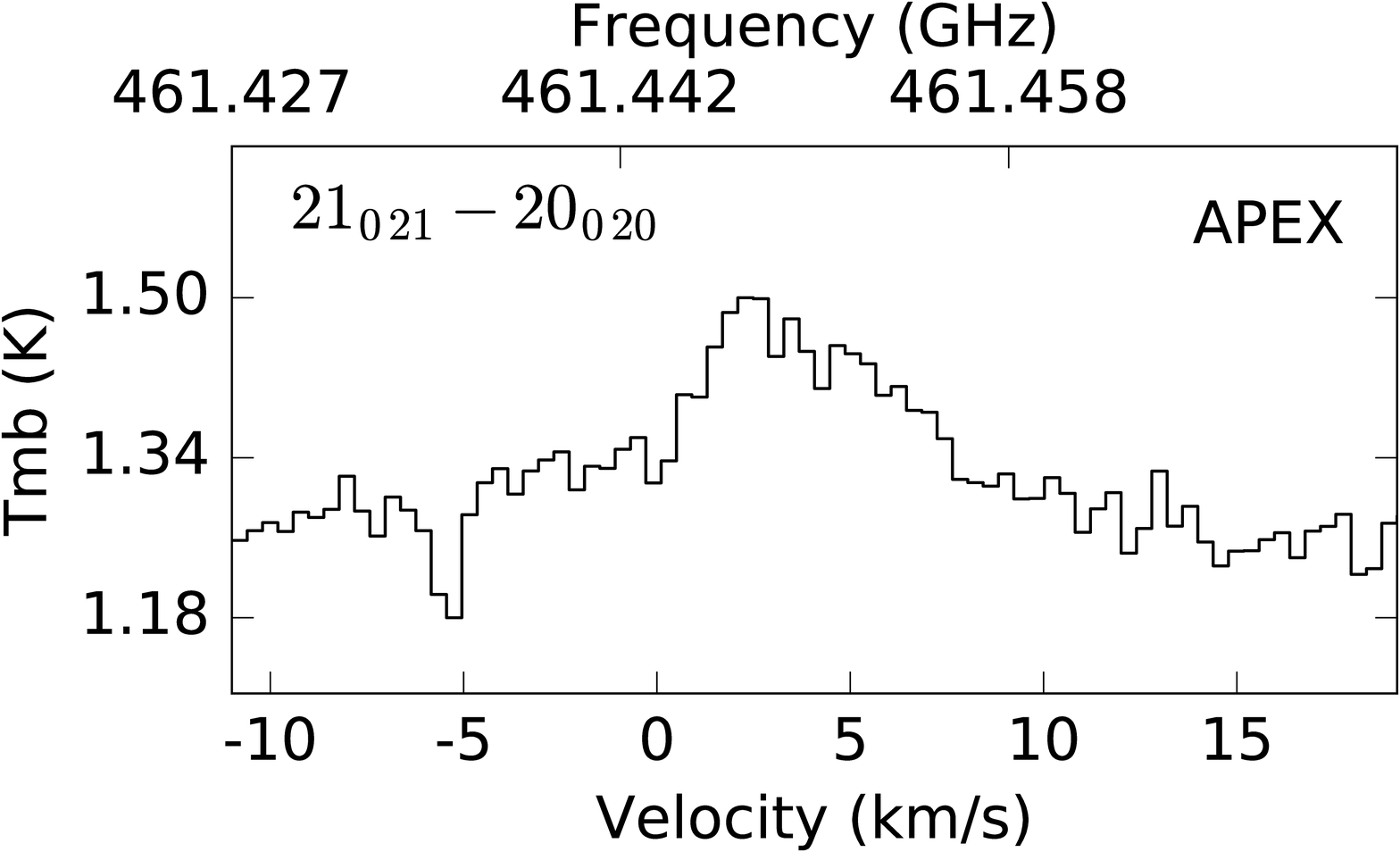} \\

\end{tabular}
\label{Obs-modHNCO}
\end{figure*}


\begin{figure*}
\centering
\setlength\tabcolsep{3.7pt}
\caption{HNCO $K_a=1$ transitions. The $19_{1 \, 18}-18_{1 \,17}$ transition could not be modelled due to the lack of its collisional rate coefficient.}
\begin{tabular}{c c c}
\includegraphics[width=0.315\textwidth, trim= 0 0 0 0, clip]{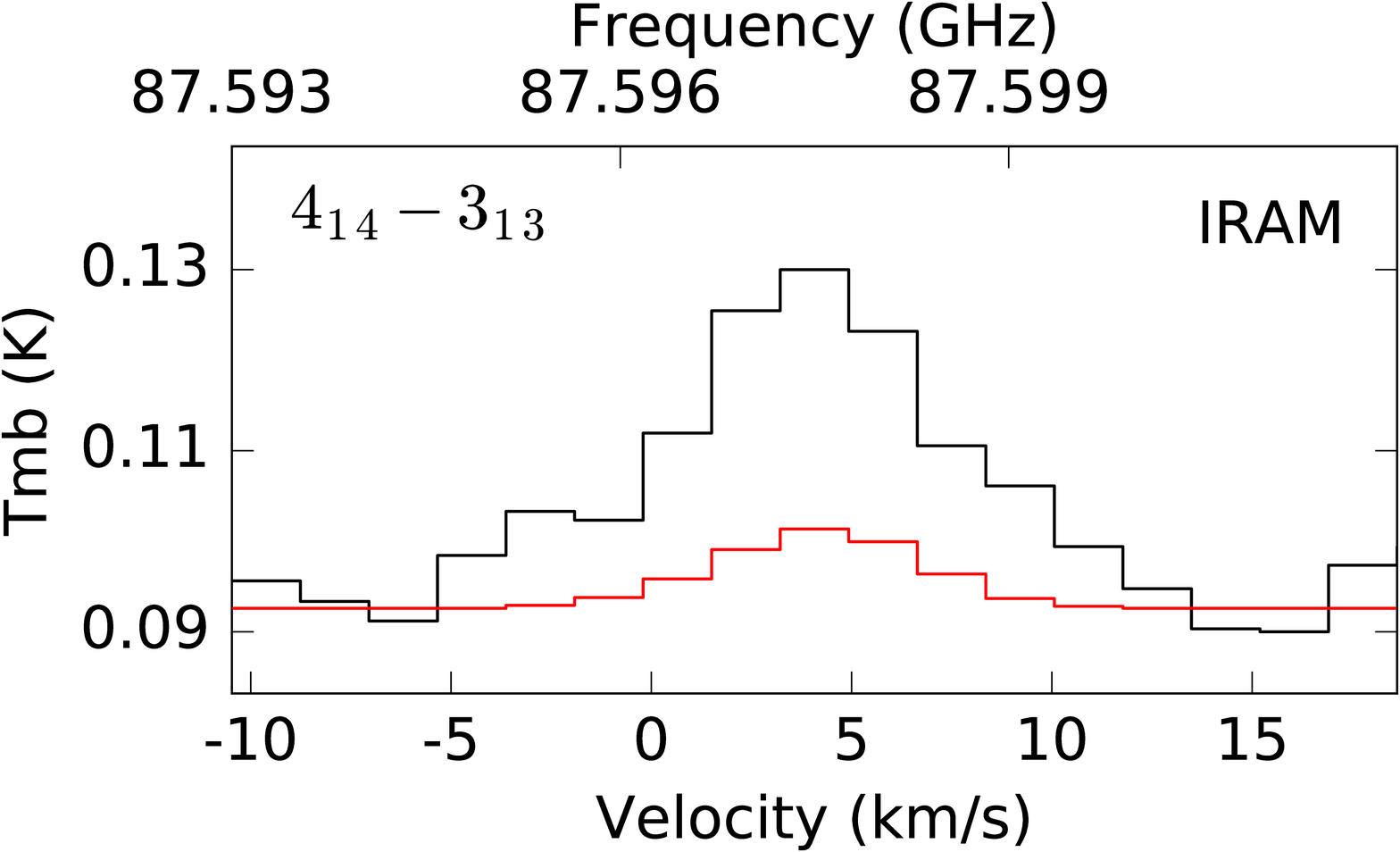} &\includegraphics[width=0.315\textwidth,trim = 0 0 0 0,clip]{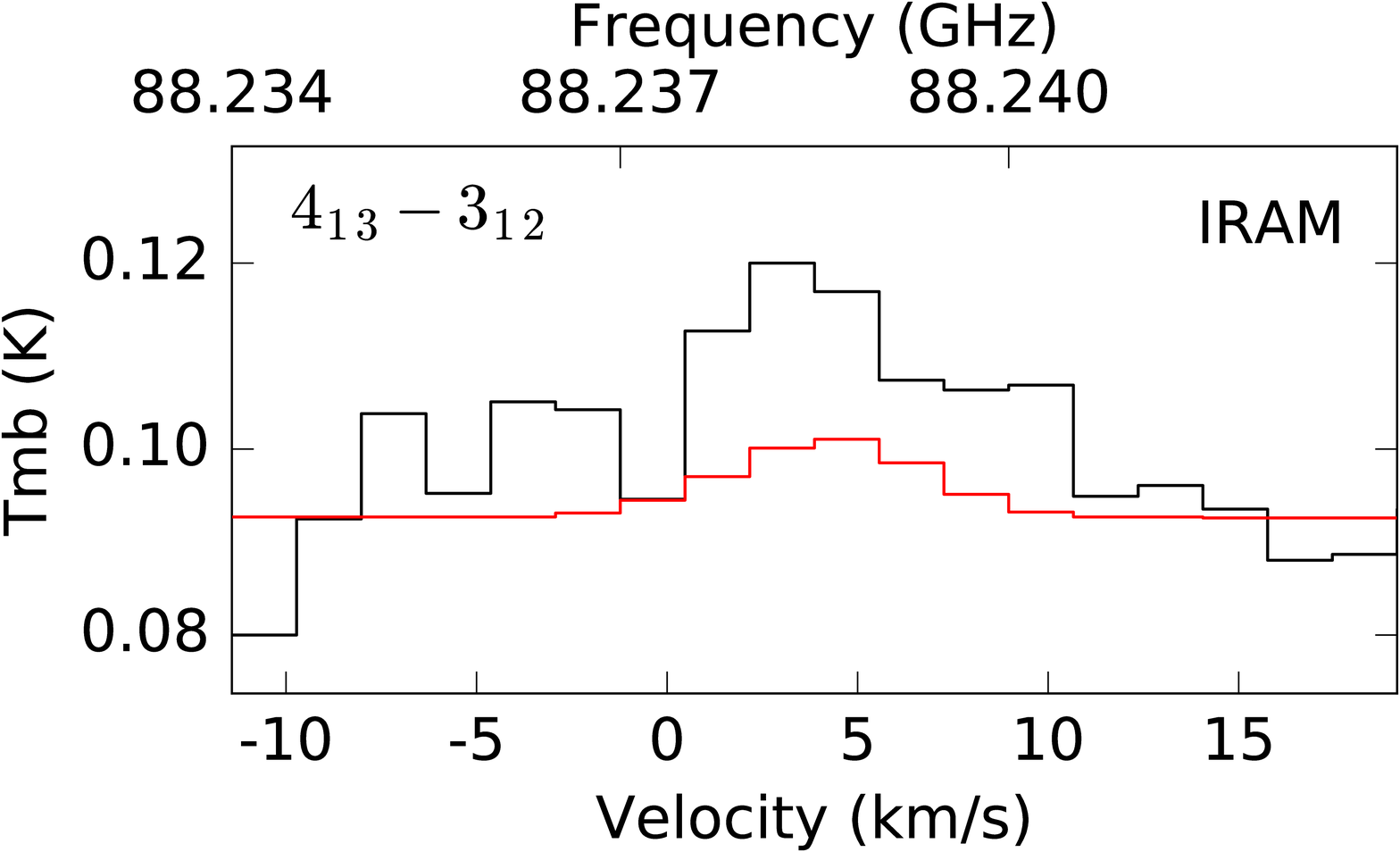}  &\includegraphics[width=0.315\textwidth,trim = 0 0 0 0,clip]{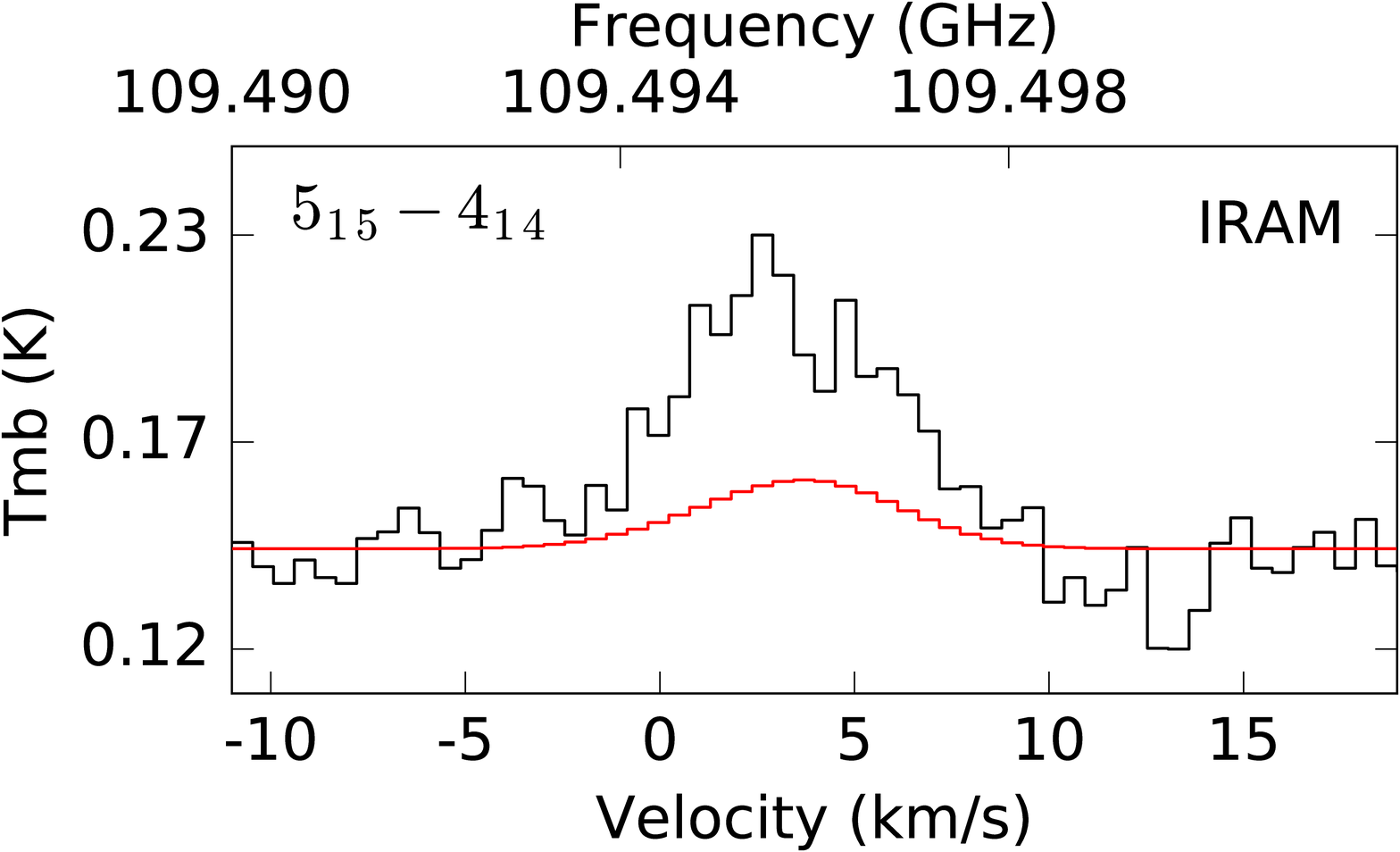}\\
\includegraphics[width=0.315\textwidth,trim = 0 0 0 0,clip]{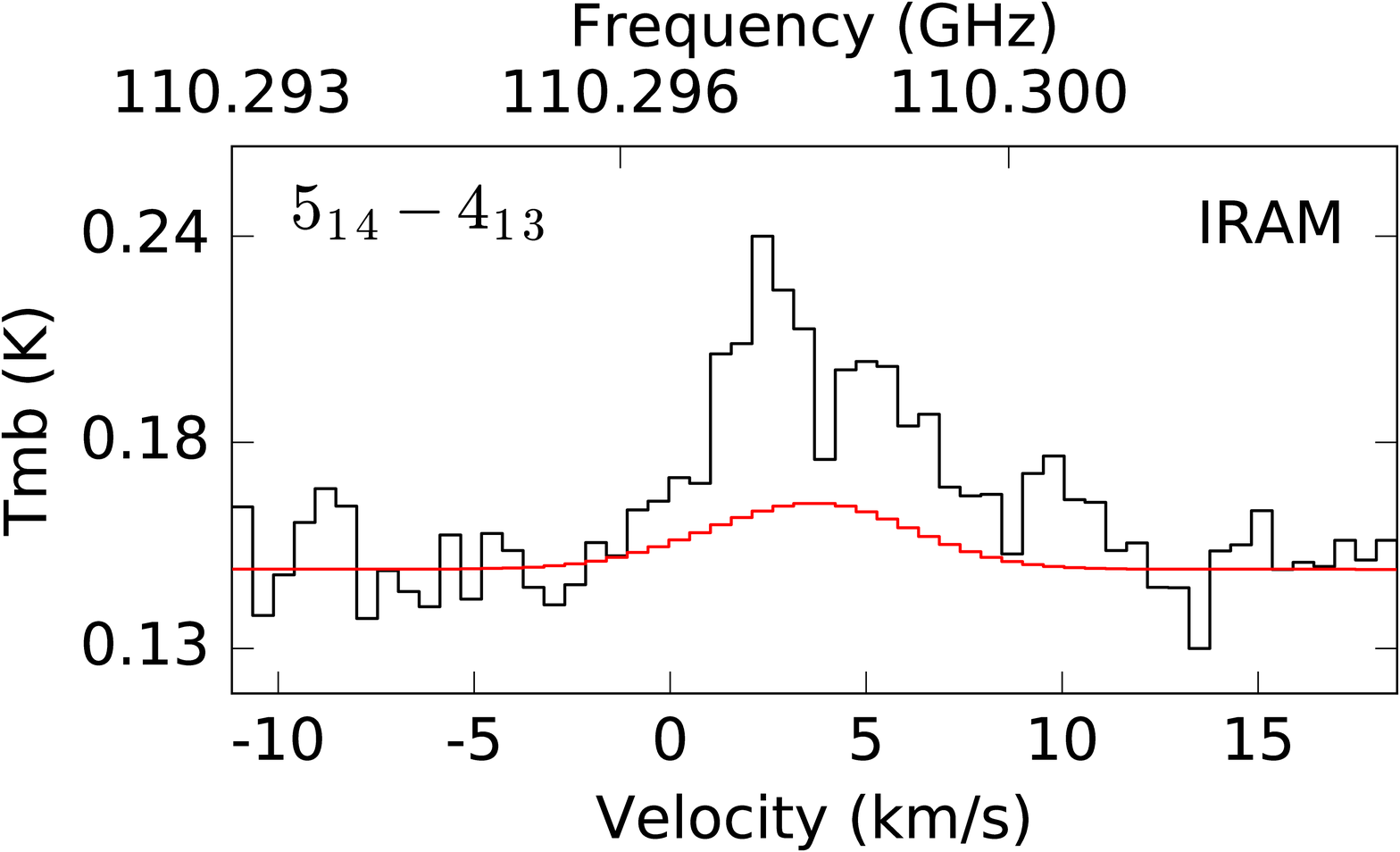}&\includegraphics[width=0.315\textwidth,trim = 0 0 0 0,clip]{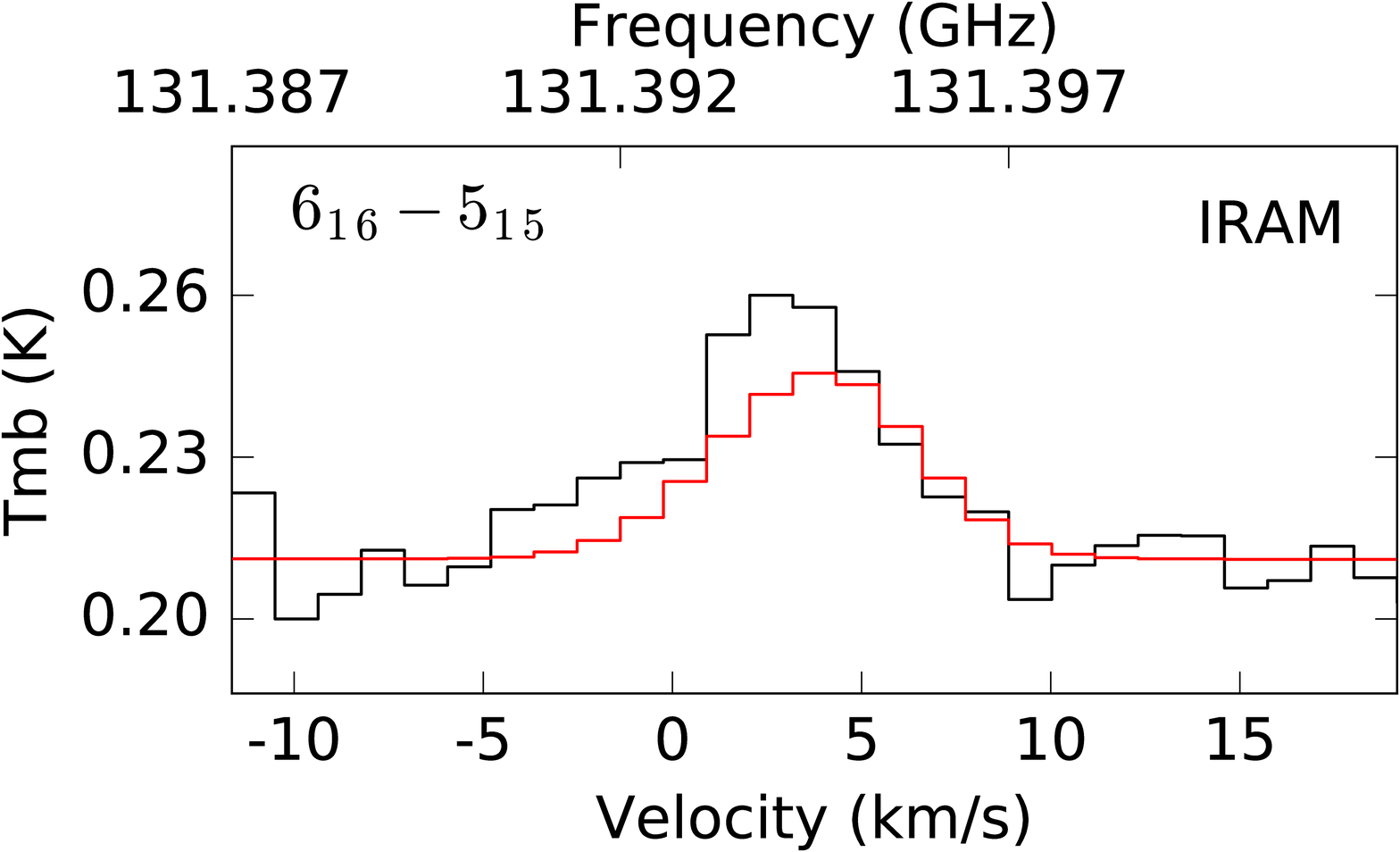}&\includegraphics[width=0.315\textwidth, trim= 0 0 0 0, clip]{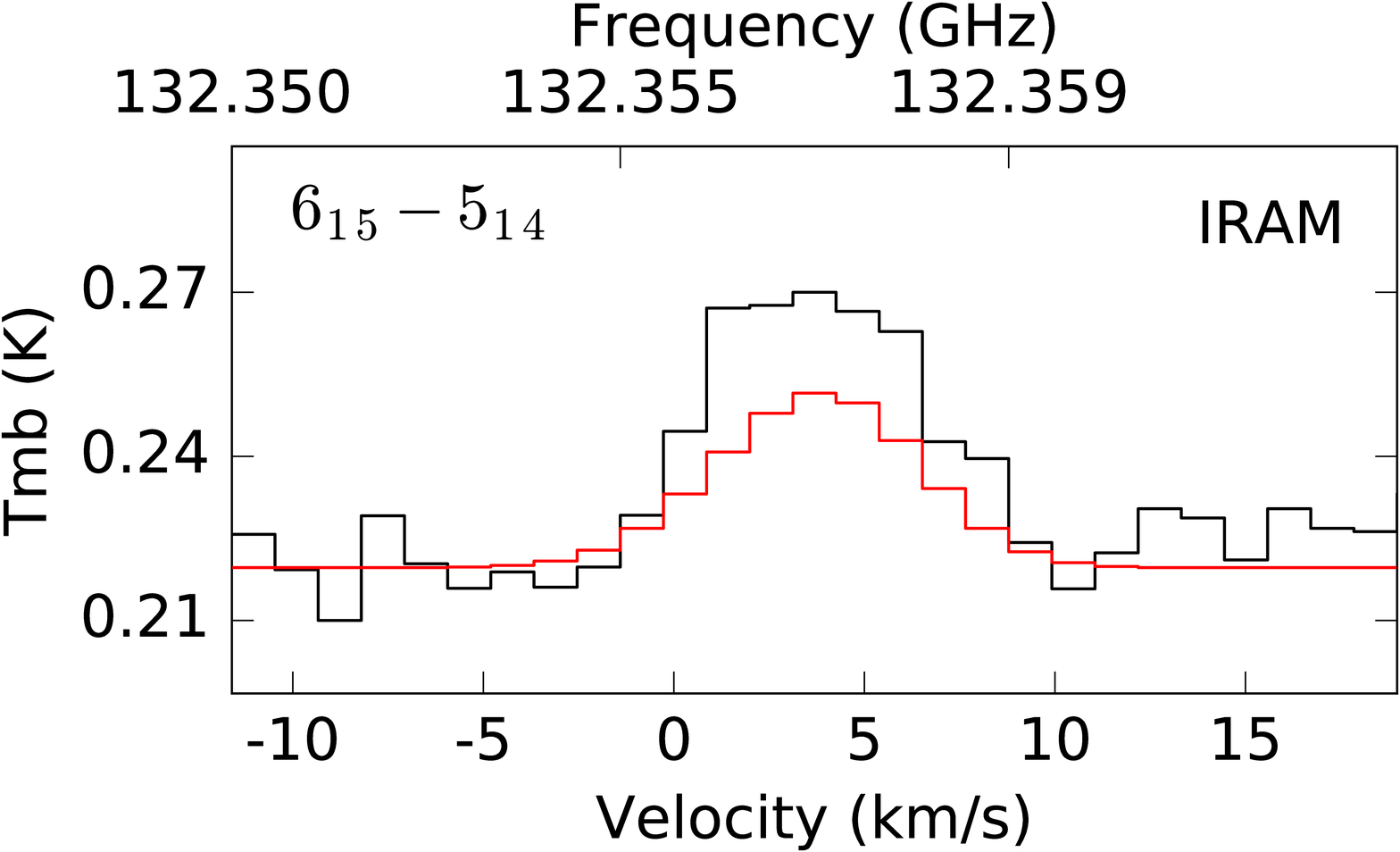} \\
\includegraphics[width=0.315\textwidth,trim = 0 0 0 0,clip]{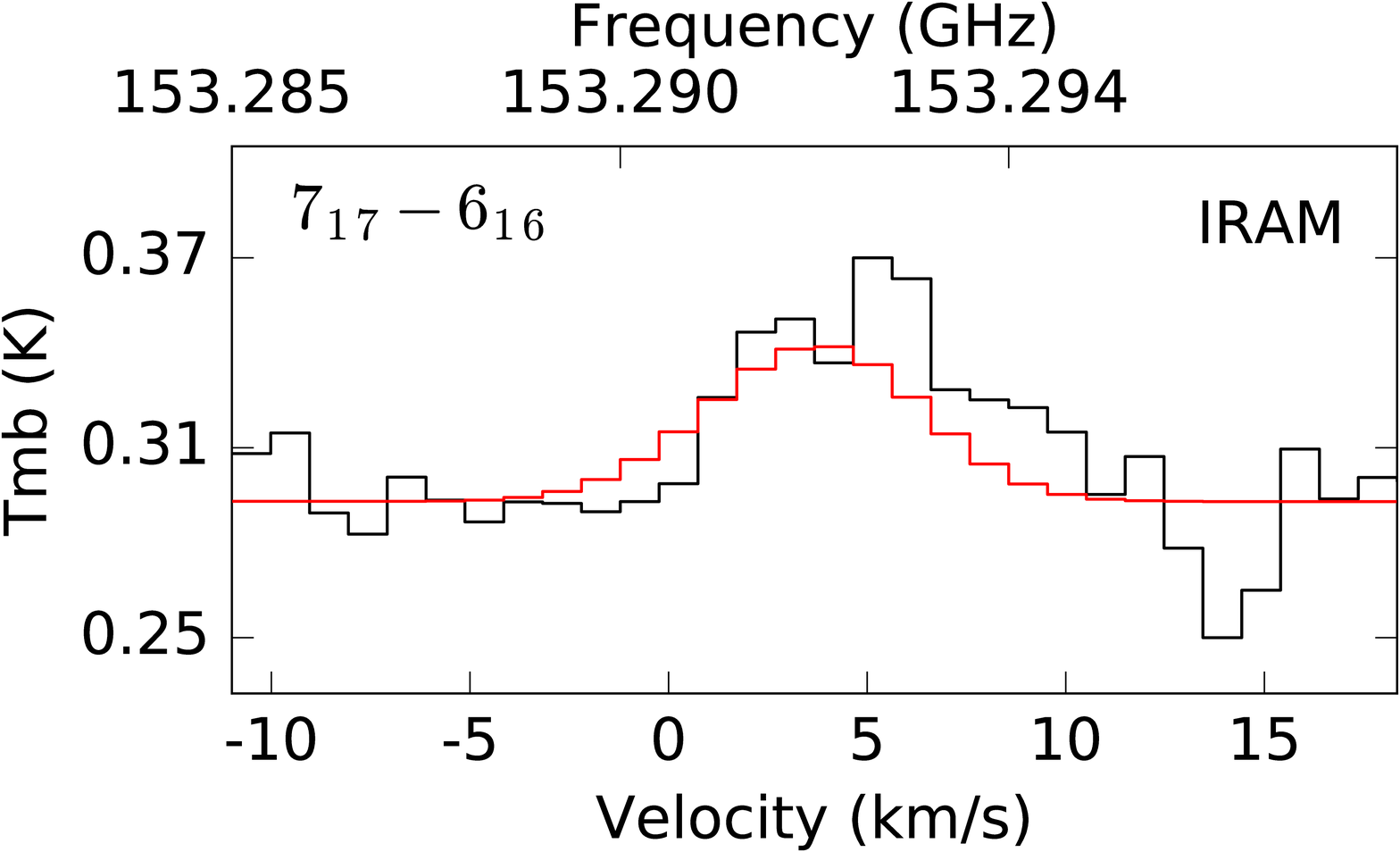}  &\includegraphics[width=0.315\textwidth,trim = 0 0 0 0,clip]{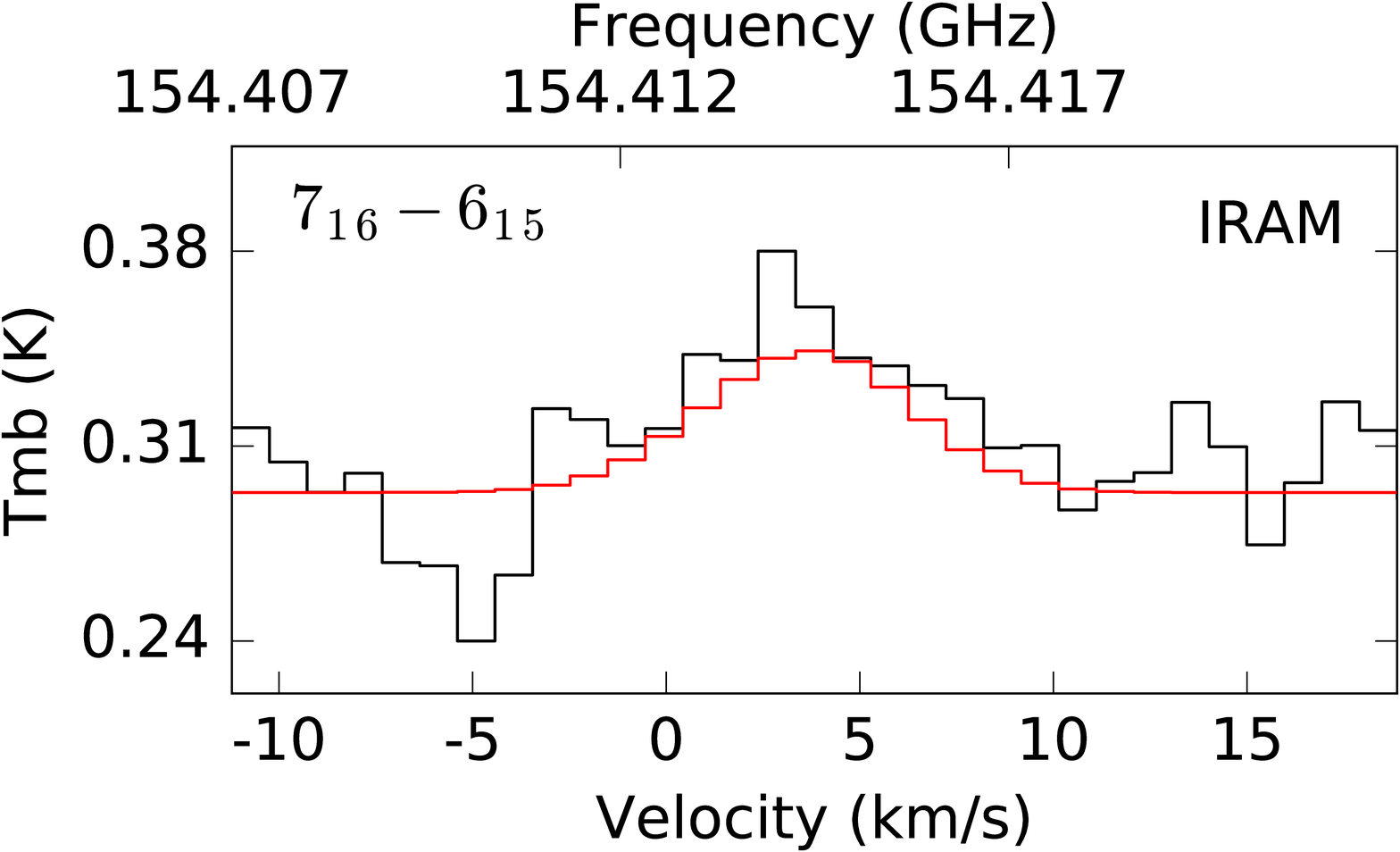} &\includegraphics[width=0.315\textwidth,trim = 0 0 0 0,clip]{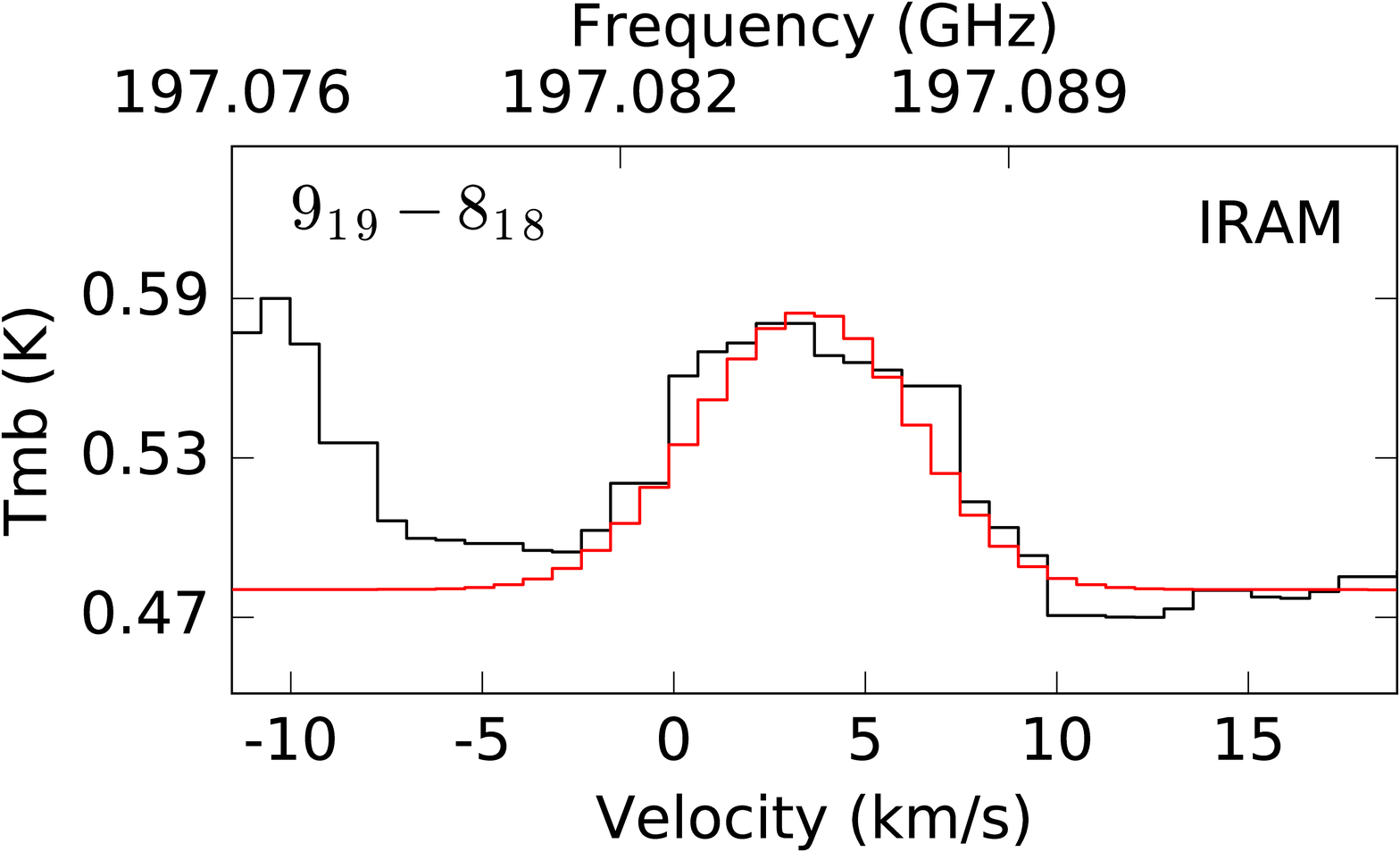}\\
\includegraphics[width=0.315\textwidth,trim = 0 0 0 0,clip]{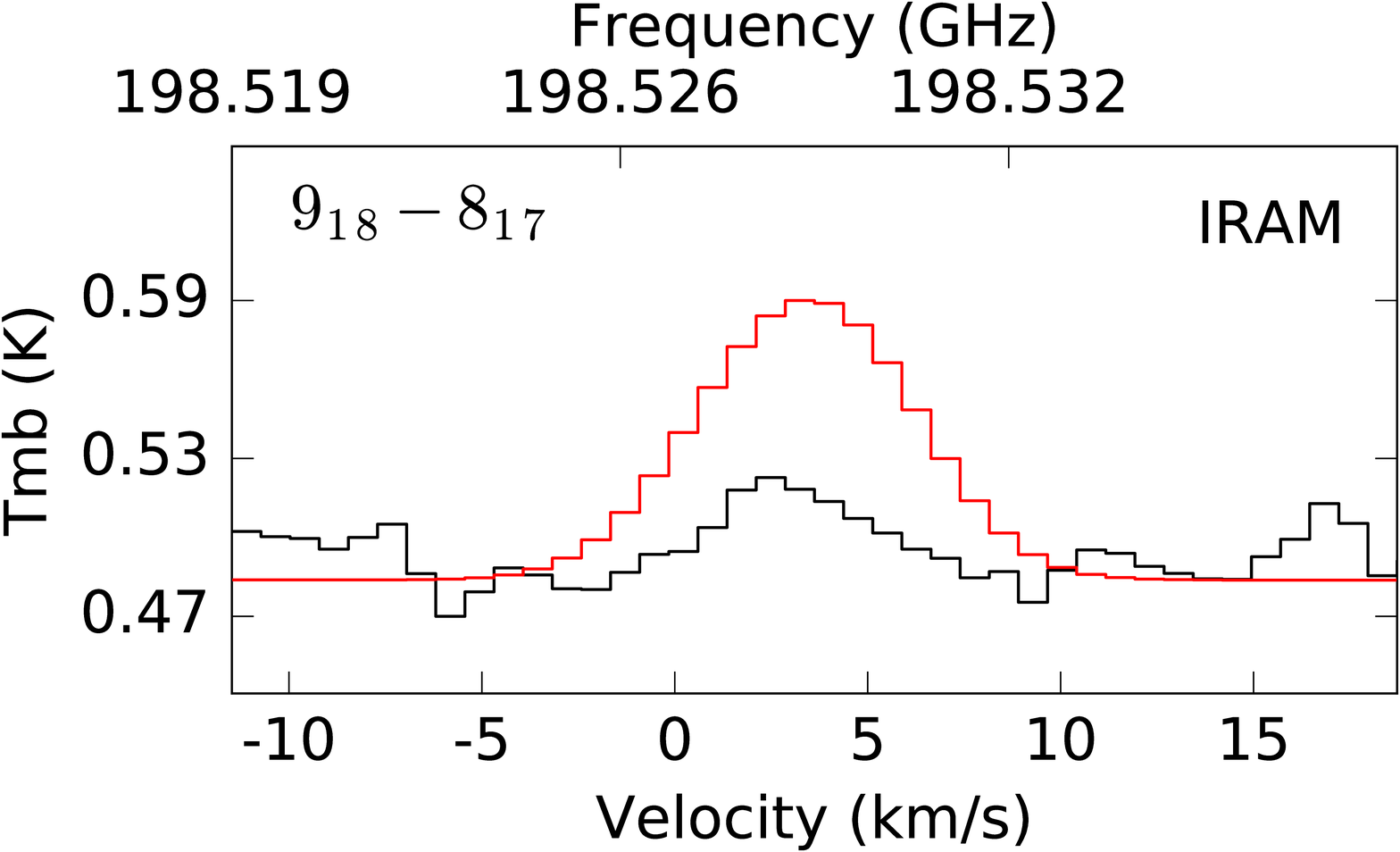} & \includegraphics[width=0.315\textwidth, trim= 0 0 0 0, clip]{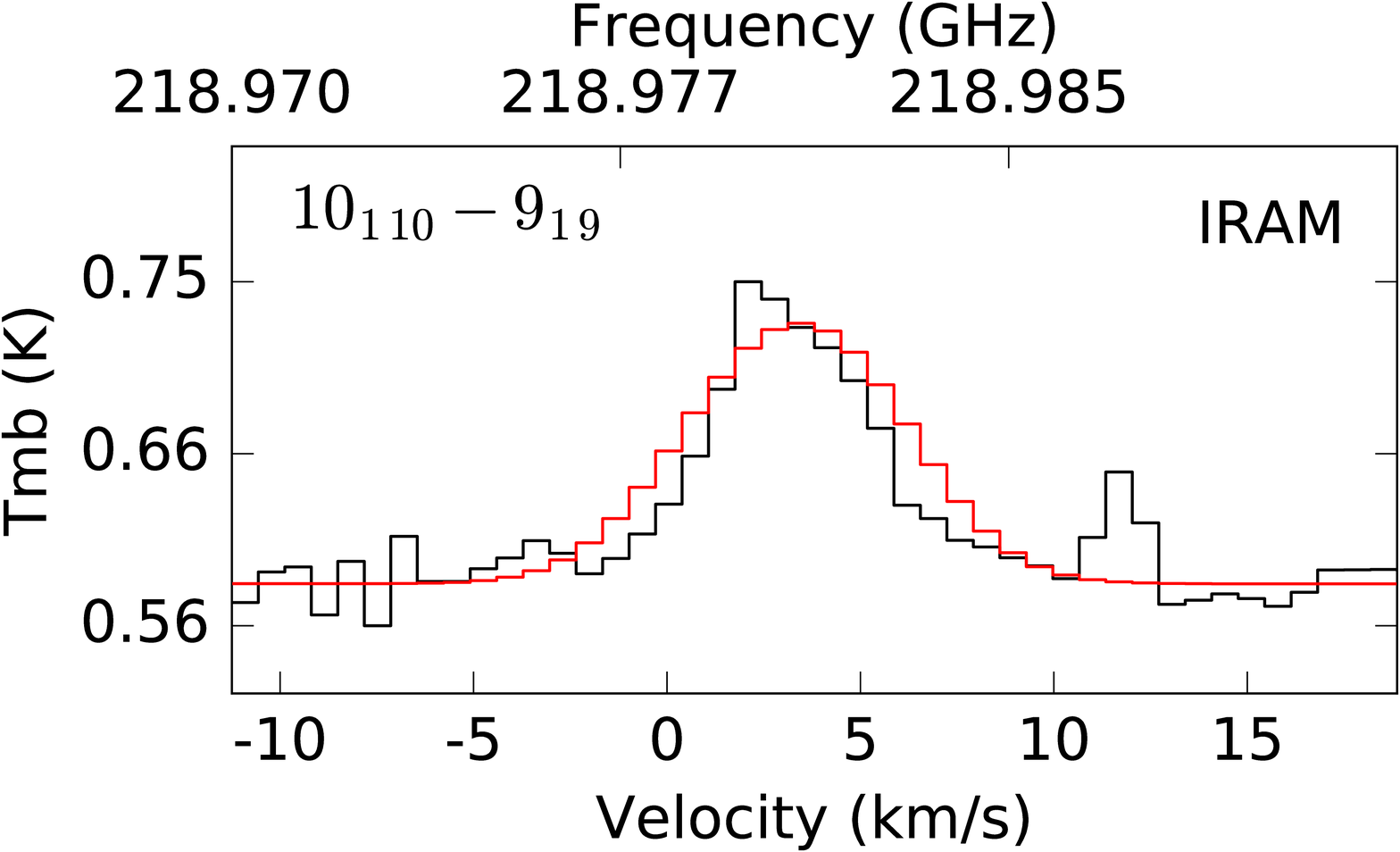} &\includegraphics[width=0.315\textwidth,trim = 0 0 0 0,clip]{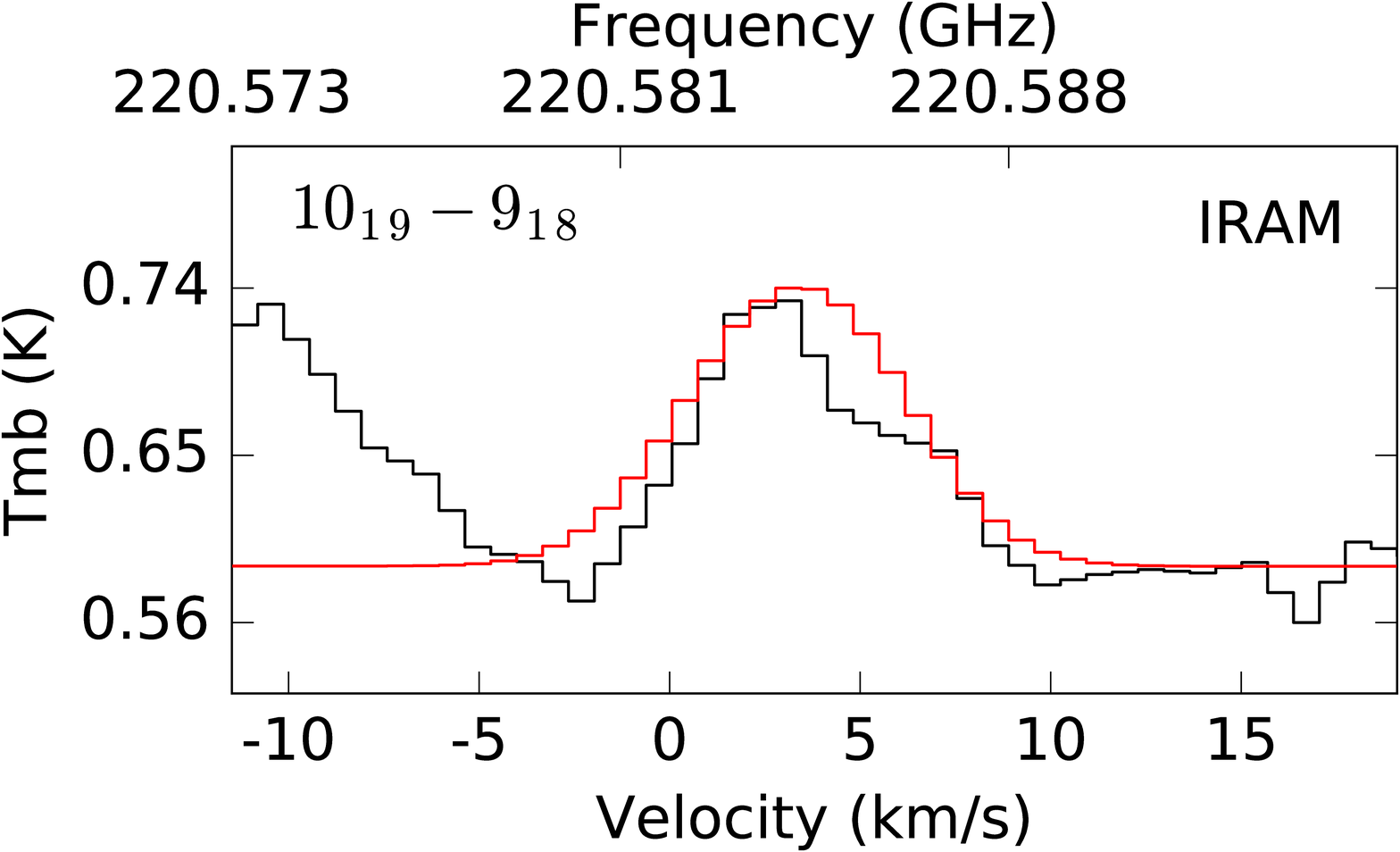}\\
\includegraphics[width=0.315\textwidth,trim = 0 0 0 0,clip]{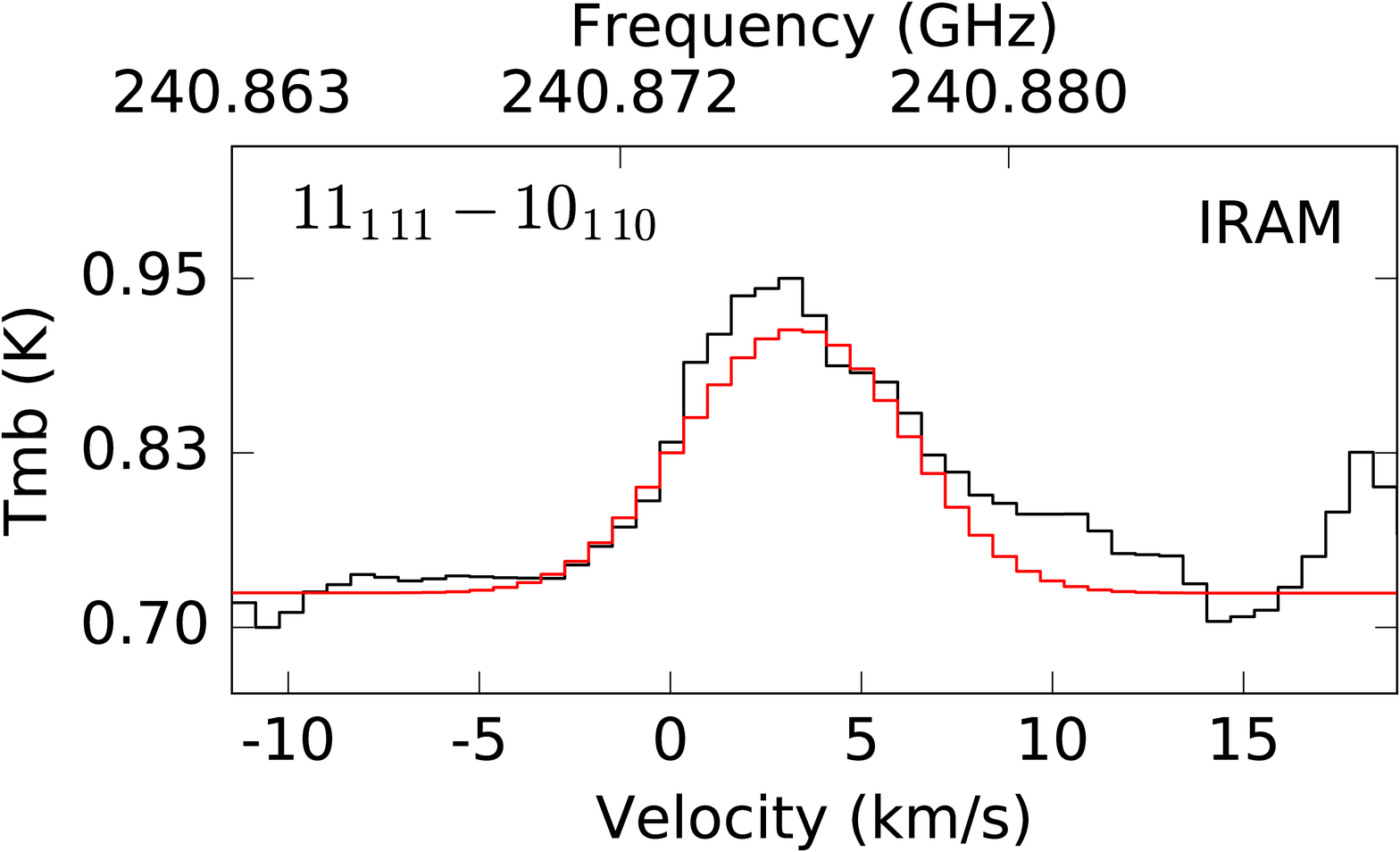} & \includegraphics[width=0.315\textwidth, trim= 0 0 0 0, clip]{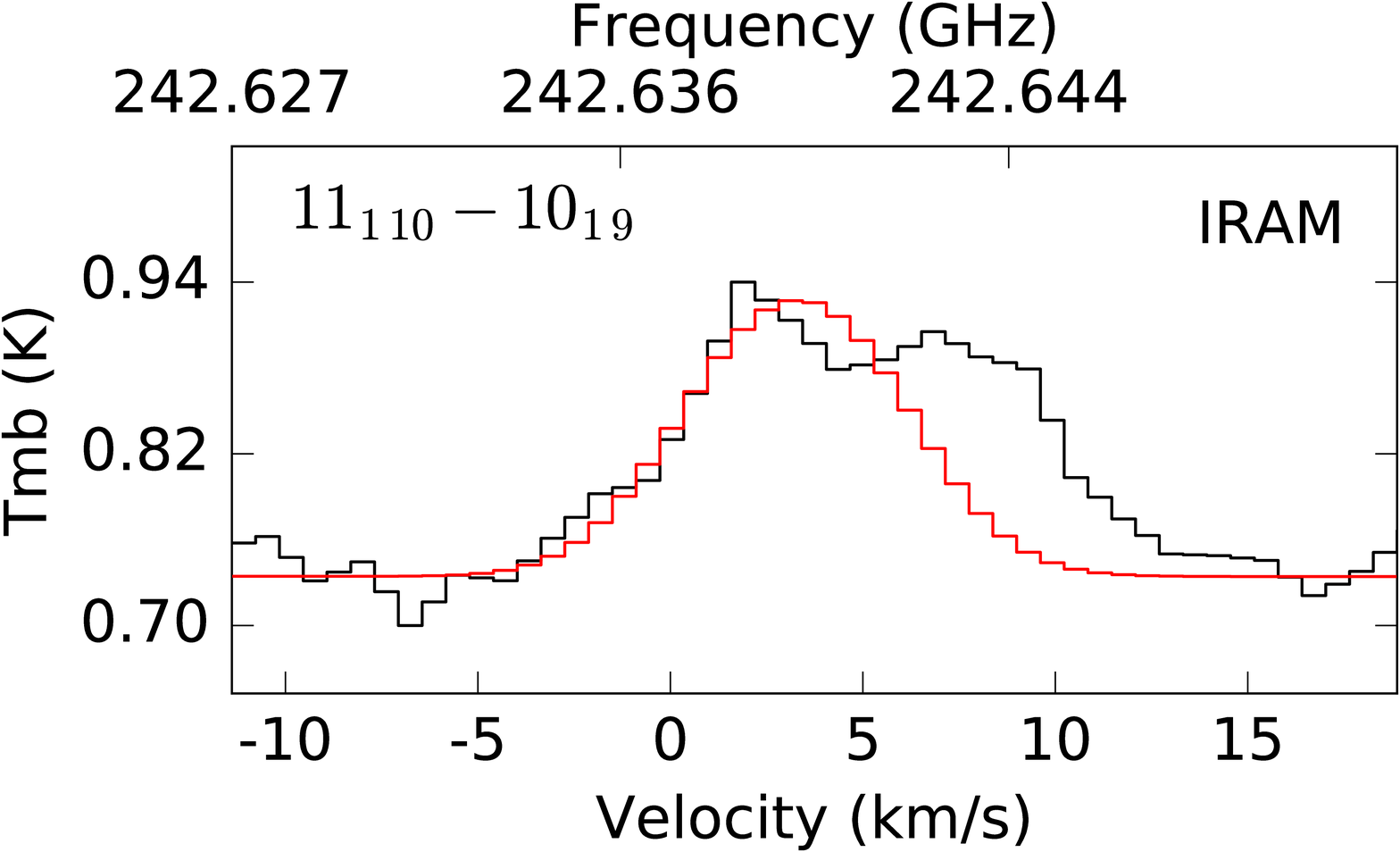} &\includegraphics[width=0.315\textwidth,trim = 0 0 0 0,clip]{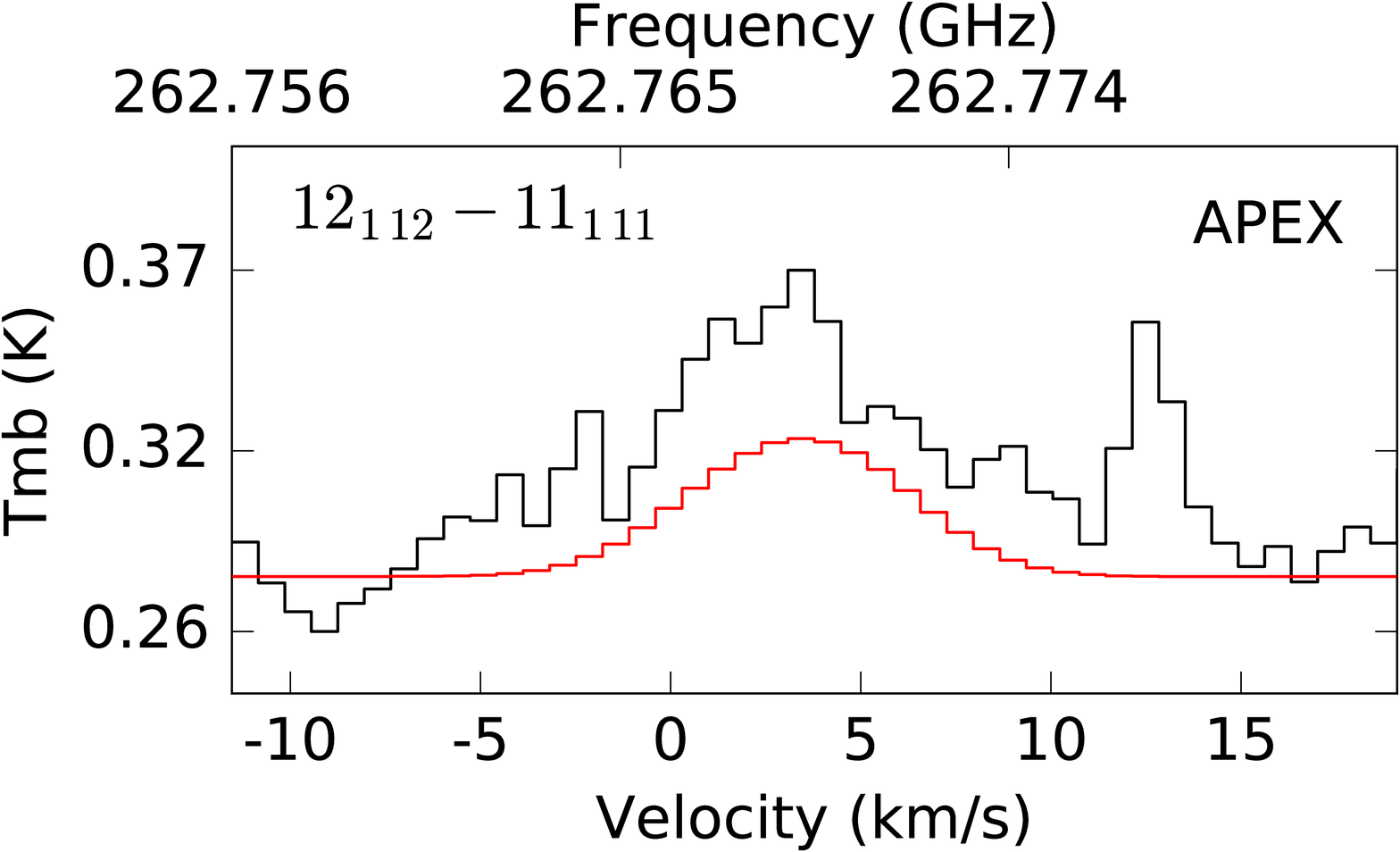}\\
\includegraphics[width=0.315\textwidth, trim= 0 0 0 0, clip]{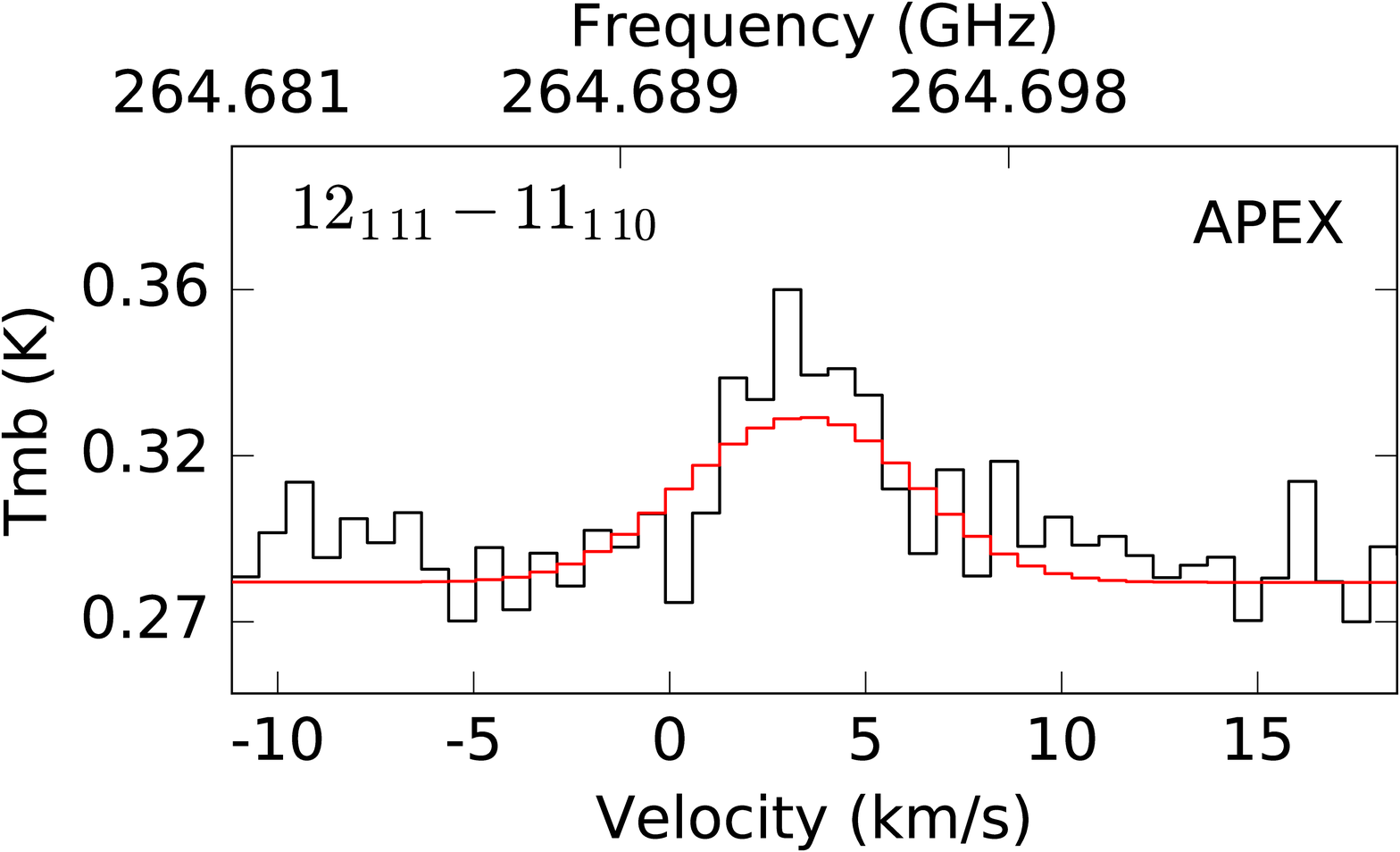} &\includegraphics[width=0.315\textwidth,trim = 0 0 0 0,clip]{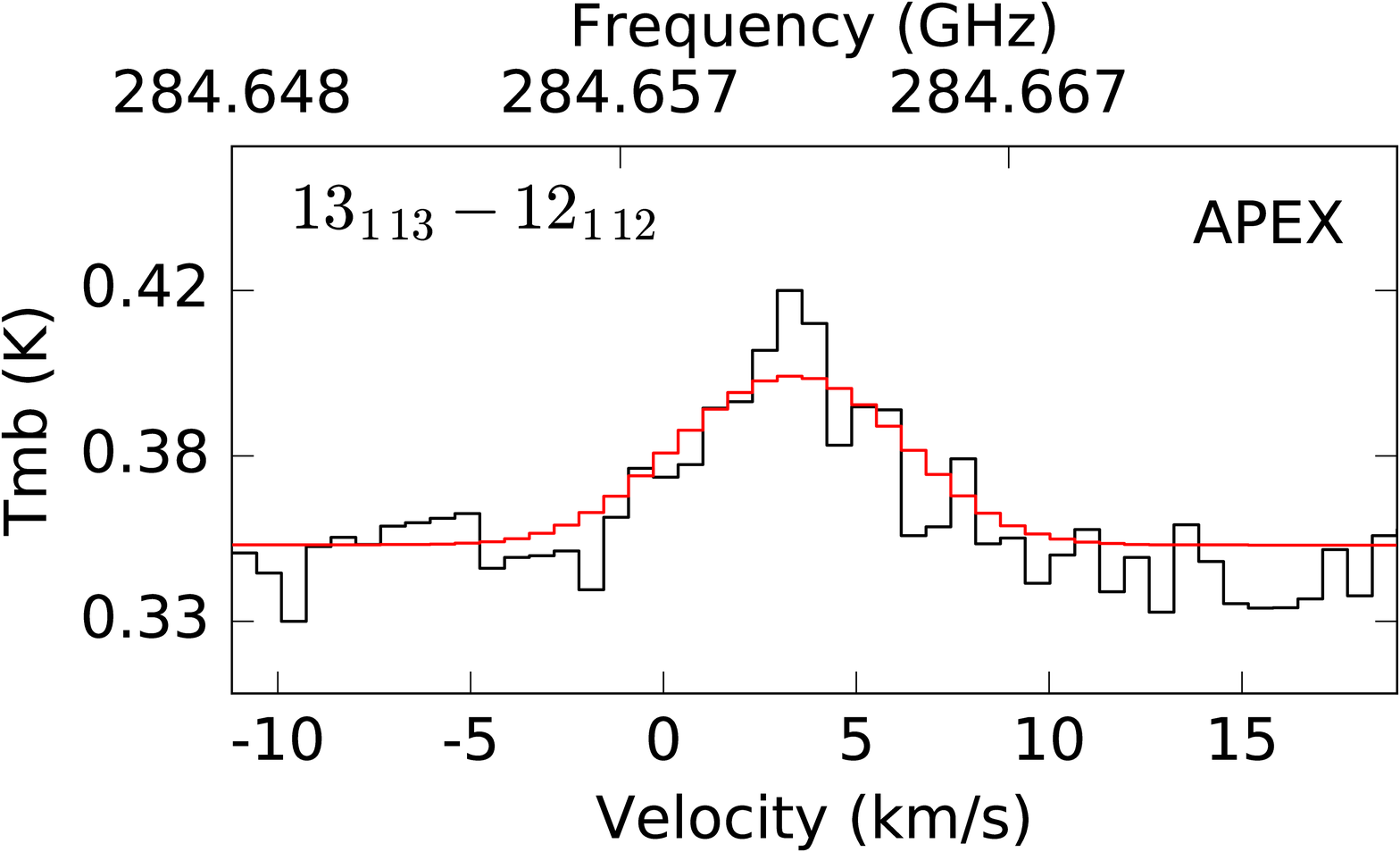}  &\includegraphics[width=0.315\textwidth,trim = 0 0 0 0,clip]{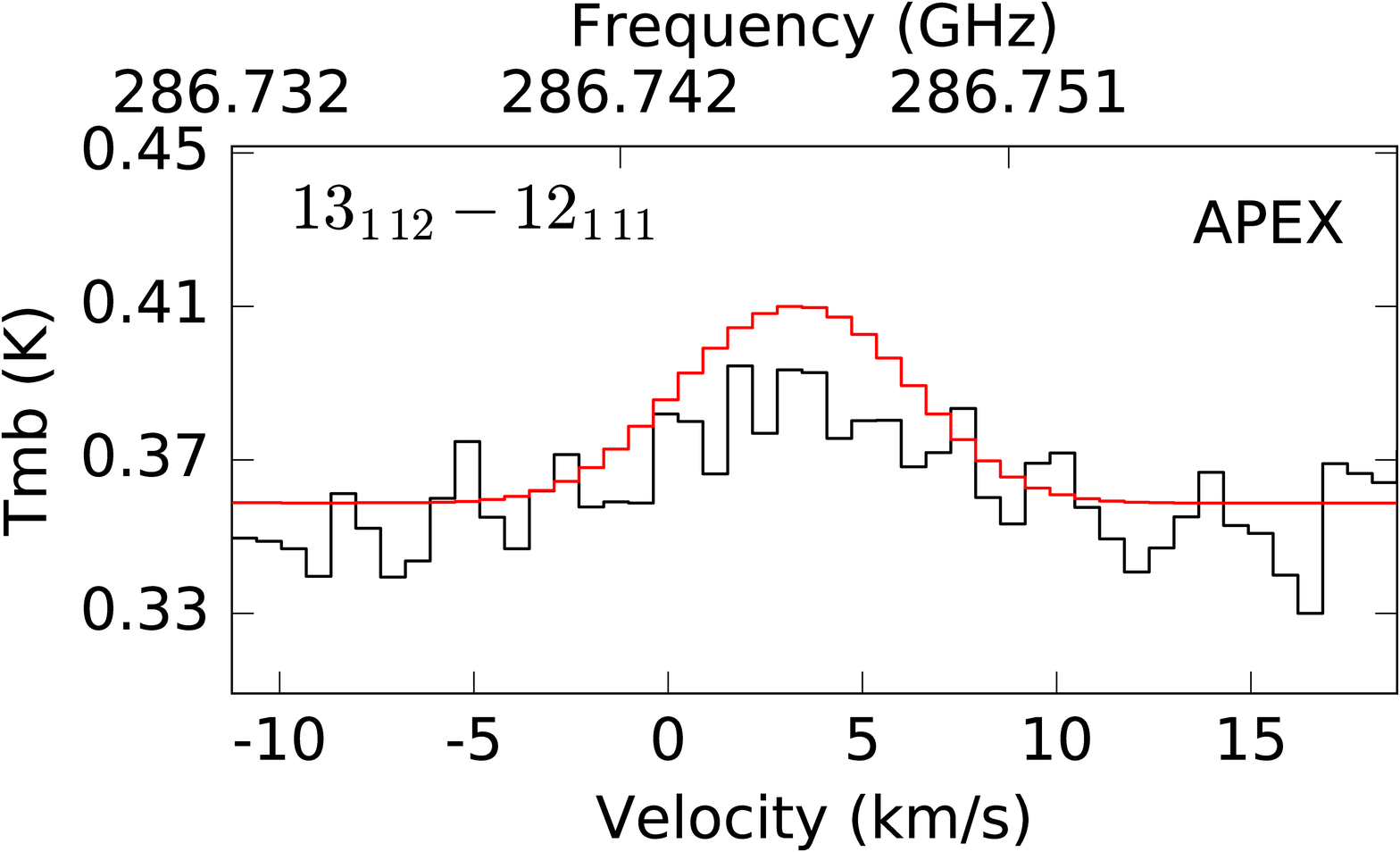}\\

\end{tabular}
\label{imagenes-cont1}
\end{figure*}

\begin{figure*}
\centering
\setlength\tabcolsep{3.7pt}
\contcaption{}
\label{imagenes-cont2}
\begin{tabular}{c c c}
\includegraphics[width=0.315\textwidth,trim = 0 0 0 0,clip]{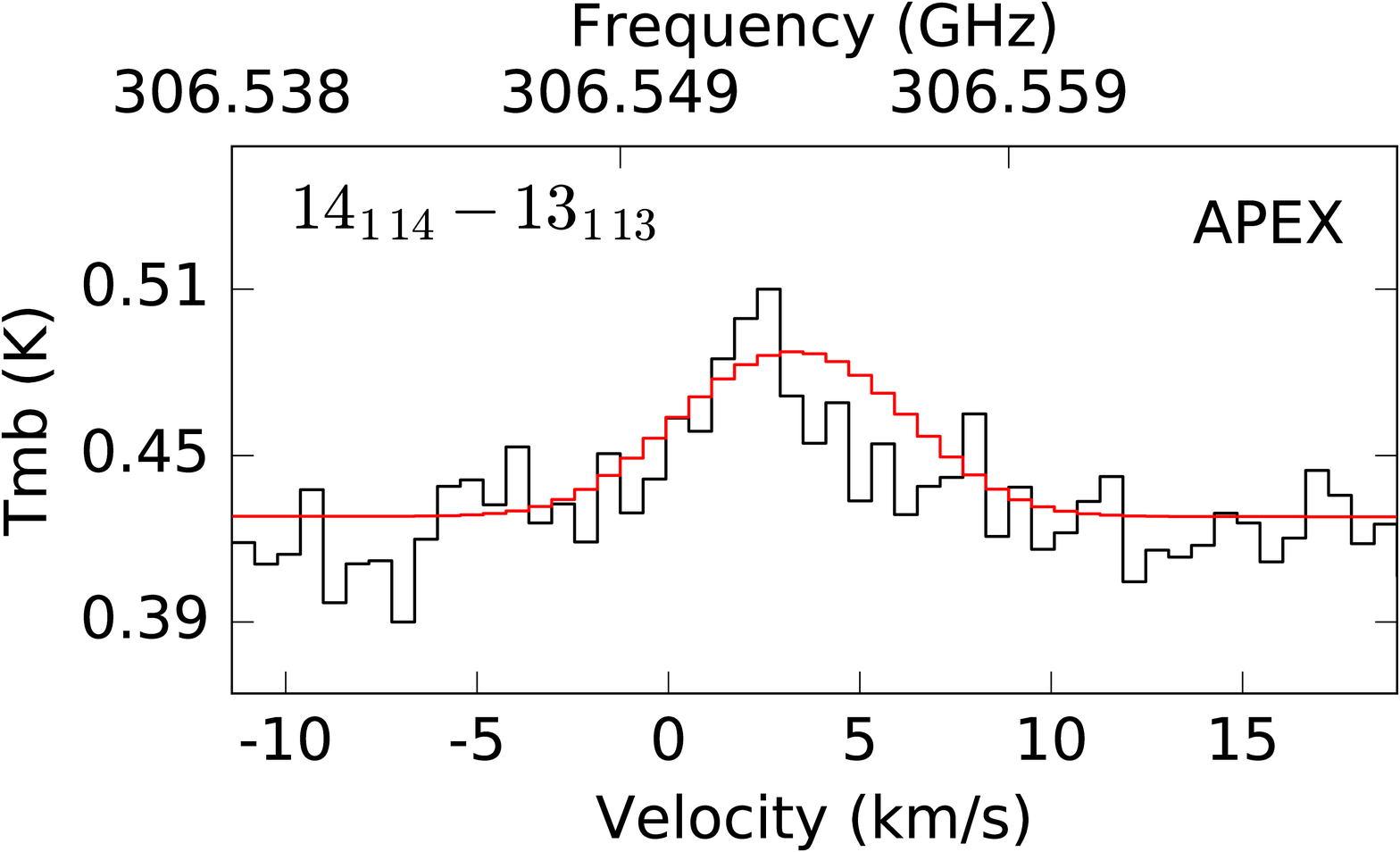}&\includegraphics[width=0.315\textwidth,trim = 0 0 0 0,clip]{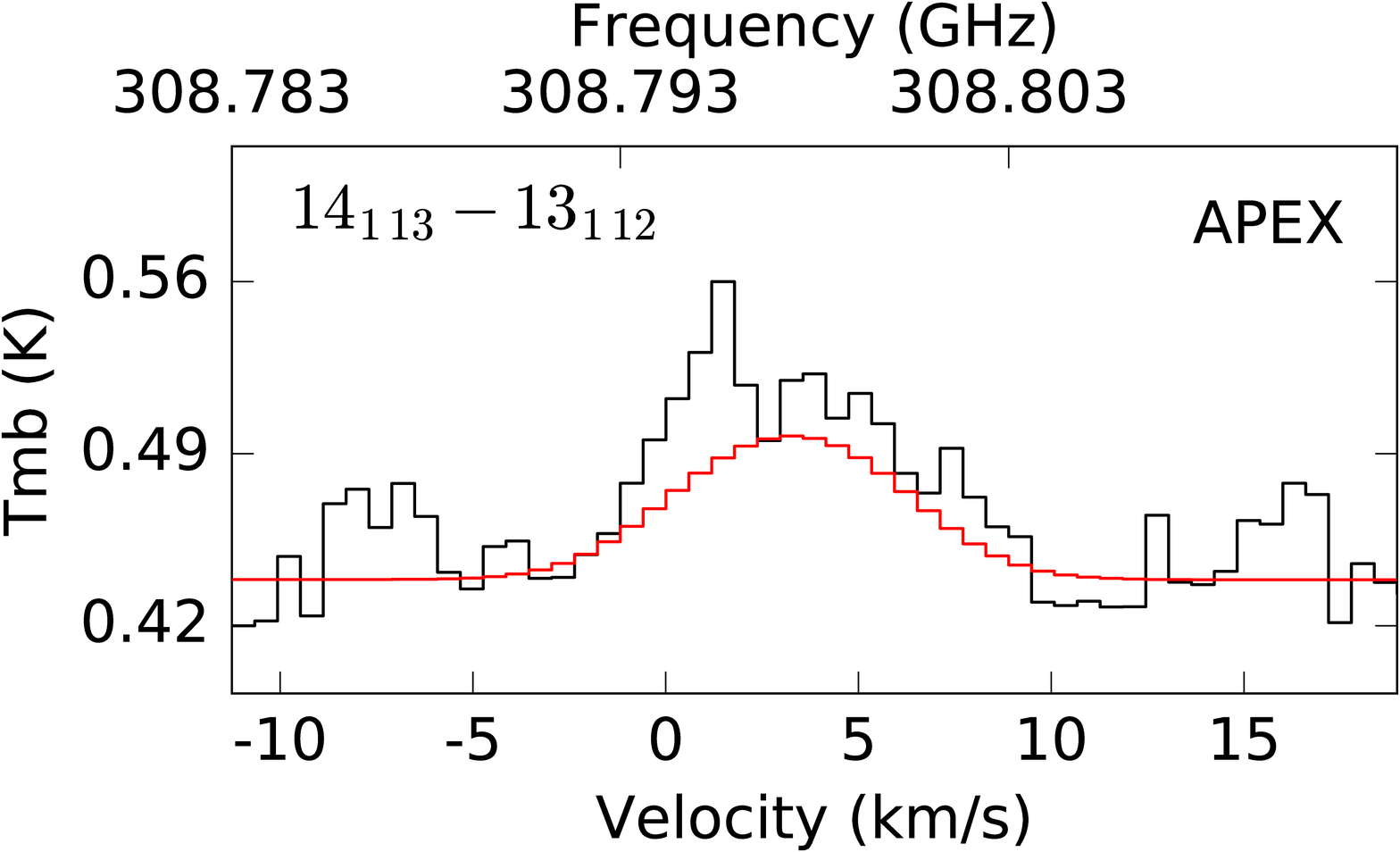}&\includegraphics[width=0.315\textwidth, trim= 0 0 0 0, clip]{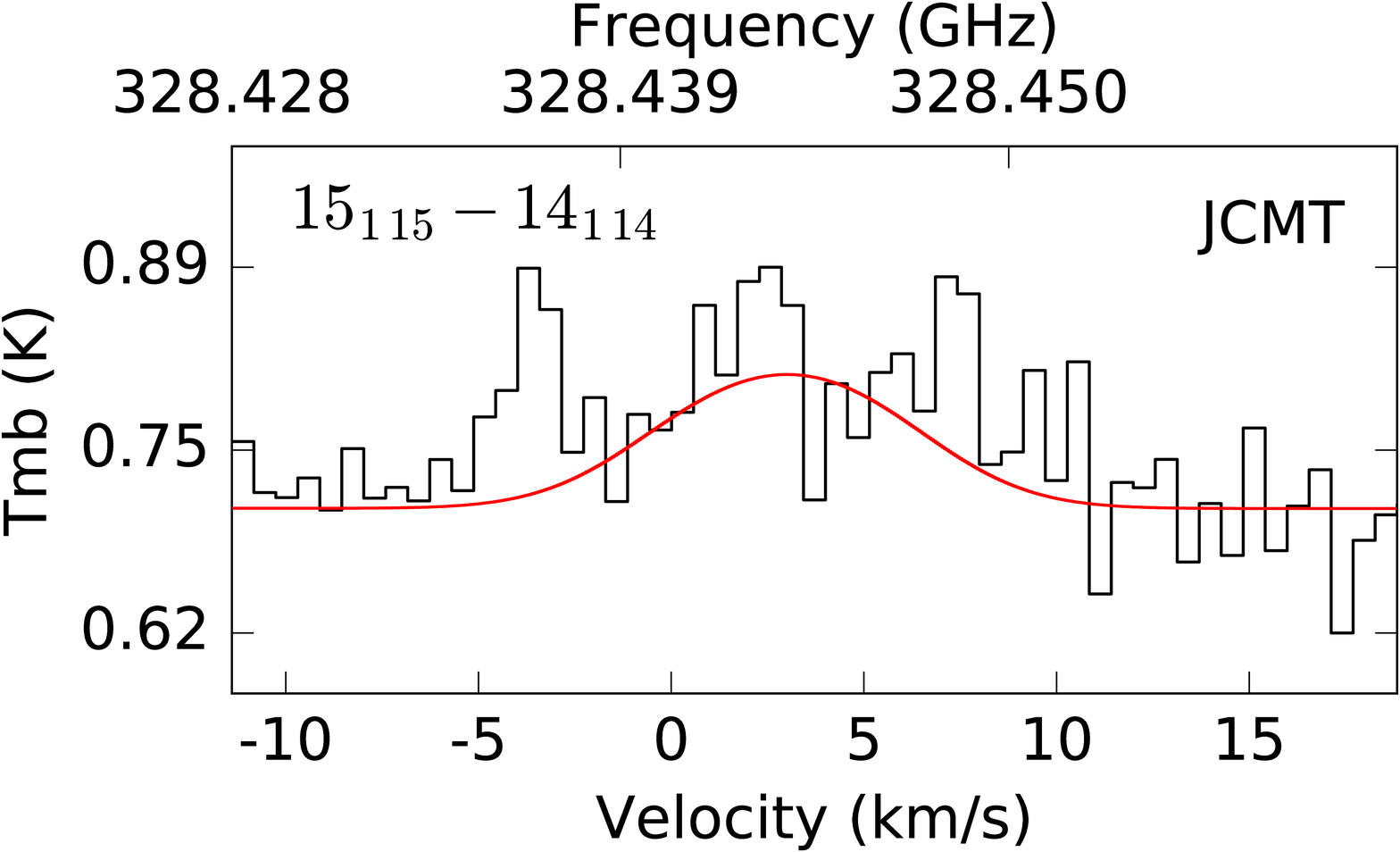}\\
\includegraphics[width=0.315\textwidth,trim = 0 0 0 0,clip]{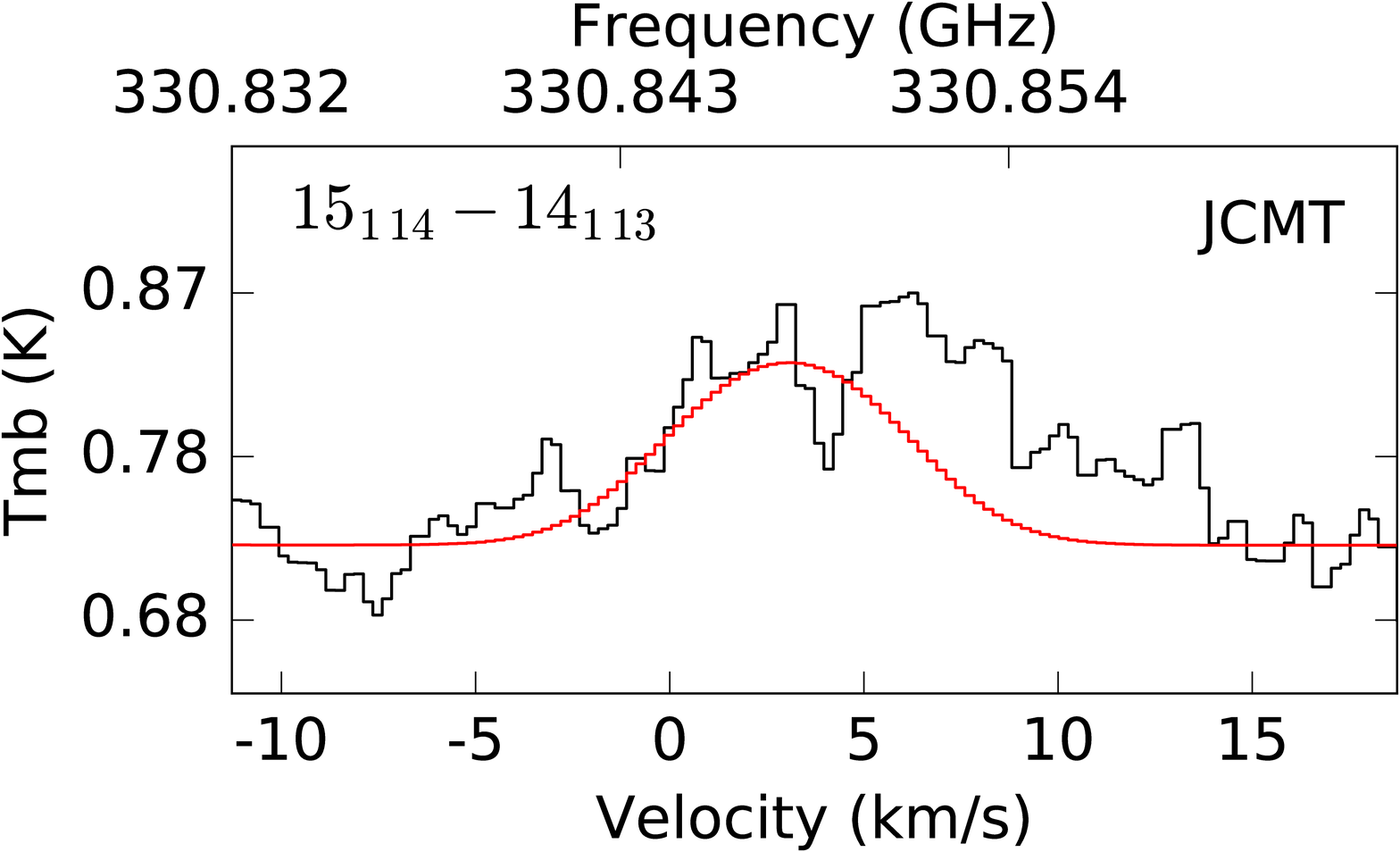}  &\includegraphics[width=0.315\textwidth,trim = 0 0 0 0,clip]{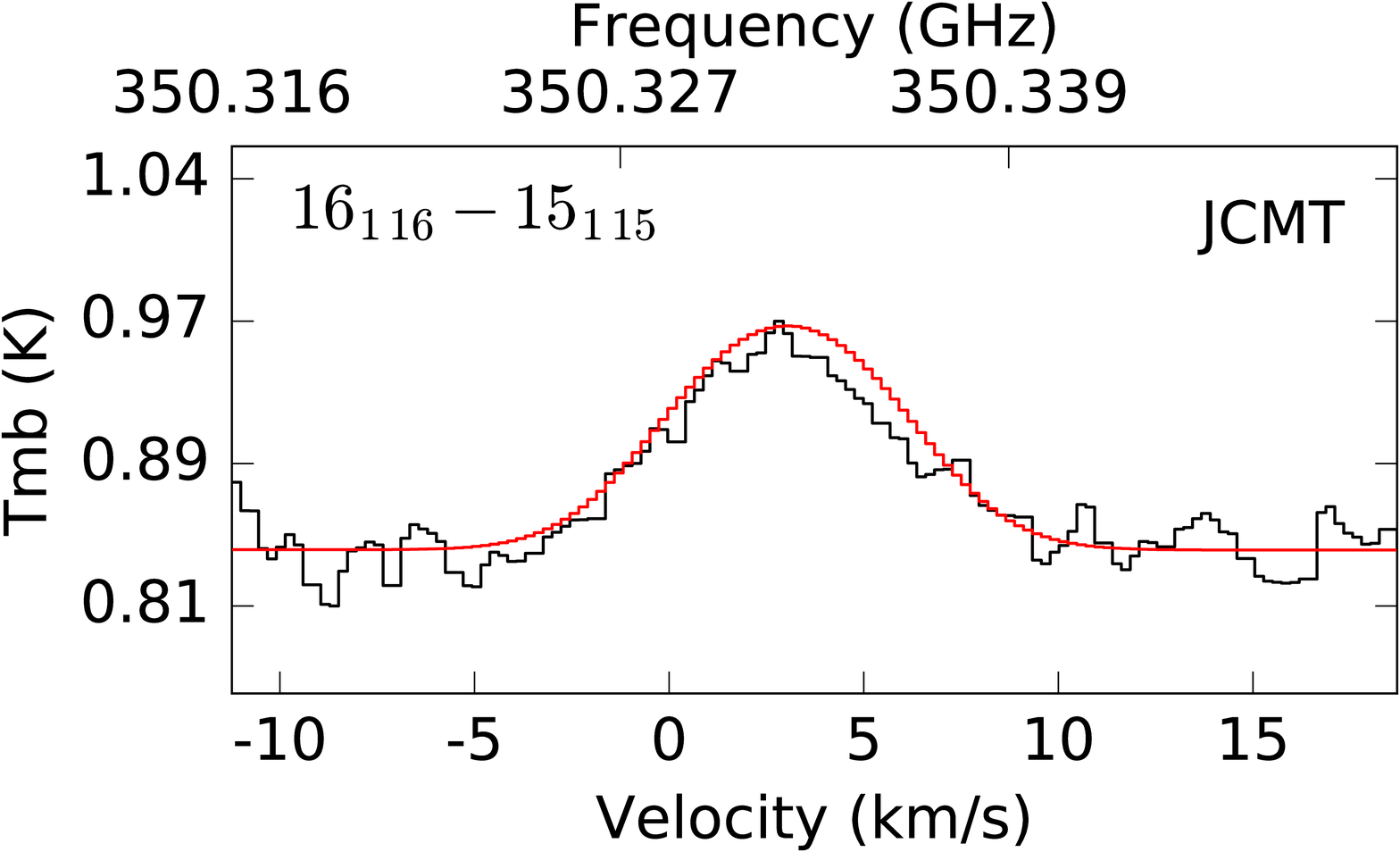} &\includegraphics[width=0.315\textwidth,trim = 0 0 0 0,clip]{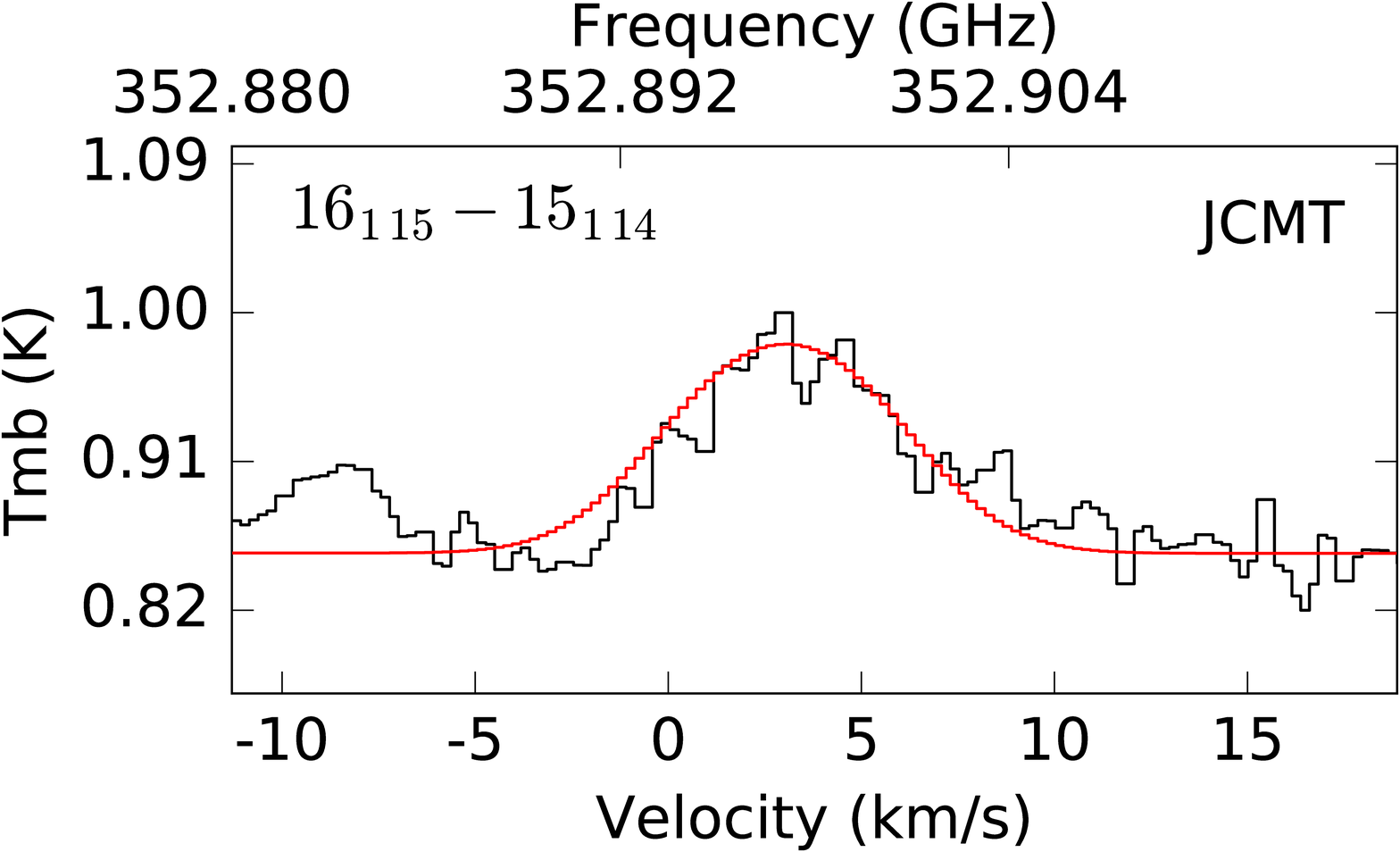}\\
\includegraphics[width=0.315\textwidth,trim = 0 0 0 0,clip]{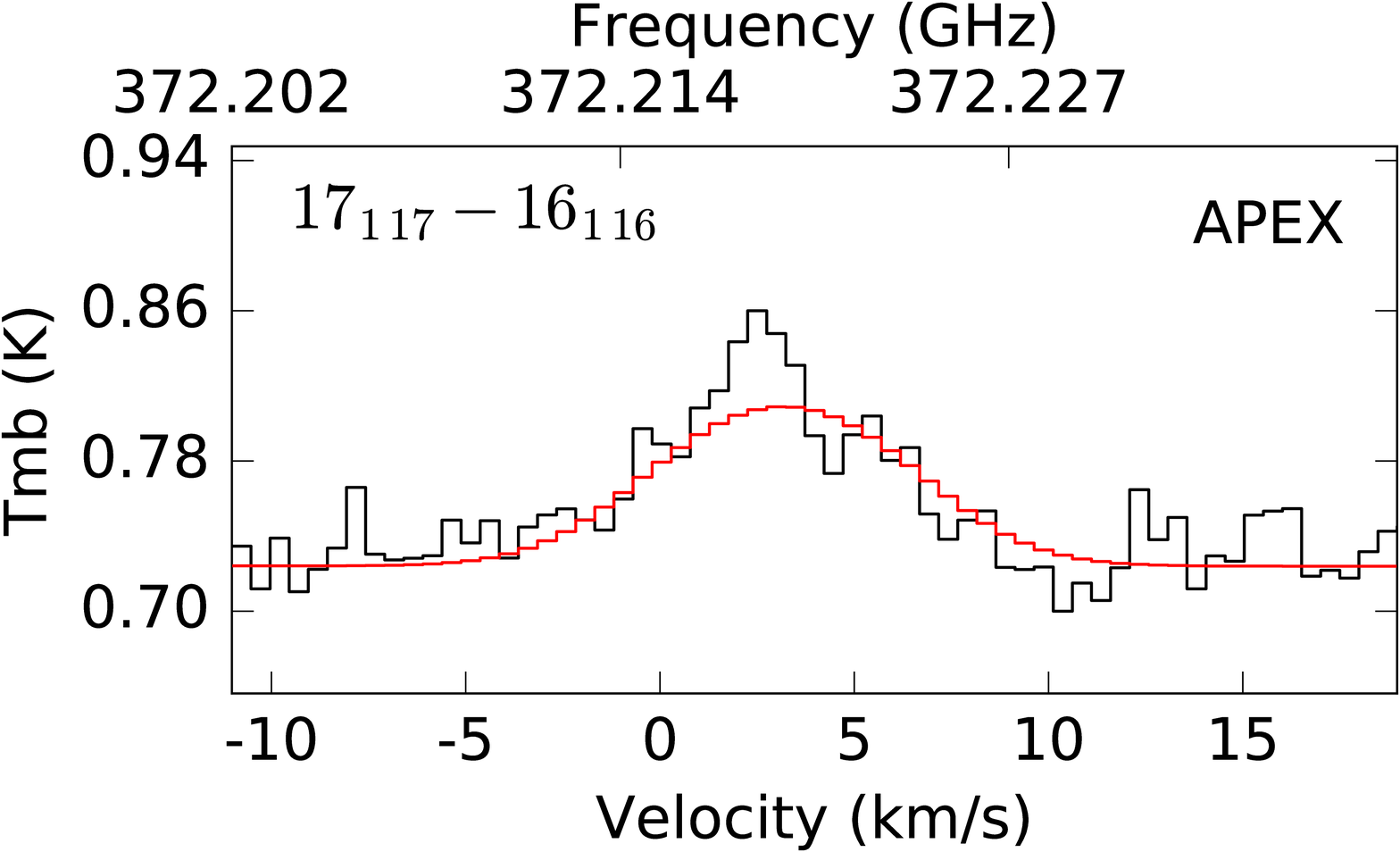} & \includegraphics[width=0.315\textwidth, trim= 0 0 0 0, clip]{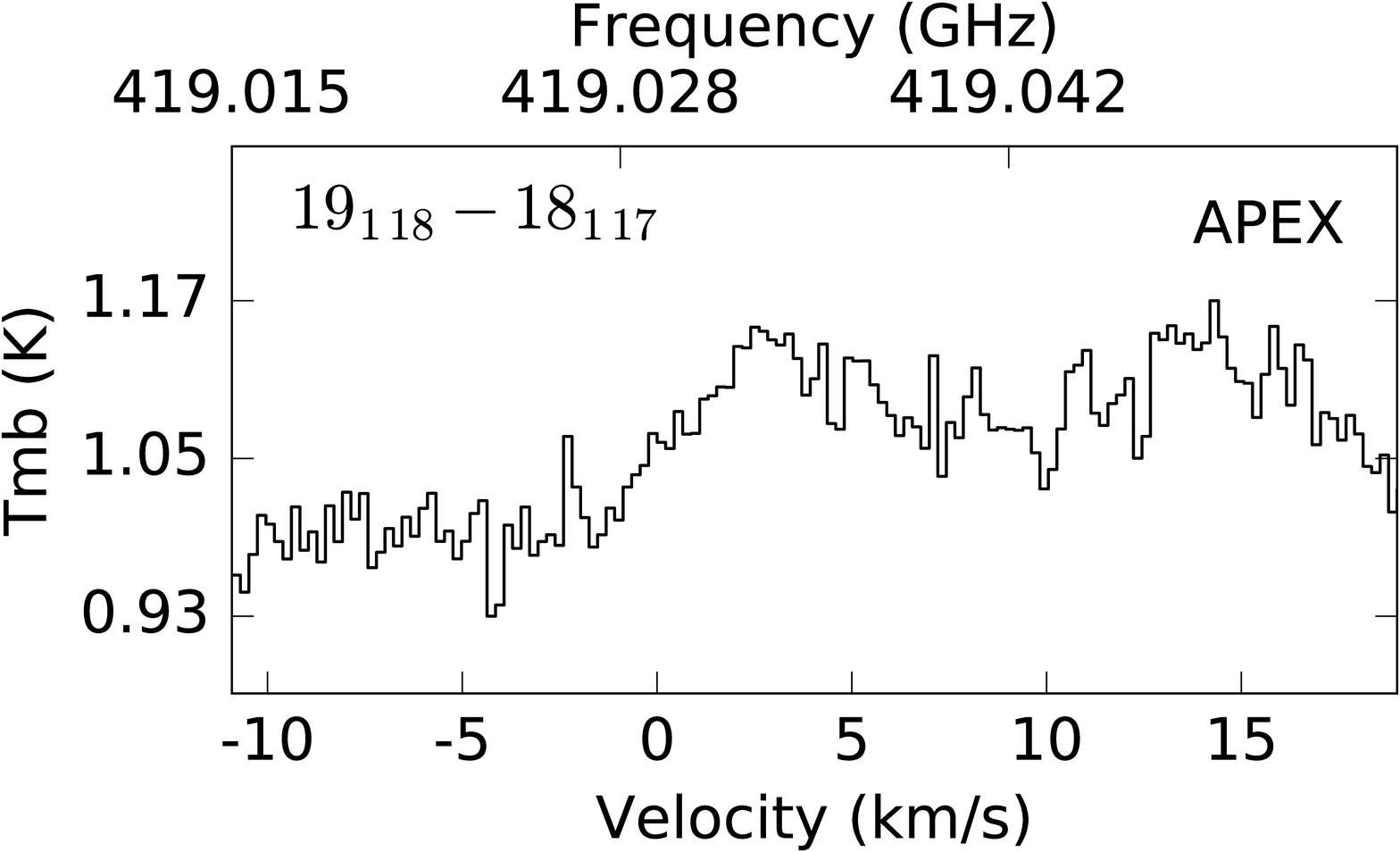} &\includegraphics[width=0.315\textwidth,trim = 0 0 0 0,clip]{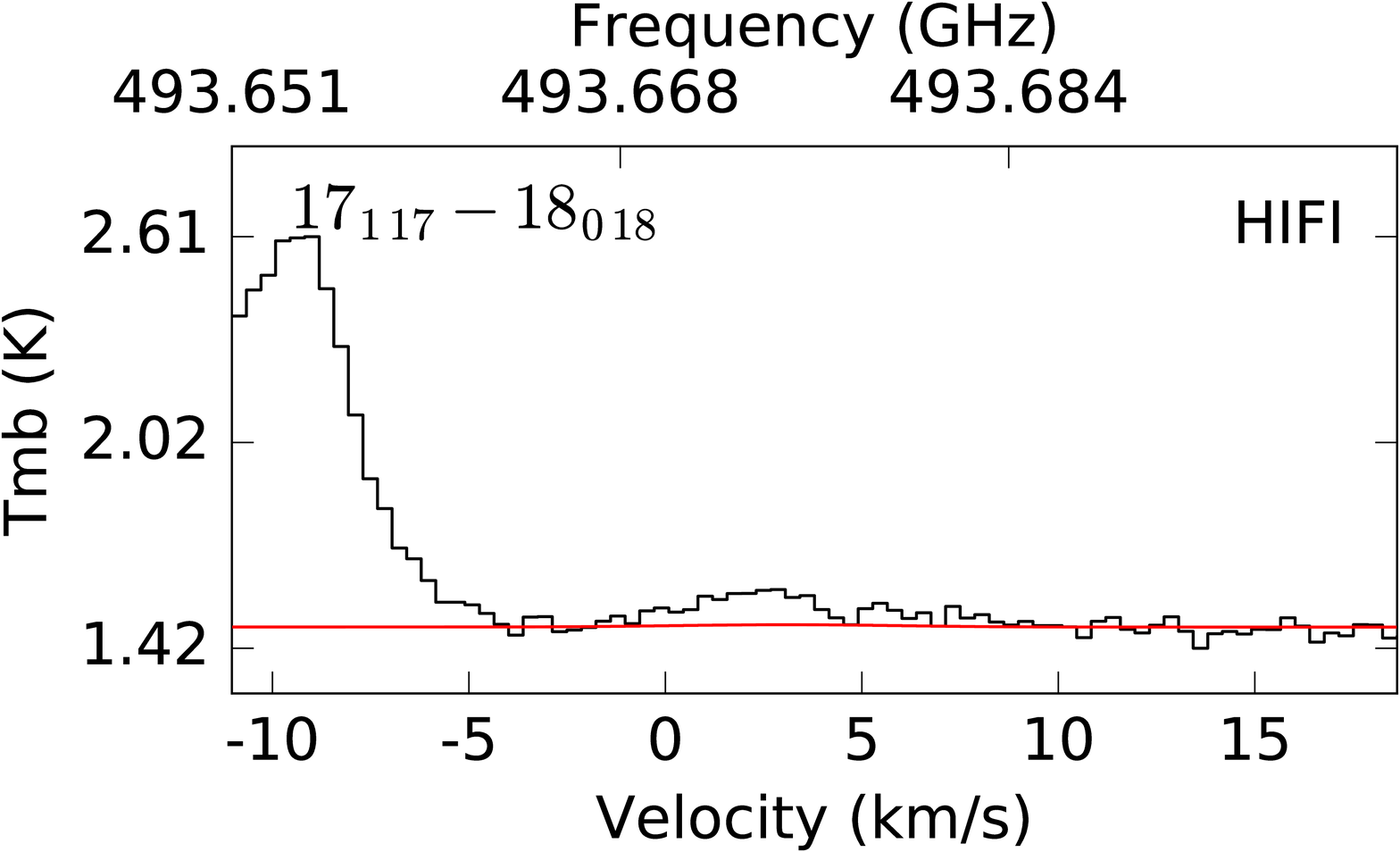}\\  
\includegraphics[width=0.315\textwidth,trim = 0 0 0 0,clip]{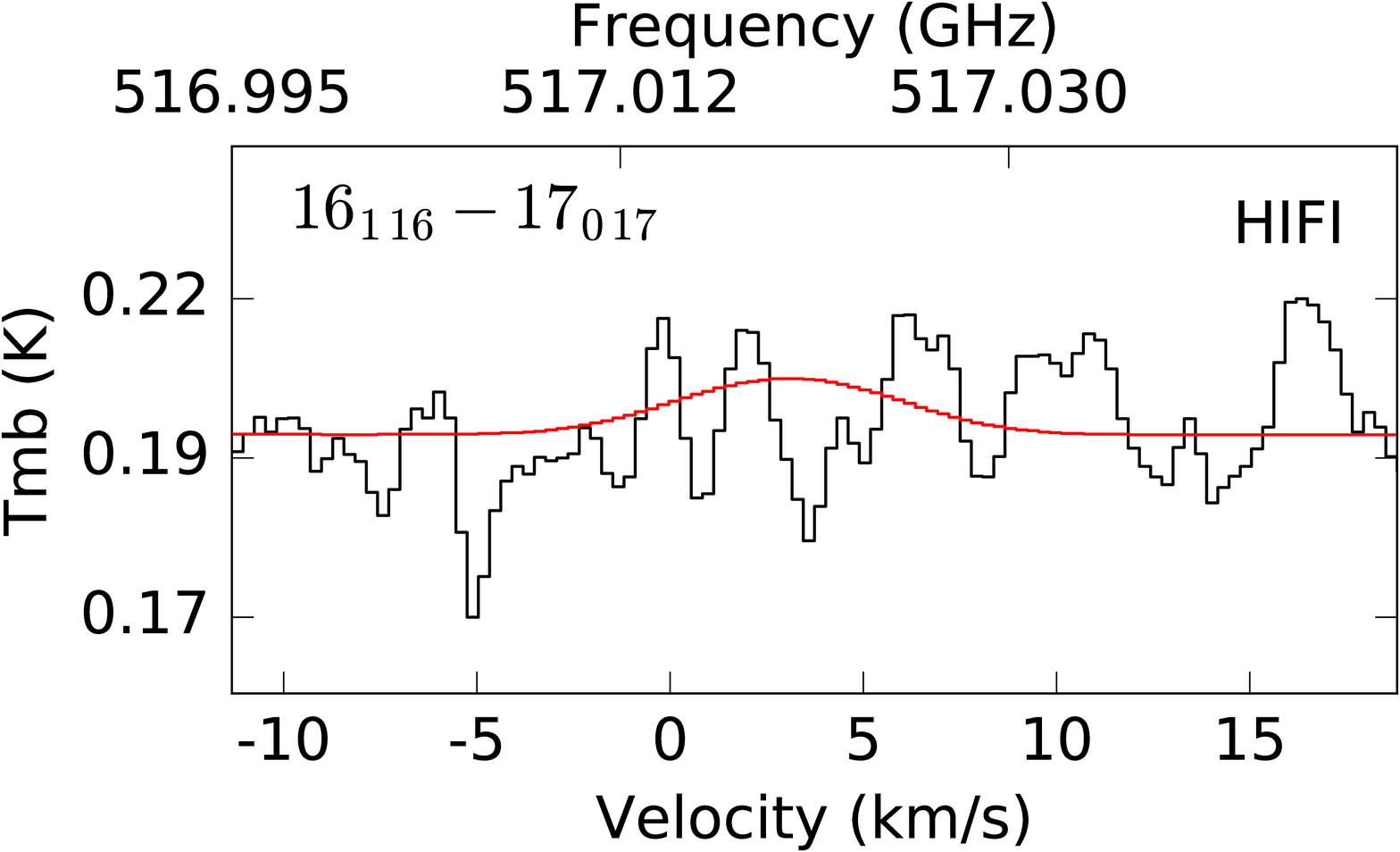} &\includegraphics[width=0.315\textwidth,trim = 0 0 0 0,clip]{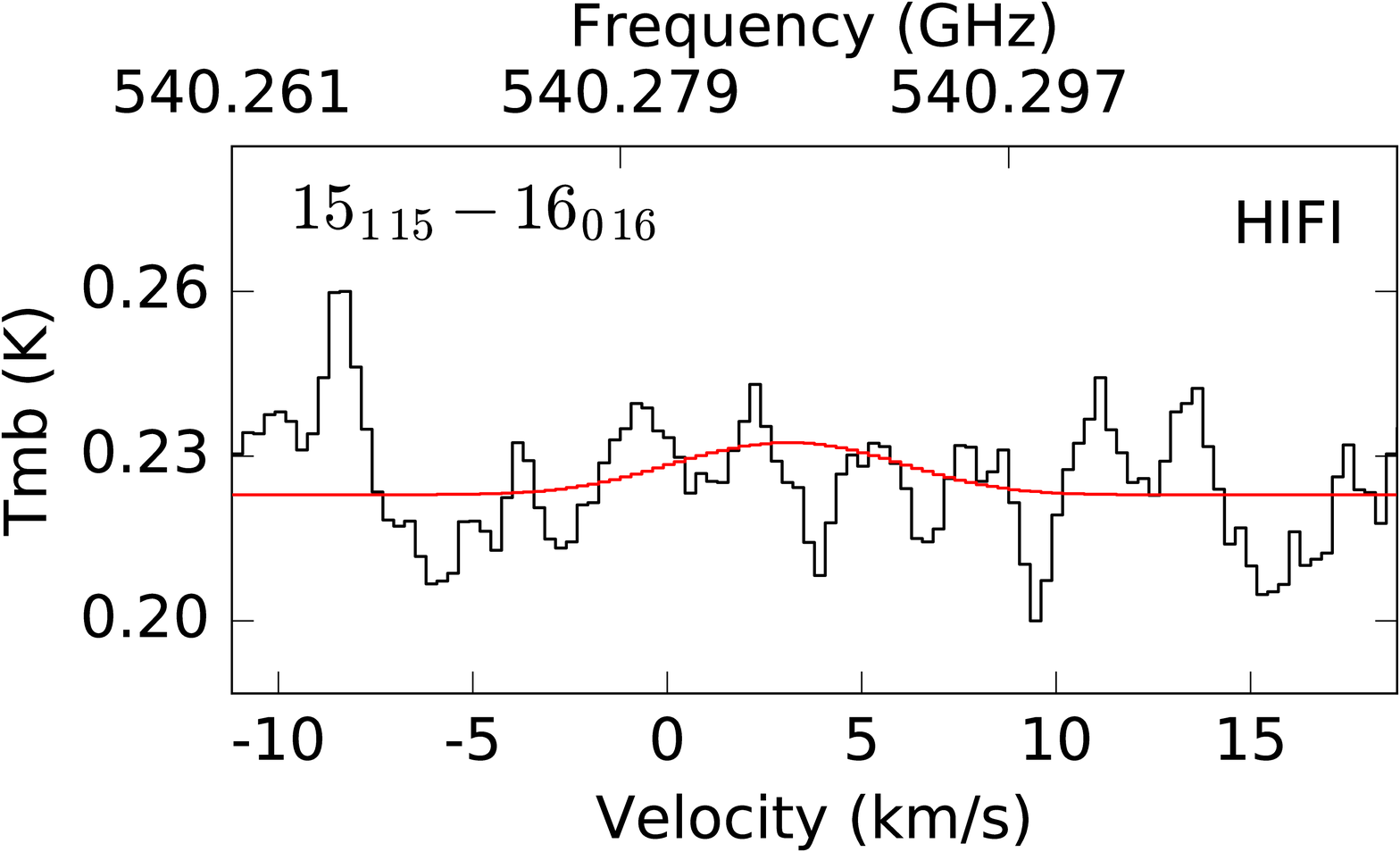}&\includegraphics[width=0.315\textwidth,trim = 0 0 0 0,clip]{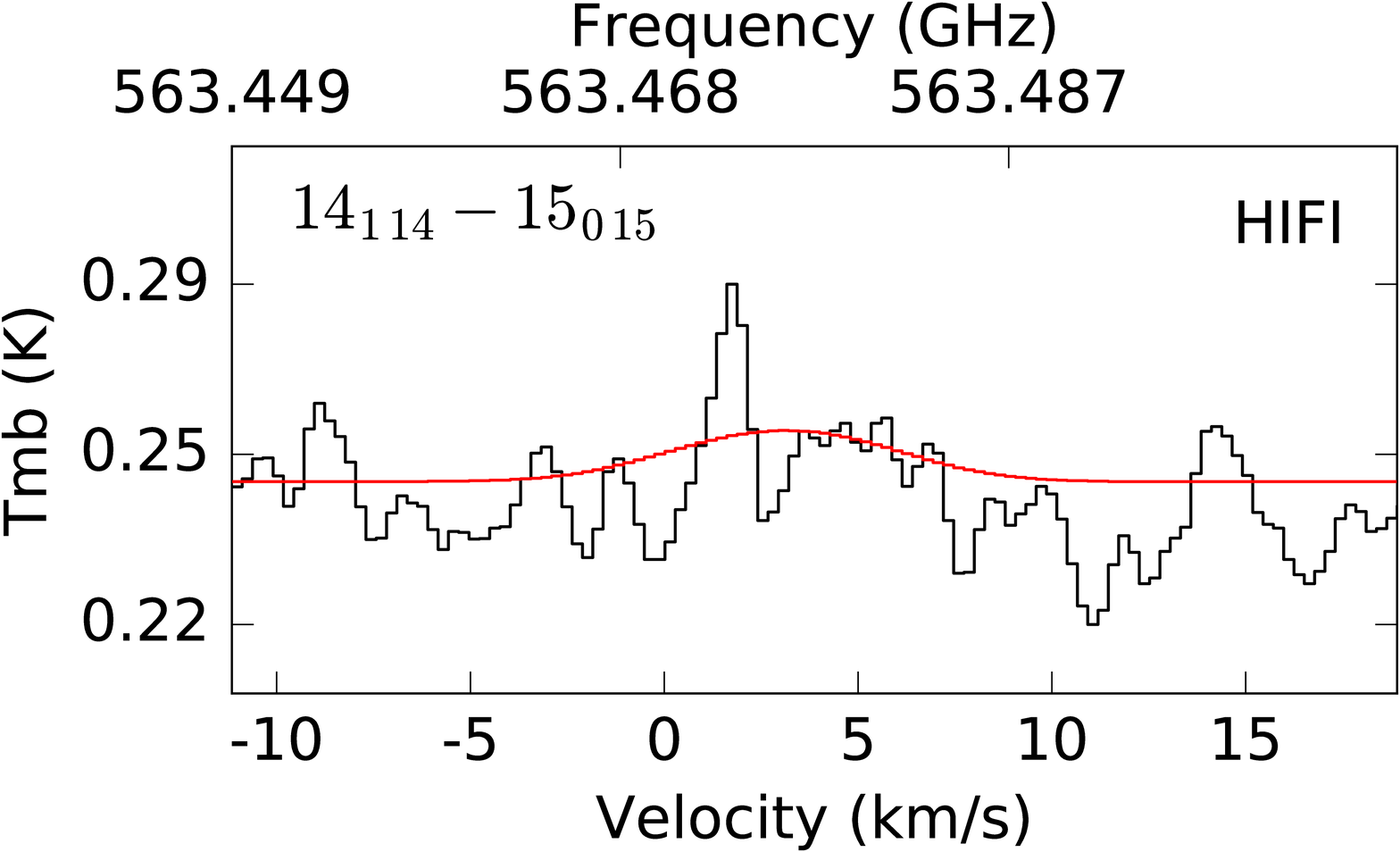} \\
\includegraphics[width=0.315\textwidth, trim= 0 0 0 0, clip]{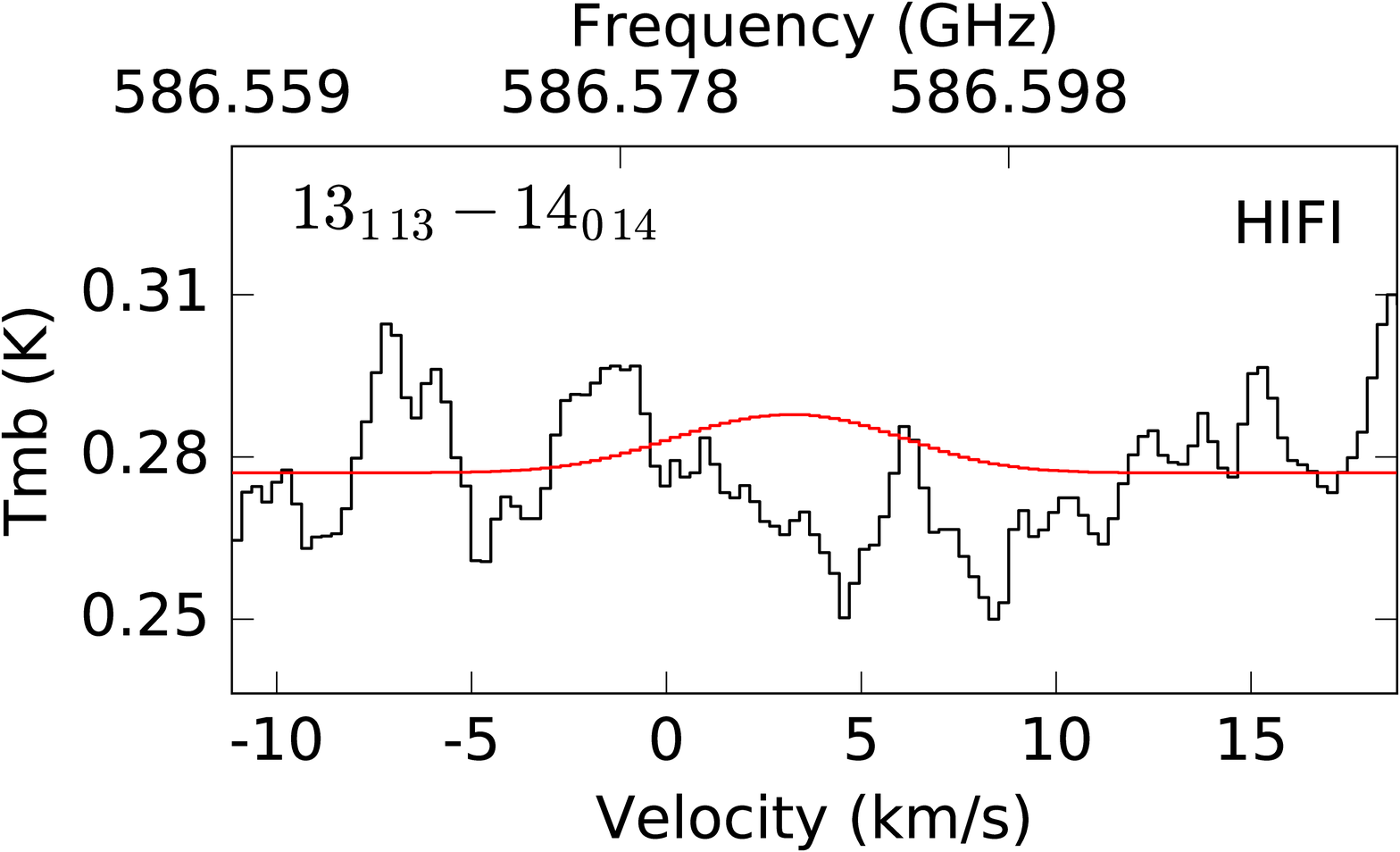} &\includegraphics[width=0.315\textwidth,trim = 0 0 0 0,clip]{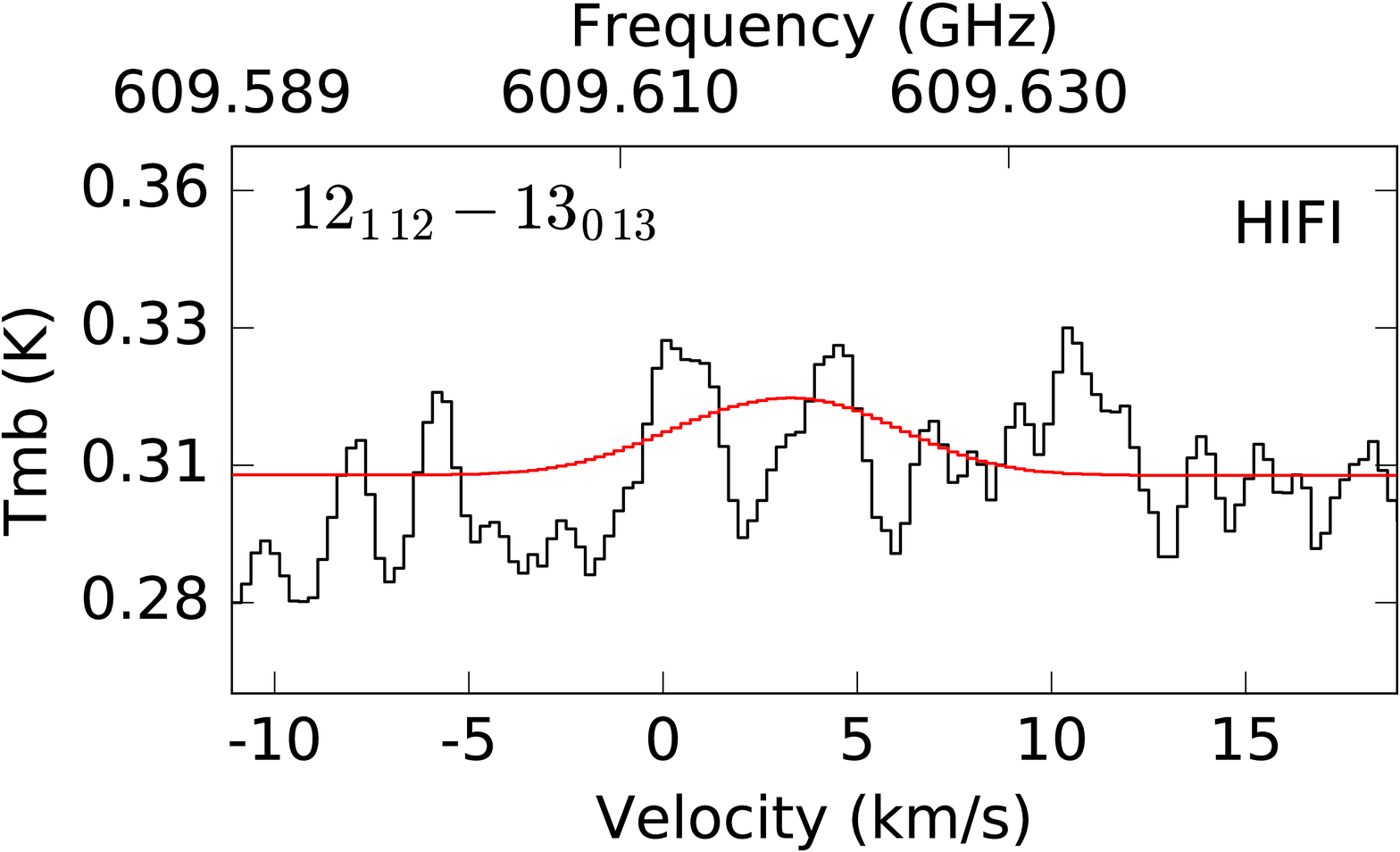}  &\includegraphics[width=0.315\textwidth,trim = 0 0 0 0,clip]{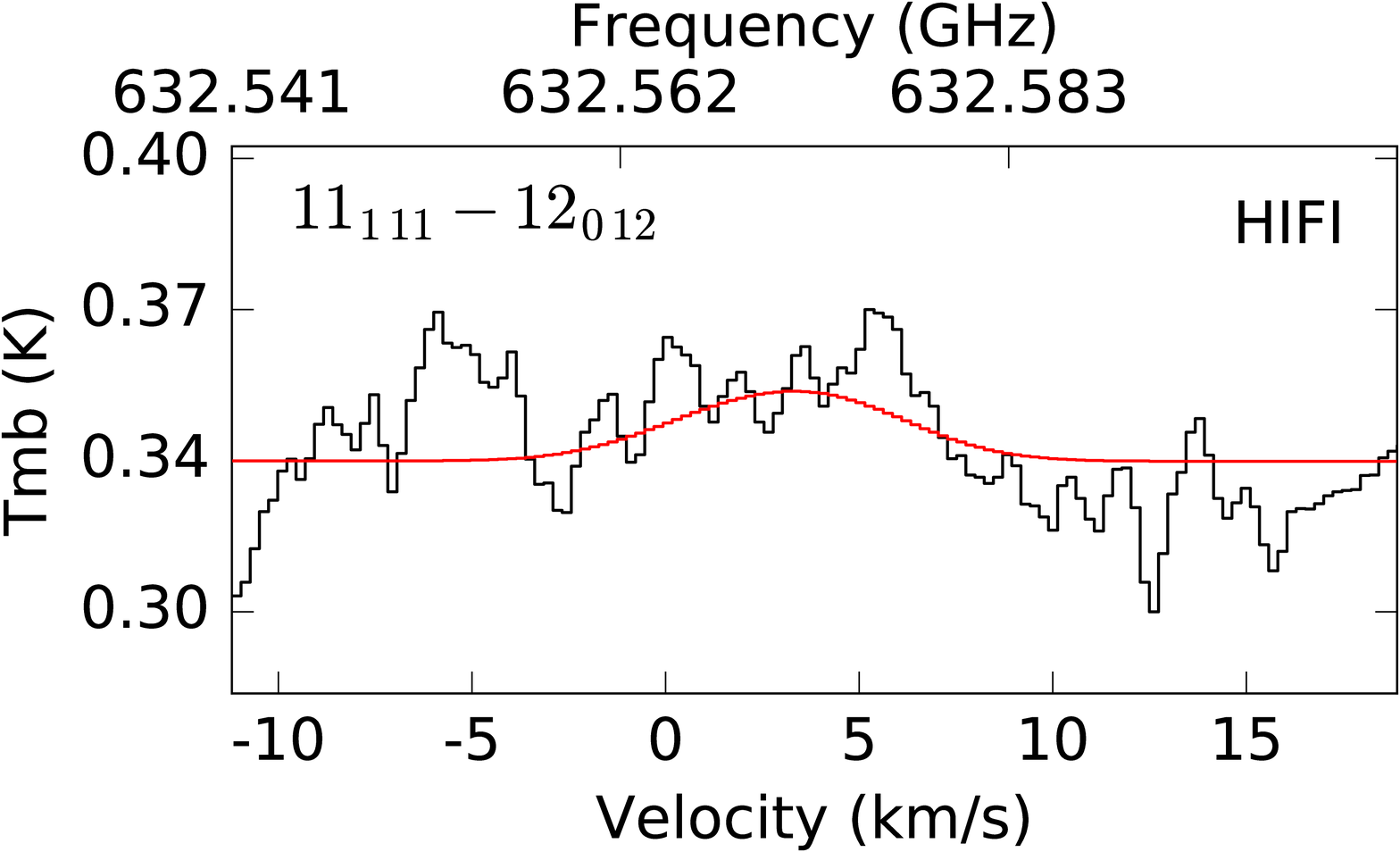}\\ 
\includegraphics[width=0.315\textwidth,trim = 0 0 0 0,clip]{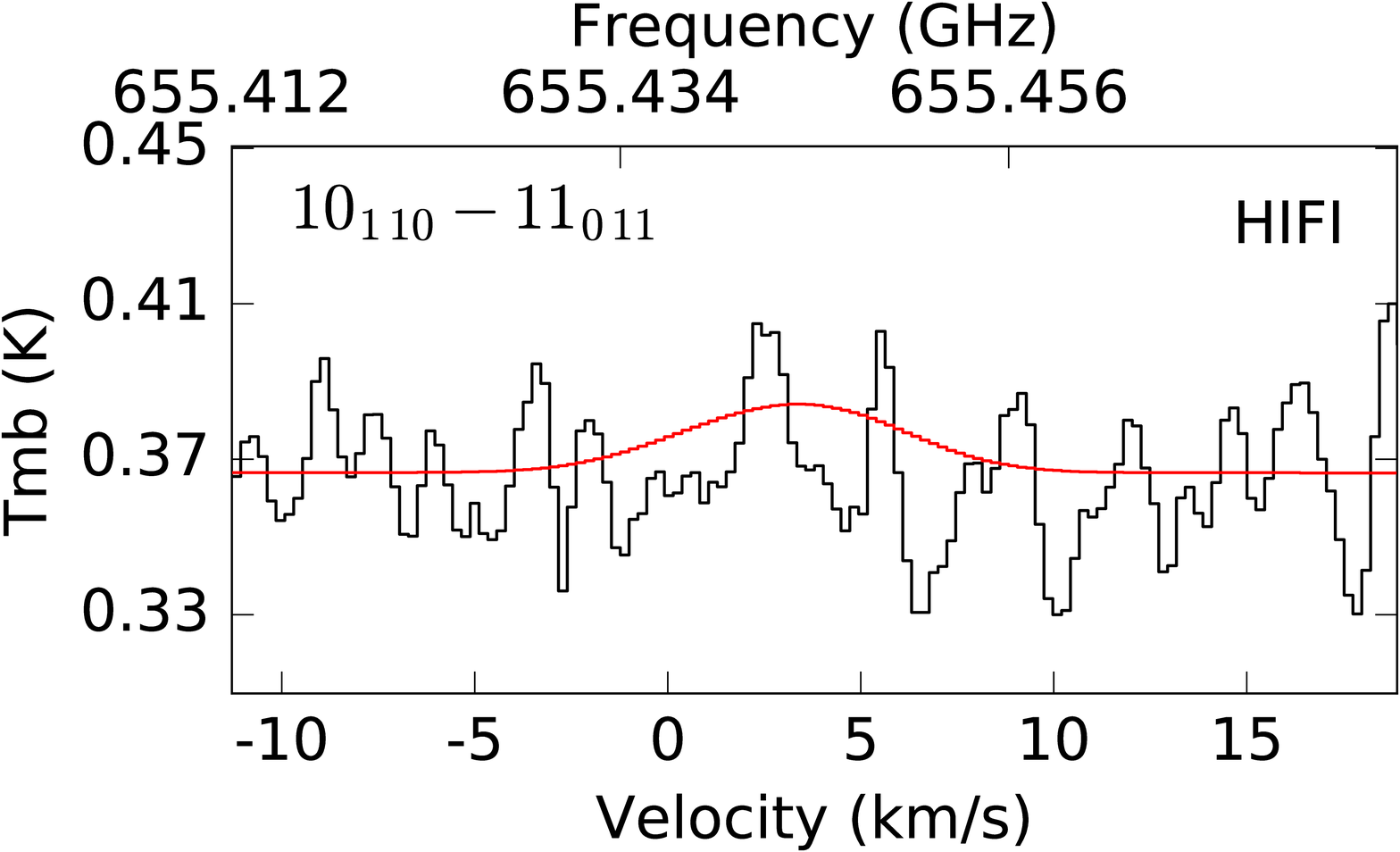}&\includegraphics[width=0.315\textwidth,trim = 0 0 0 0,clip]{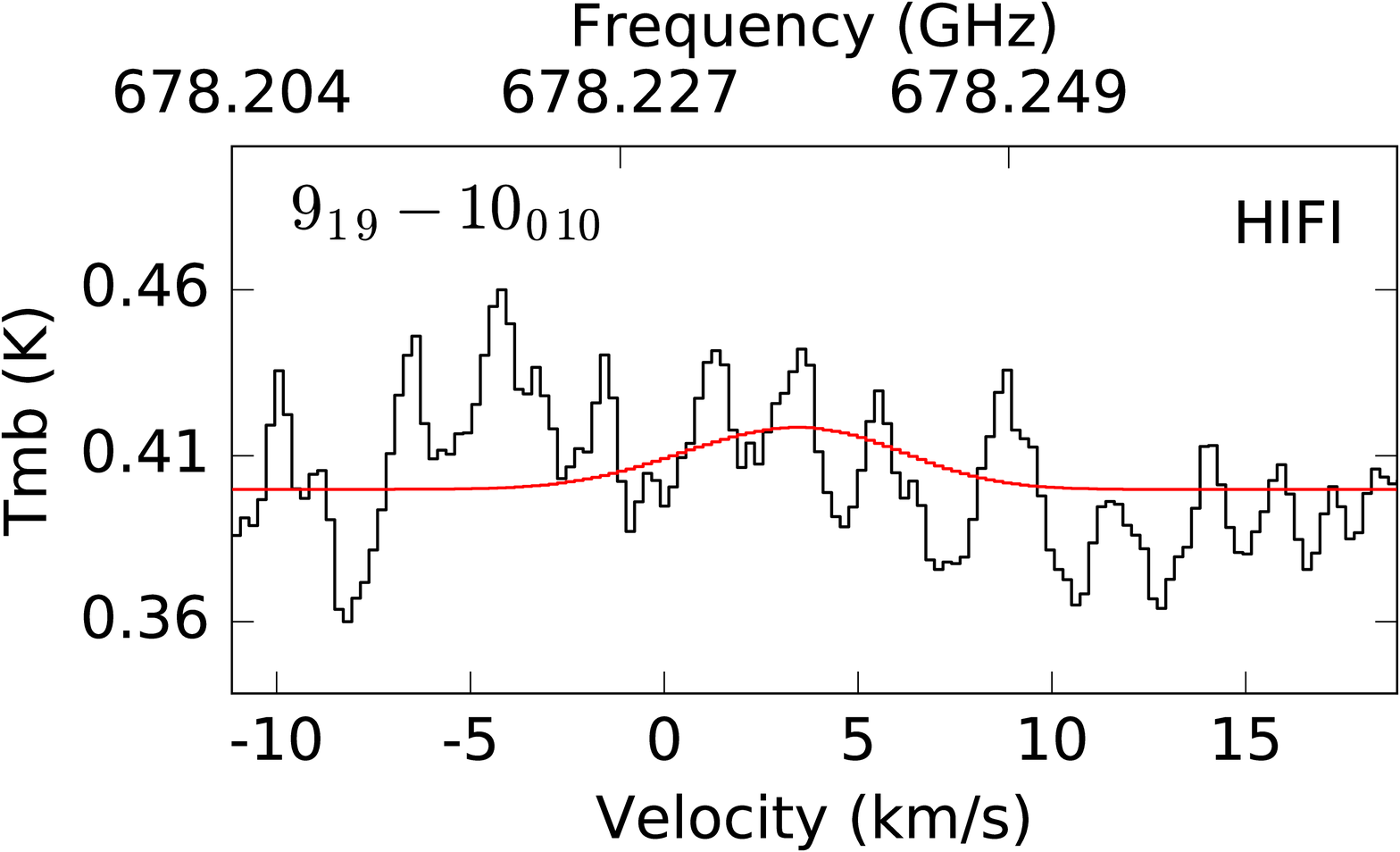}&\includegraphics[width=0.315\textwidth, trim= 0 0 0 0, clip]{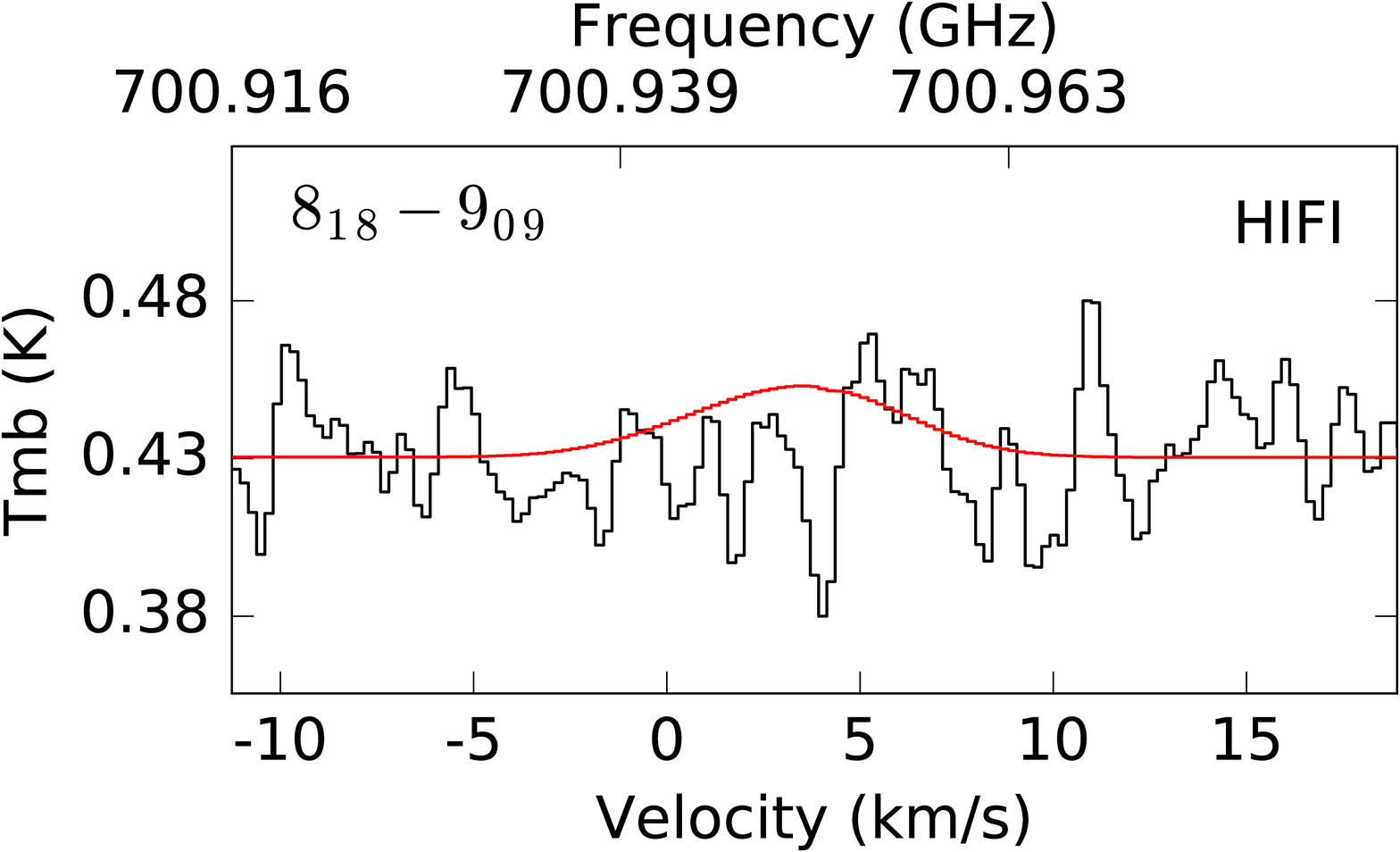}\\

\end{tabular}
\label{imagenes-cont3}
\end{figure*}

\begin{figure*}
\centering
\setlength\tabcolsep{3.7pt}
\contcaption{}
\label{imagenes-cont2}
\begin{tabular}{c c c}
\includegraphics[width=0.315\textwidth,trim = 0 0 0 0,clip]{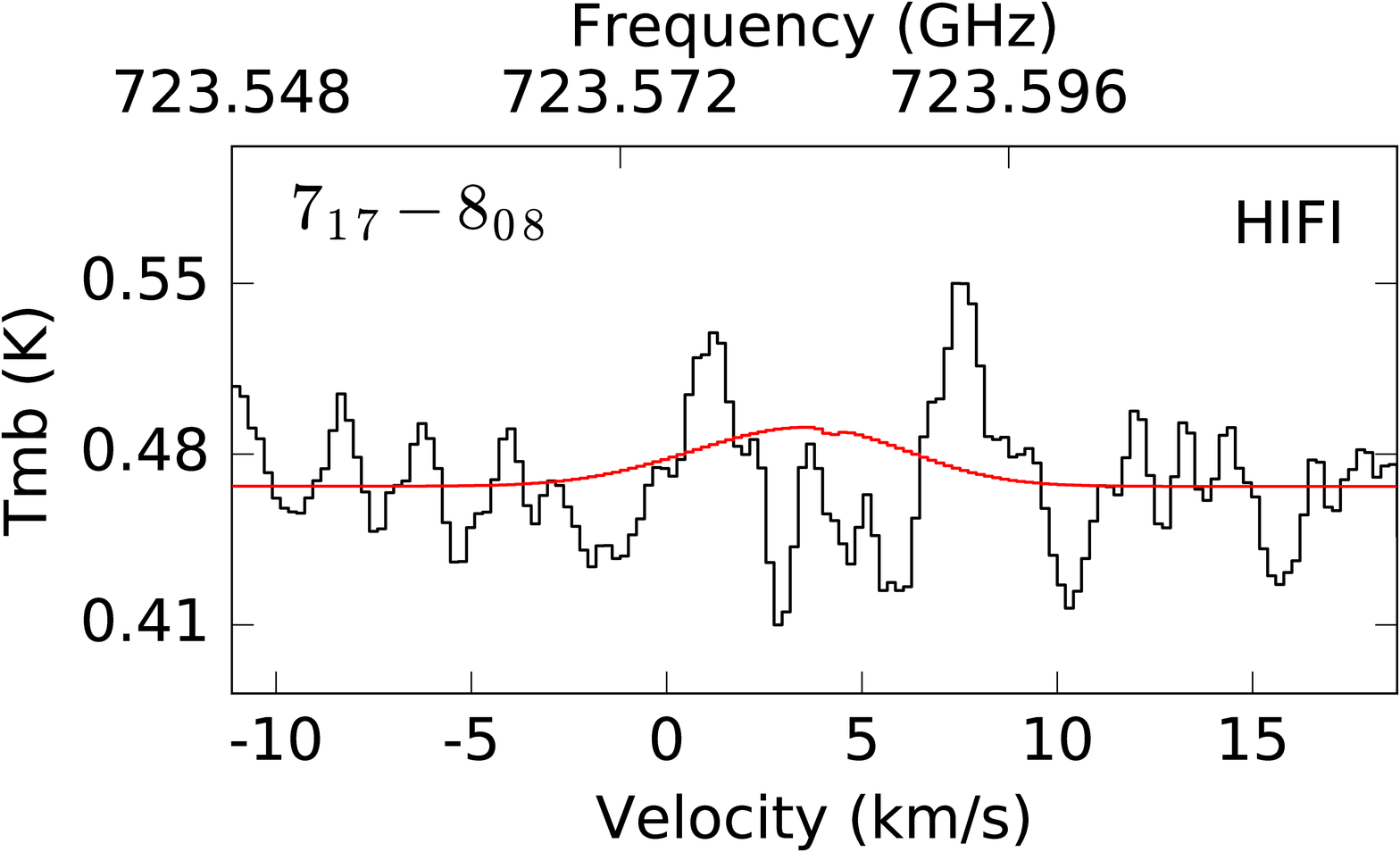}  &\includegraphics[width=0.315\textwidth,trim = 0 0 0 0,clip]{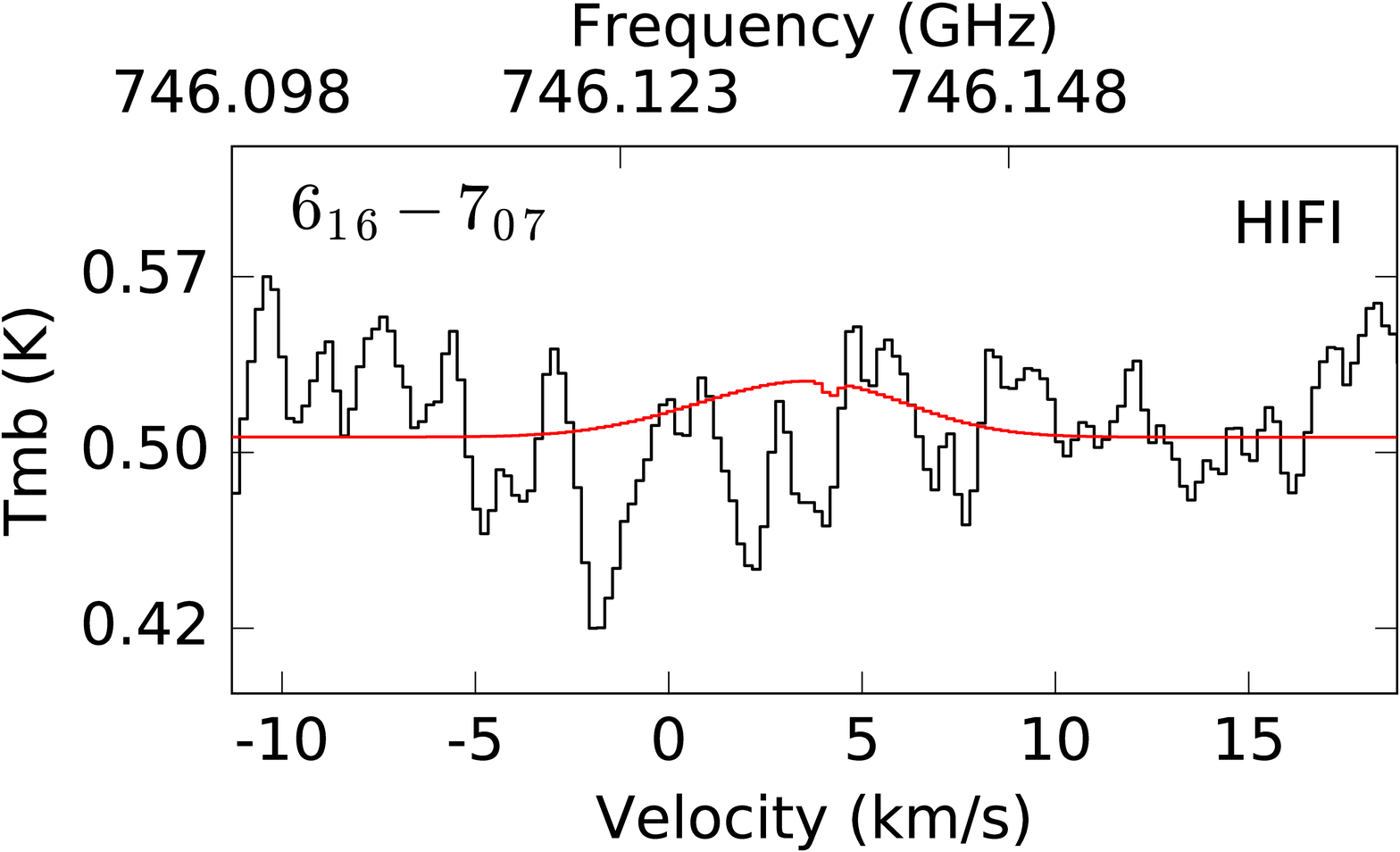} &\includegraphics[width=0.315\textwidth,trim = 0 0 0 0,clip]{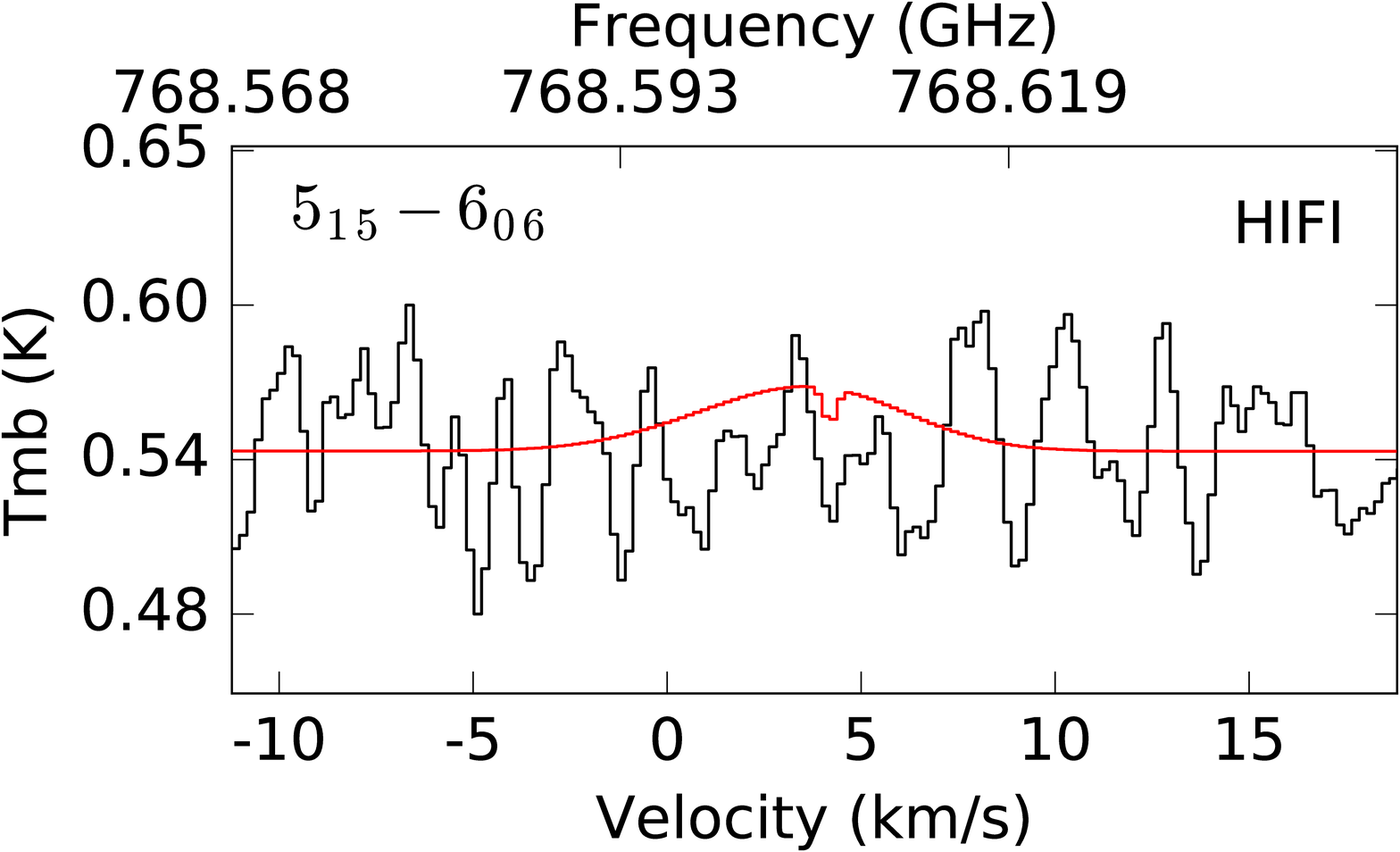}\\
\includegraphics[width=0.315\textwidth,trim = 0 0 0 0,clip]{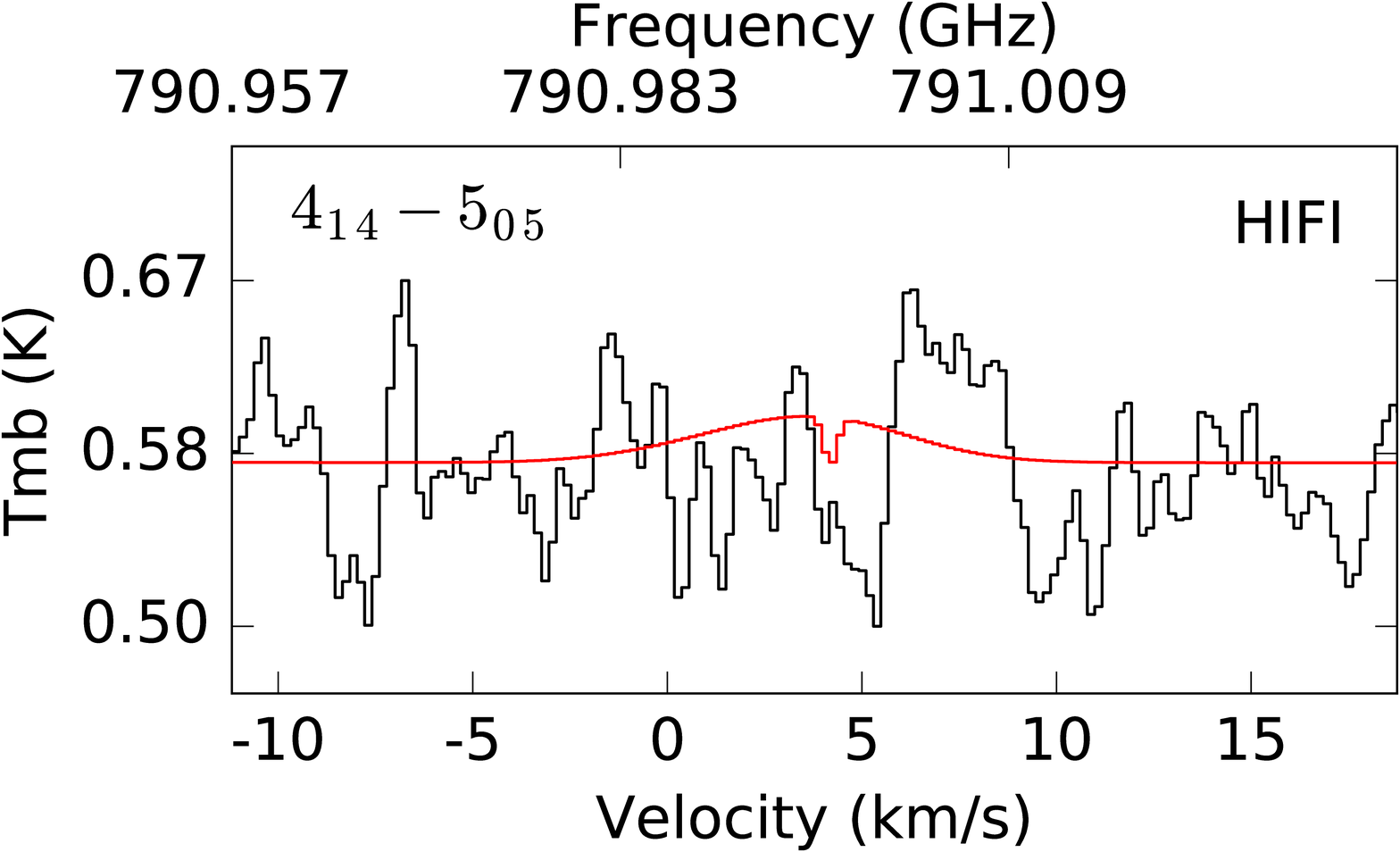}&\includegraphics[width=0.315\textwidth, trim= 0 0 0 0, clip]{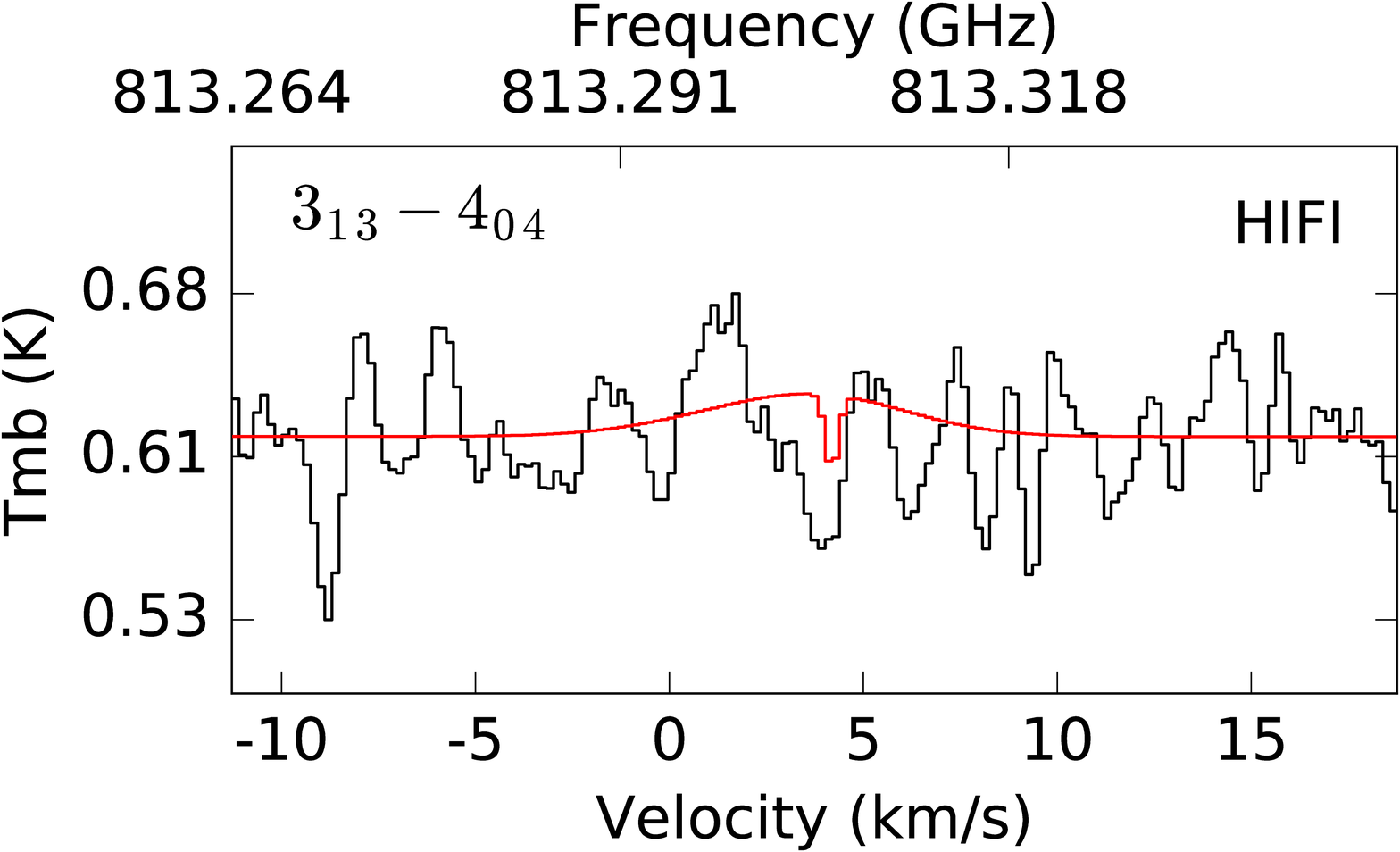} &\includegraphics[width=0.315\textwidth,trim = 0 0 0 0,clip]{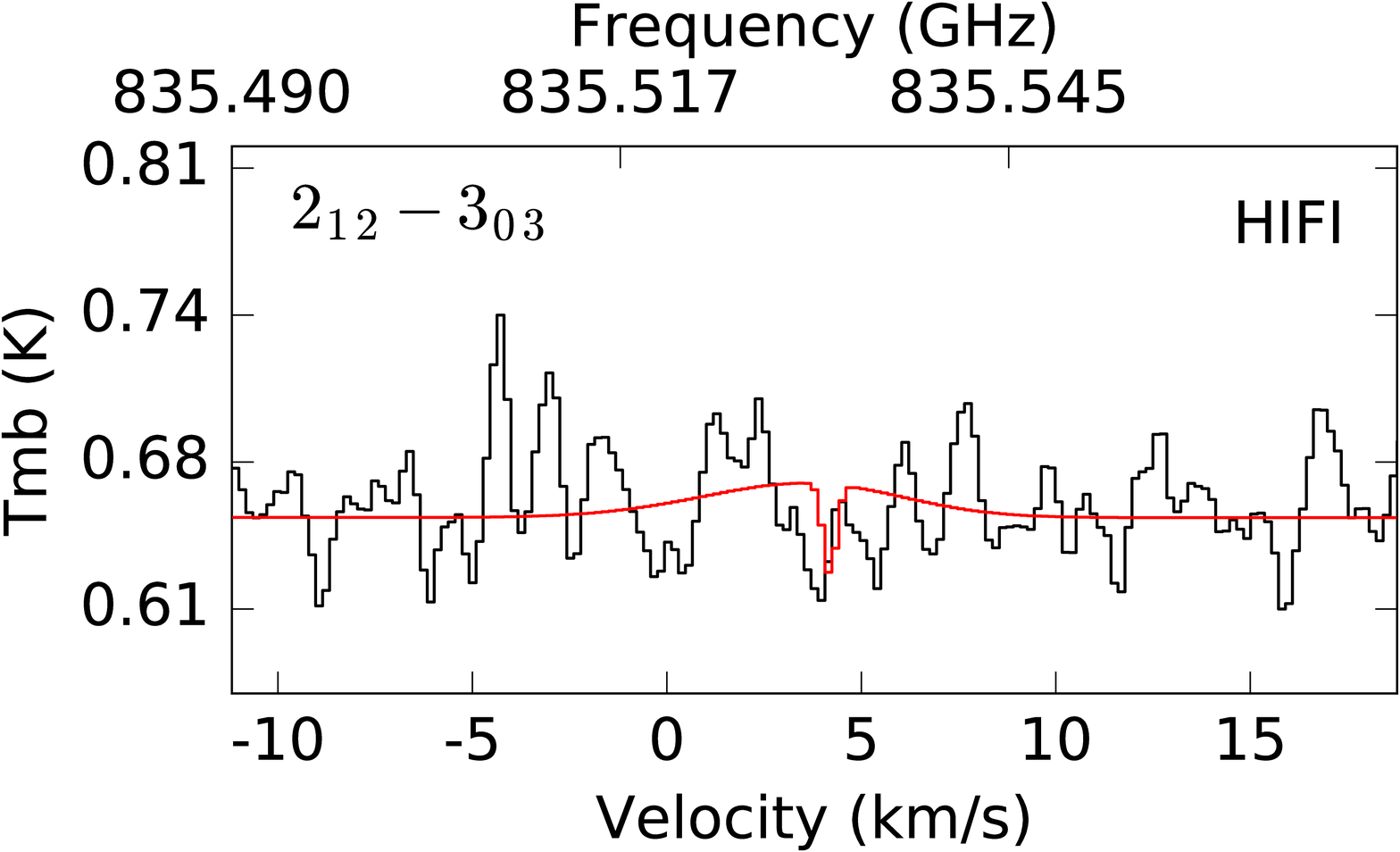}\\  
\includegraphics[width=0.315\textwidth,trim = 0 0 0 0,clip]{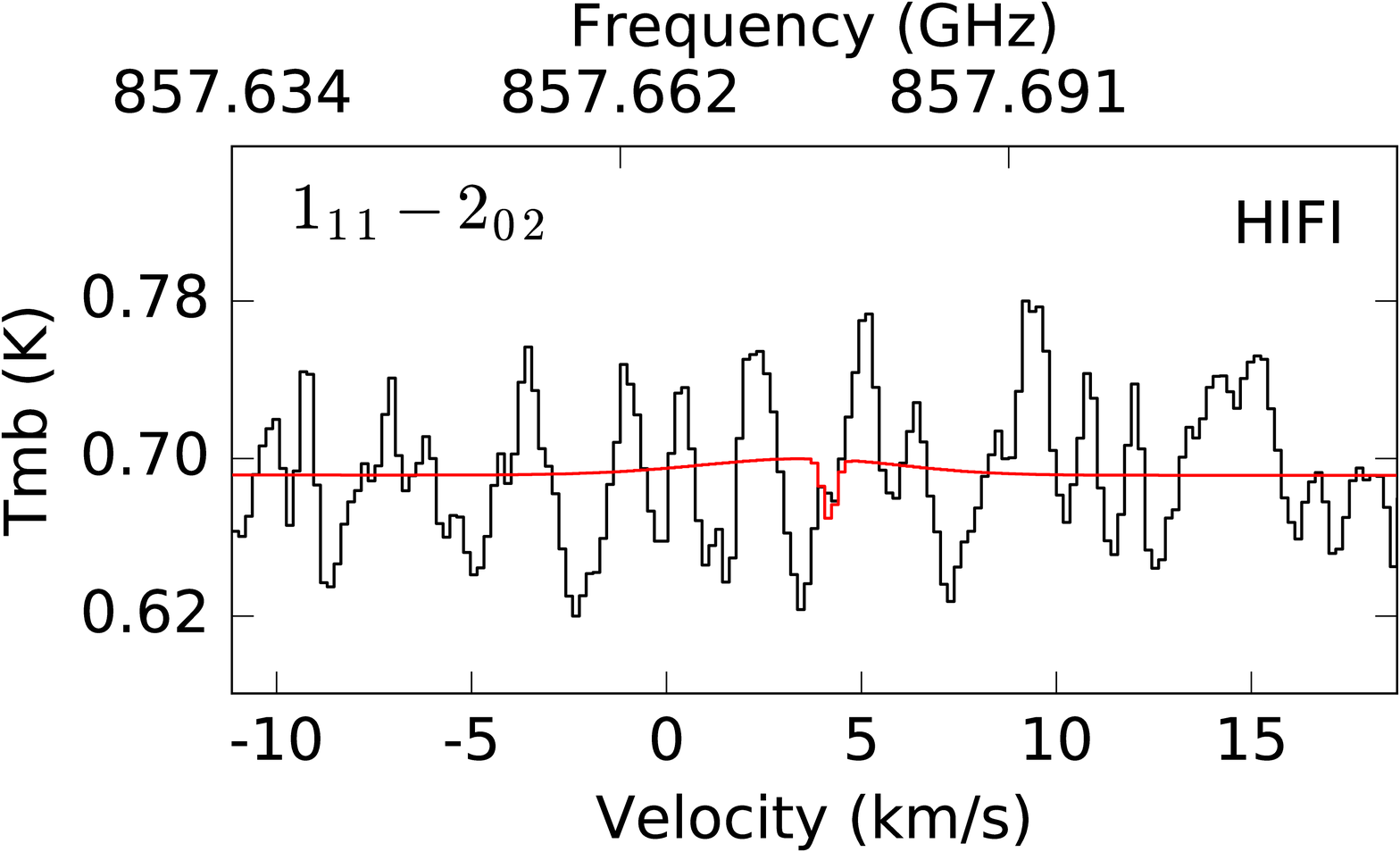} &\includegraphics[width=0.315\textwidth,trim = 0 0 0 0,clip]{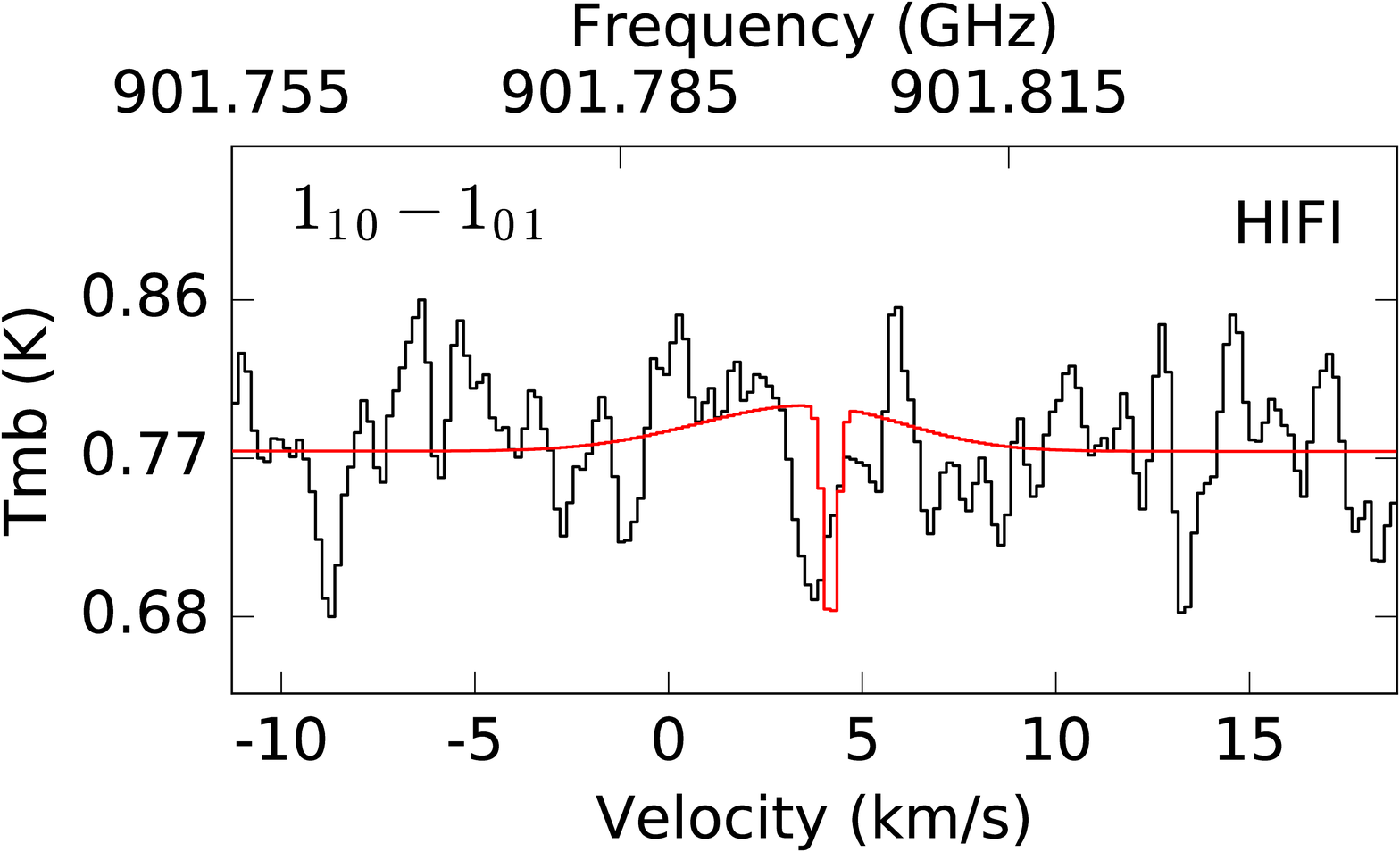}&\includegraphics[width=0.315\textwidth,trim = 0 0 0 0,clip]{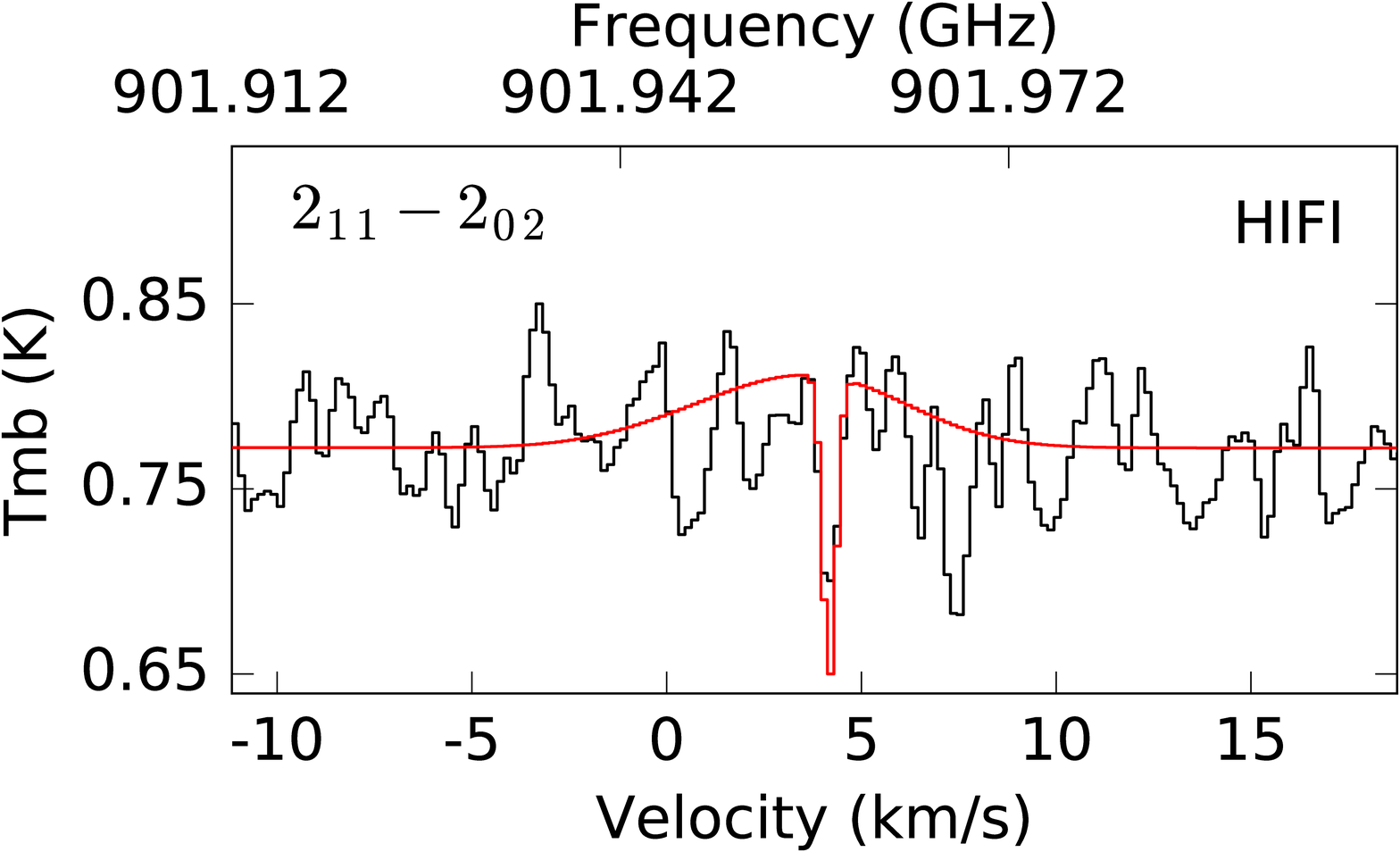} \\
\includegraphics[width=0.315\textwidth, trim= 0 0 0 0, clip]{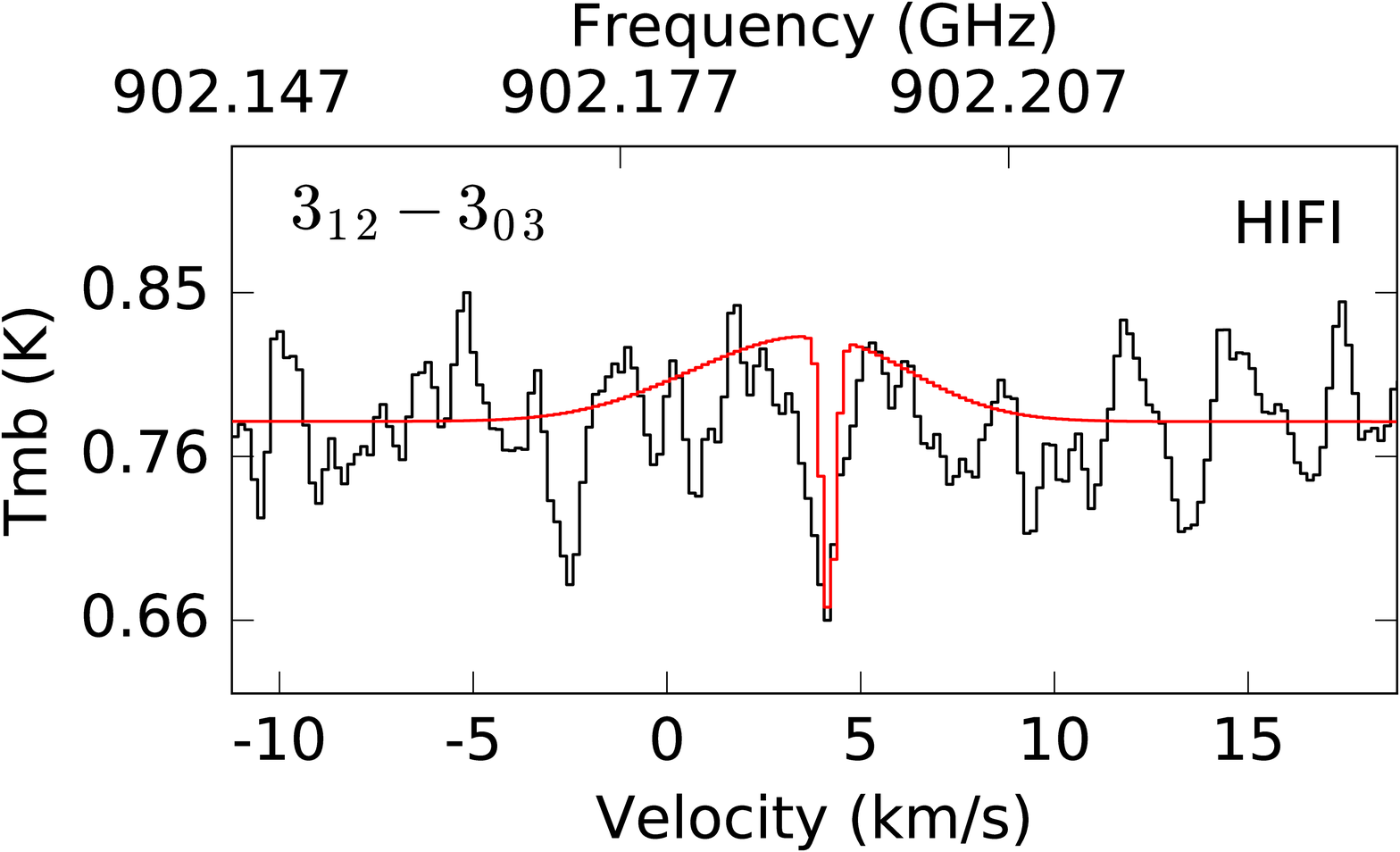} &\includegraphics[width=0.315\textwidth,trim = 0 0 0 0,clip]{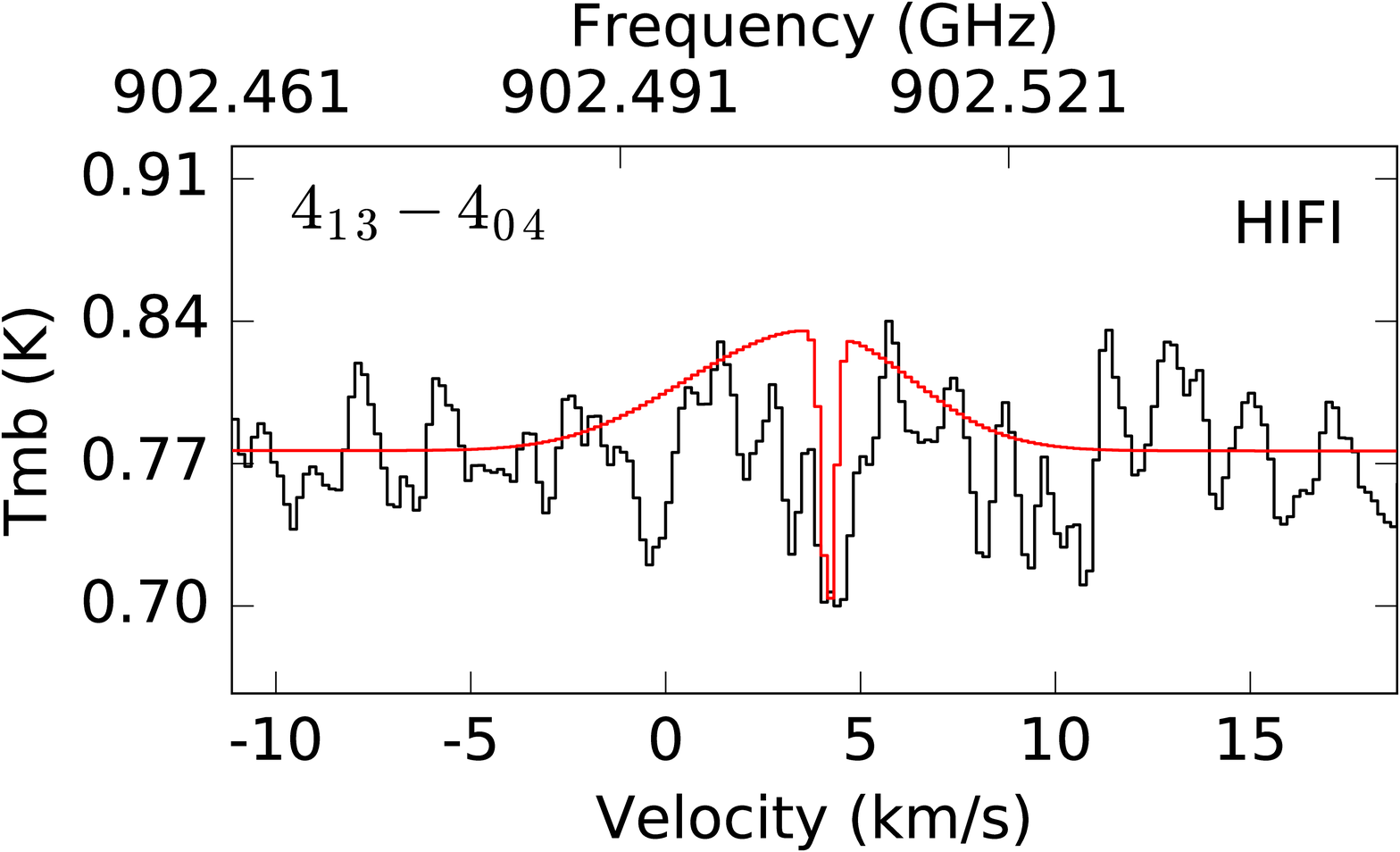}  &\includegraphics[width=0.315\textwidth,trim = 0 0 0 0,clip]{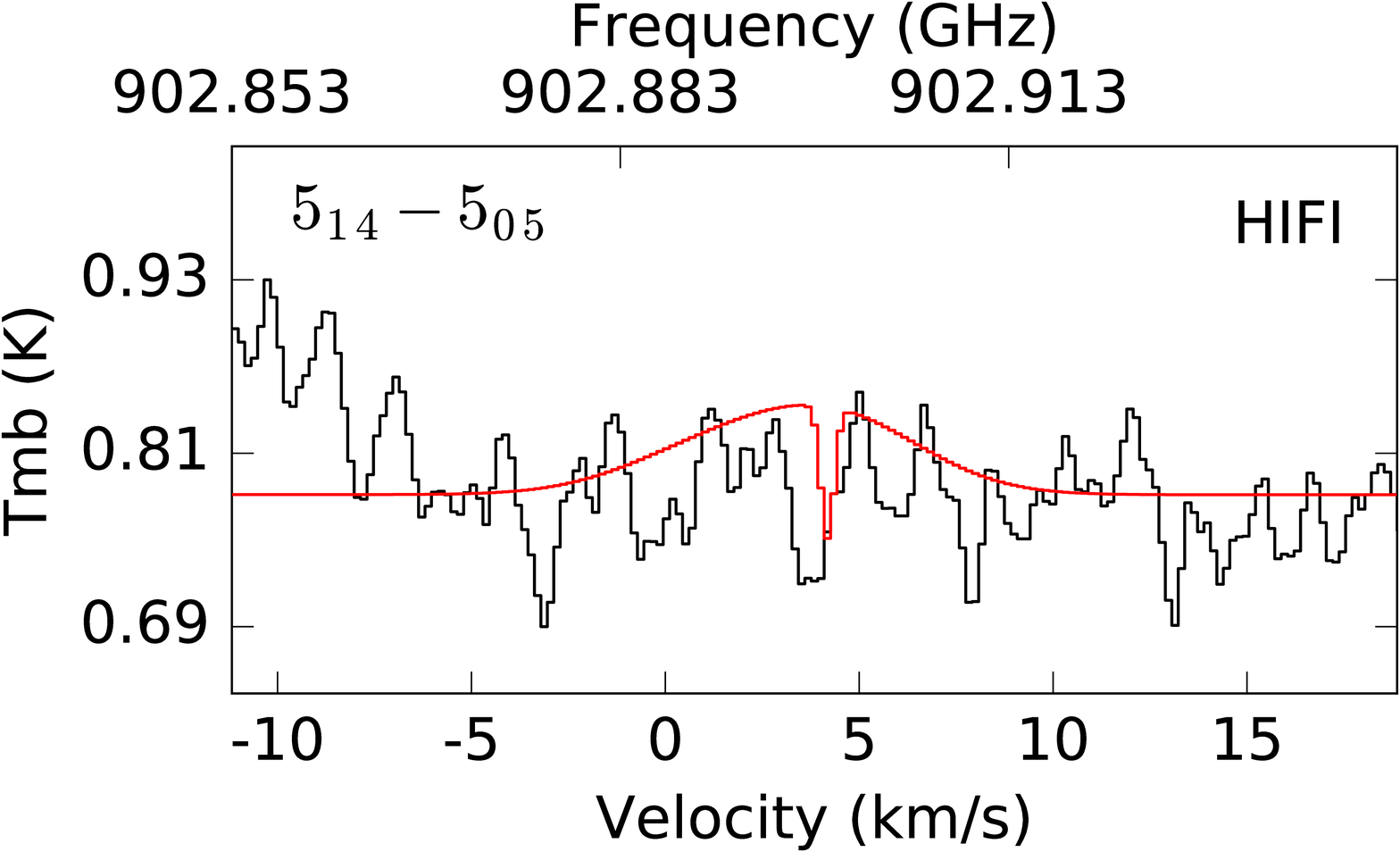}\\ 
\includegraphics[width=0.315\textwidth,trim = 0 0 0 0,clip]{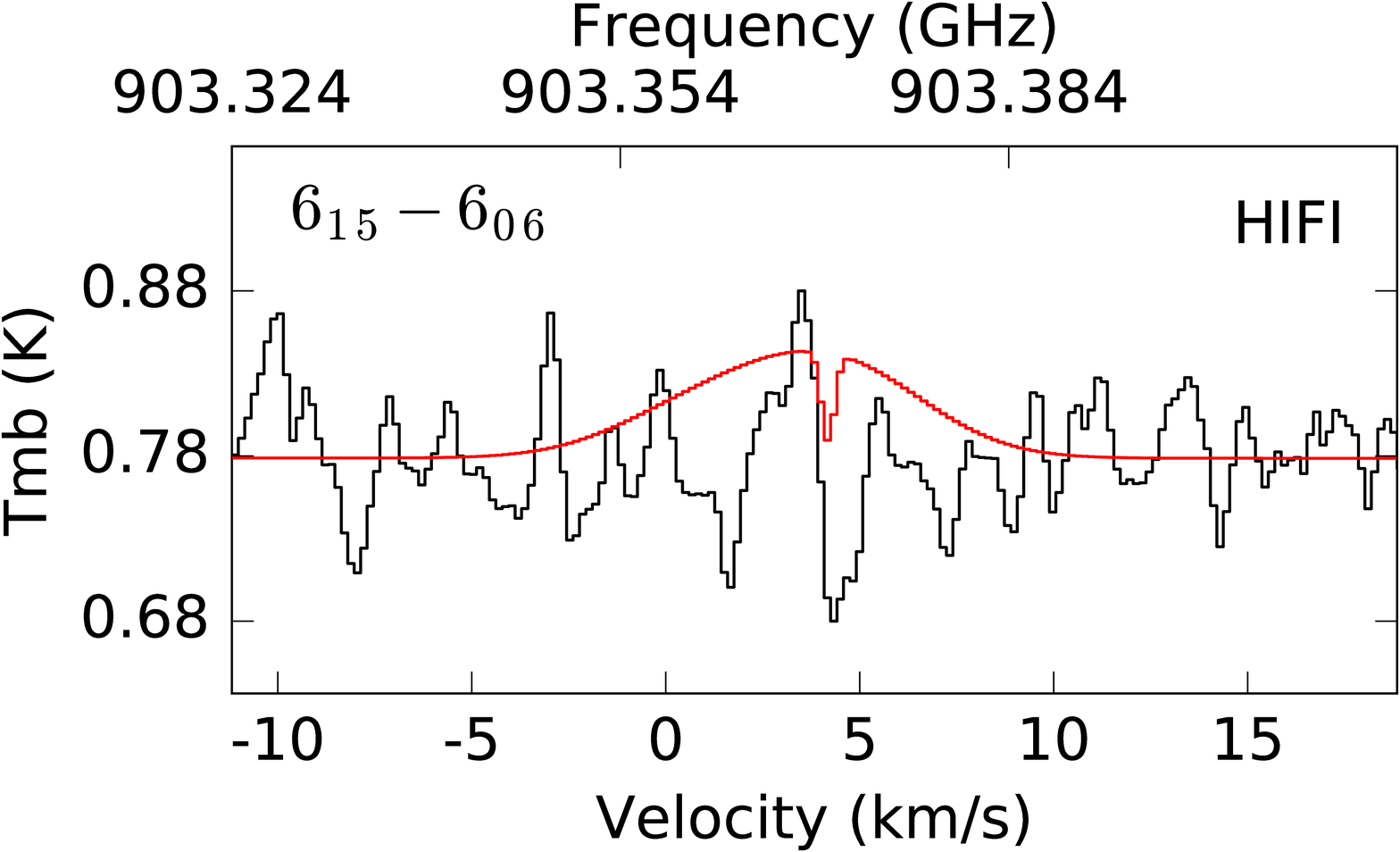}&\includegraphics[width=0.315\textwidth,trim = 0 0 0 0,clip]{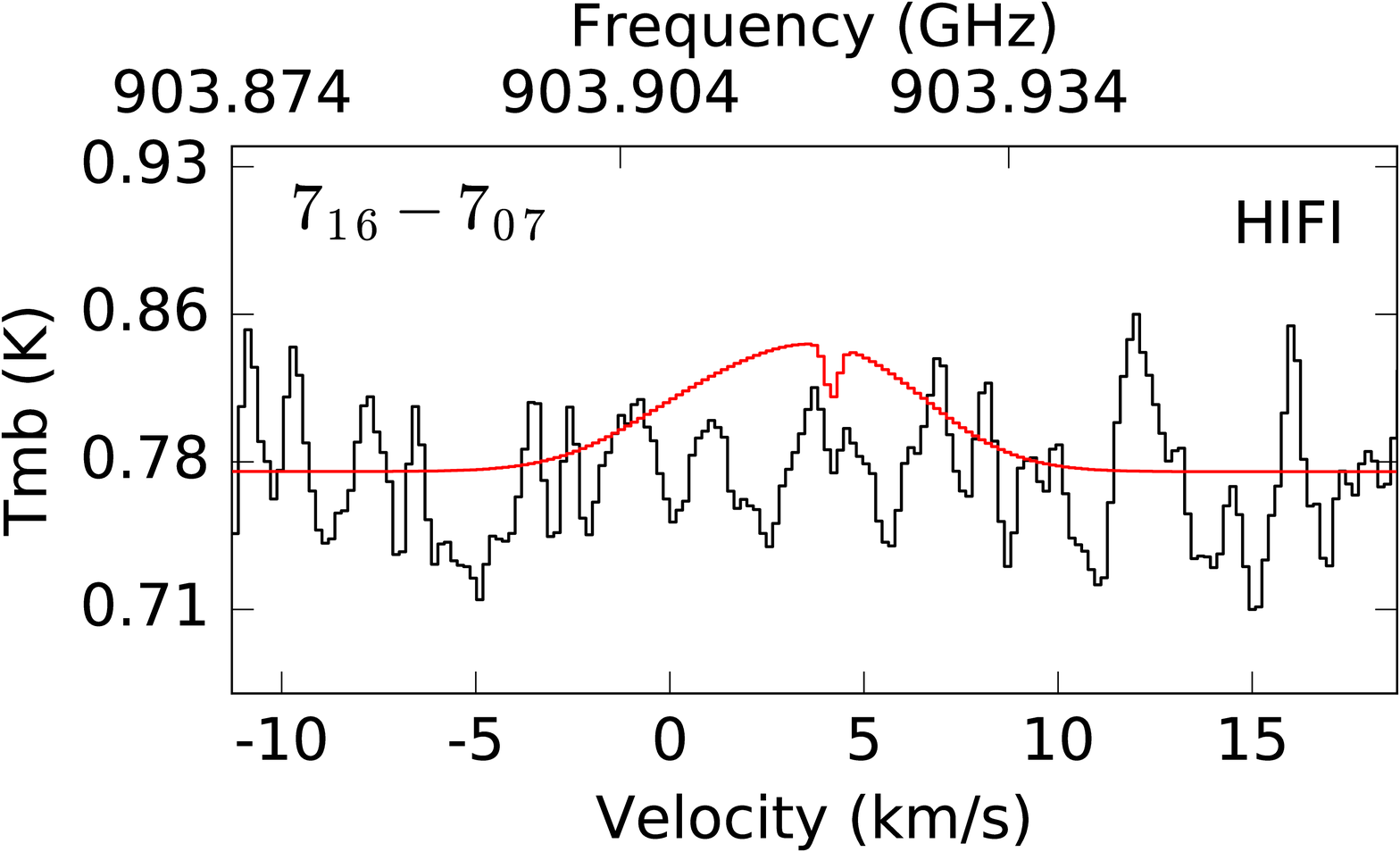} &\includegraphics[width=0.315\textwidth, trim= 0 0 0 0, clip]{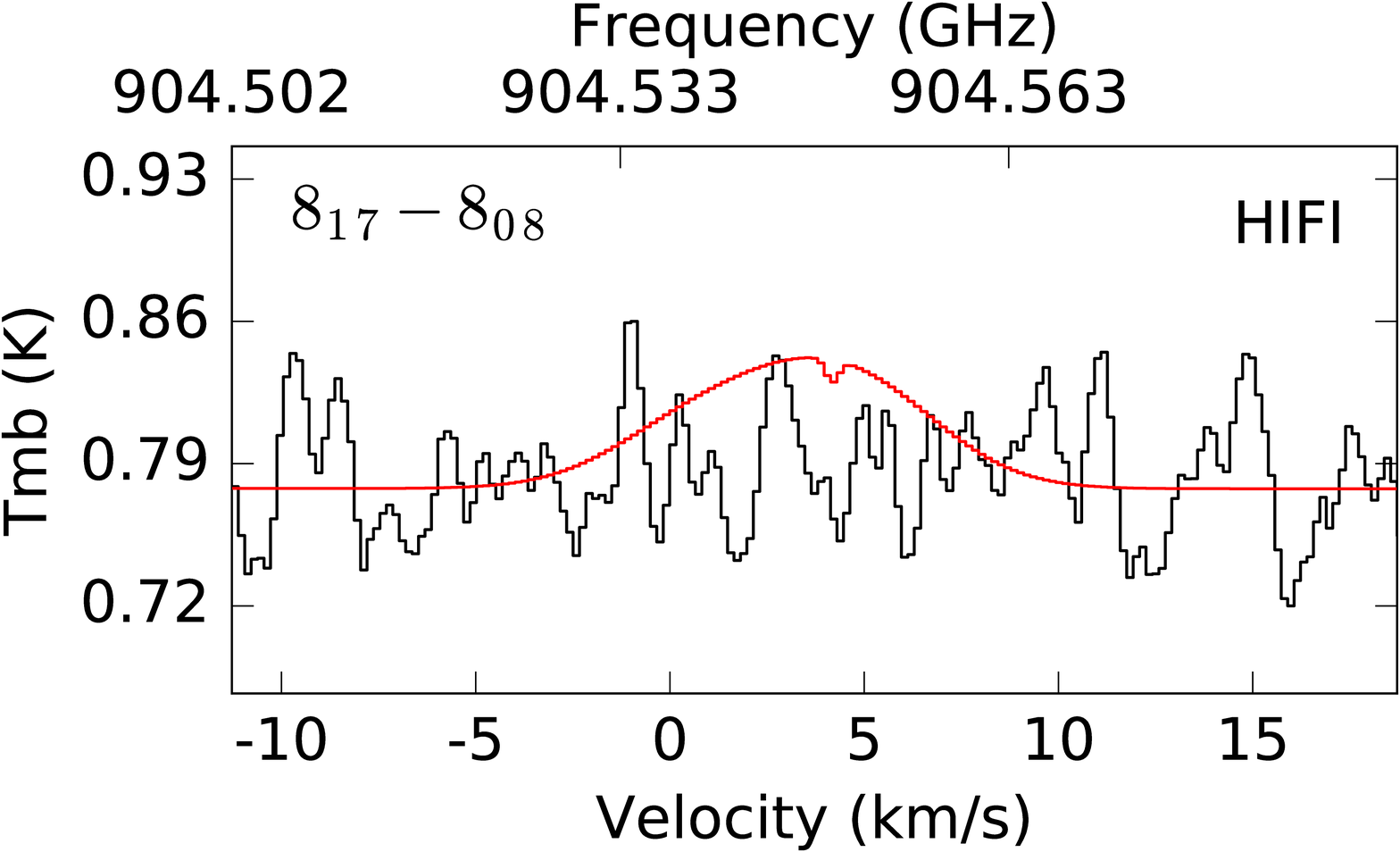}\\
 \includegraphics[width=0.315\textwidth,trim = 0 0 0 0,clip]{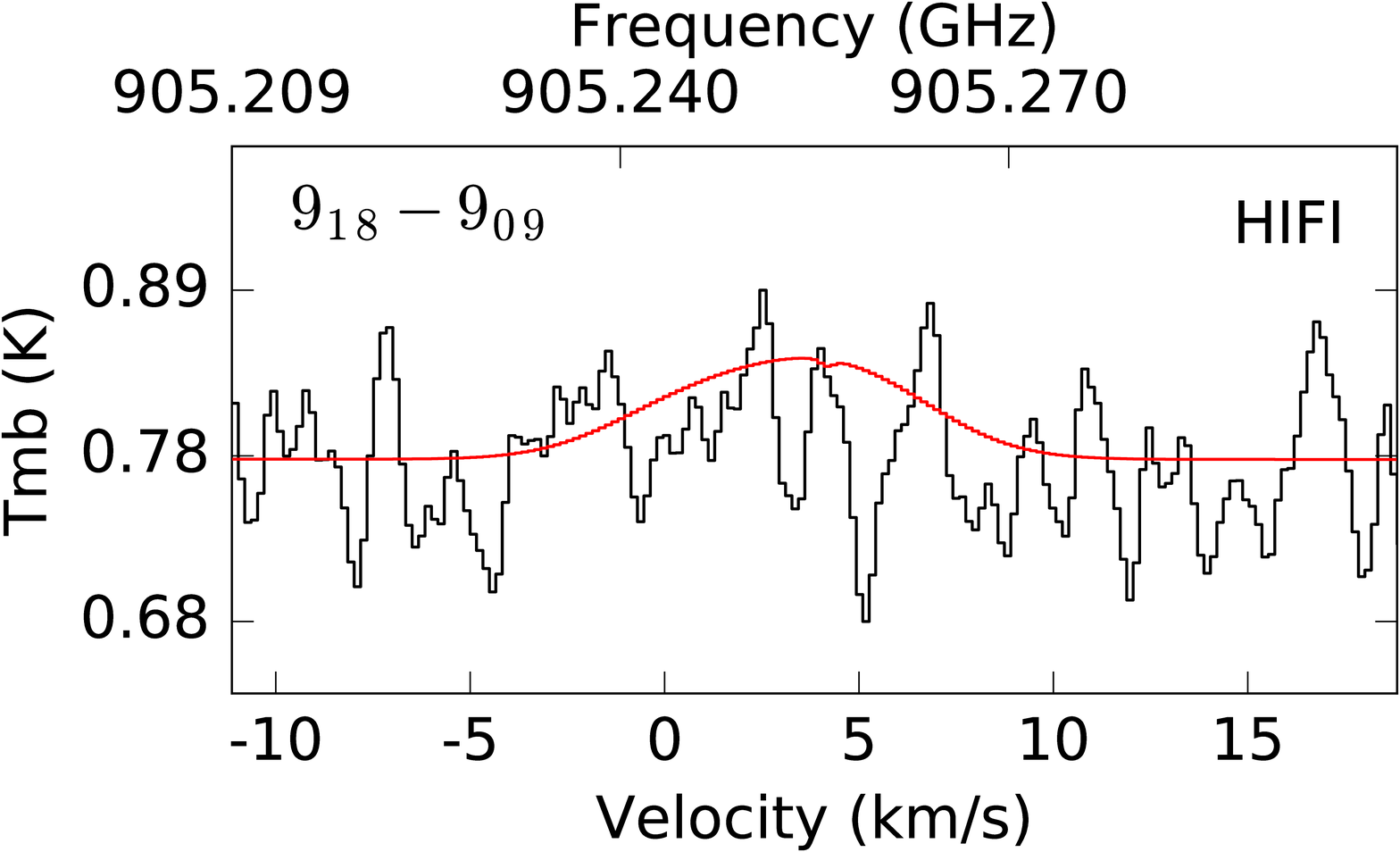}  &\includegraphics[width=0.315\textwidth,trim = 0 0 0 0,clip]{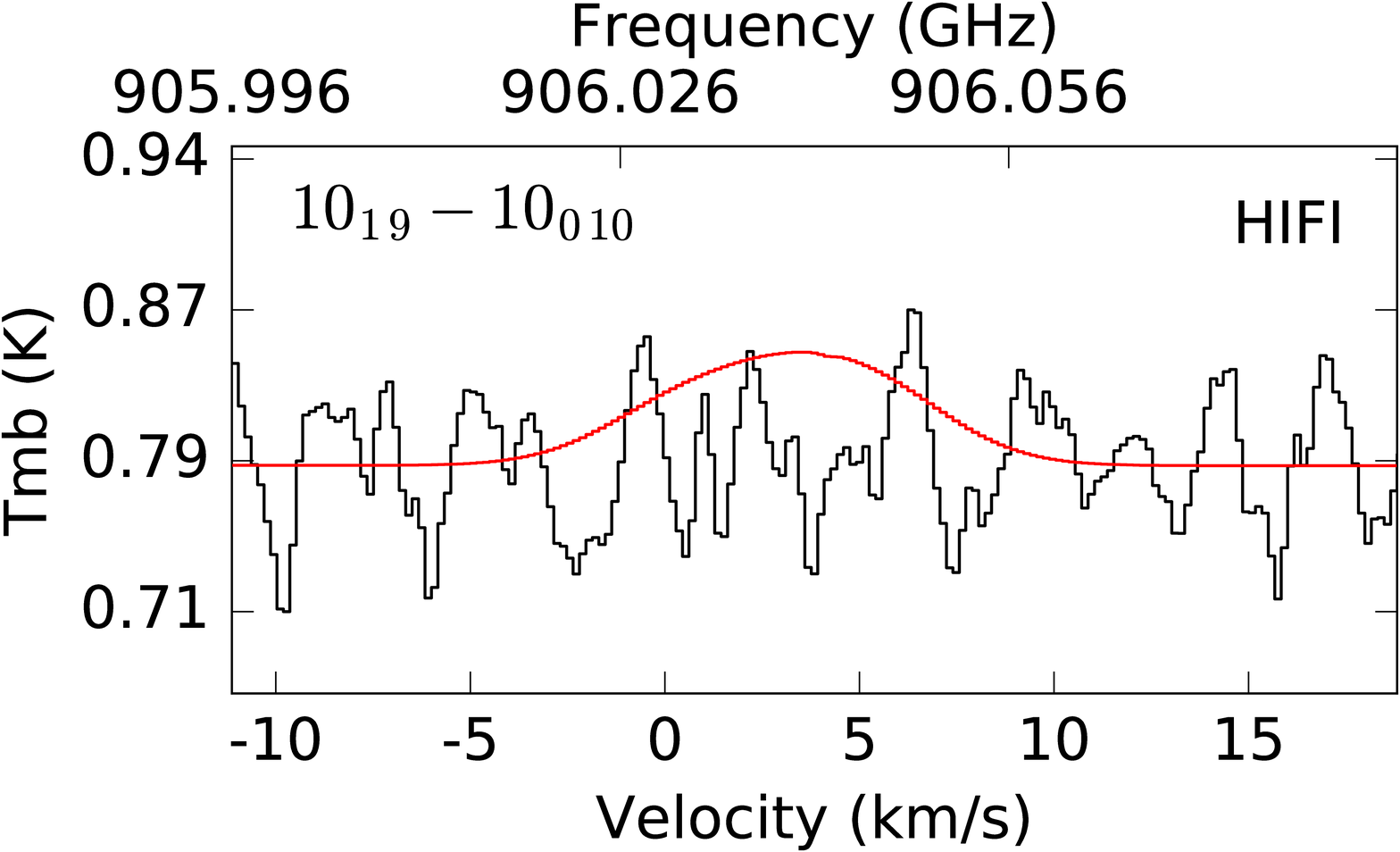} &\includegraphics[width=0.315\textwidth,trim = 0 0 0 0,clip]{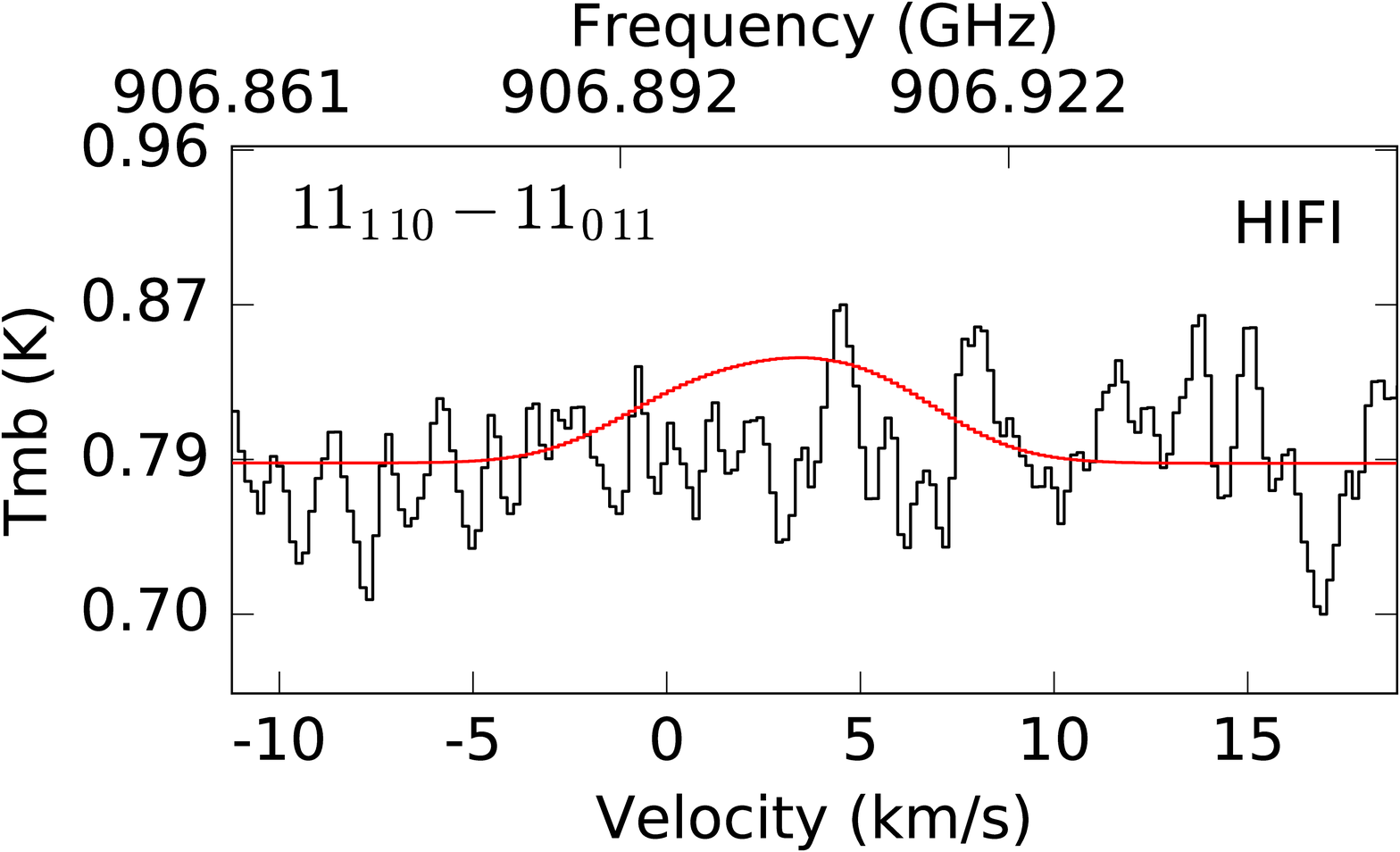}\\
\end{tabular}
\label{imagenes-cont3}
\end{figure*}

\begin{figure*}
\centering
\setlength\tabcolsep{3.7pt}
\contcaption{}
\label{imagenes-cont2}
\begin{tabular}{c c c}
 \includegraphics[width=0.315\textwidth,trim = 0 0 0 0,clip]{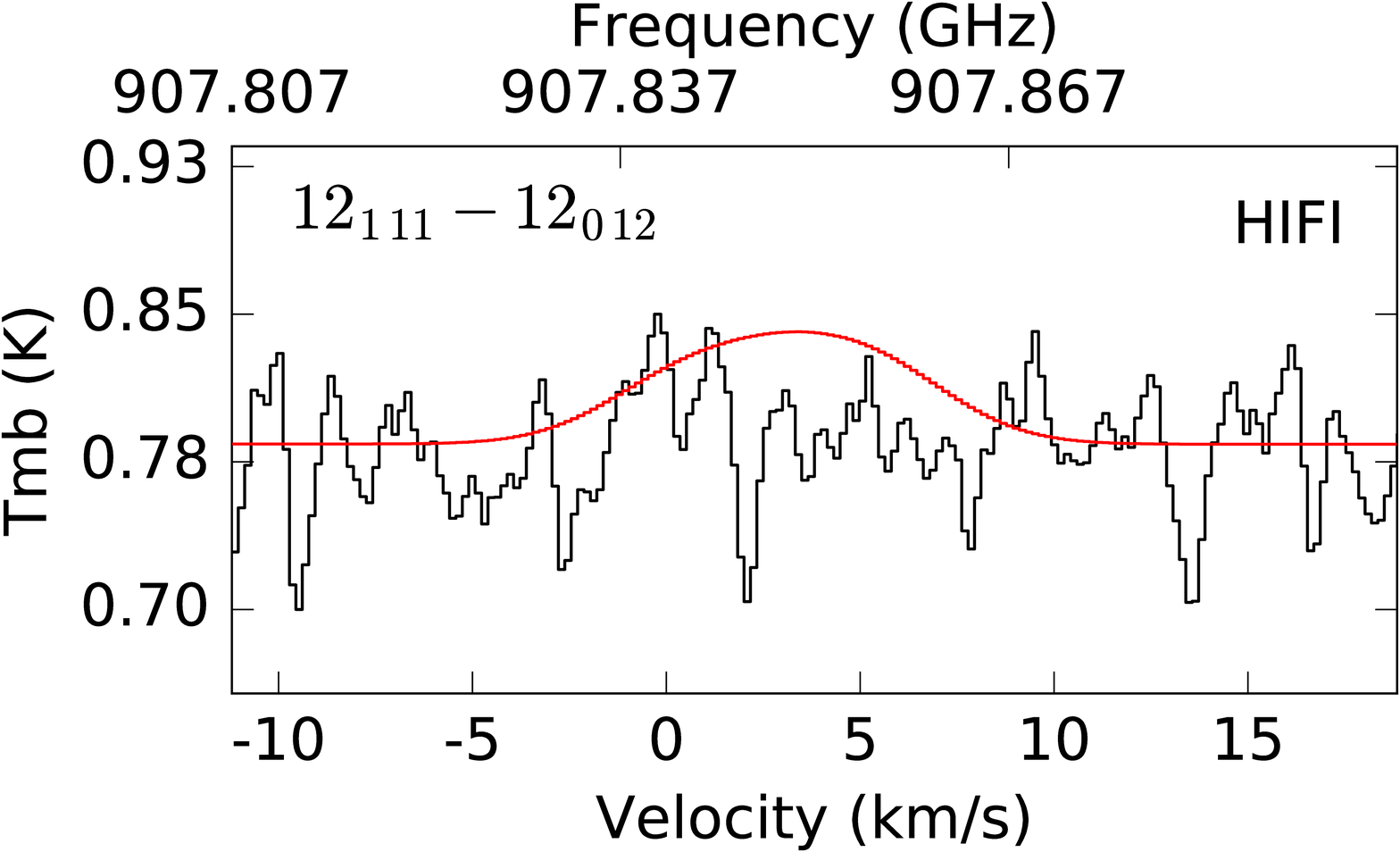} &\includegraphics[width=0.315\textwidth, trim= 0 0 0 0, clip]{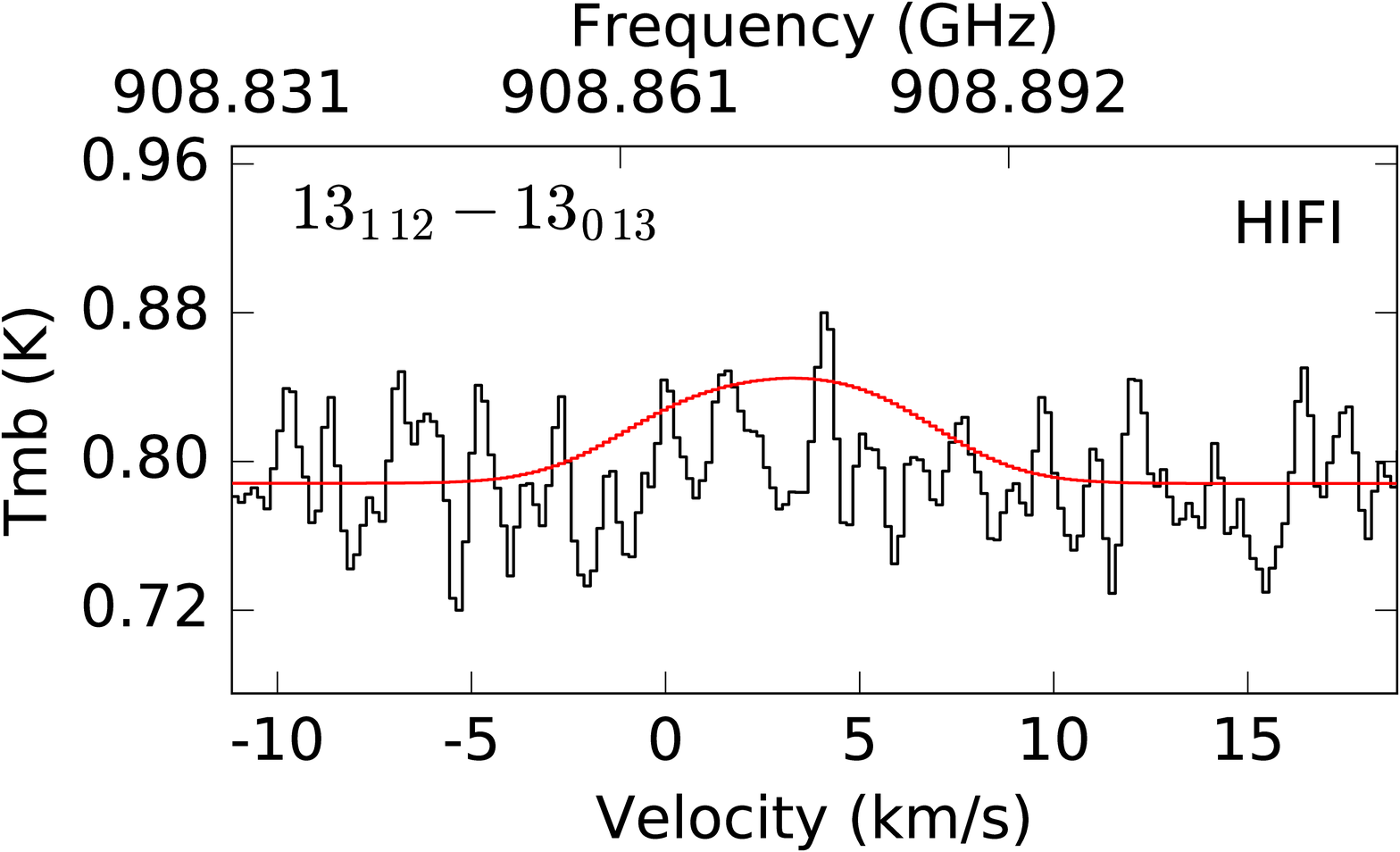} &\includegraphics[width=0.315\textwidth,trim = 0 0 0 0,clip]{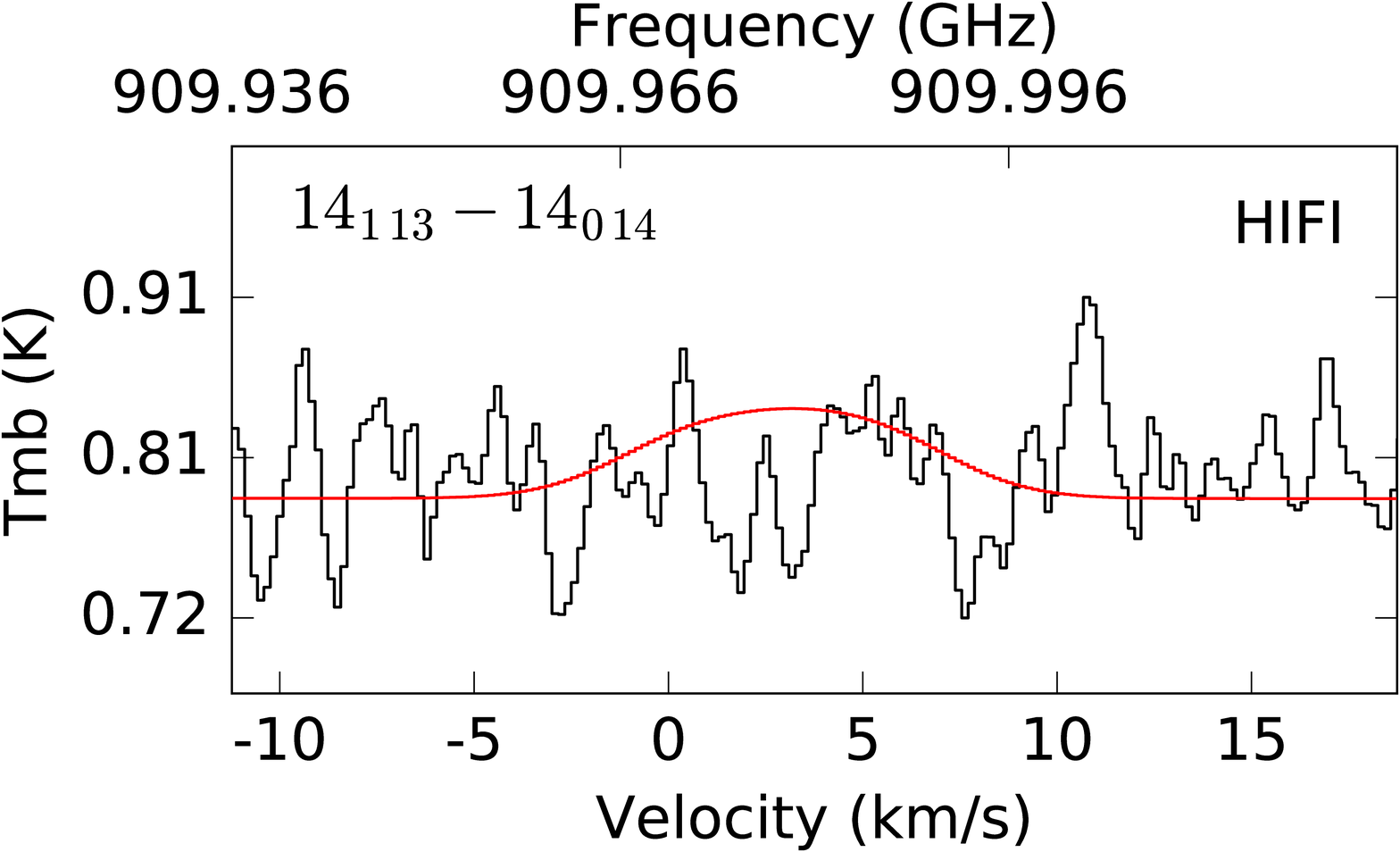}\\   
\includegraphics[width=0.315\textwidth,trim = 0 0 0 0,clip]{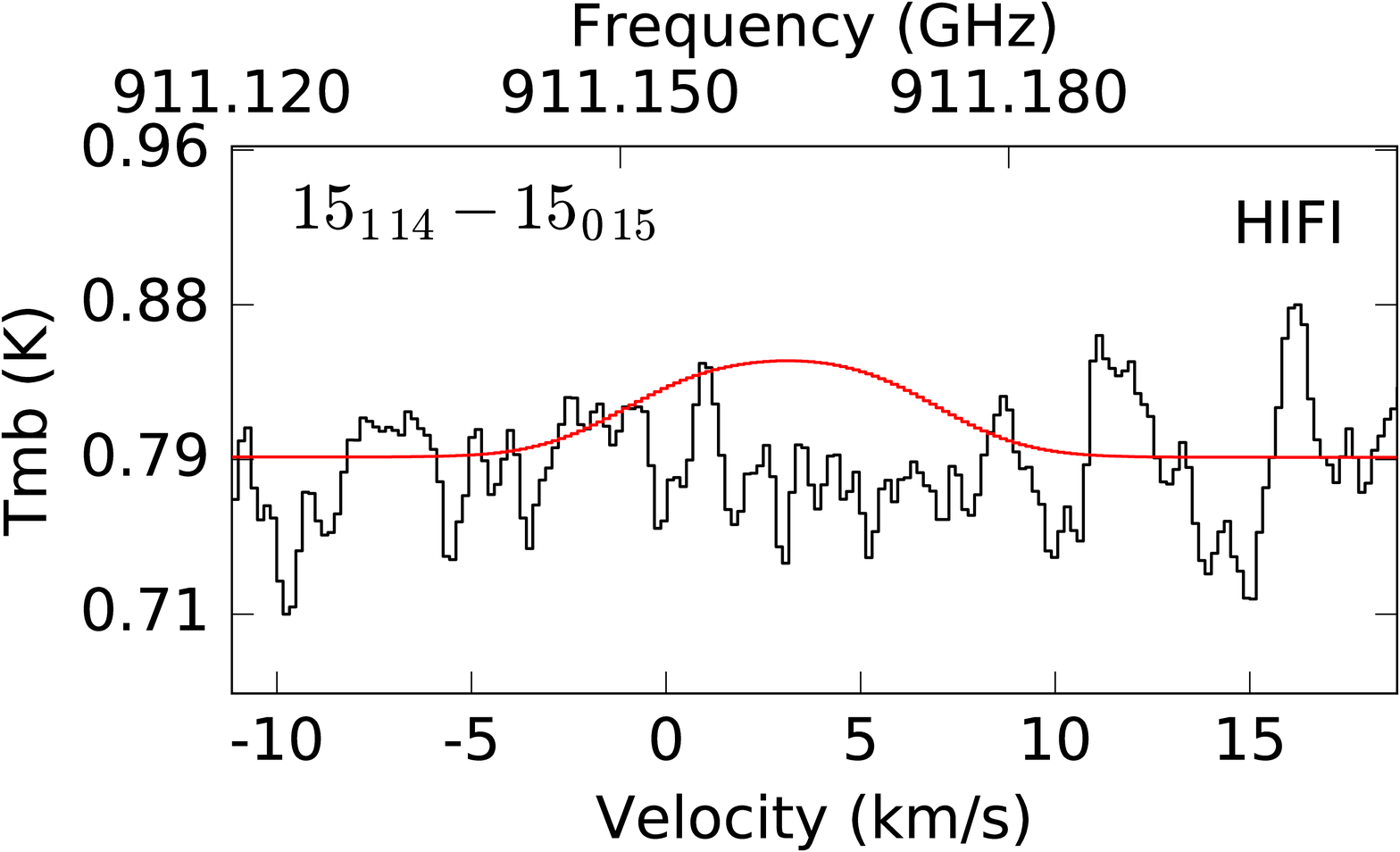} &\includegraphics[width=0.315\textwidth,trim = 0 0 0 0,clip]{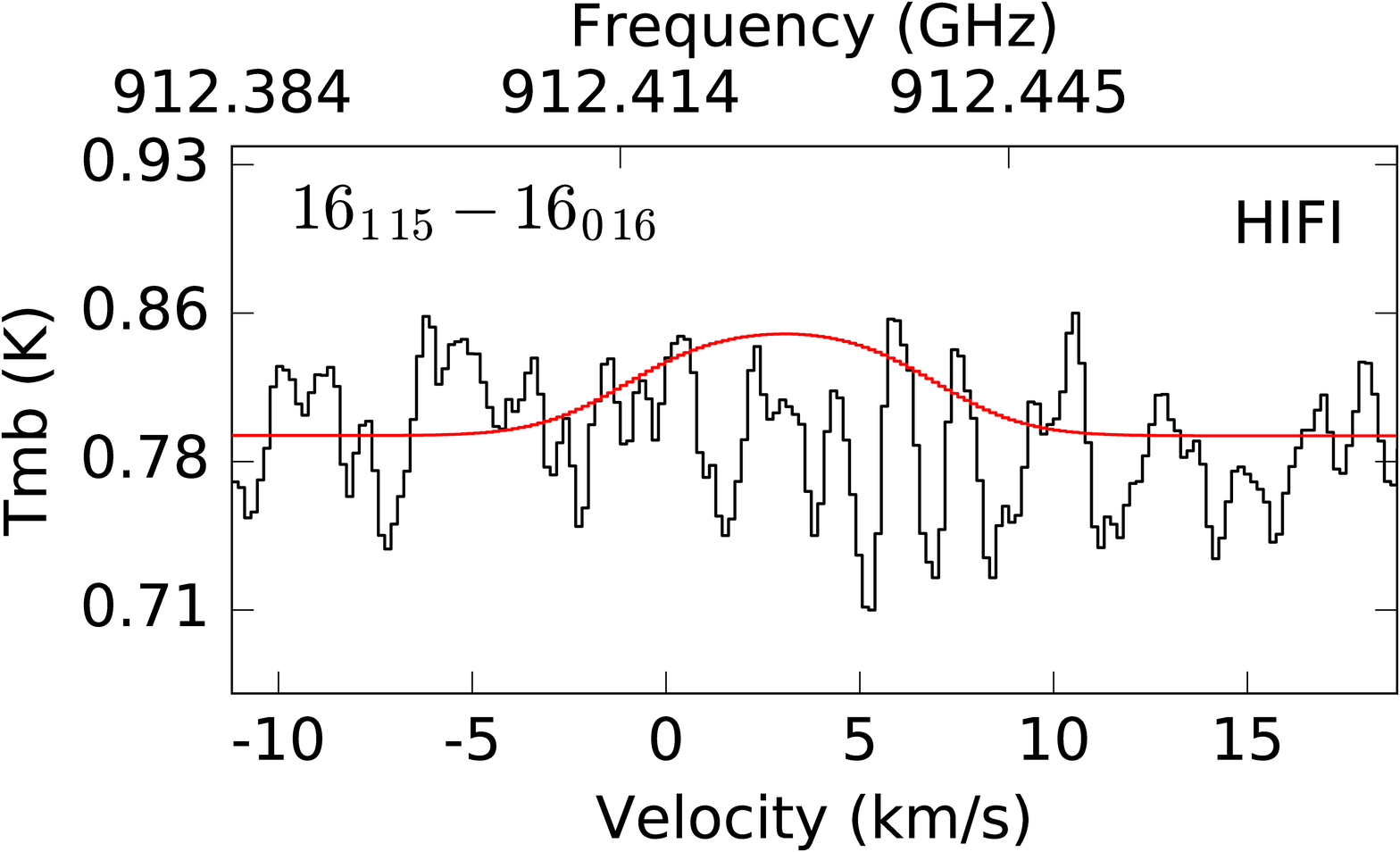}&\includegraphics[width=0.315\textwidth,trim = 0 0 0 0,clip]{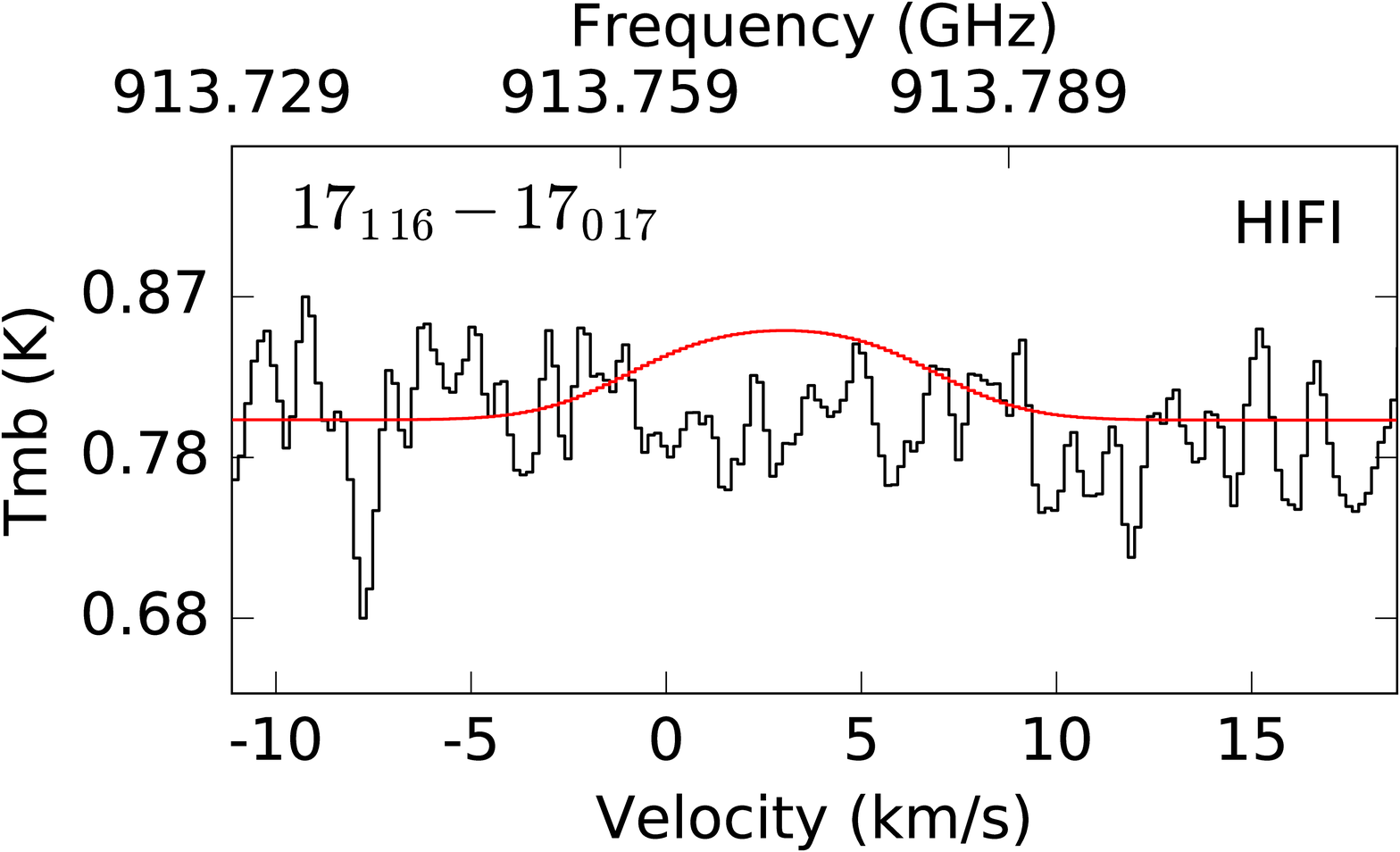} \\
\includegraphics[width=0.315\textwidth, trim= 0 0 0 0, clip]{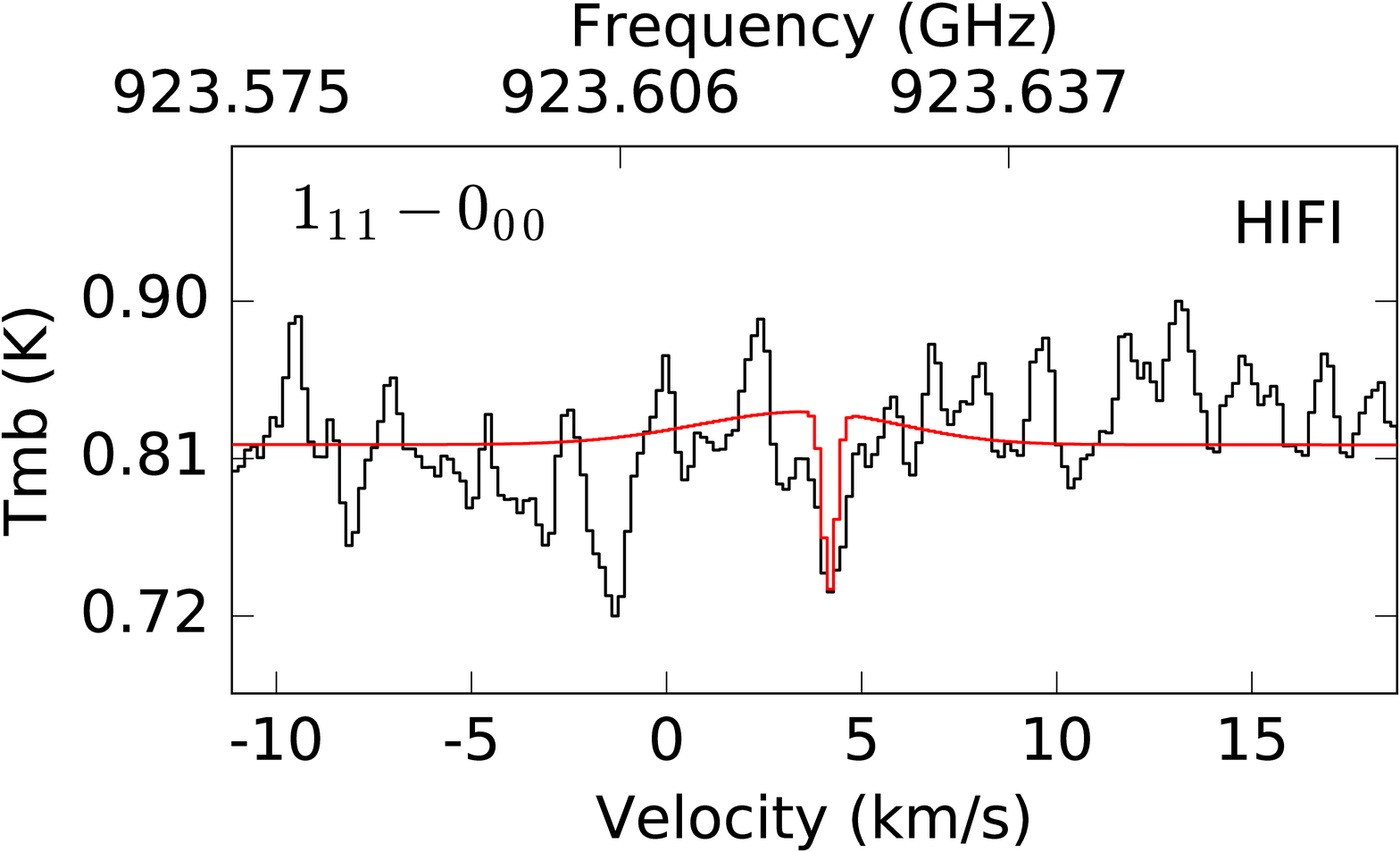} &\includegraphics[width=0.315\textwidth,trim = 0 0 0 0,clip]{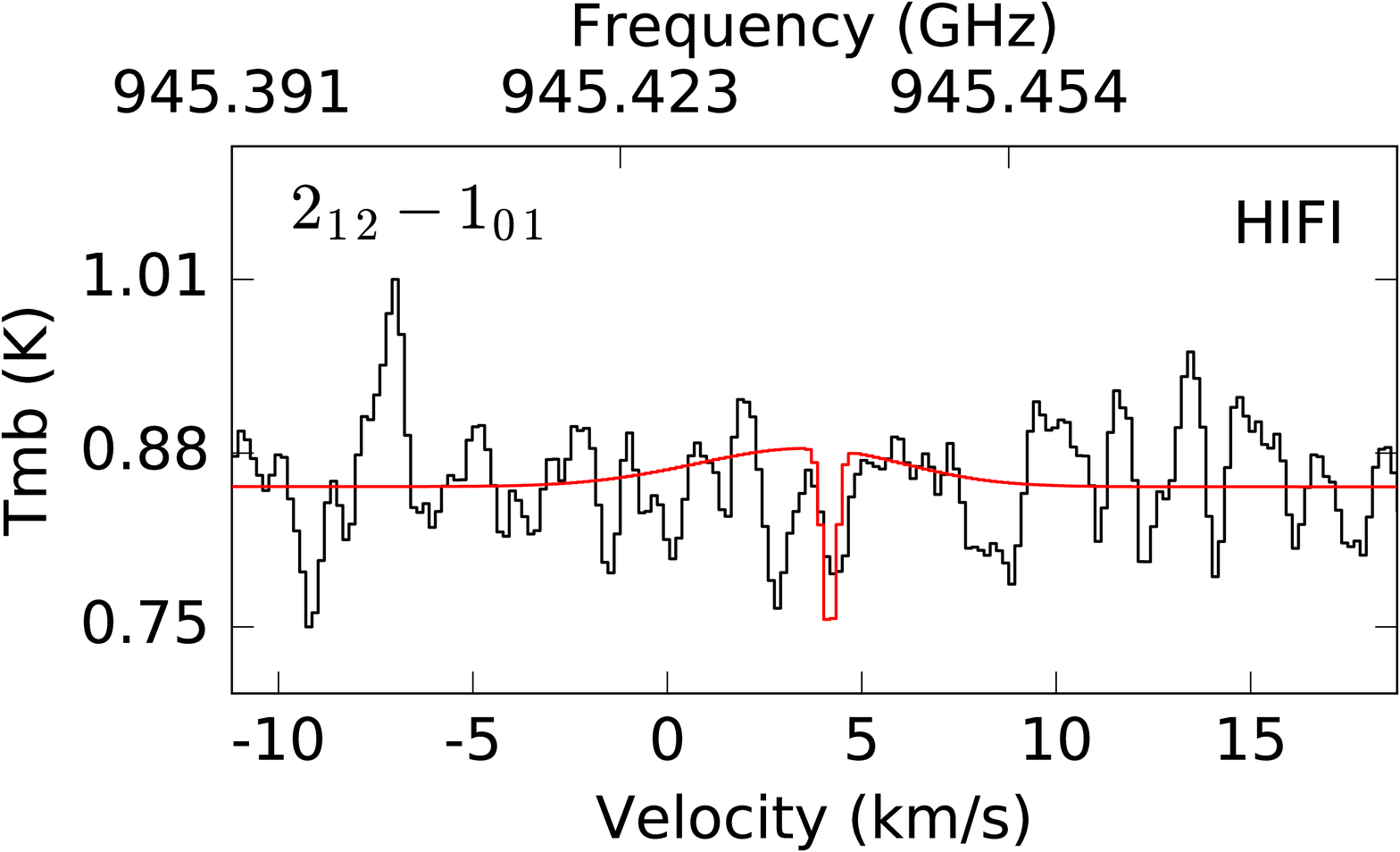}  &\includegraphics[width=0.315\textwidth,trim = 0 0 0 0,clip]{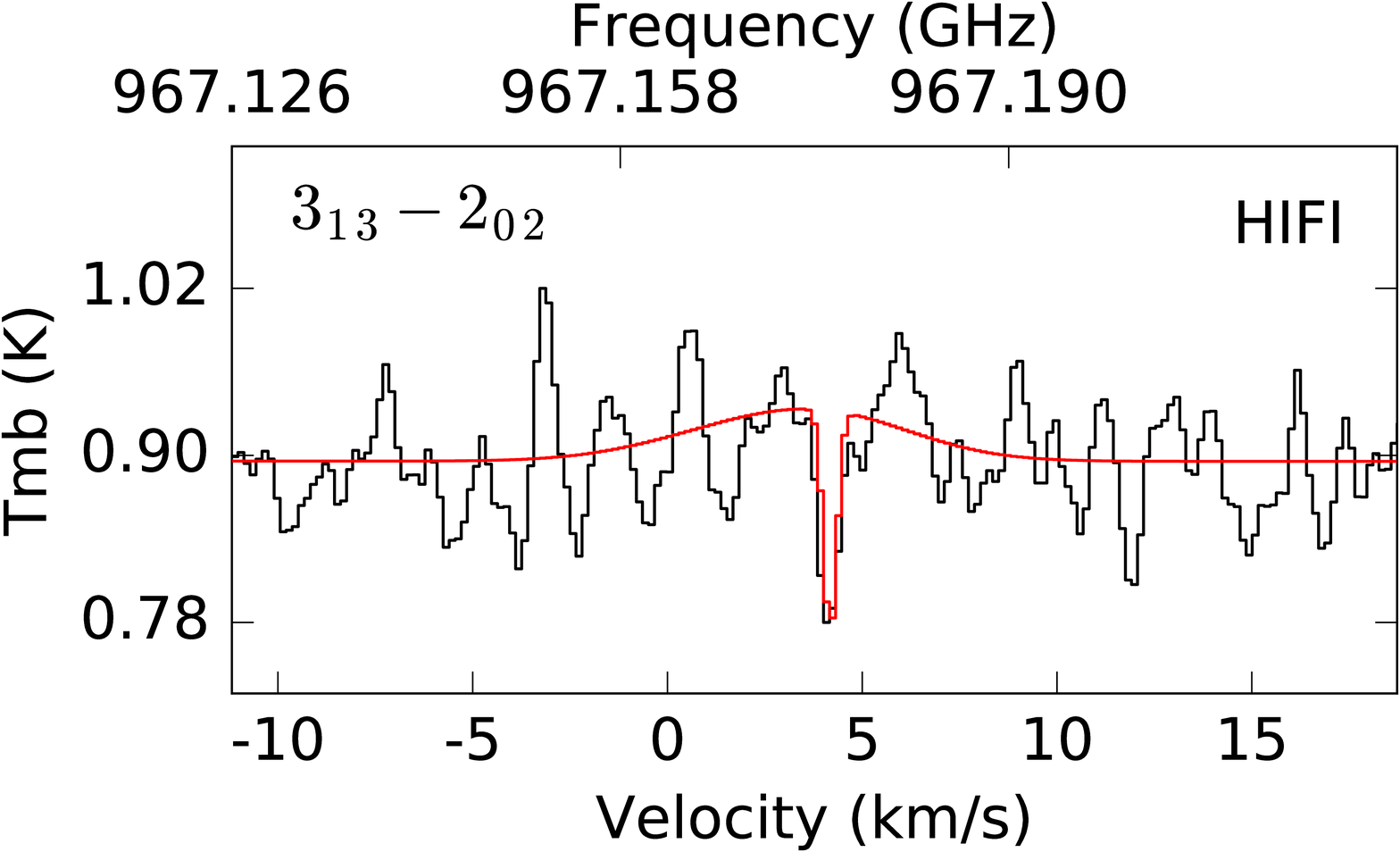}\\
\includegraphics[width=0.315\textwidth,trim = 0 0 0 0,clip]{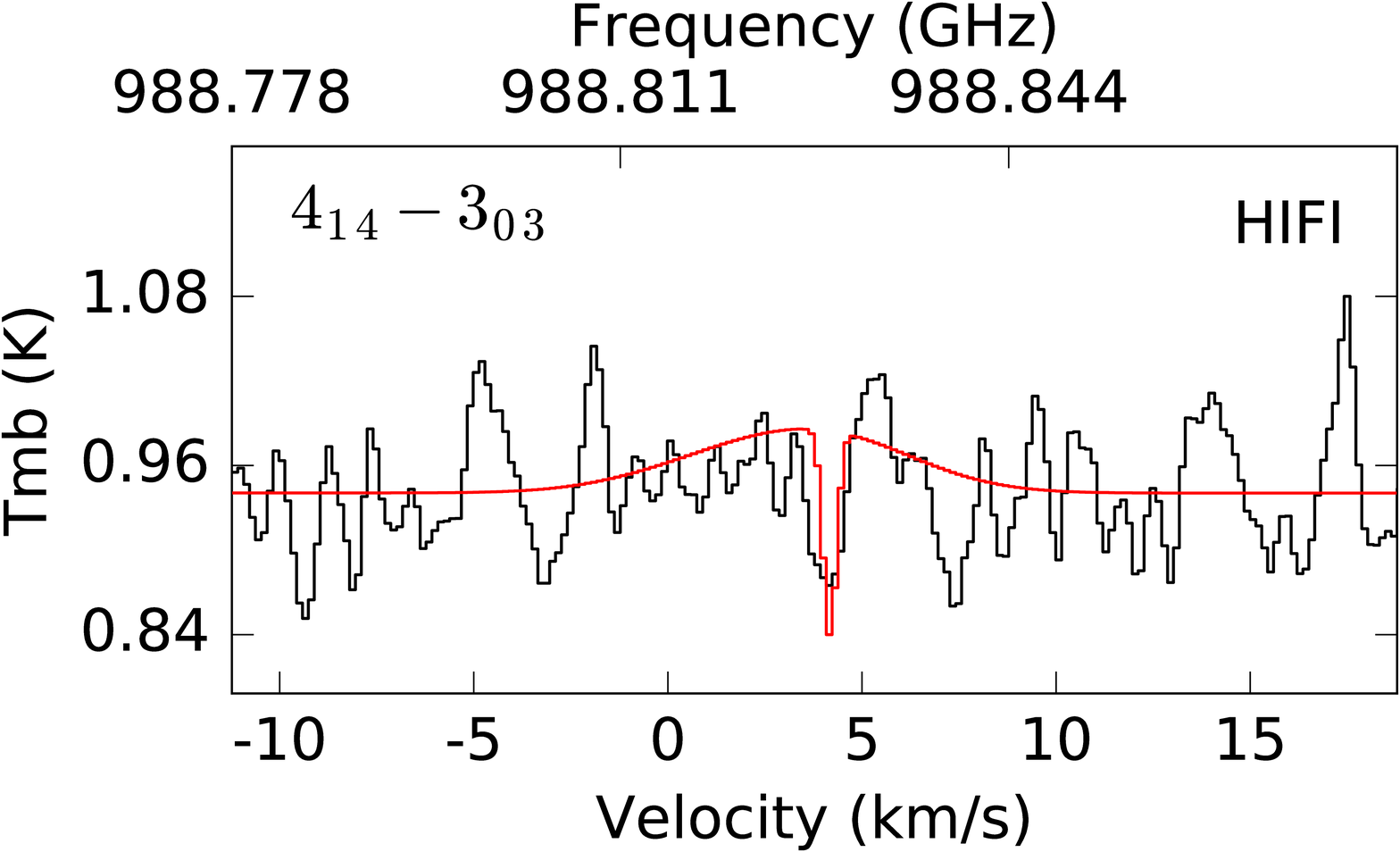}&\includegraphics[width=0.315\textwidth,trim = 0 0 0 0,clip]{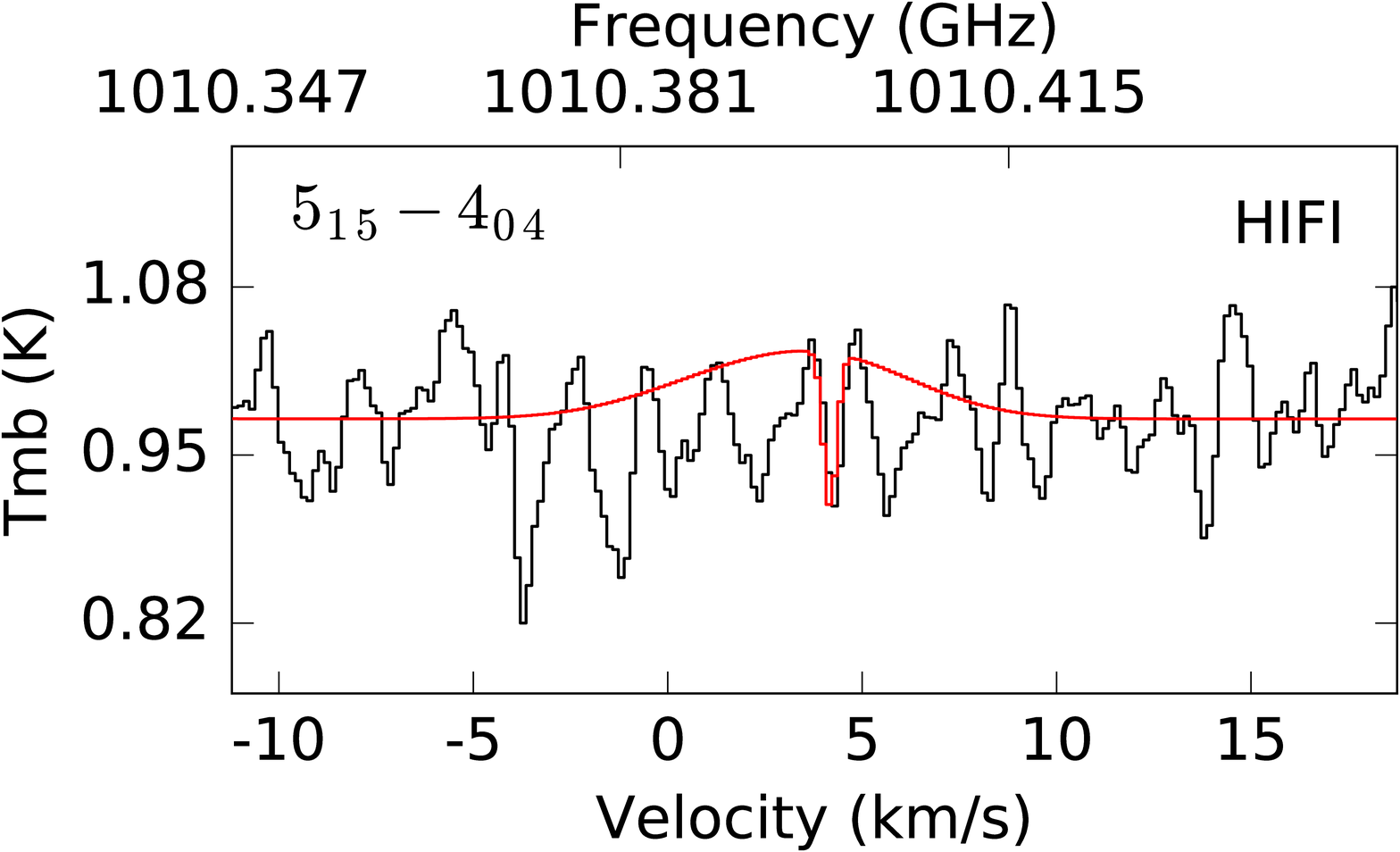} &\includegraphics[width=0.315\textwidth, trim= 0 0 0 0, clip]{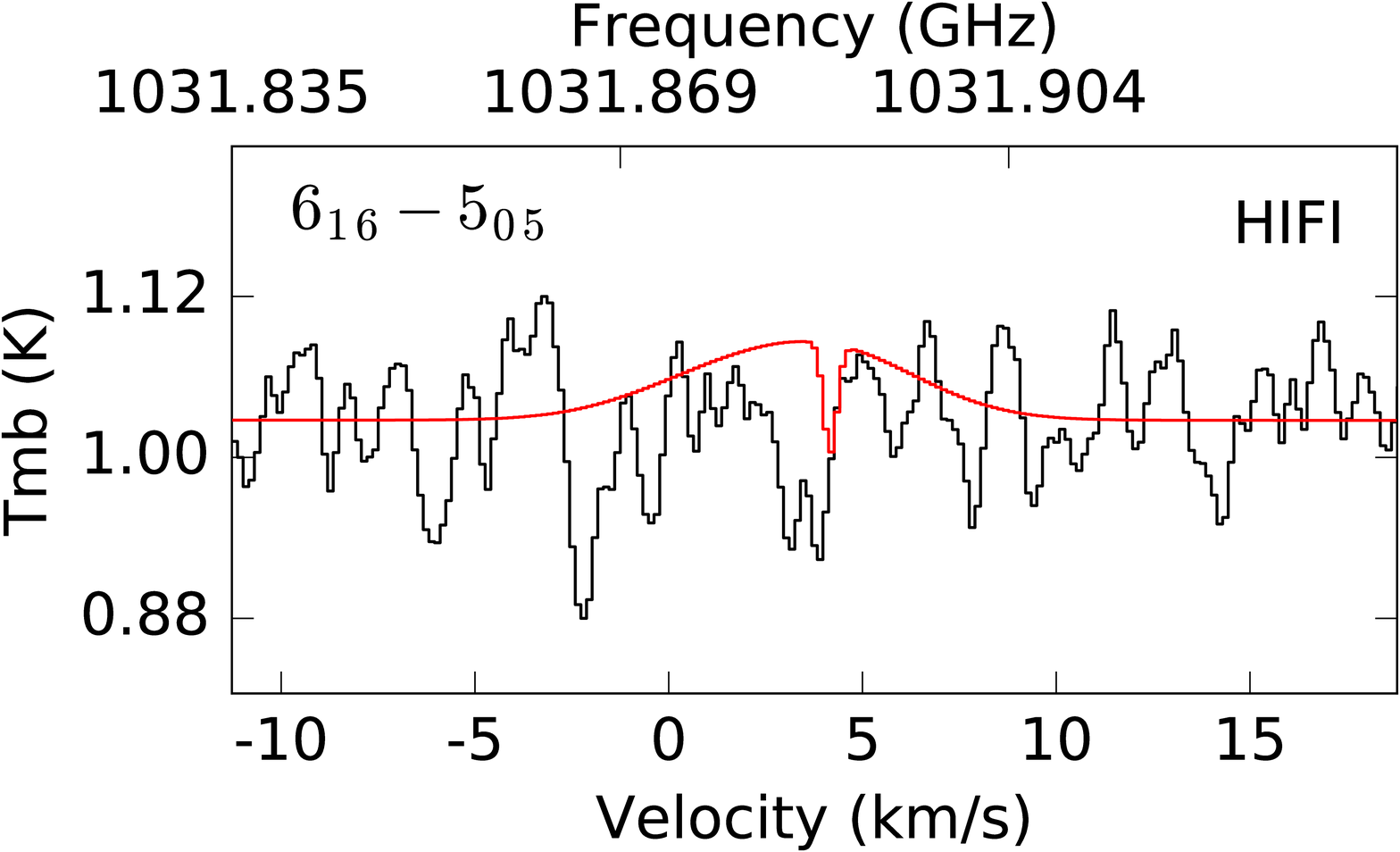} \\\includegraphics[width=0.315\textwidth,trim = 0 0 0 0,clip]{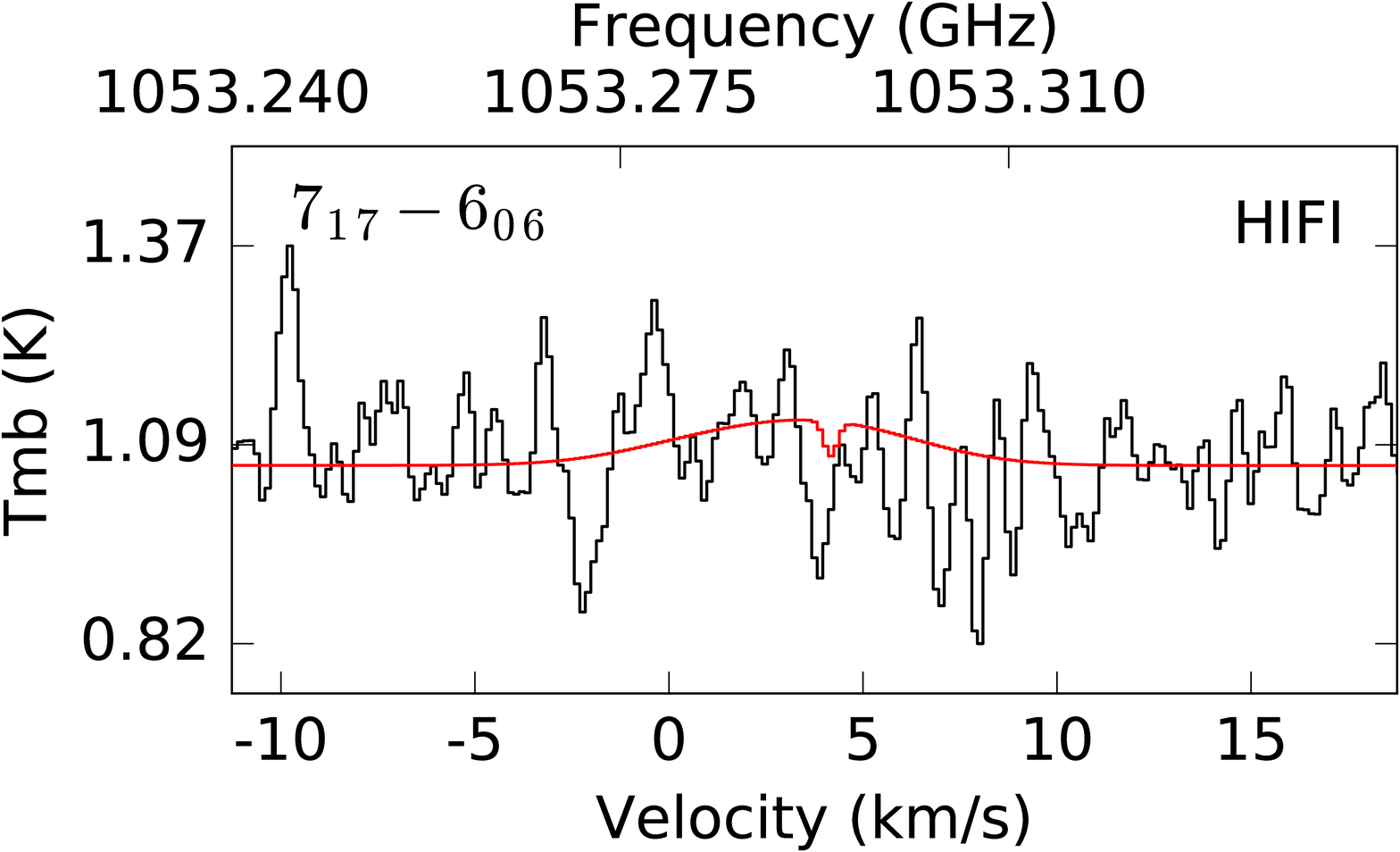}  &\includegraphics[width=0.315\textwidth,trim = 0 0 0 0,clip]{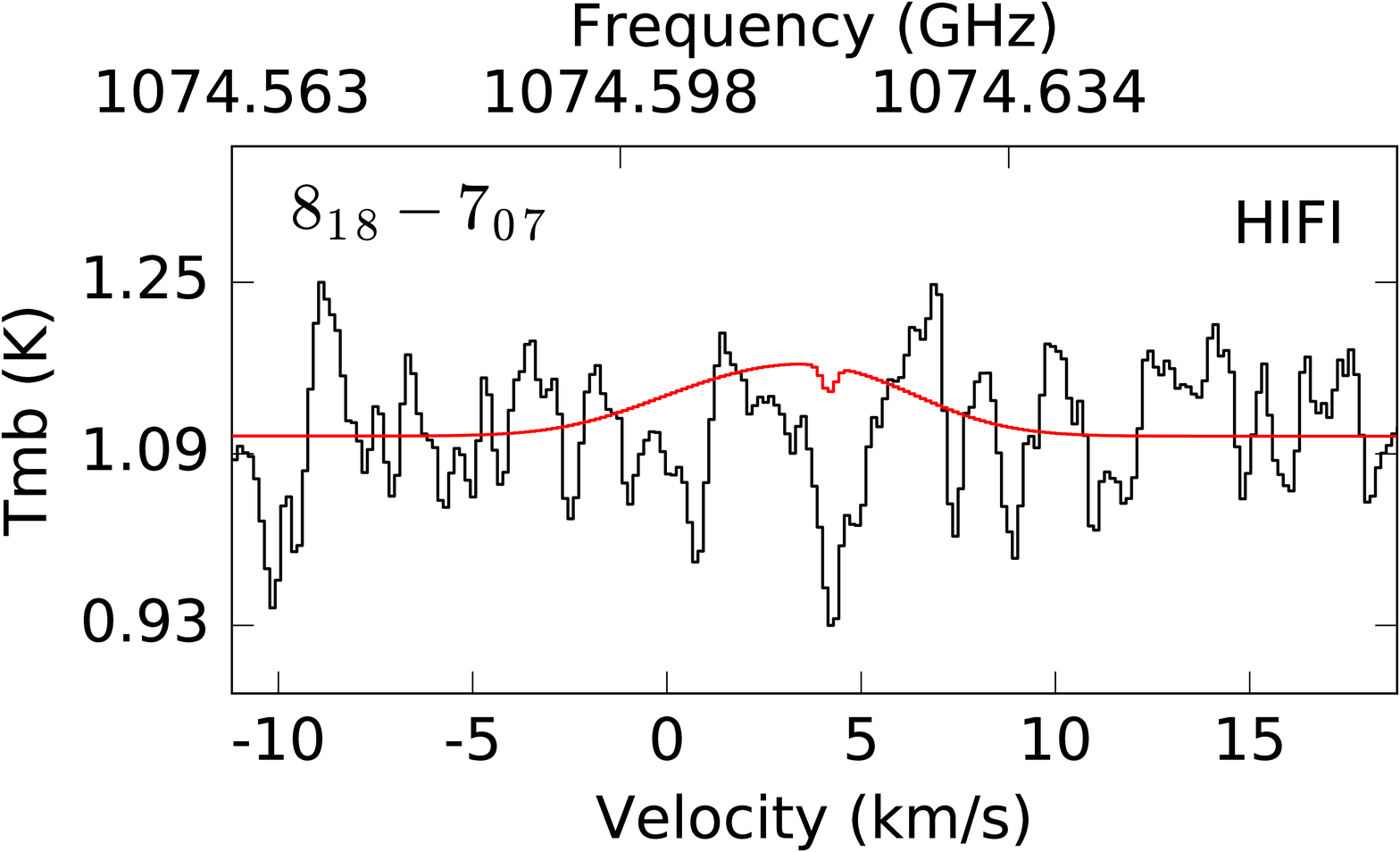} &\includegraphics[width=0.315\textwidth,trim = 0 0 0 0,clip]{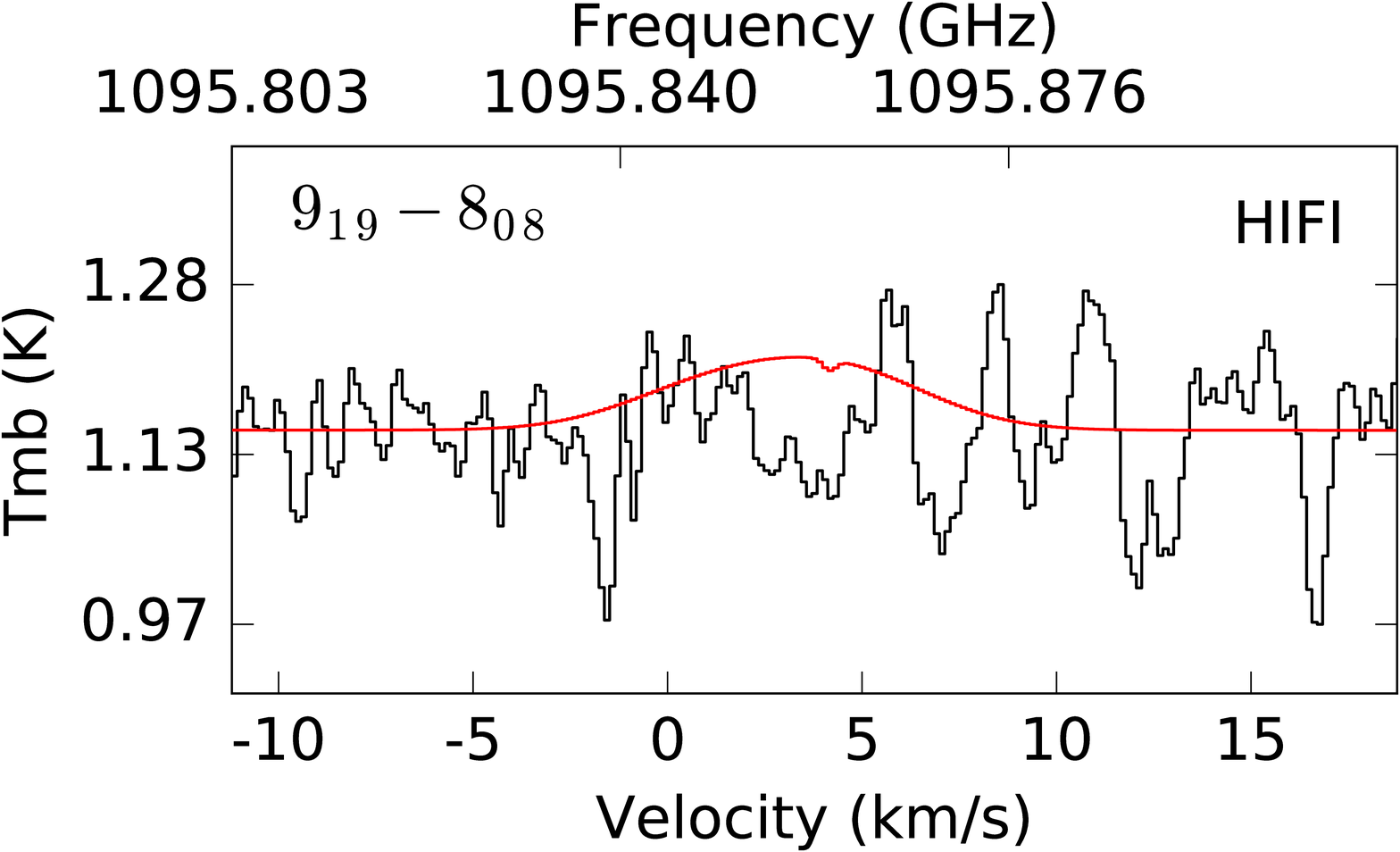}\\
\includegraphics[width=0.315\textwidth,trim = 0 0 0 0,clip]{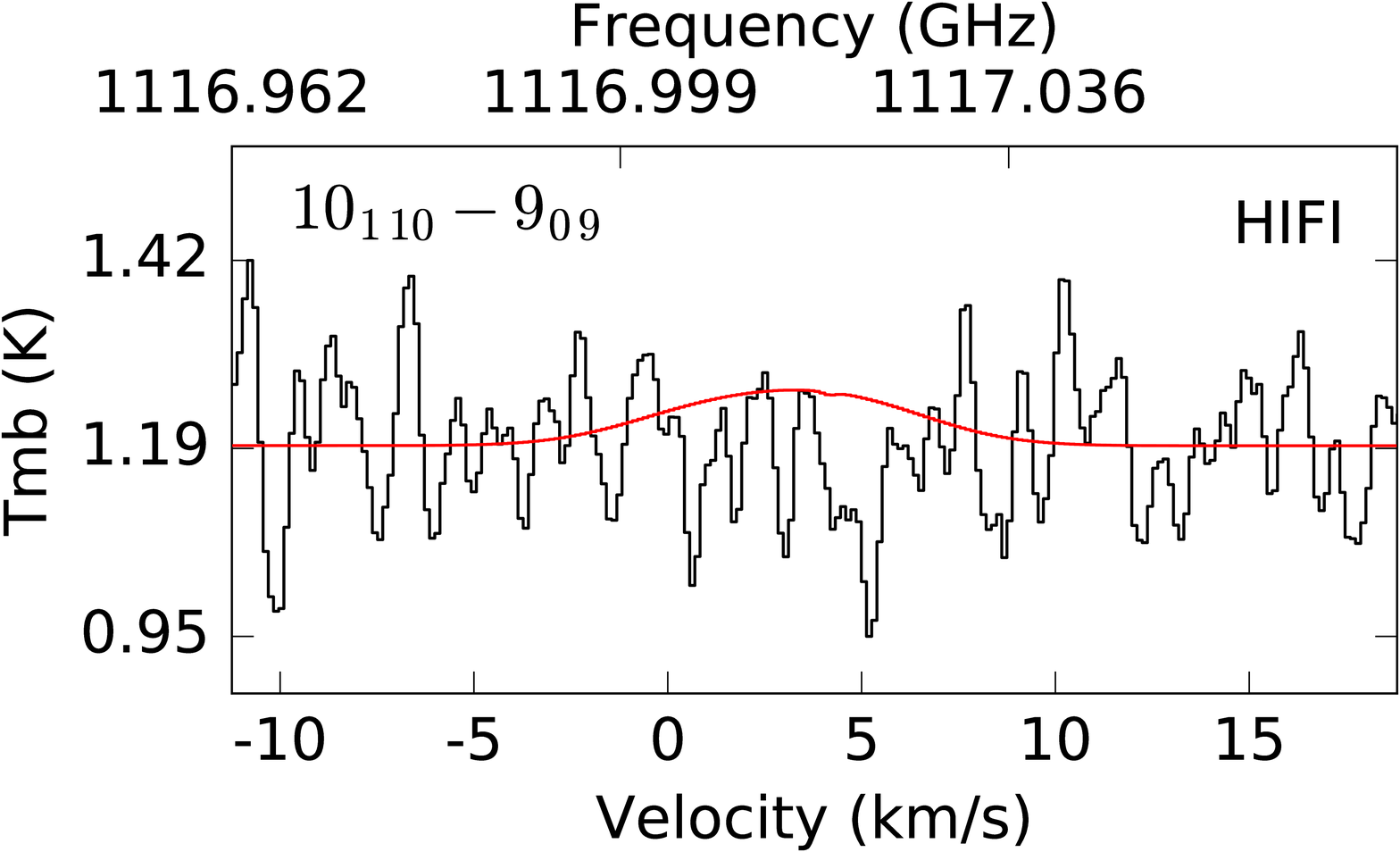} &\includegraphics[width=0.315\textwidth, trim= 0 0 0 0, clip]{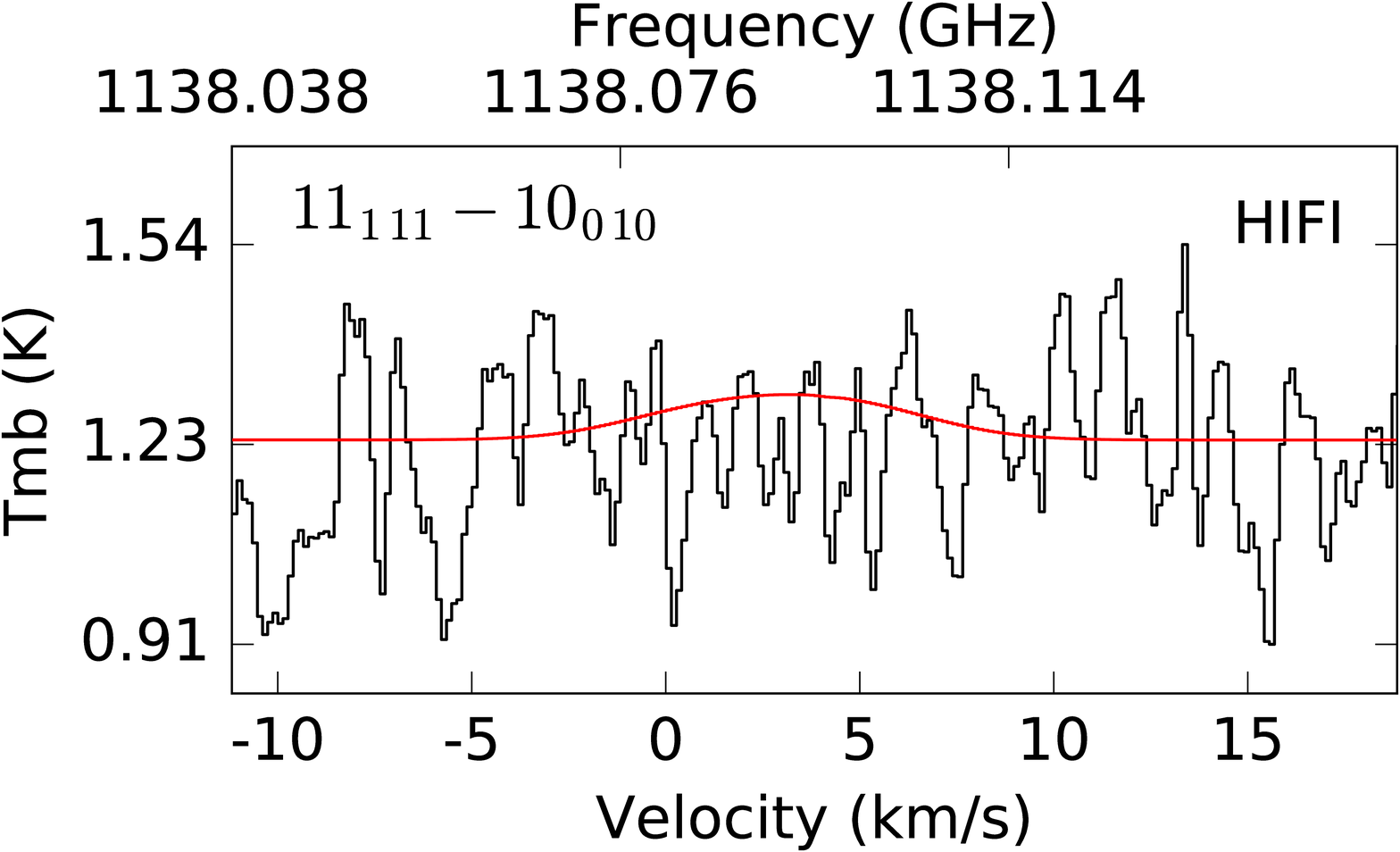} &\includegraphics[width=0.315\textwidth,trim = 0 0 0 0,clip]{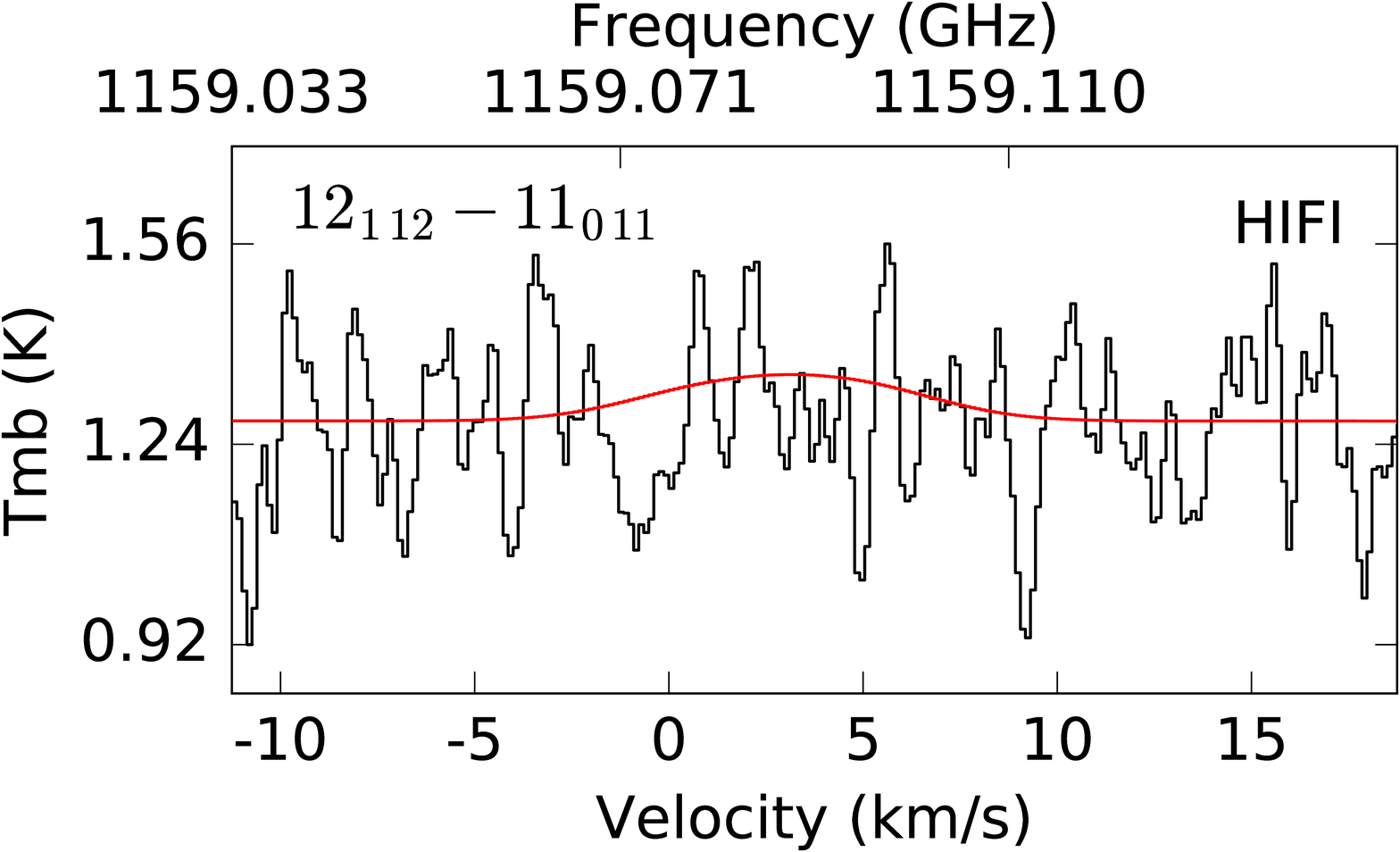}\\

\end{tabular}
\label{imagenes-cont3}
\end{figure*}

\begin{figure*}
\setlength\tabcolsep{3.7pt}
\contcaption{}
\label{imagenes-cont2}
\begin{tabular}{l l l}
\includegraphics[width=0.315\textwidth,trim = 0 0 0 0,clip]{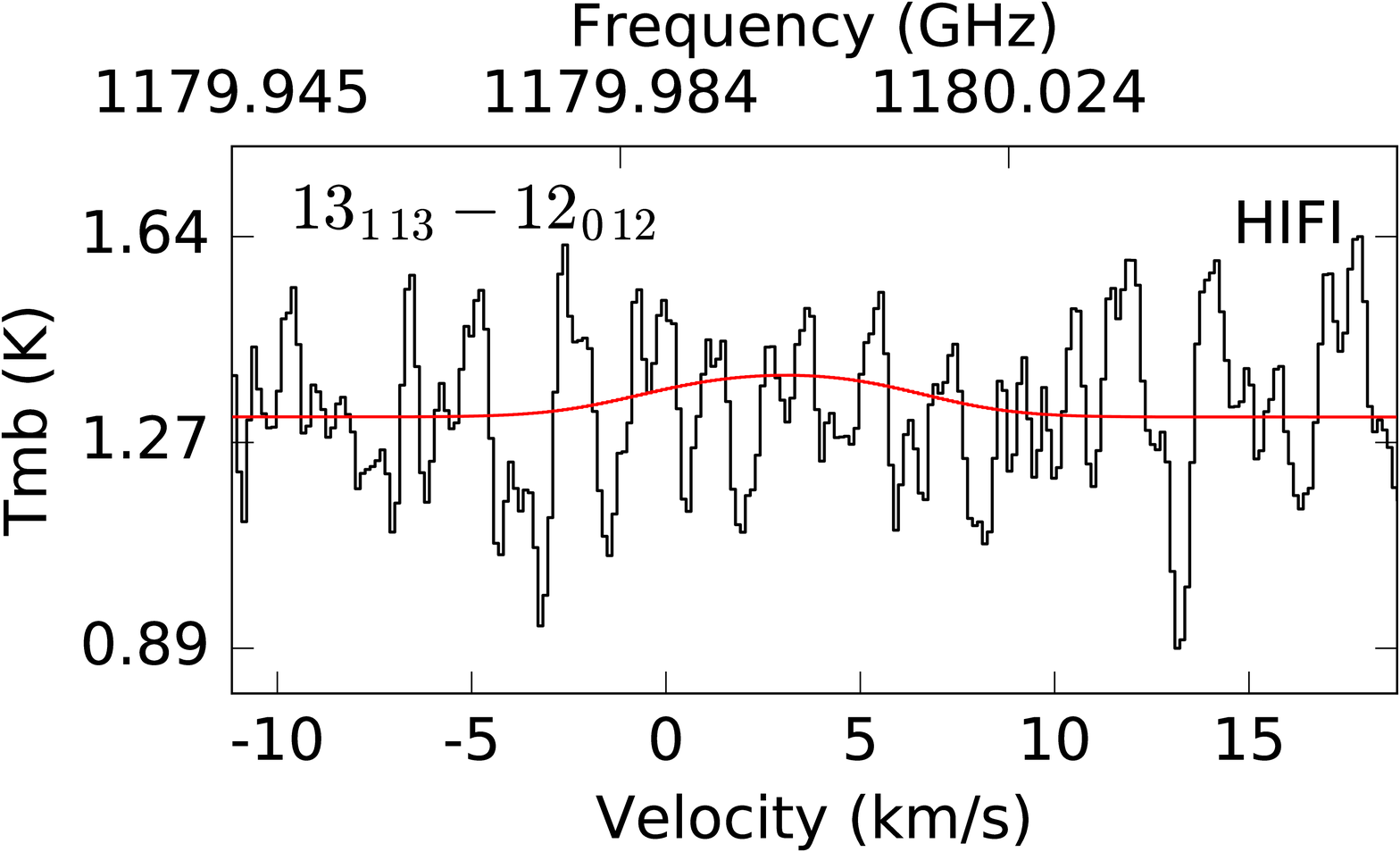}&&\\
\end{tabular}
\label{imagenes-cont3}
\end{figure*}

\begin{figure*}
\centering
\setlength\tabcolsep{3.7pt}
\caption{HNCO $K_a=2$ transitions. In all cases, two transitions very close in frequency are always contained in the same line profile (see Table 1). The transitions with $J\geq 9$ were not modelled due to the lack of their corresponding collisional rate coefficients.}
\begin{tabular}{c c c}
\includegraphics[width=0.315\textwidth, trim= 0 0 0 0, clip]{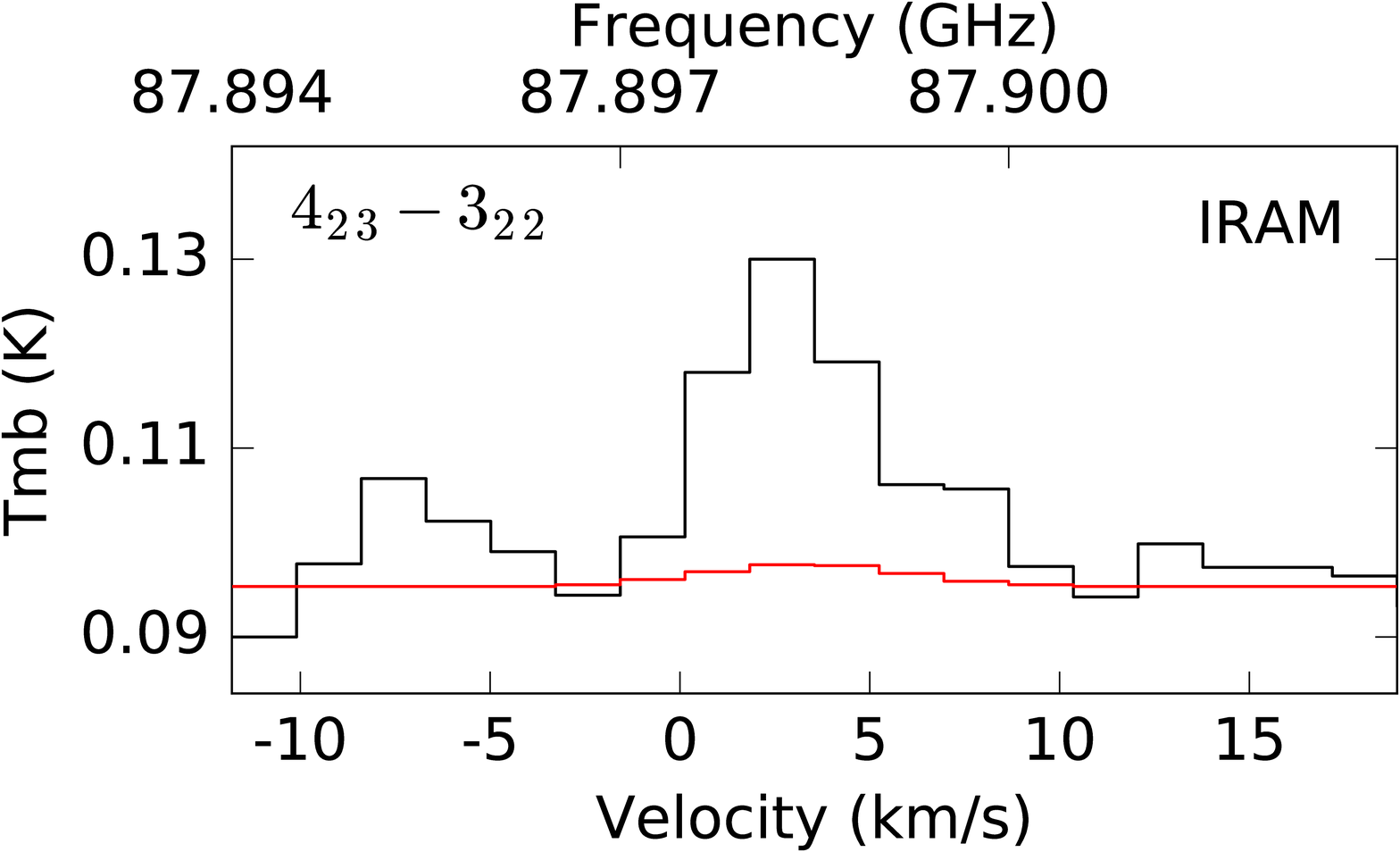} &\includegraphics[width=0.315\textwidth,trim = 0 0 0 0,clip]{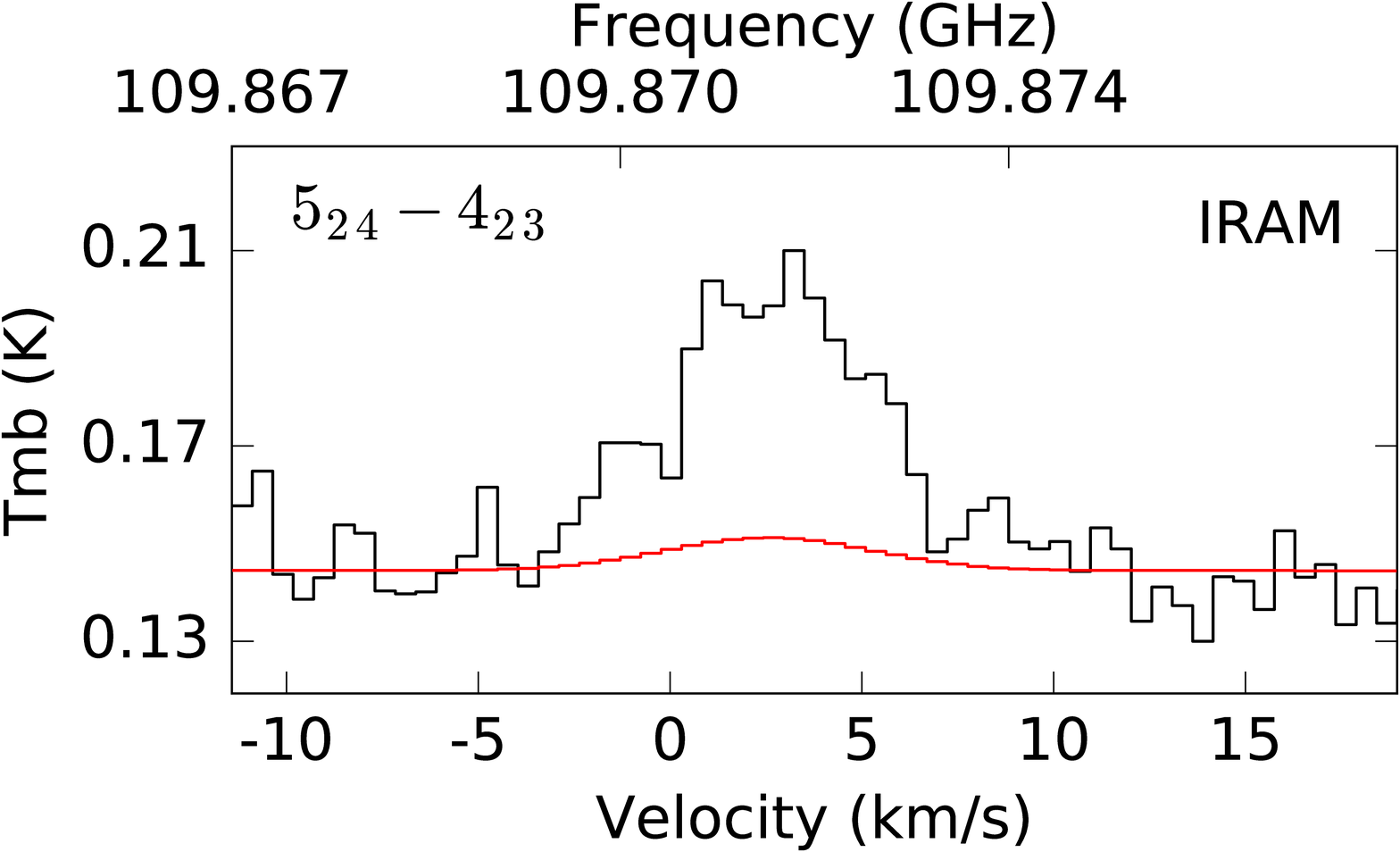}  &\includegraphics[width=0.315\textwidth,trim = 0 0 0 0,clip]{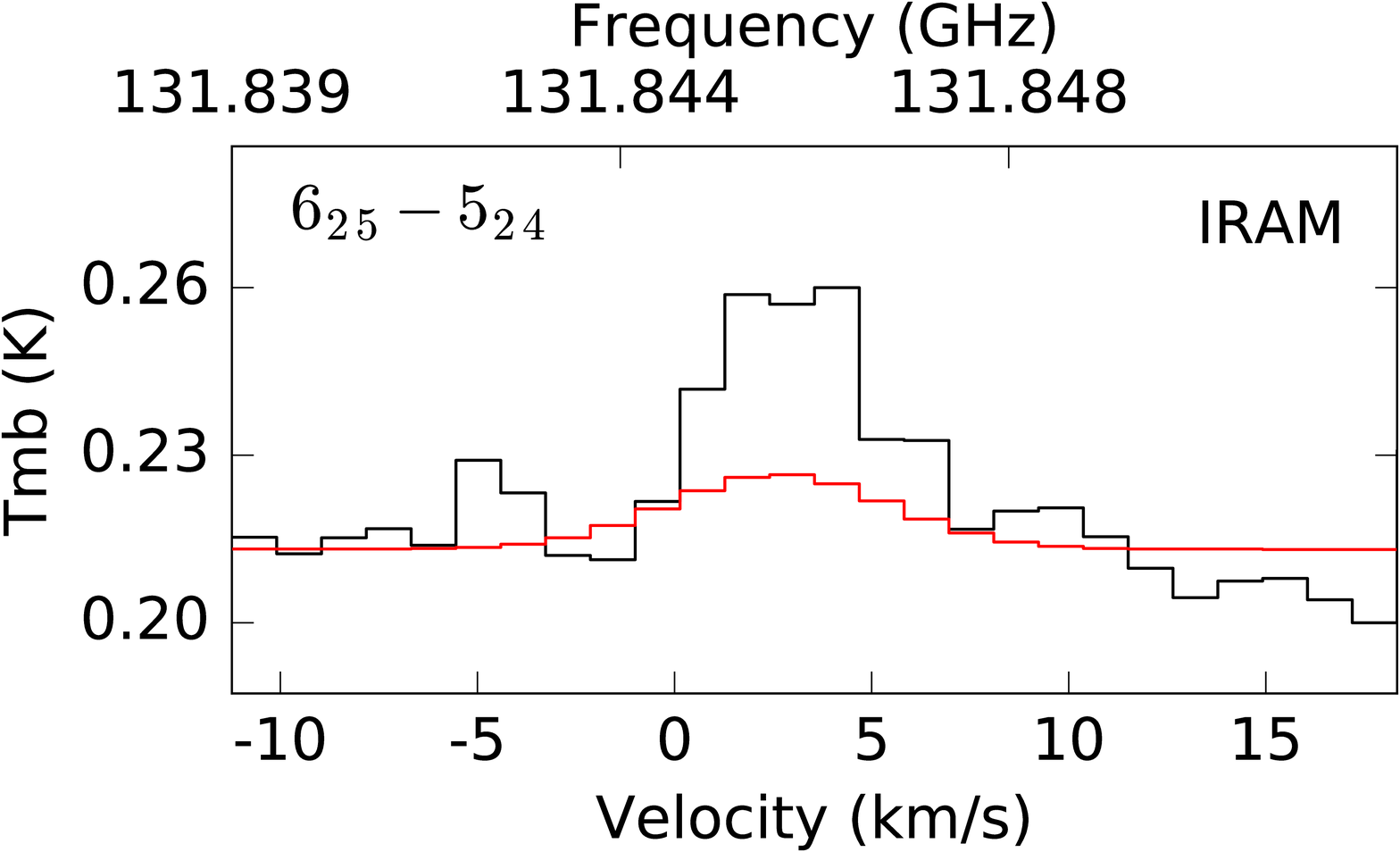}\\ 
\includegraphics[width=0.315\textwidth,trim = 0 0 0 0,clip]{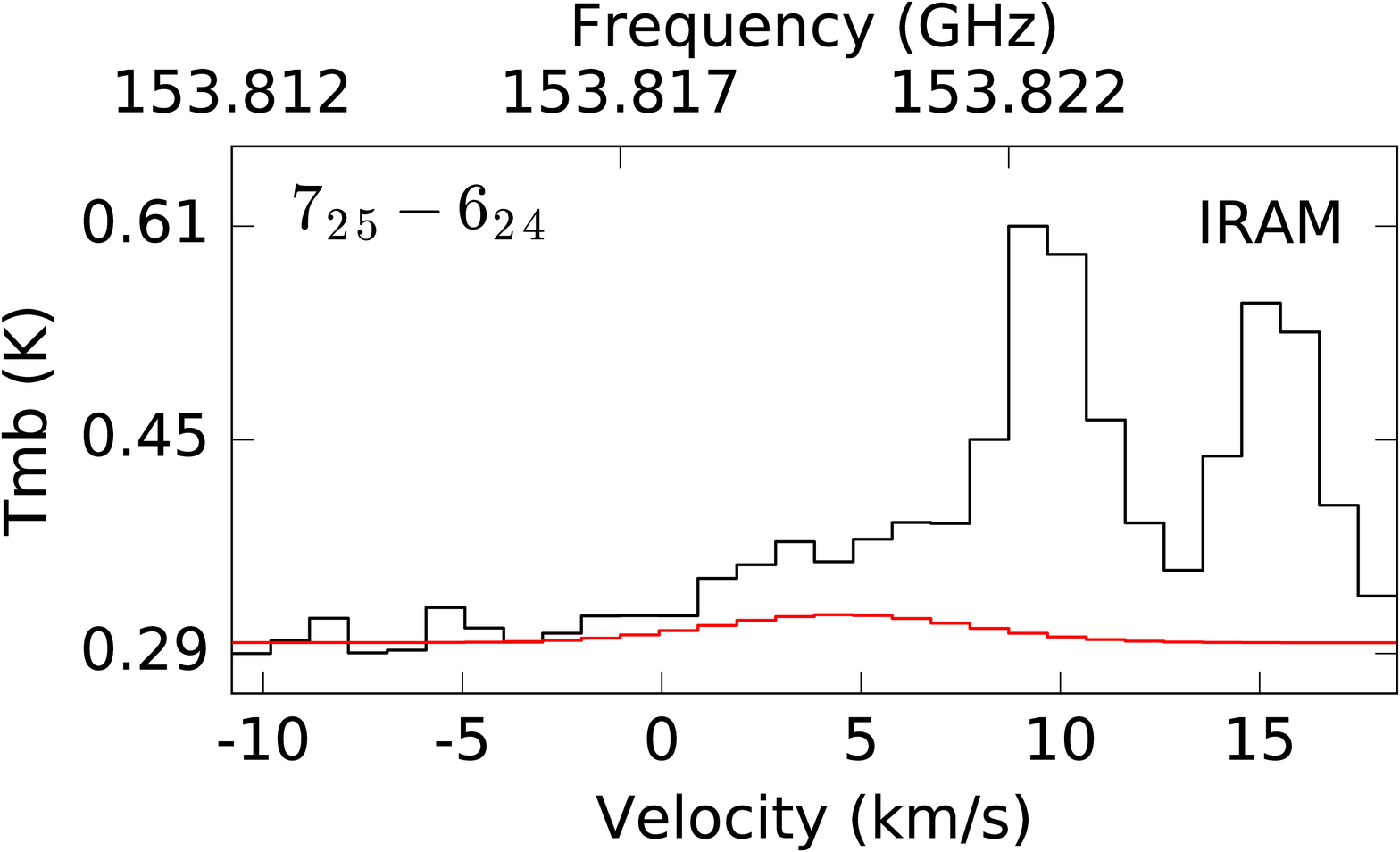} &\includegraphics[width=0.315\textwidth,trim = 0 0 0 0,clip]{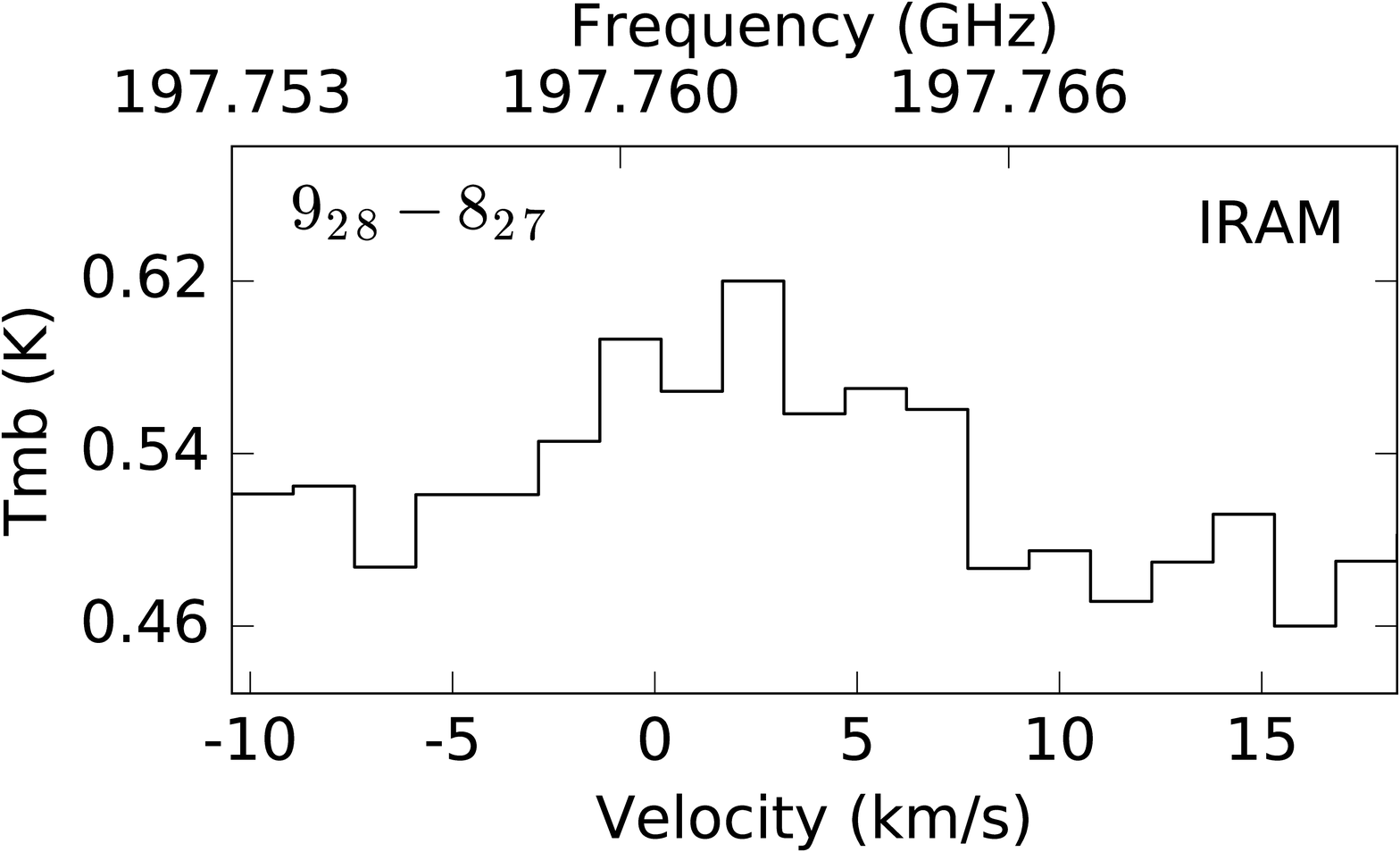}&\includegraphics[width=0.315\textwidth, trim= 0 0 0 0, clip]{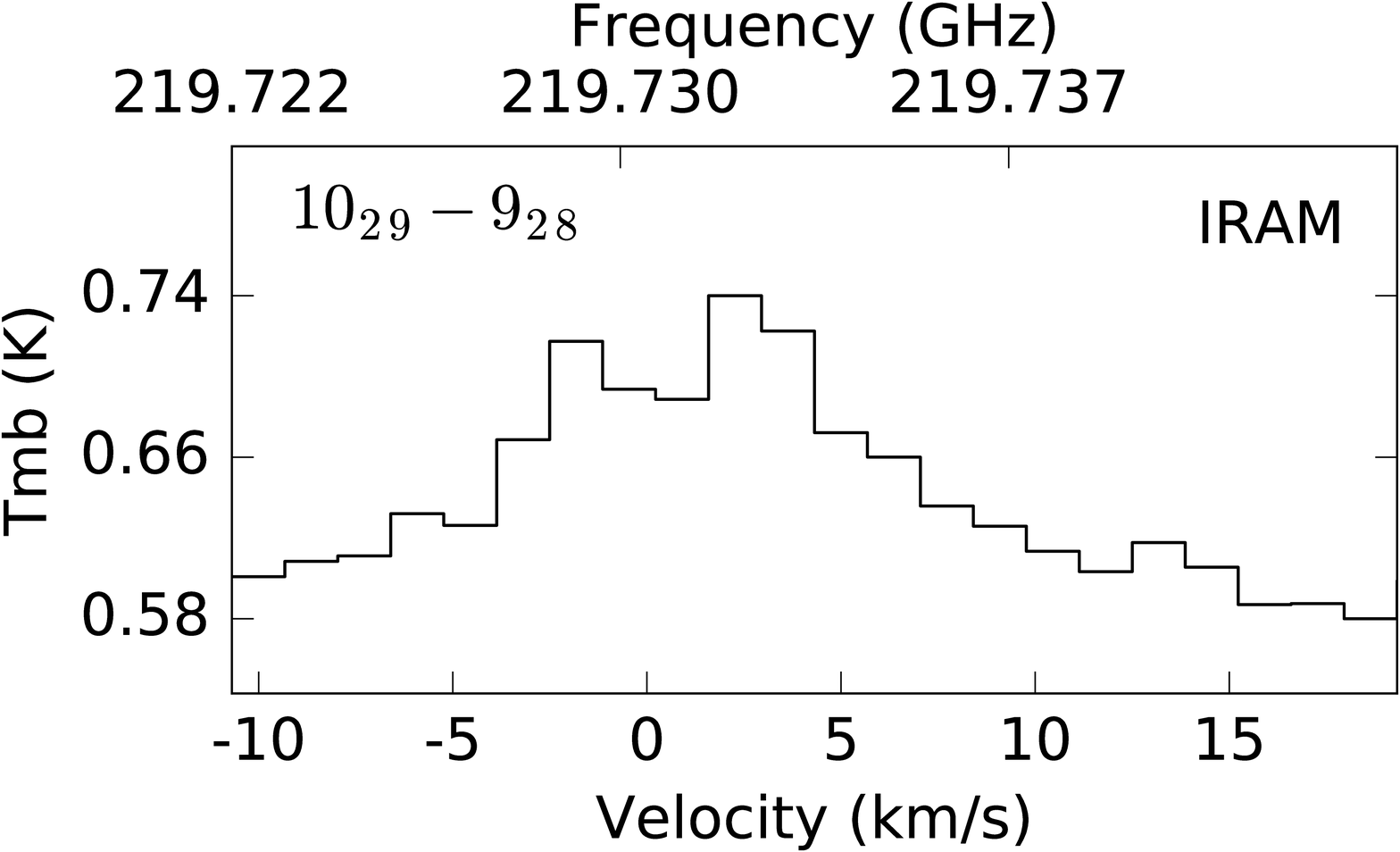} \\\includegraphics[width=0.315\textwidth,trim = 0 0 0 0,clip]{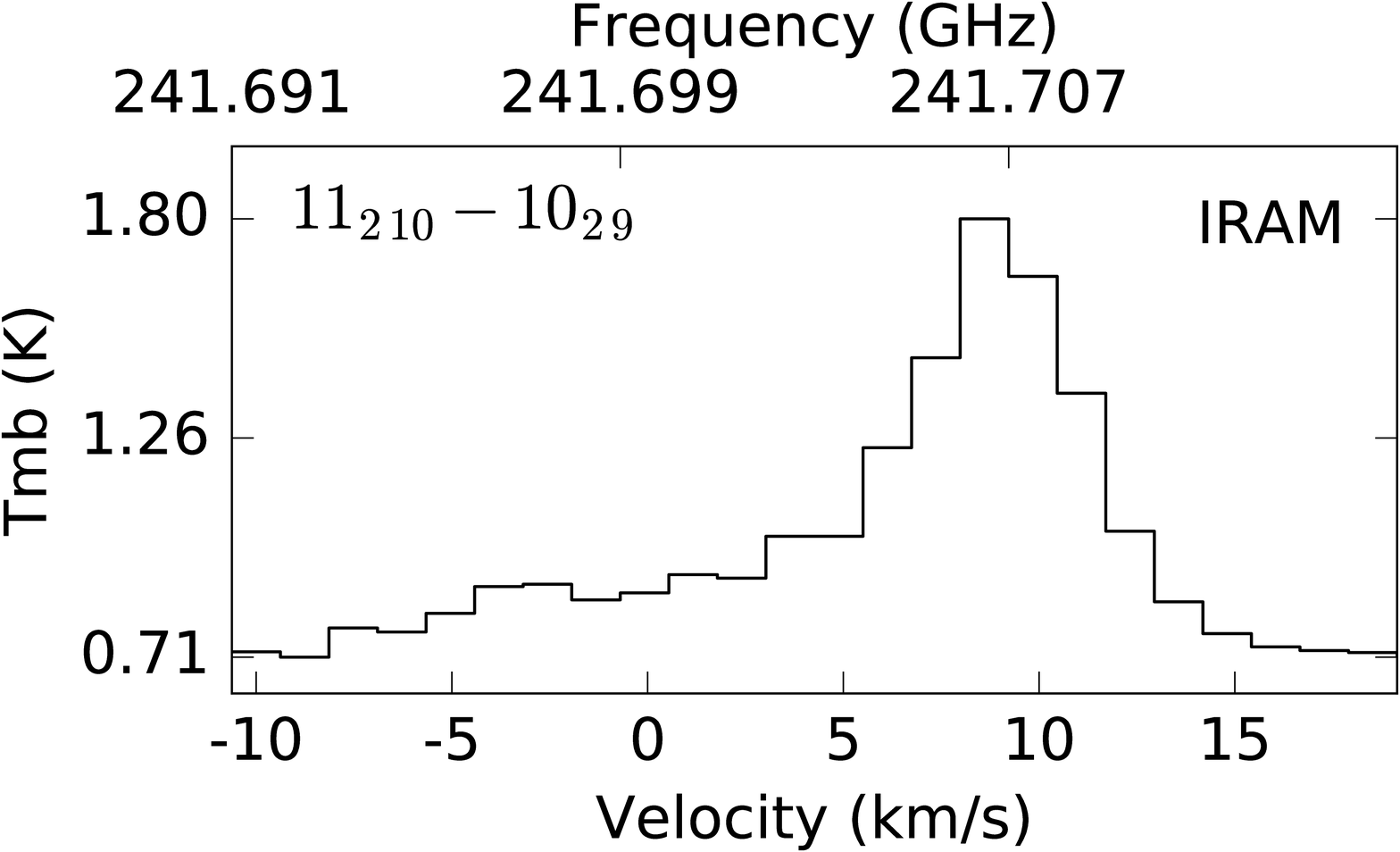}  &\includegraphics[width=0.315\textwidth,trim = 0 0 0 0,clip]{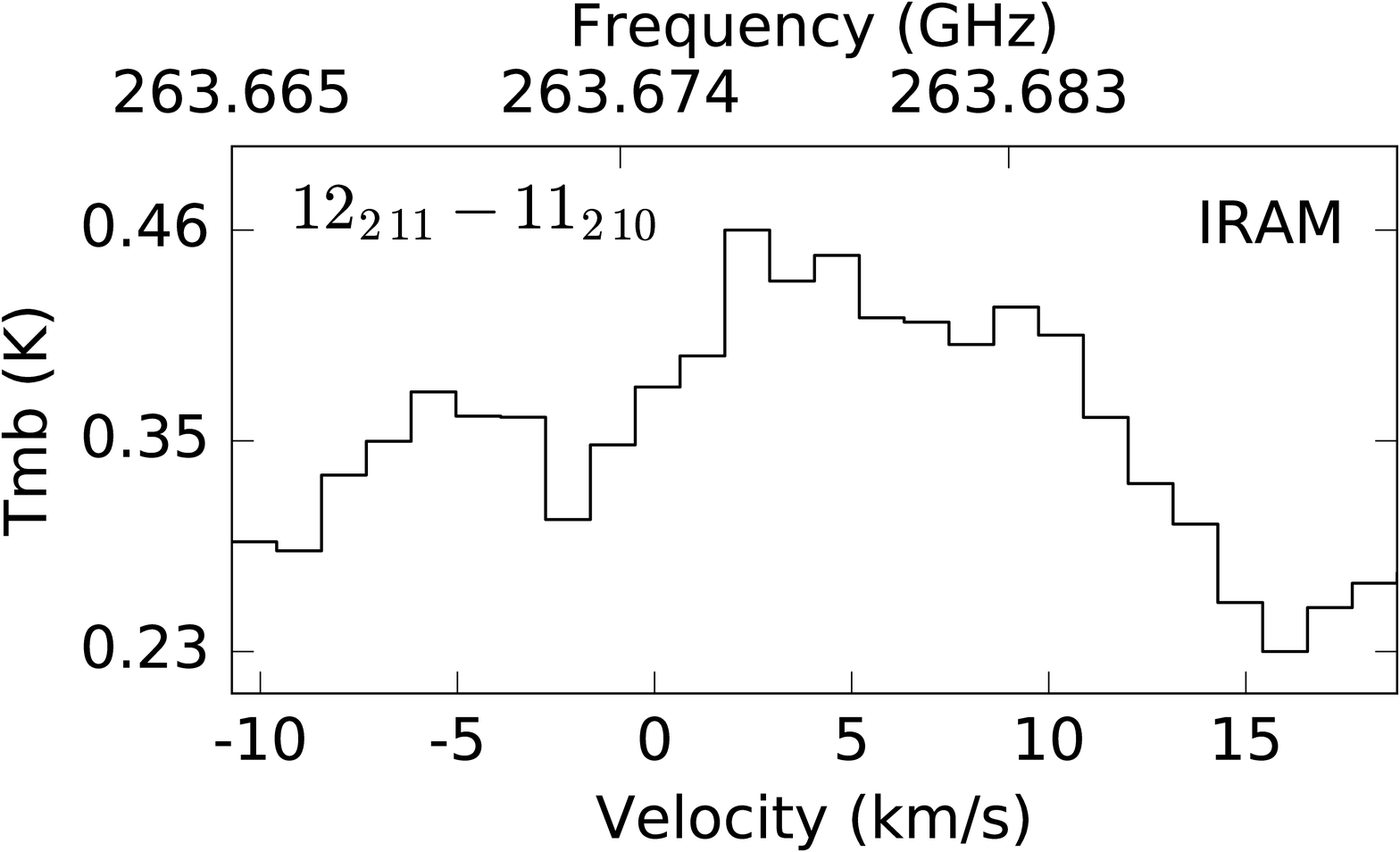} &\\
\end{tabular}
\label{imagenes-cont4}
\end{figure*}



\bsp	
\label{lastpage}
\end{document}